\documentclass[rmp,aps,nofootinbib,endfloats]{revtex4-1}
\usepackage{graphics}
\usepackage{epsfig}

\def\mrm{\mathrm}
\def\ra{\rightarrow}
\newcommand{\dzero}{D\O}

\newcommand{\gev}{\ensuremath{\mathrm{Ge\kern -0.1em V}}}
\newcommand{\mev}{\ensuremath{\mathrm{Me\kern -0.1em V}}}
\newcommand{\gevc}{\ensuremath{\mathrm{Ge\kern -0.1em V}/c}}
\newcommand{\mevc}{\ensuremath{\mathrm{Me\kern -0.1em V}/c}}
\newcommand{\gevcc}{\ensuremath{\mathrm{Ge\kern -0.1em V}/c^{2}}}
\newcommand{\mevcc}{\ensuremath{\mathrm{Me\kern -0.1em V}/c^{2}}}
\newcommand{\evcc}{\ensuremath{\mathrm{e\kern -0.1em V}/c^{2}}}

\newcommand{\pb}{\ensuremath{\mathrm{pb}^{-1}}}
\newcommand{\fb}{\ensuremath{\mathrm{fb}^{-1}}}

\newcommand{\pt} {\mbox{$p_T$}}

\newcommand{\mt}{\ensuremath{m_{t}}}
\newcommand{\mw}{\ensuremath{m_{W}}}
\newcommand{\mz}{\ensuremath{m_{Z}}}
\newcommand{\mh}{\ensuremath{m_{H}}}
\newcommand{\st}{\ensuremath{\tilde{t}}}
\newcommand{\tchi}{\ensuremath{\tilde{\chi}}}

\newcommand{\ttbar}{\ensuremath{t\overline{t}}}
\newcommand{\mttbar}{\ensuremath{m_{t\overline{t}}}}
\newcommand{\ljt}{\ensuremath{\ell\nu q q^{\prime} b \overline{b}}}
\newcommand{\had}{\ensuremath{q q^{\prime} b q q^{\prime} \overline{b}}}
\newcommand{\dil}{\ensuremath{\ell^{+}\nu b\ell^{-}\overline{\nu}\overline{b}}}
\newcommand{\ttljt}{\ensuremath{\ttbar\ra\ljt}}
\newcommand{\ttdil}{\ensuremath{\ttbar\ra\dil}}
\newcommand{\tthad}{\ensuremath{\ttbar\ra\had}}
\newcommand{\ppbar}{\ensuremath{p\overline{p}}}
\newcommand{\qqbar}{\ensuremath{q\overline{q}}}
\newcommand{\sigmattbar}{\ensuremath{\sigma_{\ttbar}}}

\newcommand{\zg}{\ensuremath{Z/\gamma}}
\newcommand{\Zboson}{\ensuremath{Z}-boson}

\newcommand{\Wboson}{\ensuremath{W}-boson}

\newcommand{\Zpjets}{\zg +jets}
\newcommand{\Wpjets}{\ensuremath{W}+jets}

\newcommand{\met}{\ensuremath{E_{T}\!\!\!\!\!\!\!\backslash\;\;}}
\newcommand{\stat}{\ensuremath{\mathrm{stat}}}
\newcommand{\syst}{\ensuremath{\mathrm{syst}}}
\newcommand{\statsyst}{\ensuremath{\mathrm{stat}+\mathrm{syst}}}
\newcommand{\lumi}{\ensuremath{\mathrm{lumi}}}
\newcommand{\theo}{\ensuremath{\mathrm{theory}}}
\newcommand{\fttbar}{f_{\ttbar}}
\newcommand{\mtreco}{\ensuremath{m_{t}^{\mrm{reco}}}}

\newcommand{\pythia}{P{\sc ythia}}
\newcommand{\herwig}{H{\sc erwig}}
\newcommand{\alpgen}{A{\sc lpgen}}
\newcommand{\madevent}{M{\sc adevent}}
\newcommand{\mcnlo}{MC@NLO}
\newcommand{\tauola}{T{\sc auola}}
\newcommand{\evtgen}{E{\sc vtgen}}
\newcommand{\singletop}{S{\sc ingletop}}

\begin{document}
\title{ Top Quark Physics at the Tevatron using {\boldmath $t\bar{t}$} Events}

\author{Fr\'ed\'eric D\'eliot}
\email{frederic.deliot@cea.fr}
\affiliation{Institut de Recherche sur les lois Fondamentale de l'Univers, 
Service de Physique des Particules, CEA Saclay - Bat 141, F-91191 
Gif-sur-Yvette Cedex, France}
\author{Douglas A. Glenzinski}
\email{douglasg@fnal.gov}
\affiliation{Fermilab, Particle Physics Division, P.O. Box 500, Batavia, IL 
60510, USA}

\begin{abstract}  
We review the field of top-quark physics using \ttbar\ events with an emphasis on experimental 
techniques.  The role of the top quark in the Standard Model of particle 
physics is summarized and the basic phenomenology of top-quark production and decay is introduced.  
We discuss how contributions from physics beyond the Standard Model could affect the top-quark properties or event samples.  
The many measurements made at the Fermilab Tevatron, which test the Standard Model predictions or probe for direct evidence 
of new physics using the top-quark event samples, are reviewed here. 
\end{abstract}                                                                 

\date{\today}
\maketitle

\tableofcontents

\newpage

\newpage

% ============================================================================
% --- INPUT THE VARIOUS SECTIONS HERE
% ============================================================================

% ======================================================================
\section{Introduction}
\label{sec:intro}
% ======================================================================

The Standard Model of particle physics (SM) contains six quarks that can be arranged into three generations of two quarks each, and six leptons, similarly arranged.  The top quark is the weak-isospin partner to the bottom quark, which together constitute the third generation of quarks.  The existence of three quark generations was postulated as early as 1973 by Kobayashi and Maskawa since mixing among three generations (and no fewer) could provide a mechanism for CP violation~\cite{KM1973}.  Experimentally the multi-generation quark model found firm footing in 1974 with the discovery of the charm quark~\cite{JPsi1974BR,JPsi1974ST}, the weak-isospin partner to the strange quark in the second generation of quarks.  The discovery of the third generation tau lepton in 1975~\cite{Tau1975} implied the existence of a third generation of quarks.  The bottom quark was discovered in 1977~\cite{Bottom1977} and the search for its weak-isospin partner began in earnest.  It was nearly two decades before the top quark was directly observed by experiment~\cite{Top1994,Top1995c,Top1995d} owing to its surprisingly large mass.  Of course the immediate consequence of this discovery was to further solidify the SM.  Indeed, by the time the direct observation was made there was mounting indirect evidence that the top quark existed and that it had to be heavy~\cite{Schaile1992}. But it could be argued that this indirect evidence was model dependent and that an important consequence of the top-quark discovery was the definitive elimination of ``top-less'' theories that might otherwise still be viable.  In fact, the principal interest in the top quark arises because it offers a window into new physics, into physics Beyond the Standard Model (BSM).  Thanks to its very large mass (it weighs about as much as a gold atom) BSM contributions to physics involving top quarks can occur in a wide variety of ways - the production mechanisms can be affected, the decay widths can be altered, its intrinsic properties changed, or the experimental signature mimicked by a new particle of similar mass.  A thorough exploration of all these possibilities is what constitutes the field of ``Top Quark Physics'', pioneered over the last 15 years by the experiments at the Fermilab Tevatron.   

In recent years there have been several reviews of top-quark physics~\cite{Quadt:2007jk,Incandela:2009, Wicke:2010, Wagner:2010, Pleier:2009, Demina:2008, Kehoe:2008}.  In this article we will concentrate on what has been experimentally established by the Tevatron experiments and the analysis techniques required to do so.  The emphasis of this article will be on the $\ppbar\ra\ttbar$ process responsible for most of the progress to date.  While the $\ppbar\ra tq$ process has been recently observed~\cite{Aaltonen:2010st, Abazov:2009st, Aaltonen:2009st} the available samples are still rather small, which limits their sensitivity relative to what can be accomplished with the \ttbar\ sample.  Recent reviews of the emerging field of top-quark studies using the $tq$ events are available~\cite{Heinson:2010, Heinson:2011}.
The study of top quarks is expected to evolve over the next years using data from CERN's Large Hadron Collider (LHC).  Where appropriate we comment on differences between the Tevatron studies and the LHC possibilities.
% and on the expected evolution of top quark physics in the coming years as the LHC reaches full energy and luminosity.

% ======================================================================
\section{Standard Model Production and Decay}
\label{sec:proddecay}
% ======================================================================

We begin by introducing some concepts and phenomenology to which we will refer throughout this article.  We will not summarize the Standard Model theory in any detail.  This has already been done many times and interested readers should consult the references~\cite{Quigg2009,Kronfeld:2010bx} for in-depth discussions. 

Simply put, high energy physics endeavors to enumerate the elementary particles which constitute all known matter and to characterize the interactions which govern their behavior.  The best guess to date is called the Standard Model of particle physics.  The SM is a quantum gauge theory based on the $SU(3)$ gauge symmetry of quantum chromodynamics (QCD) and the $SU(2)\times U(1)$ gauge symmetry of quantum electrodynamics (QED).
%It's been around a long time and is very successful at describing HEP
%experimental data, often to great precision.

In the SM there are twelve elementary particles - six quarks and six leptons, each arranged into three generations as shown here
\begin{equation}
\nonumber
\begin{array}{ccc}
    \left( \begin{array}{c} u \\ d \end{array} \right) \:\:
  & \left( \begin{array}{c} c \\ s \end{array} \right) \:\:
  & \left( \begin{array}{c} t \\ b \end{array} \right) \vspace*{0.125in} \\

    \left( \begin{array}{c} \nu_e \\ e \end{array}\right) \:\:
  & \left( \begin{array}{c} \nu_\mu \\ \mu \end{array}\right) \:\:
  & \left( \begin{array}{c} \nu_\tau \\ \tau \end{array}\right). \\
\end{array}
\end{equation}
The up-type quarks, up ($u$), charm ($c$), and top ($t$), have electric charge $+\frac{2}{3}$ while the down-type quarks, down ($d$), strange ($s$), and bottom ($b$), have electric charge $-\frac{1}{3}$.  The neutrinos, $\nu_e$, $\nu_\mu$, and $\nu_\tau$ are the electric-charge-neutral lepton partners to the electron ($e$), muon ($\mu$) and tau ($\tau$) leptons, respectively.  The $e$, $\mu$, and $\tau$ all carry electric charge $-1$. The measured masses of these particles are given in 
Tab.~\ref{tab:SMmasses}.  All these particles have anti-particles with the same mass and quantum numbers except for the opposite electric charge, the opposite parity, and being singlets rather than doublets of the $SU(2)$ group.  Often anti-particles are represented with the same symbol as the particle but with a bar; for example a quark is represented as $q$ and an anti-quark as $\overline{q}$. An important subtlety worth noting is that the weak eigenstates of the down-type quarks are mixtures of the mass eigenstates.  This mixing is represented by the CKM matrix,
\begin{equation}
\nonumber
  \left( \begin{array}{c} d^{\prime} \\ s^{\prime} \\ b^{\prime} \end{array}\right) =
  \left( \begin{array}{ccc}
    V_{ud} & V_{us} & V_{ub} \\
    V_{cd} & V_{cs} & V_{cb} \\
    V_{td} & V_{ts} & V_{tb}
  \end{array}\right)
  \left( \begin{array}{c}
    d \\ s \\ b
  \end{array}\right). 
\label{eq:ckm}
\end{equation}

\begin{table}
\caption{The measured masses of the elementary particles in the Standard Model.  
For quarks other than the top, it is estimated from hadron masses. 
For the neutrino masses only limits exist.  Those listed are at the $90\%$ confidence level~\cite{PDG}.}
\label{tab:SMmasses}
\centering
\begin{tabular}{cccc}\hline
  \multicolumn{4}{c}{quark masses}                               \\ \hline
  $m_u \sim 0.002$ & $m_c = 1.27 $ & $\mt = 171.2$ & (\gevcc )  \\
  $m_d \sim 0.005$ & $m_s = 0.104$ & $m_b = 4.20$  & (\gevcc )  \\ \hline 
  \multicolumn{4}{c}{lepton masses}                              \\ \hline
  \multicolumn{4}{c}{$m_\nu < 2\times10^{-6}$ \evcc , $\sum_{i} m_{\nu_{i}}^{2} > 2\times 10^{-17} (\evcc )^{2}$}                       \\
  $m_e = 0.511$ & $m_\mu = 105.6$ & $m_\tau = 1776.8$ & (\mevcc )\\ \hline
\end{tabular}
\end{table}
 
\begin{figure}
\centering
  \epsfxsize=0.6in\epsfbox{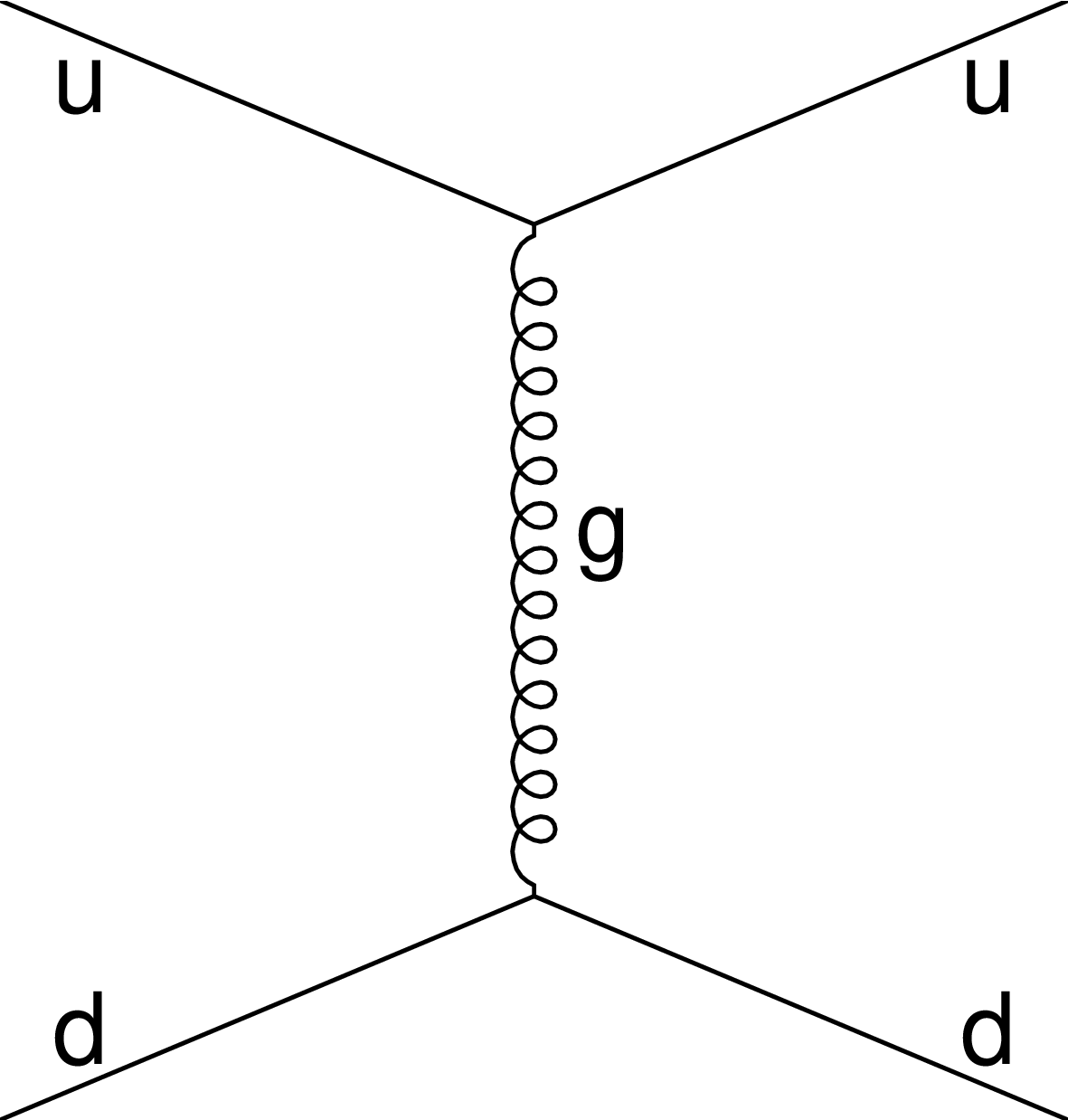}\hspace{0.25in}
  \epsfxsize=0.6in\epsfbox{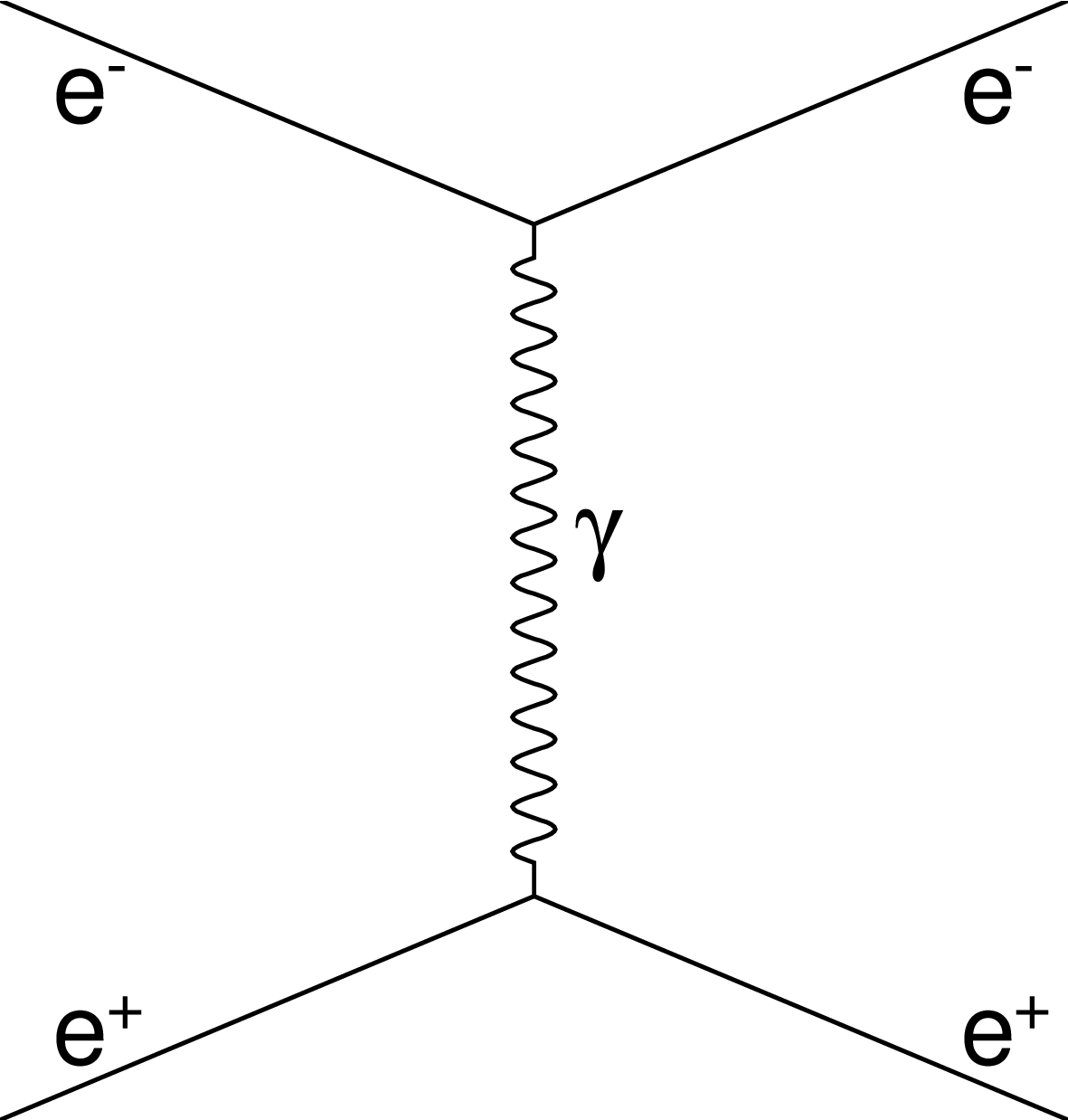}\hspace{0.25in}
  \epsfxsize=0.6in\epsfbox{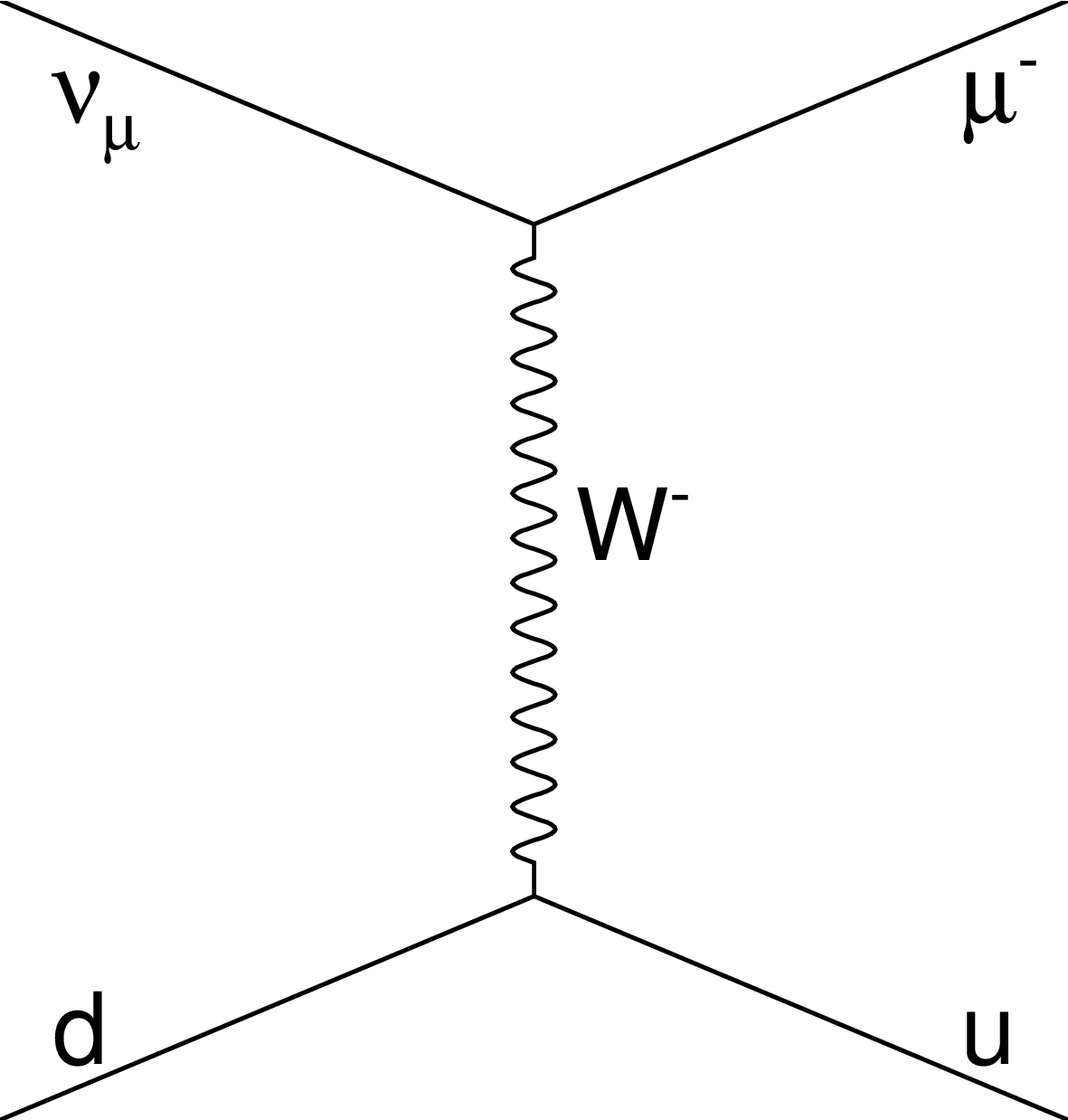}
\caption{\label{fig:interactions}Example of Feynman diagrams depicting SM interactions for the strong (left), electromagnetic (middle), and weak (right) forces.}
\end{figure}

The interactions of the elementary particles are governed by three forces, the strong force of QCD, and the electromagnetic and weak forces of QED.  These forces are mediated by the exchange of bosons.  The strong force is mediated by the exchange of gluons ($g$), the electromagnetic force by the exchange of photons ($\gamma$), and the weak force by the exchange of the charged $W$ bosons ($W^\pm$) or the neutral $Z$ bosons ($Z^0$).  The strengths of these interactions are characterized by coupling constants with these approximate relative strengths $\alpha_{\mrm{strong}} : \alpha_{\mrm{em}} :\alpha_{\mrm{weak}} \approx \frac{1}{10} : \frac{1}{100} : \frac{1}{10000}$.  While the gluon and photon are massless, the $W$ and $Z$ bosons are massive and have been measured to be $\mw=80.40\:\gevcc$ and $\mz=91.188\:\gevcc$~\cite{PDG}.  Figure~\ref{fig:interactions} shows Feynman diagrams for an example of strong, electromagnetic, and weak interactions.

Not all of the elementary particles participate in all three interaction forces.  The photons couple to the electric charge, so neutrinos, for example, do not participate in the electromagnetic interaction.  The gluons couple to the color charge.  Only quarks carry color charge and participate in the strong interaction. All the quarks and leptons can participate in the weak interaction.  The gauge bosons can interact subject to these same constraints.  Gluons carry color and can self-interact.  Similarly $W$ and $Z$ bosons can interact with each other.  The electrically charged $W$ boson can couple to the photon.

In the simplest form of the SM, the $SU(3)\times SU(2)\times U(1)$ symmetries predict that all the elementary particles and the gauge bosons are massless, which is clearly discrepant with experimental observations.  
The easiest way to remedy this is to introduce a scalar field whose vacuum state breaks the $SU(2)\times U(1)$ electroweak symmetry.
With a non-zero vacuum expectation value, $v=246\:\gev$, this field, called the Higgs field, imparts mass to the $W$ and $Z$ bosons via gauge interactions, and to the quarks and leptons via Yukawa interactions.
The electroweak gauge interactions establish a well defined relationship between the Higgs vacuum expectation value and the masses of the $W$ and $Z$ bosons.  The quark and lepton masses are linearly proportional to a Yukawa coupling, $Y_{q,\ell}$, and the Higgs vacuum expectation value.  From their measured masses we can infer that the top-quark Yukawa coupling is of order $Y_{t}\sim 1$, while the others are significantly smaller ranging from $Y_{e}\sim 10^{-6}$ for the electron to $Y_{b}\sim 10^{-2}$ for the $b$-quark.  

The physical manifestation of the Higgs field is a scalar boson, $H^0$.  
The Higgs boson is the only SM particle that has not been experimentally observed.  While the mass of the Higgs boson is not predicted by the SM, it can be inferred since quantum loops including the Higgs boson induce corrections to several experimental observables at a level comparable to the precision with which these observables are measured.  A global fit to these so called ``precision electroweak observables'' predicts that the mass of the Higgs boson lies in the range of about 
$40 < \mh < 160\:\gevcc$ at the $95\%$ confidence level~\cite{PDG, gfitter}.  Direct searches for the Higgs boson at LEP and the Tevatron have ruled-out the masses $\mh < 115\:\gevcc$ and $158 < \mh < 175\:\gevcc$ at the $95\%$ confidence level.  It should be noted that all these constraints are valid only in the context of the SM.  The electroweak symmetry may well be broken in some other way.  
%%%[moved this to the newphy section] The search to discover and 
%%%describe the physics that breaks electroweak symmetry is among the 
%%%most ardently pursued by high energy experimentalists and theorists 
%%%alike.  Is it the SM Higgs, or is it some new physics, something 
%%%beyond the SM?  The principal aim of top-quark physics is to help 
%%%answer these questions. By probing the top-quark sample new physics 
%%%effects may reveal themselves in a variety of ways as discussed in 
%%%Sec.~\ref{sec:newphys}. 
% add citation to GFitter?
% add citation to direct searches?
% add something more about Yukawa couplings?

A peculiarity of QCD worth noting is that quarks can only be observed as bound state hadrons.  There are two types of hadrons: mesons are quark-anti-quark bound states, and baryons are three-quark bound states. An important experimental consequence of this is that quarks produced in a high energy \ppbar\ interaction manifest themselves as collimated streams of hadrons called ``jets''.  The energy and direction of a jet are correlated to the energy and direction of its parent quark.  The process by which the quark evolves into a jet is called ``hadronization'' and consists of a parton shower, which can be perturbatively calculated, and a fragmentation process, which is a non-perturbative process modeled using Monte-Carlo techniques.  As will be seen later in this review, the presence of jets in the final state gives rise to several experimental challenges.

Most hadrons are unstable and decay to lighter particles.  The probability of decay occurring at a time $t$ is described by an exponential, $P\sim e^{-t/\lambda}$, where the lifetime, $\lambda$, is characteristic of the decaying particle.  For example, hadrons containing a $b$-quark decay with a lifetime of about 1.5~ps.  This is long enough that their flight distance in the lab frame is large (of order mm) compared to the experimental resolution - a fact exploited to identify $b$-hadrons in jets.  The ability to identify 
$b$-jets is important in reducing background contributions for many of the \ttbar\ measurements we will discuss.

\begin{figure}
\centering
  \epsfxsize=0.6in\epsfbox{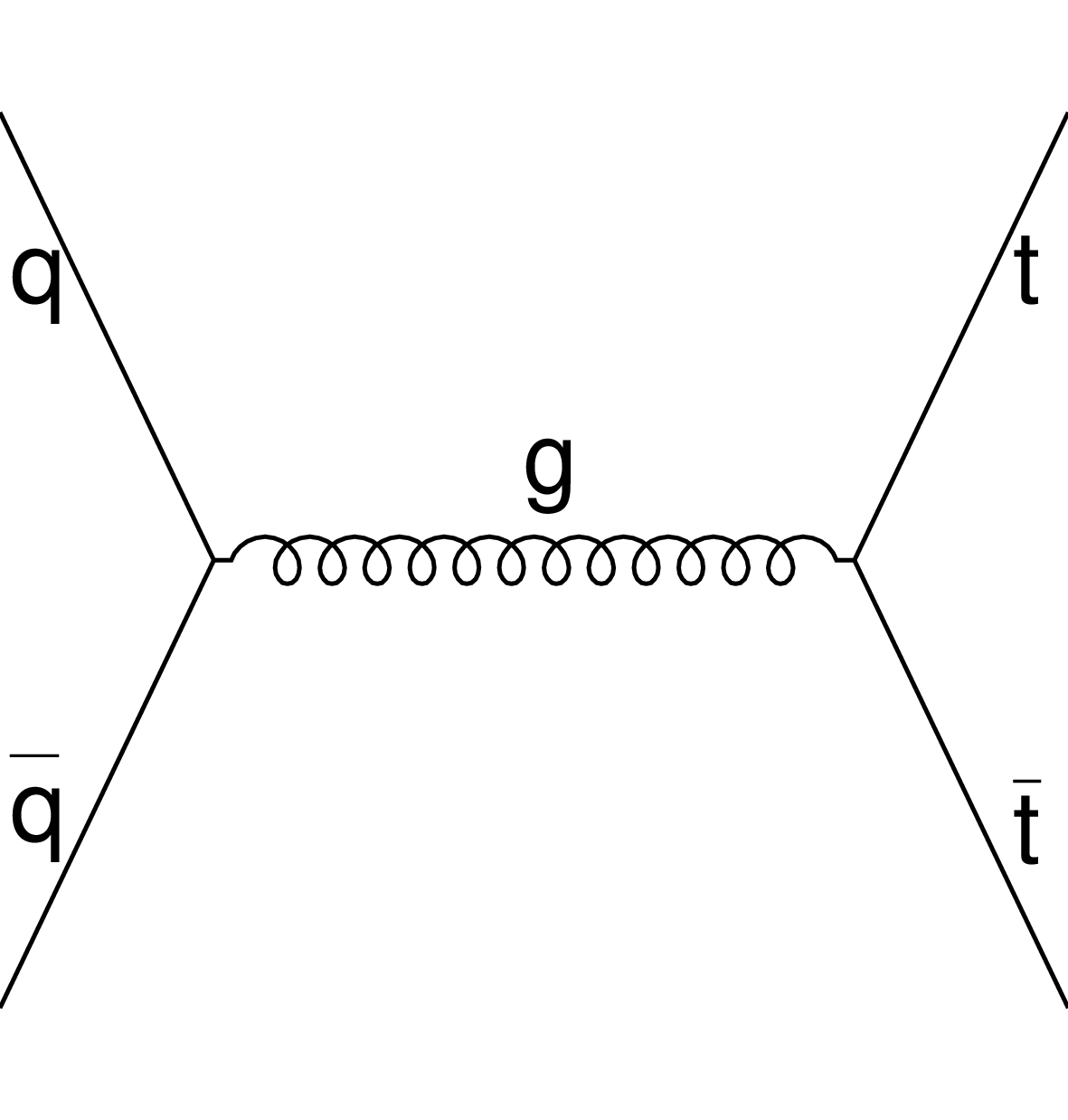}\hspace{0.25in}
  \epsfxsize=0.6in\epsfbox{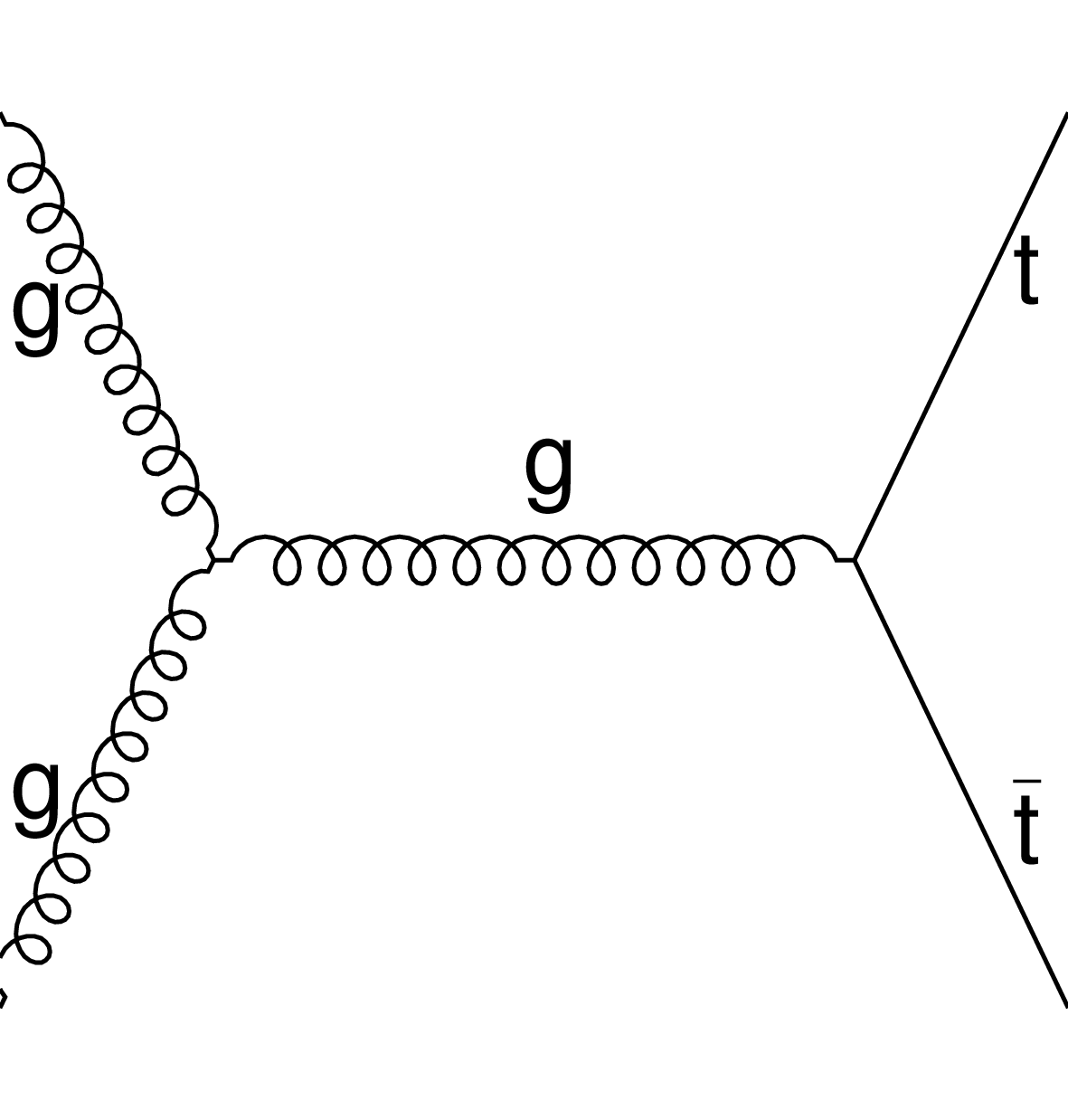}\hspace{0.25in}
  \epsfxsize=0.6in\epsfbox{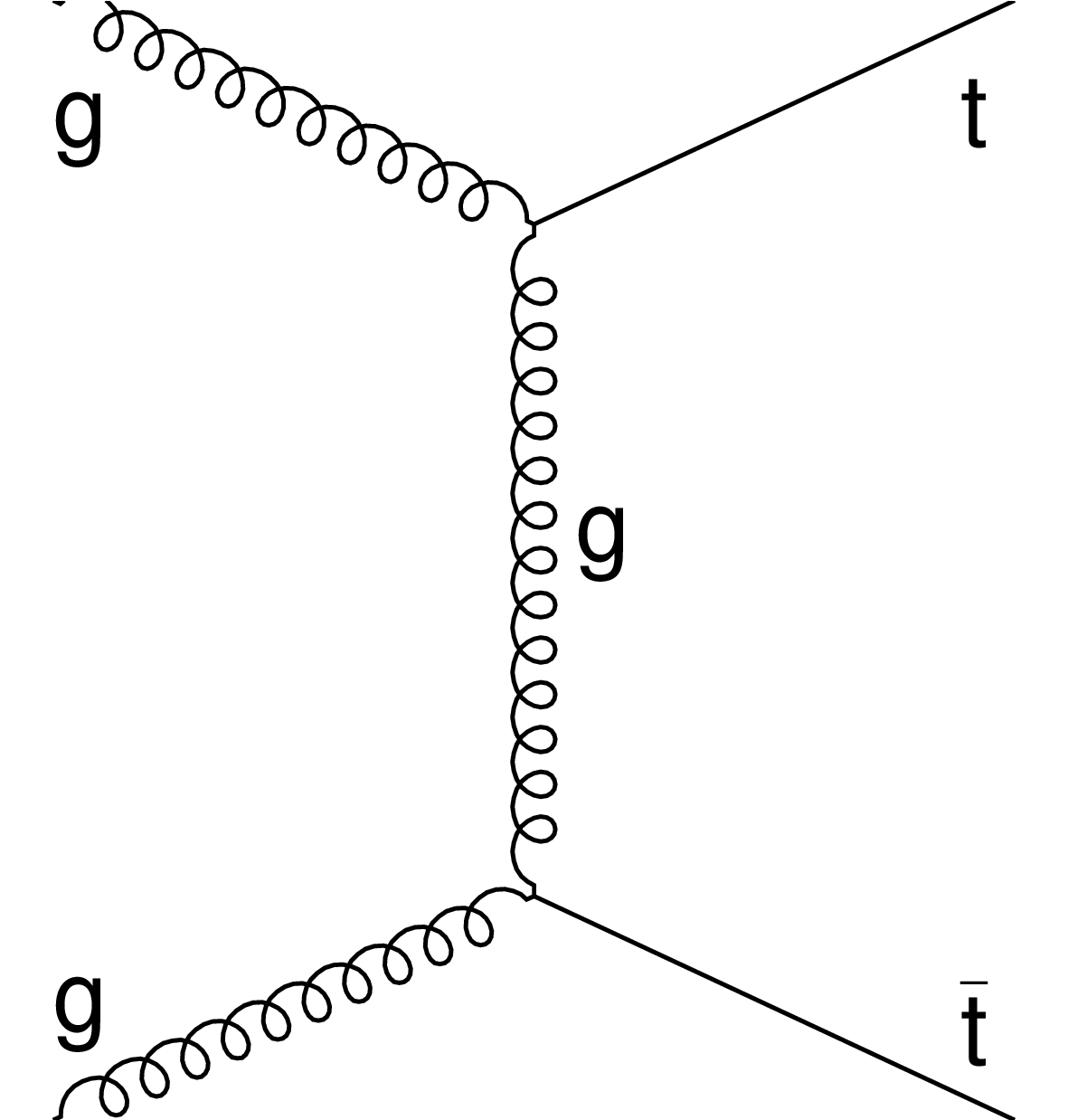}
\caption{\label{fig:ttbarproduction}Feynman diagrams of 
  $\ppbar\ra\ttbar$ for the $q\overline{q}$ annihilation process (left) and 
  the gluon-gluon fusion process (middle and right).}
\end{figure}

From 1989-2009 top quarks could only be produced at the Tevatron, a proton ($p$) anti-proton ($\overline{p}$) collider with a center-of-mass energy of $\sqrt{s} = 1.96$~TeV located at Fermilab.  For \ppbar\ collisions at this energy top quarks are predominantly produced in pairs (\ttbar ) through the strong interaction diagrams shown in Fig.~\ref{fig:ttbarproduction}.  The most recent production cross section calculations are accomplished at next-to-leading-order (NLO) and include logarithmic corrections that account for contributions from higher order soft gluon radiation~\cite{Ahrens:2010zv, Langenfeld:2009wd, Cacciari:2008zb, Kidonakis:2008mu}.  For a top-quark mass of $172$~\gevcc\ the production cross section is calculated in~\cite{Langenfeld:2009wd} to be
$\sigma( \ppbar\ra\ttbar ) = 7.57^{+0.50}_{-0.70}$~pb for 
$\sqrt{s} = 1.96$~TeV.  The uncertainty includes contributions from varying the factorization and renormalization scales, and from varying the parton distribution functions (PDF).  The PDFs describe the momentum distribution of the quarks and gluons that make up the protons and anti-protons. They are empirically determined from fits to deep-inelastic-scattering (DIS) cross section measurements performed by a variety of experiments.  The fits to the DIS data are performed by a number of different collaborations and are made available as software packages.   Common choices are the MRST~\cite{mrst} and CTEQ~\cite{cteq} PDF fits.  The cross section above uses the CTEQ6.6~\cite{Nadolsky:2008zw} PDF parameterization and changes by approximately $\mp 0.24$~pb for a $\pm 1.0$~\gevcc\ change in the top-quark mass.

\begin{figure}
\centering
  \epsfxsize=0.9in\epsfbox{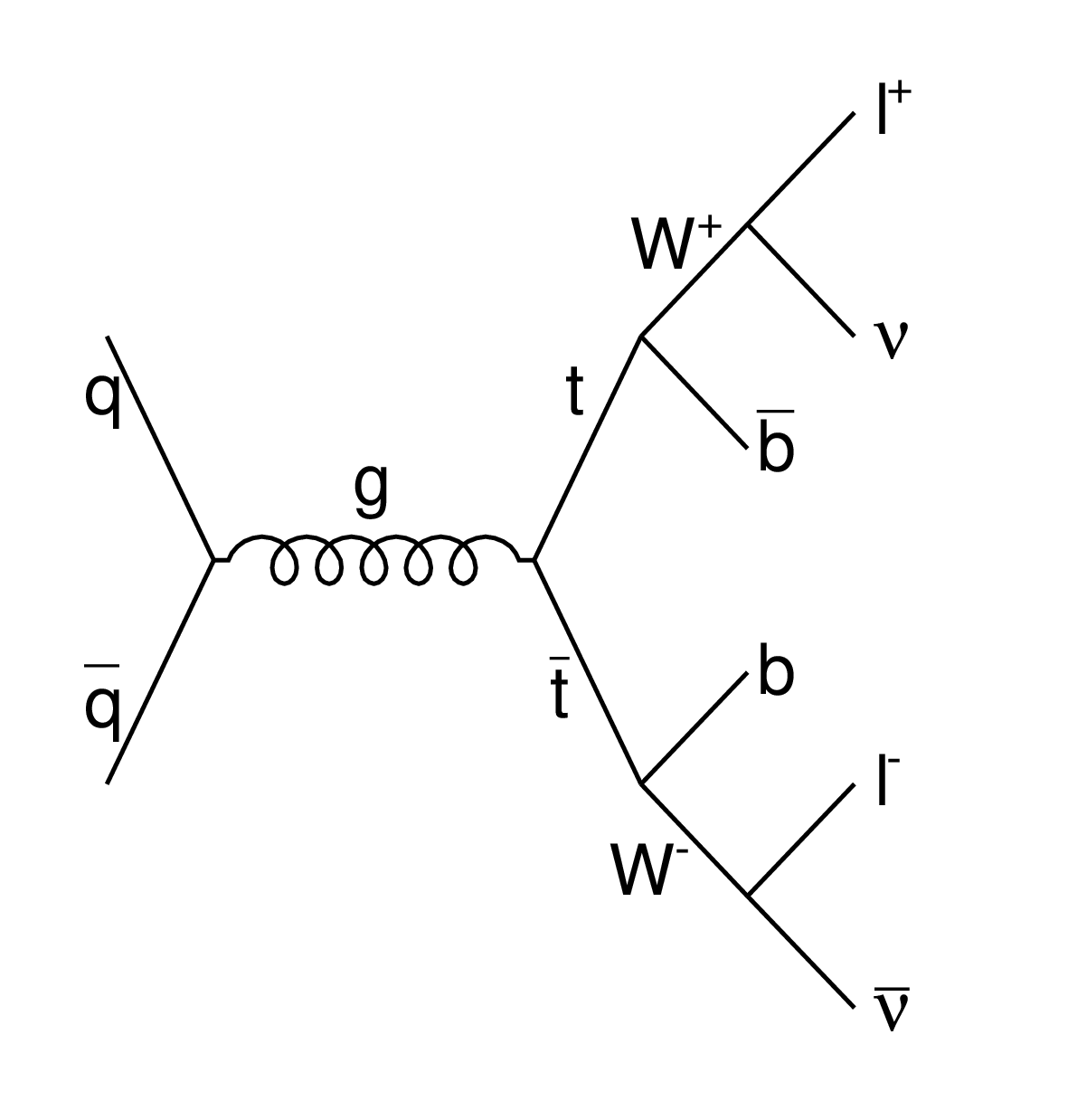}\hspace{0.25in}
  \epsfxsize=0.9in\epsfbox{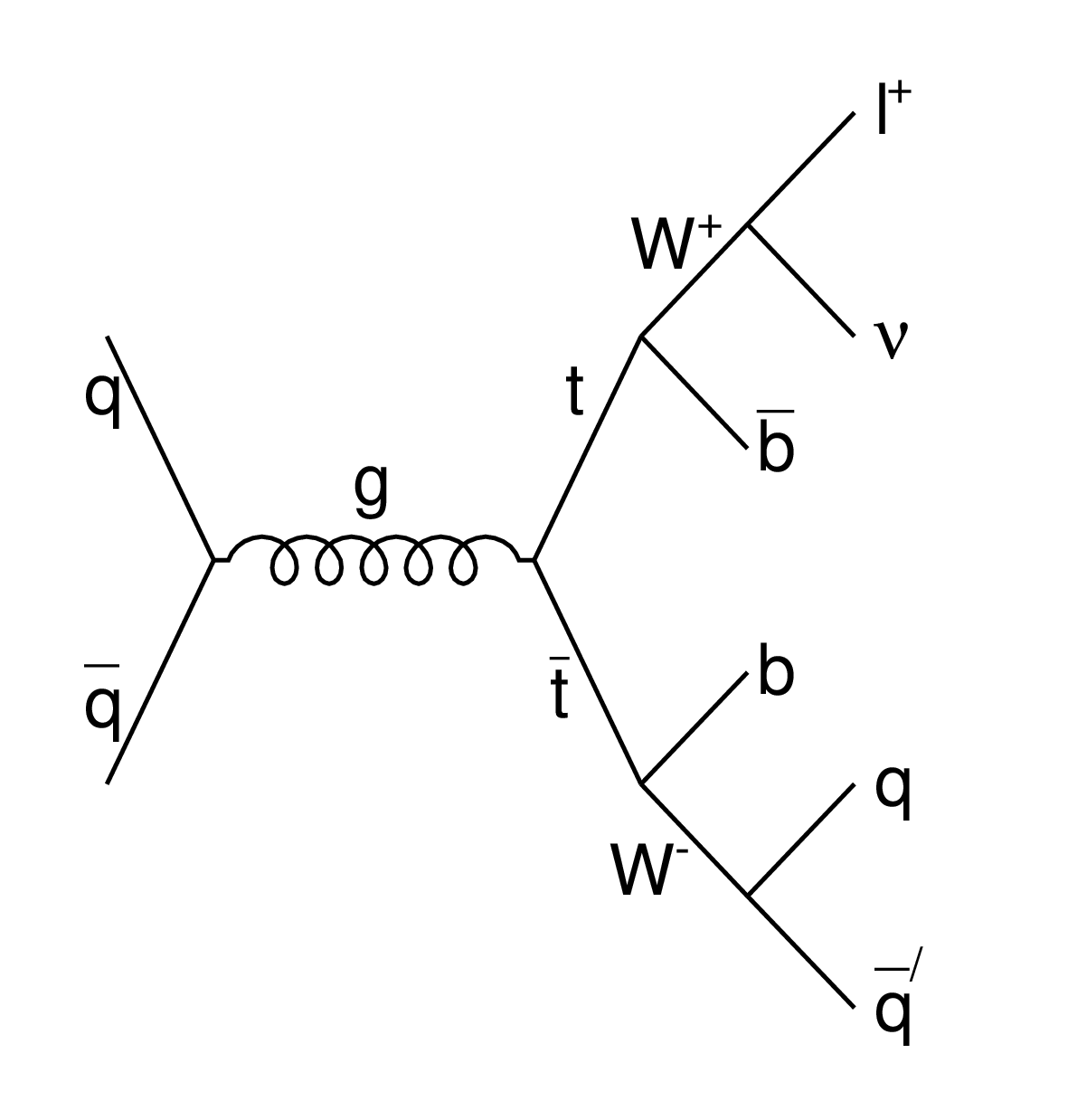}\hspace{0.25in}
  \epsfxsize=0.9in\epsfbox{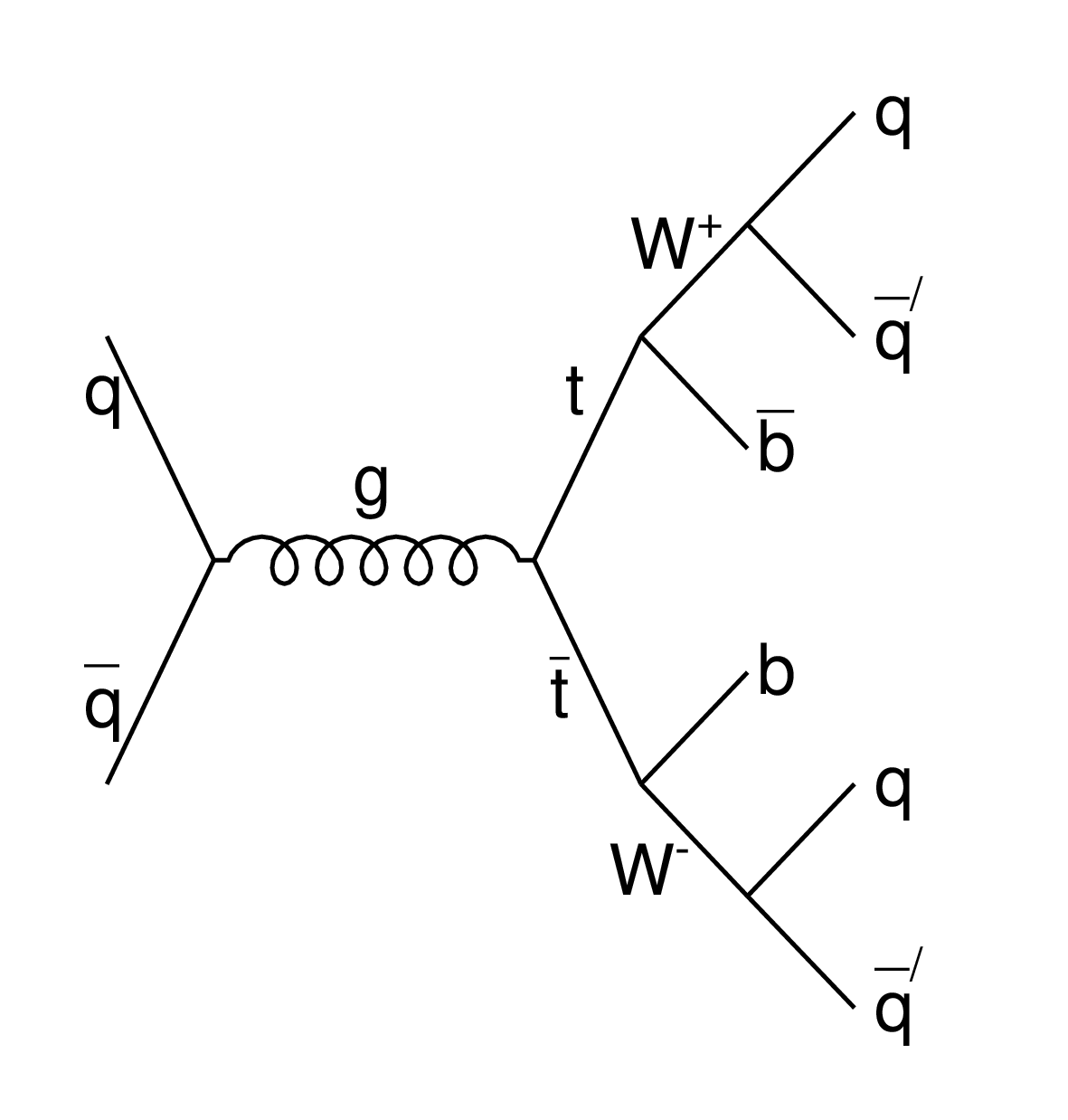}
\caption{\label{fig:ttbardecay}Feynman diagrams of 
  $\ppbar\ra\ttbar$ production with the \ttbar\ decaying to the dilepton (left), lepton+jets (middle), and all hadronic (right) final states.}
\end{figure}

In the SM the top-quark branching fraction is completely dominated by $t\ra W^{+}b$ with the decays $t\ra W^{+}q$ ($q=s,\: d$) contributing at the $< 1\%$ level.  Thus for 
\ttbar\ events the experimental final state is determined by the decay of the $W$ bosons.  The leptonic decays ($\ell = e,\: \mu ,\: \tau$) of the $W$ boson have a branching fraction of ${\cal B}(W\ra \ell\nu)=10.8\%$ each, while the hadronic decays have a total branching fraction of ${\cal B}(W\ra q q^{\prime})=67.7\%$~\cite{PDG}.  For \ttbar\ events there are then three possible final states.  The dilepton final state ($dil$) corresponds to both $W$ bosons decaying leptonically, \ttdil , and occurs 10\% of the time.  The lepton+jets final state ($ljt$) corresponds to one $W$ decaying leptonically and the other hadronically, \ttljt , and occurs 44\% of the time.  The all hadronic final state ($had$) corresponds to both $W$ bosons decaying hadronically, \tthad , and occurs 46\% of the time.  Figure~\ref{fig:ttbardecay} shows the Feynman diagrams corresponding to these three final states, which, experimentally, are treated separately since the contributing sources of background and the dominant detector effects differ among them.  Since the experiments cannot efficiently reconstruct decays of the tau lepton it should be understood in what follows that $\ell=e,\: \mu$ only unless otherwise noted.

The decay width of the top quark is also well specified in the SM.  Including radiative QCD and QED corrections the decay width for a top-quark mass of 172~\gevcc\ is $\Gamma_{t} = 1.4$~\gevcc \cite{Kuhn1993, Kuhn1994}.  The theoretical uncertainty is at the level of a few percent and is dominated by uncertainties in $\alpha_s$ and in missing higher order QCD corrections.  The large decay width of the top quark has important experimental consequences.  Since $\Gamma_{t}$ is large relative to the hadronization scale of QCD, $\Lambda_{QCD} = 250$~MeV, the top quark decays before quark/anti-quark bound states are formed. Essentially, the top quark is produced and decays as a free quark - it is unique among the quarks in this respect.  As a result of this unique feature, many of the experimental techniques used to explore the properties of the lighter quarks are not useful in exploring the top quark so that new methodologies had to be developed.  
%%%The techniques described below have been pioneered over the last 15
%%%years at the CDF and \dzero\ experiments at the Fermilab Tevatron.

% ======================================================================
\section{New Physics Affecting Top-Quark Properties}
\label{sec:newphys}
% ======================================================================

It is important to note that the SM has several shortcomings and is widely regarded as a low energy approximation of a more complete description of particle physics~\cite{Quigg2009}.  There are numerous hypothesized theories that address these shortcomings and offer this more complete description.  The theories differ primarily in the manner by which they break the electroweak symmetry and impart mass to the elementary particles.
The search to discover and describe the physics that accomplishes this is among the most ardently pursued by high energy experimentalists and theorists alike.  Is it the SM Higgs, or is it some new physics, something Beyond the Standard Model (BSM)?  The principal aim of top-quark physics is to help answer these questions.  With a Yukawa coupling of order unity, the top quark strongly couples to the dynamics of electroweak symmetry breaking and is thus expected to be a good place to probe for BSM effects.  In addition, because of its large mass, top-quark observable can receive large quantum loop corrections from possible new particle contributions - significantly larger than for the lighter quarks and leptons.  The strategy for probing for BSM effects using the top quark employs both direct searches for new particles and indirect searches by looking for deviations of experimental observables from the SM expectations.

New physics contributions can affect top-quark observables directly in two ways \cite{Hill:1993hs}.  First non-standard top-quark production can appear through intermediate heavy states such as new gauge bosons that decay into a \ttbar\ pair or into a final state mimicking the SM \ttbar\ signature.  Secondly the top quark can decay into exotic particles such as a charged Higgs boson, $t \to H^+ b$, in models with multiple Higgs doublets.  In the first case, BSM effects distort the SM top-quark production observables, while in the second case BSM effects produce discrepancies among the different decay observables.  It is thus important to measure the top-quark properties using as many decay final states as possible and to compare them.

There are many examples of new physics models that affect the production or decay observables in \ttbar\ samples. For example, new heavy states that carry color charge can modify the \ttbar\ production observables. They can be color singlet scalar or vector particles or color non-singlet particles. Examples of models that predict color singlet scalar particles are the supersymmetric extension of the SM, SUSY~\cite{Nilles:1983ge,Haber:1984rc,Martin:1997ns,Chung:2005a}, or more generally models with multiple Higgs fields, MHDM~\cite{Grossman:1994jb}. Examples of models that predict color singlet vector particles include technicolor~\cite{Eichten:1979ah,Eichten:1986eq}, topcolor \cite{Hill:1991at}, or top assisted technicolor \cite{Hill:1994hp}. Color non-singlet resonances arise in these BSM theories~\cite{Simmons:1996fz,Chivukula:1996yr,Choudhury:2007ux}. 
Compactified extra dimension models can also affect \ttbar\ production and decay.  These models produce Kaluza-Klein (KK) modes that either decay preferentially into \ttbar\ pairs or into final states that mimic the SM \ttbar\ signature.
The flat extra dimension TeV$^{-1}$ models produce KK modes that can decay to \ttbar\ pairs~\cite{Antoniadis:1994yi,Dienes:1998vh,Dienes:1998vg,Antoniadis:1999bq,Rizzo:1999br,Cheung:2001mq}.
Universal extra dimension models pair produce a KK mode that each decay as $\mrm{KK}\ra t\gamma$ so that they mimic the SM signature but also include additional photons in the final state~\cite{Appelquist:2000nn,Cheng:2002ab,Rizzo:2001sd}.
Warped Randall-Sundrum extra dimension models predict KK gravitons that can decay to \ttbar\ \cite{Randall:1999ee,Davoudiasl:2000wi}, while bulk
Randall-Sundrum models with fermions and gauge bosons in the bulk predict in addition KK top quarks that would be produced in pairs and would decay to $Wb$ as in the SM~\cite{Davoudiasl:1999tf,Grossman:1999ra,Pomarol:1999ad,Chang:1999nh,Randall:2001gb,Huber:2000fh,Randall:2001gc,Csaki:2002gy,Hewett:2002fe,Agashe:2003zs}.
Some BSM theories predict a fourth family of (heavy) quarks and leptons~\cite{Frampton:1999xi,Holdom:2009rf}. Such extensions can still be compatible with electroweak precision tests~\cite{Kribs:2007nz,Eberhardt:2010bm}.  These additional generations can appear for instance in grand-unified theories~\cite{Langacker:1980js} but some of the models cited above also predict new particles that look like heavier quarks.  Similarly, Little Higgs models predict new heavy vector-like quarks~\cite{Han:2003gf,Dobrescu:1997nm}.  All of these can produce decay final states that mimic the SM \ttbar\ signatures.

New physics contributions can also affect top-quark decays or other properties of the \ttbar\ sample.  For example, some SUSY and MHDM theories produce top-quark decays into new particles~\cite{Chung:2005a,Grossman:1994jb}.  Other models modify the SM top-quark branching fractions.  For example flavor changing neutral current (FCNC) decays, $t\ra cX^0$, or $t\ra uX^0$ where $X^0$ is any neutral particle like a photon or a $Z$ boson, can be modified by BSM physics.  Since these processes are extremely suppressed in the SM even small new physics effects can yield large changes in FCNC branching fractions~\cite{AguilarSaavedra:2004wm}.  The \ttbar\ sample can also be probed for the presence of more general anomalous couplings~\cite{Kane:1991bg} affecting the kinematic properties of top-quark decays and appearing as discrepancies with the SM predictions.  These might affect the \ttbar\ forward-backward charge asymmetry, the \ttbar\ spin correlations, or the polarization fractions of $W$ bosons from top-quark decays~\cite{Cao:2009uz,Shu:2009xf,Cheung:2009ch,Djouadi:2009nb}.  The measured intrinsic properties of the top quark might also be affected.  For example some BSM theories predict particles with electric charge $-4/3$ or $+5/3$ that decay into top quarks~\cite{Contino:2008hi}.

Over the last decade the Tevatron experiments have undertaken a broad top physics program aimed at exploring these possibilities both by comparing measured top-quark properties to SM predictions, as discussed in Sec.~\ref{sec:topprop}, and by directly searching for new particles in the top-quark event samples, as discussed in Sec.~\ref{sec:topsearch}.

% ======================================================================
\section{Experimental Apparatus}
\label{sec:detec}
% ======================================================================

The \ppbar\ collision data at the Tevatron are collected and analyzed by two collaborations, CDF and \dzero .  For both the CDF and \dzero\ experiments the detector apparatus is designed to be a general purpose collider detector with the capability of reconstructing charged particle trajectories and jets, identifying electrons, muons, photons, and $b$-quark jets, and measuring the energy and momentum of as many of the final state particles as possible.  Functionally the two detectors behave very similarly, although they differ significantly in their technical details. They both have a cylindrical geometry centered on the \ppbar\ beam line and employ a coordinate system with the $z$ axis parallel to the beam line and pointing in the direction of the proton beam.  The angle $\theta$ is the polar angle relative to the $z$ axis and the angle $\phi$ is the azimuthal angle in the plane transverse to the $z$ axis.  The energy, $E$, and momenta, $\vec{p}$, of electrons, muons, photons, and jets are estimated using information the calorimeters and spectrometers.  
It is convenient to define the transverse energy, $E_{T}=E\sin\theta$, and the transverse momentum, $p_{T}=\left|\vec{p}\right|\sin\theta$.
Since the \ppbar\ beams are traveling along the $z$ axis the initial state is known to satisfy $\left( E_{T}, p_{T} \right) = \left( 0, 0 \right)$ while the $\left( E_{z}, p_{z} \right)$ components of the initial state are unknown event-by-event - and in general are non-zero - due to the ambiguities introduced by the parton distribution functions.  Because the collision events are boosted along the $z$-axis hadron collider experiments often use pseudo-rapidity, $\eta = -\ln\tan\frac{\theta}{2}$, in place of $\cos\theta$ when defining angular regions since differences in pseudo-rapidity, $\Delta\eta$, are invariant under these boosts.

\begin{figure} 
\centering
  \epsfysize=4in\epsfbox{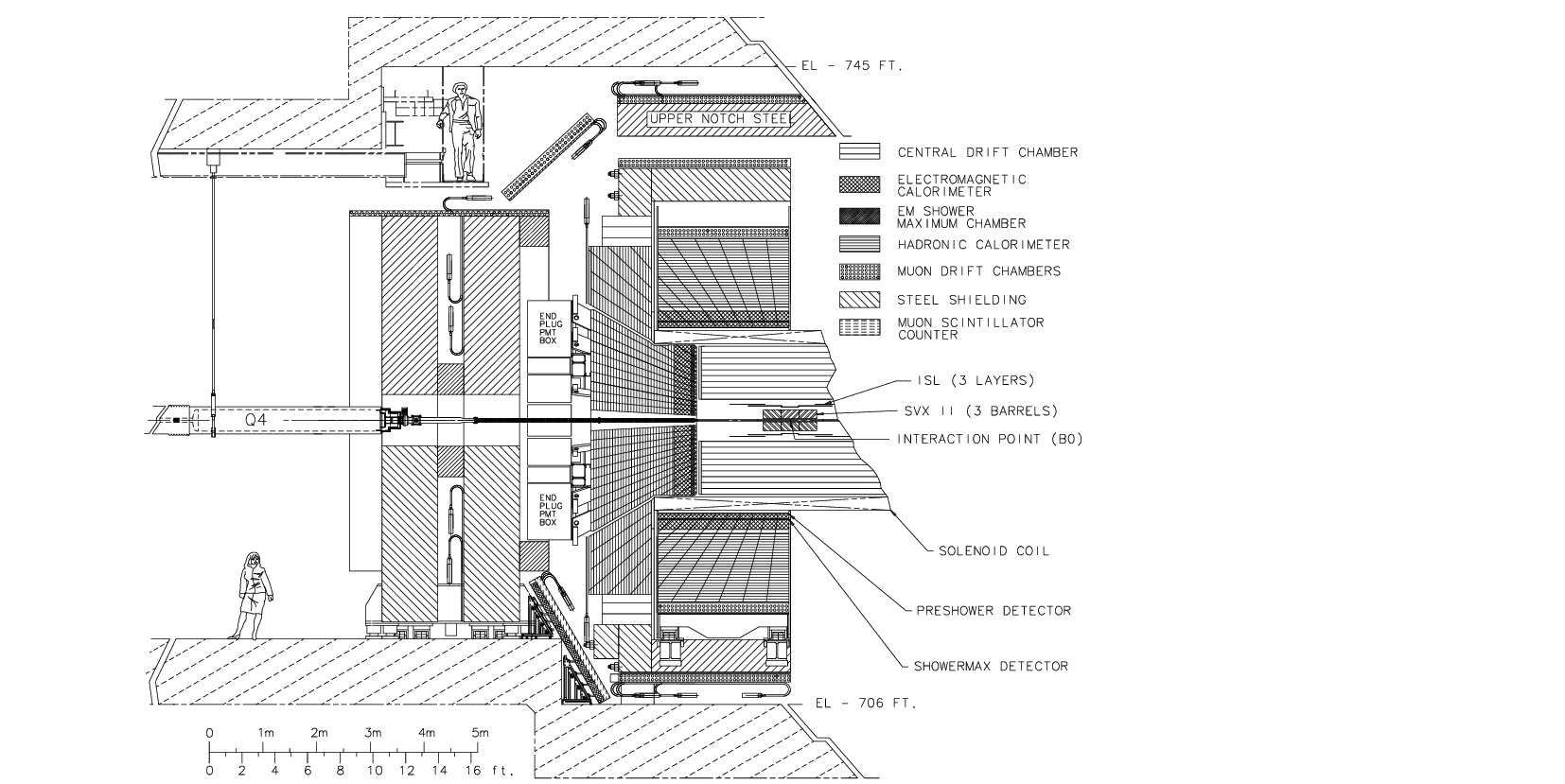}\vspace*{0.25in} 
  \epsfysize=4in\epsfbox{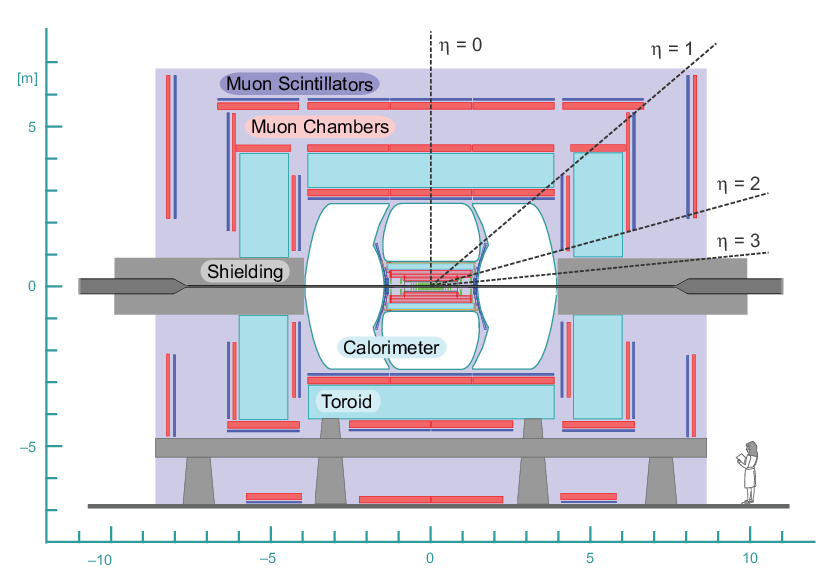}\vspace*{0.25in} 
\caption{\label{fig:detecs}A cut away view of the CDF (top) and \dzero\ (bottom) detectors.} 
\end{figure}

The CDF and \dzero\ detectors are shown in Fig.~\ref{fig:detecs} and are described in detail in references~\cite{cdf} and~\cite{dzero} and only a generic description is provided here.  The detectors are divided into a central region covering (approximately) $\left|\eta\right| < 1.3$ and a forward region covering (approximately) $1.3 < \left|\eta\right| < 3.5$.  In the central region charged particle trajectories are reconstructed as tracks using a set of custom built components.
At the inner most radii, immediately surrounding the \ppbar\ beam line, are precision vertex systems comprised of several concentric layers of silicon microstrip detectors with a per layer charged particle position resolution of around $10\:\mu\mrm{m}$ in the transverse plane.  These detectors are important in reconstructing the primary \ppbar\ interaction vertex as well as identifying displaced vertices from the decay of long lived particles (e.g. $b$-hadrons).  The silicon systems are surrounded by large radii, multi-layered, low mass tracking chambers which achieve position resolutions of around $150\:\mu\mrm{m}$ per hit.  The silicon and tracking systems are inside superconducting solenoids, which  provide magnetic fields parallel to the beam line and induce curvature in the trajectories of charged particles from which their momentum can be deduced. The transverse momentum of charged particles is typically measured with a resolution of 
$\sigma_{p_{T}} / p_{T} < 0.01~p_{T}$.  Outside the solenoid are the calorimeters with an electromagnetic compartment in front of a hadronic compartment.  The electromagnetic (EM) compartment consists of roughly $20$ radiation lengths or more of material and fully contains $e^{\pm}$ and $\gamma$ showers.  The showers induced by pions, kaons, and other long lived hadrons are initiated in the EM calorimeter but continue through and are contained in the hadronic compartment.  The energies of electrons and photons are typically measured with a resolution of approximately $\sigma_{E} / E = 14\% / \sqrt{E}$ while the energies of jets are typically measured with a resolution of approximately $\sigma_{E} / E = 10-15\% / \sqrt{E}$.  The calorimeters are segmented into towers with a typical granularity of $\eta\times\phi = 0.1\times0.1$. 
Beyond the hadronic calorimeters are the muon chambers, small scintillator plus tracking systems which serve to identify muons, the only charged particles able to traverse the calorimeters without being absorbed. In the forward region there is some limited tracking capability from the silicon system, but with a much worse $p_{T}$ resolution than that achieved in the central region.   There are also EM and hadronic calorimeters together with muon chambers. In the far forward region, $\left| \eta \right| > 3.5$, is a luminosity monitor used to measure the \ppbar\ collision rate in the detector during data taking.   In total each experiment has about one million read-out channels.  

The data acquisition system digitizes the read-out channels and records them to permanent storage (tape) for later analysis.
The Tevatron delivers a collision to each experiment every 396~ns, which corresponds to about $2.5$M collisions per second.  This is far too many collisions to record them all and each experiment employs a three level trigger system to identify the most interesting events to keep for later analysis.  The level one trigger is hardware based and employs simple thresholds and a coarse granularity to identify events with high energy depositions in the calorimeter or segments in the tracking system consistent with having originated from a high momentum particle.  The level one system reduces the initial event rate by about two orders of magnitude.  The level two trigger system employs some custom hardware and processing to match objects identified at level one and to achieve improved resolution with finer granularity.  For example, at level two a high momentum track may be matched to a high energy deposition in the EM calorimeter or to a track segment in the muon chamber, consistent with what is expected from an electron or muon, respectively.  The level two system reduces the event rate by another one order of magnitude.  The level three system employs a computing farm and custom software to reconstruct the full event and confirm the level two decision with a resolution and granularity that nearly matches that achieved after final calibrations and offline event reconstruction.  The level three system achieves an additional order of magnitude reduction in the event rate so that each experiment writes the most interesting events to tape at a rate of about $100$~Hz.

% ======================================================================
\section{Event Reconstruction}
\label{sec:reco}
% ======================================================================

The events which are written to tape are later processed on large scale computing farms using custom software to perform event reconstruction. This offline processing uses final calibration constants and alignments and more sophisticated algorithms relative to what is used in the level three trigger in order to achieve improved efficiencies and resolutions.  The reconstruction software combines information from the tracking, calorimeter, and muon chamber systems to provide particle identification.  Charged particles reconstructed in the tracking systems are extrapolated to the calorimeters and muon chambers.  Tracks matched to energy depositions in the EM calorimeter or to muon chamber hits are taken to be electron or muon candidates, respectively.  An EM energy deposition without a track match is taken to be a photon candidate.  A contiguous set of calorimeter towers with energy depositions above a predetermined noise threshold and matched to a collimated set of tracks in the tracking chamber is taken to be a jet originating from the hadronization of an initial quark or gluon state.
A jet which contains tracks that form a secondary vertex significantly displaced from the primary interaction vertex is taken to be a candidate $b$-jet.  The presence of a neutrino in the final state is inferred from a momentum imbalance in the transverse plane since neutrinos traverse the entire detector volume without interacting.  In the events of interest the final state particles are all relativistic, so the momentum imbalance is calculated using the calorimeter energies and is called ``missing $E_{T}$'' or MET and is defined as $\met = - \sum_{i} E_{T}^{i}\hat{i}$, where the index $i$ runs over all calorimeter towers, $E_{T}^{i}$ is the transverse energy measured in the $i$th tower and $\hat{i}$ is the unit vector pointing from the primary event vertex to the center of the $i$th tower in the transverse plane.  The resolution of the \met\ calculation can be improved by correcting for muons and jets in the event.
%%%The main features of the particle identification described here
%%%are shown in Fig.~\ref{fig:pid}.

The \ttbar\ all hadronic final state (\tthad ) is characterized by a large jet multiplicity, and the presence of 2 $b$-quark jets.  Analyses in this final state exploit the jet reconstruction and $b$-jet identification capabilities of the detector to identify events of interest.  
The \ttbar\ lepton+jets (\ttljt ) final state contains one high energy lepton, large \met, high jet multiplicity, and 2 $b$-quark jets.
The \ttbar\ dilepton (\ttdil ) final state has two high energy isolated leptons and large \met\ together with 2 $b$-quark jets.  
Analyses using the dilepton or lepton+jets final state exploit the full range of detector capabilities using electron, muon, and $b$-jet identification, jet reconstruction, and missing energy estimates to identify candidate \ttbar\ events.  Each analysis begins by choosing the subset of recorded data satisfying a particular set of trigger criteria.  This eliminates the need to run the full analysis software over those events whose topologies and kinematics significantly differ from those expected of \ttbar\ events.  The trigger criteria are designed to exploit a given characteristic of at least one of the \ttbar\ final states and to identify those events with high efficiency.  For each event satisfying the trigger criteria the primary interaction vertex is identified, leptons and jets are reconstructed, an estimate of \met\ is made, and $b$-jets are identified using a dedicated algorithm.  The event sample at this stage is dominated by background events originating from non-\ttbar\ processes.  Additional final state dependent selection criteria are used to suppress these background contributions and improve the purity of the sample.

\subsection{Trigger Criteria}
\label{sec:trig}

The $had$ data set is collected using trigger criteria requiring at least four jets each with raw (uncorrected) energy of $E_{T}>15$~GeV in events with large total transverse energy as determined by summing over all calorimeters towers (e.g. $\sum_{i} E_{T}^{i} > 175$~GeV).
For each 1~pb$^{-1}$ of collision data these triggers collect about 8-10 thousand events, depending on the details of the trigger selection, dominated by QCD production of multi-jet final states events (e.g. $\ppbar\ra qqqq$ or $\ppbar\ra qqgg$).

The $dil$ and $ljt$ data sets are collected using several different sets of trigger criteria.  The two most important sets of trigger criteria focus on identifying the electron or muon from the \Wboson\ decays\footnote{Because of the additional experimental challenges they introduce, \ttbar\ events with one of the $W$ bosons decaying to a tau lepton are not often used.  Since the tau lepton quickly decays to a multi-particle final state the $W\ra\tau\nu$ decay is difficult to trigger on in an inclusive manner.  Those leptonic tau decays, $\tau\ra e\nu\nu$ and $\tau\ra\mu\nu\nu$, satisfying the inclusive muon and electron trigger criteria are included in the data set described here.}.  These leptons are typically high energy and isolated.  The first set of trigger criteria is an inclusive high-$E_{T}$ electron data set requiring a candidate electron with $E_{T} > 20$~GeV.  The electron is identified at the trigger level as a high energy cluster in the EM calorimeter matched to a high momentum track in the tracking chamber.  Additional requirements are made on the ratio of energy deposition in the EM calorimeter to the energy deposition in the hadronic calorimeter, and on the shape of the EM shower in the calorimeter.  The electron candidate must be isolated, with few surrounding tracks and/or only small additional energy deposits in neighboring calorimeter towers.  The second set of trigger criteria is an inclusive high-$p_{T}$ muon data set requiring a candidate muon with $p_{T} > 20$~GeV/c.  The muon is identified at the trigger level as a track segment reconstructed in the muon chambers matched to a high momentum track reconstructed in the tracking chamber.   The muon candidate is additionally required to 
%%%match to small energy depositions in the calorimeter (since muons at %%%%these energies are minimum ionizing), and to 
be isolated.   Because these first two sets of trigger criteria make tracking requirements their acceptance is limited to the central detector region.   The number of \ttbar\ events collected can be significantly increased by including triggers which collect events in the forward detectors.  In order to keep the purity of the trigger sample reasonable these forward triggers are less inclusive.  They
usually require some high energy jets (e.g. $E_{T}^{\mrm{jet}} > 10$~GeV) in addition to a high energy EM calorimeter cluster in the forward region (e.g. $E_{T}^{EM} > 20$~GeV) or large \met\ (e.g. $\met > 35$~GeV).  For each 1~pb$^{-1}$ of collision data these triggers collect approximately 1-2 thousand $\ppbar\ra W+\mrm{jets}$ and $\ppbar\ra \zg+\mrm{jets}$ events, depending on the details of the trigger selection.

\subsection{Primary Interaction Vertex}
\label{sec:pvtx}

The counter-rotating proton and anti-proton beams at the Tevatron are  brought into collision near the centers of the CDF and \dzero\ detectors.  The exact location of the collision varies for each beam crossing.  The shape of the luminous region is determined using tracks in di-jet events ($\ppbar\ra q\overline{q}$), which are predominantly promptly produced and thus originate from the collision vertex.  The tracks are constrained to a common origin and the distribution of the $(x,y,z)$ position of the resulting vertex is determined using millions of events\footnote{For a vector of magnitude $v$ we define 
$ ( v_{x},\: v_{y},\: v_{z} ) = v ( \sin\theta\cos\phi,\:\sin\theta\sin\phi,\:\cos\theta ) $.}.   
The distribution of the collision positions along the $z$ axis can be approximated as a Gaussian with a mean near $z=0$ and an RMS of approximately $30$~cm.  
%
%%%In the transverse plane the beam is well described by a two dimensional %%%Gaussian whose mean varies as a function of $z$ but is typically within %%%$\mathcal{O}(1)\:\:\mrm{mm}$ of the $z$ axis.   The RMS width of the %%%beam in the transverse plane also varies with $z$ and increases from %%%about $30\:\:\mu m$ near $z=0$ to about $40\:\:\mu m$ for $\left| z%%%\right| > 40\:\:\mrm{cm}$.  The mean $x$ and $y$ as a function of $z$ are %%%well described by linear functions whose parameters are determined run-%%%by-run and stored in a beamline database. 
%
In order to achieve the best jet energy resolutions and highest $b$-jet identification efficiency it is necessary to reconstruct the location of the interaction vertex for each event.  
For $dil$ and $ljt$ events the $z$ coordinate of the highest energy lepton track can be used together with the known shape of the luminous region, as determined from the di-jet data as described above, to estimate the location of the interaction vertex with a resolution of a few hundred microns in $z$ and about $35\:\:\mu m$ in $x$ and $y$.  
These resolutions are improved by a factor of three to five by using the tracks in the event itself.  In order to avoid bias from long lived $b$-hadron decays in $b$-jets, only tracks whose distance-of-closest-approach (also known as the impact parameter, $d_0$) to the beamline is comparable to the $d_0$ resolution are used.  These tracks are then constrained to originate from a common point, whose coordinates are taken as the position of the \ttbar\ interaction vertex. An iterative procedure is employed by eliminating tracks with large $d_0$ relative to the fitted vertex to form a new vertex.  The process is repeated until no large impact parameter tracks are included in the vertex fit.   For $had$ events there is no lepton to seed the identification of the \ttbar\ vertex.  Again, tracks in the event are used and are constrained to originate from a common origin and an iterative procedure is used to remove the bias from tracks originating from $b$-hadron decay.  Depending on event topology, resolutions of $5-20\:\:\mu\mrm{m}$ in the transverse plane and of $100\:\:\mu\mrm{m}$ along $z$ are achieved on an event-by-event basis.

Since the \ppbar\ inelastic cross section is very large there is often more than one hard scatter interaction per beam-beam crossing. The mean number of expected hard scatters is linearly dependent on the instantaneous luminosity of the colliding beams.  For the Tevatron parameters the mean number of expected hard scatters per beam-beam crossing is about $3.5$ at an instantaneous luminosity of $\mathcal{L}_{\mrm{inst}} = 1.0\times 10^{32}\:\mrm{cm}^{-2}\mrm{s}^{-1}$. The actual number of hard scatters per event is Poisson distributed about the mean.  These additional interactions are typically separated in $z$. Since the RMS of the luminous region along the $z$ axis is large relative to the resolution with which a single interaction vertex is reconstructed, the experiments can efficiently reconstruct and readily resolve multiple interaction vertices in a single event.  This is important since the typical Tevatron stores have an initial instantaneous luminosity that exceed $3\times 10^{32}\:\mrm{cm}^{-2}\mrm{s}^{-1}$ and an average instantaneous luminosity over the duration of a store, which is typically about 15 hours, of about  
$1.5\times 10^{32}\:\mrm{cm}^{-2}\mrm{s}^{-1}$.   For events with multiple reconstructed interaction vertices it is necessary to identify which of the vertices corresponds to the \ttbar\ interaction vertex.  For the leptonic final states the vertex closest to the high energy lepton is used.  For the $had$ final state, the scalar sum $p_{T}$ of all tracks associated with each vertex is calculated and the vertex with the largest $\sum_{i} \left| p_{T}^{i} \right|$ is used.  The \ttbar\ interaction vertex is correctly identified nearly 100\% of the time.

\subsection{Electrons and Muons}
\label{sec:emu}

The reconstruction of high energy electrons and muons from \Wboson\ decays is important in identifying the signal \ttbar\ events.  The leptons from $Z\ra e^{+}e^{-}$ and $Z\ra \mu^{+}\mu^{-}$ decays are kinematically very similar to the $W$ decay leptons and are thus used as data control samples to measure efficiencies and backgrounds for the lepton reconstruction and selection criteria employed in the \ttbar\ analyses.   The reconstruction of lower energy electrons and muons from semi-leptonic $b$-hadron decays is sometimes used to identify $b$-jets.

Electron reconstruction begins by identifying energy depositions in the EM calorimeter above a specified threshold, typically around $2$~GeV.  In the central detector region, tracks from the tracking chamber with momentum above a few GeV/$c$ are extrapolated from the tracker outer radius to the front face of the EM calorimeter.   If the extrapolated track position is consistent with the energy-weighted centroid of the calorimeter cluster within uncertainties, including contributions from multiple scattering in the material between the tracker and the calorimeter, the resulting EM calorimeter and track match are taken to be a candidate electron.  To reduce contributions from hadrons mimicking electrons, the shape of the shower in the calorimeter, and the ratio of track $p_{T}$ to calorimeter $E_{T}$ are required to be consistent with that expected of electrons as determined from $Z\ra e^{+}e^{-}$ events and test beam data.  The energy deposited in the hadronic calorimeter tower immediately behind the candidate electron is required to be small relative to the EM energy deposit.  Electrons consistent with having originated from $\gamma\ra e^{+}e^{-}$ conversions in the detector material are identified with dedicated algorithms and removed from further consideration.  In the forward detector region where there is very limited coverage from the full tracking chamber, electrons are identified using stricter criteria on the EM cluster energy, shower shape, and hadronic to EM energy deposition ratios.  The purity of the resulting forward electron sample can be significantly improved by searching for hits in the tracking system consistent with expectations derived from the $E_T$ and position of the EM cluster and the location of the event primary interaction vertex. 

Muon reconstruction begins by identifying track segments in the muon chambers.  Tracks from the tracking chamber with momentum above several GeV/$c$ are extrapolated from the tracker outer radius to the front face of the muon chamber.  If the extrapolated track position is consistent with the position of the muon chamber track segment within uncertainties, including contributions from multiple scattering in the material between the tracker and the muon chamber, the resulting segment and track match are taken to be a candidate muon. Through going cosmic ray muons are identified using dedicated algorithms and removed. Hadrons whose shower are not fully contained in the calorimeter and other intervening material are said to ``punch-through'' to the muon chambers and can mimic a muon signature.  To reduce contributions from punch-through hadrons, stricter segment-track matching and track $p_T$ criteria together with track quality criteria can be employed.  The track quality criteria usually include requirements on the number hits used to form the track, the impact parameter, and the track fit $\chi^2$.  These track quality criteria also help reduce contributions from hadron decays-in-flight (e.g. $K^{\pm}\ra\mu^{\pm}\nu$).  The energy depositions in the calorimeter towers traversed by the muon can be required to be consistent with that expected from a muon as determined using $Z\ra \mu^{+}\mu^{-}$ and $J/\psi\ra \mu^{+}\mu^{-}$ data control samples.

The leptons from $W$- and \Zboson\ decays are typically well separated from other objects in the event.  To further suppress hadronic backgrounds isolation criteria are employed.  Calorimeter isolation requires that the energy deposited in an annulus around the calorimeter towers associated with the lepton candidate be small.  Tracking isolation requires that the scalar sum $p_T$ of all tracks in a cone around the lepton candidate be small.  The requirements are sometimes absolute (e.g. $E(\mrm{annulus}) < 3\:\:\mrm{GeV}$) and sometimes relative (e.g. $E(\mrm{annulus})/E(\mrm{electron}) < 0.1$).

Samples of $Z\ra \ell^{+}\ell^{-}$ events collected with the inclusive electron and muon triggers described above, are used to measure the trigger, reconstruction, and identification efficiencies for electrons and muons, respectively.   The ``tag-and-probe'' method is employed where the lepton leg which fired the inclusive trigger is the ``tag'' leg, while the second leg is unbiased and used as a ``probe'' to measure the efficiency of interest.   The efficiencies are measured with total uncertainties of $<1\%$ relative and are characterized as a function of $p_{T}$, $\eta$, instantaneous luminosity, and number of reconstructed interaction vertices.  A more detailed description of these measurements can be found in~\cite{CDFWZxsec}. 

The absolute momentum scale of the spectrometer is determined using $J/\psi$, $\Upsilon$, $\Upsilon^\prime$, and \Zboson\ decays to $\mu^{+}\mu^{-}$ by comparing the reconstructed invariant mass of the muon pairs to the known masses of each resonance.  By using a variety of resonances any non-linearities in the momentum scale can be identified and corrected for.  These studies result in a more precise understanding of the magnetic field uniformity, the distribution of material in the detector volume, and any small residual mis-alignments in the tracking systems. The absolute energy scale of the EM calorimeter is determined using $\pi^{0}\ra\gamma\gamma$, $J/\psi\ra e^{+}e^{-}$ and $Z\ra e^{+}e^{-}$ decays and using the $E/p$ distribution of electrons from $W\ra e\nu$ decays.  Again, by using a variety of decay processes non-linearities in the energy scale can be identified and corrected for.  In both cases the absolute scales are determined to a precision well below $1\%$ and are monitored continuously and kept constant at the $0.2-0.3\%$ level.  The observed RMS widths of the reconstructed mass distributions are dominated by the detector resolutions.  Studies of the di-muon and di-electron resonance widths help determine the momentum and energy resolutions, respectively.   The absolute scales and resolutions are characterized as a function of energy/momentum, $\eta$, instantaneous luminosity, and number of reconstructed interaction vertices.  These methodologies are described in more detail in~\cite{CDFMw, DzMw}.

\subsection{Jets}
\label{sec:jets}

Jets are reconstructed as a contiguous set of calorimeter towers each above a predetermined noise threshold.   A variety of algorithms are available to determine which towers to group together.  The algorithms differ in the criteria they use to combine towers to form jets and in the manner in which they identify and merge or separate jets which are near each other in $\eta - \phi$ space.  The behavior of the algorithms, particularly with regard to their handling of nearby jets, affects the degree to which they are sensitive to low energy "soft" QCD radiation.  Algorithms which are insensitive to soft QCD radiation are said to be ``infrared safe'' and can be compared to theoretical predictions in a very clean manner.  There are in general two classes of jet algorithms.  The $k_T$ algorithms combine calorimeter towers to form jets based on their relative transverse momentum as well as their $\eta - \phi$ separation.  They are rigorously infrared safe.  The cone algorithms combine calorimeter towers to form jets based only on their $\eta - \phi$ separation. The cone algorithms begin by identifying a set of seed towers with energy above some seed threshold, typically about $1\:\:\mrm{GeV}$.   The Midpoint algorithm employed by CDF and \dzero\ is such an algorithm with additional steps introduced in order to minimize the sensitivity to soft QCD radiation.  While rigorously the Midpoint algorithm is not completely infrared safe, in practice the residual sensitivity is experimentally inconsequential.  This is due to the fact that the energy of jets in \ttbar\ events is large (approximately $50\:\:\mrm{GeV}$ on average) compared to the seed threshold so that small fluctuations in tower-to-tower energy depositions from soft QCD radiation do not affect the results of the algorithm.  The cone algorithms offer an experimental advantage because the physical extent of each jet is well defined to have a radius of $R=\sqrt{\left(\Delta\eta\right)^{2} + \left(\Delta\phi\right)^{2}}$ while the physical extent of jets formed using a $k_T$ algorithm vary from jet to jet.  In practice the two algorithms yield very similar results~\cite{craigspaper}.  In the cone algorithms the cone radius employed can be varied.  Since the jets in \ttbar\ events are typically collimated a relatively narrow cone size is employed, $R=0.4$ for CDF and $0.5$ for \dzero.

The jet reconstruction algorithm uses raw calorimeter energies.  The total energy of the jet is estimated as the sum of energies over all towers included in the jet.  The direction of the jet is determined using an energy weighted average over all contributing calorimeter towers.  The energy and direction of the jet are then used as an estimate of the four momenta of the parent parton (ie quark or gluon).  The jet energy resolution is improved by employing a set of corrections which account for several effects.  The effects fall into two broad categories, instrumental effects, and physics effects.  The instrumental effects accounted for include corrections for non-uniform and non-linear response of each calorimeter tower as well as contributions from multiple interactions per event.  The physics effects accounted for include corrections for contributions from the underlying event, from the hadronization process, and from out-of-cone showering.  The calibration of the CDF and \dzero\ jet energy corrections are described in detail in~\cite{cdfjes} and~\cite{dzjes} and are a function of jet transverse energy and pseudo-rapidity.  The corrections vary from about 60\% for jets with transverse energy around $20\:\:\mrm{GeV}$ to about 20\% for jets with transverse energy around $100\:\:\mrm{GeV}$.  The corrections can be checked using $\ppbar\ra \gamma + \mrm{jet}$ and $\ppbar\ra Z + \mrm{jet}$ events. Given the initial constraint that $\left( E_{T}, p_{T} \right) = \left( 0,0 \right)$, it is expected that the jet transverse energy after all corrections should be equal to the transverse energy of the well measured photon or $Z$ boson (when $\zg\ra e^{+}e^{-}$ or $\mu^{+}\mu^{-}$) in the event.   
%
%An example of these checks is shown in Fig.~\ref{fig:gammajetcomparison}.  
%
The uncertainties on the jet energy scale derive from many sources, some of which are experimental (e.g. the statistics of the control sample or an inadequate understanding of the detector response) and some of which are modeling (e.g. modeling of the underlying event, hadronization, and out-of-cone showering processes) related.   Like the corrections, the uncertainties are jet $E_T$ and $\eta$ dependent and range from about $10\%$ at low energies ($< 40$~GeV) to a few $1\%$ at high energies ($> 75$~GeV).  The dominant source of uncertainty is modeling related at the lower energies and experiment related at higher energies.

The full set of jet energy corrections are designed to give an accurate estimate of the energy and momentum of the parent parton that had initiated the jet.  However, the full set of corrections are not always employed.   Frequently jets are only corrected to the particle level, which effectively corrects for all the instrumental effects as well as contributions from the underlying event and multiple interactions.  Jets corrected to this level are effectively independent of apparatus and environment and can be compared across experiments.  For the selection criteria described below, the jets are corrected to the particle level.   Later, when using the selected \ttbar\ candidate events in analyses the jets are corrected to the parton level whenever a kinematic fit is employed.

%%\begin{figure} 
%%\centering
%%  \epsfxsize=3in\epsfbox{cdfgammajetplot.eps} 
%%  \epsfxsize=3in\epsfbox{cdfgammajetVSpt.eps} 
%%\caption{\label{fig:gammajetcomparison}Checks of the CDF jet energy 
%%corrections using $\ppbar\ra\gamma + \mrm{jet}$ events.} 
%%\end{figure} 

\subsection{Missing Transverse Energy}
\label{sec:met}

We exploit the fact that the initial state satisfies
$\left( E_{T}, p_{T} \right) = \left( 0,0 \right)$ and
use an observed imbalance of momentum in the transverse plane to infer the presence of high energy neutrinos in an event.  As mentioned above, because the final state particles are all relativistic at the relevant collider energies, we use the energy measured in the calorimeters as an estimate of the total momenta of particles produced (including neutral particles).  For this reason the quantity examined is called missing transverse energy, $\met$.   The initial estimate of the missing transverse energy begins with using only calorimeter information:
$\met = - \sum_{i} E_{T}^{i}\hat{i}$, where the index $i$ runs over all calorimeter towers, $E_{T}^{i}$ is the transverse energy measured in the $i$th tower and $\hat{i}$ is the unit vector pointing from the primary event vertex to the center of the $i$th tower in the transverse plane.
Corrections are made if there are identified muons in the event since they are minimum ionizing and the calorimeter energy deposition is a poor estimate of their momenta:
$\met = - \left( \sum_{i} E_{T}^{i}\hat{i}\:\: - E_{T}^{\mu}\hat{i}_{\mu}+p_{T}^{\mu}\hat{i}_{\mu} \right)$, where $E_{T}^{\mu}$, $\hat{i}_{\mu}$, and $p_{T}^{\mu}$ are the calorimeter transverse energy, azimuthal angle unit vector, and tracker transverse momentum associated with the muon candidate.  The \met\ resolution can be further improved when jets have been reconstructed by taking advantage of the jet energy corrections described above by removing from the initial sum any calorimeter tower included in a jet and substituting the jet transverse energy (corrected to the particle level) at the jet $\phi$.  For \ttbar\ analyses the final expression is often rewritten:
$\met = -\left( \sum_{\ell} p_{T}^{\ell}\hat{i}_{\ell} + \sum_{j} E_{T}^{j,\mrm{cor}}\hat{i}_{j} + \sum_{\rm uc} E_{T}^{\rm uc}\hat{i}_{\rm uc} \right)$, where $\ell$ is the sum over identified leptons (electrons or muons) in the event, $j$ is the sum over reconstructed jets, and uc is the sum over unclustered calorimeter towers (ie. those not included in a jet).  For electrons (muons), $p_{T}^{\ell}$ is estimated from the EM calorimeter (spectrometer).  For jets, $E_{T}^{j,\mrm{cor}}$ is the jet transverse energy corrected to the particle level.  For the unclustered calorimeter towers the raw calorimeter energy is used.  This expression is equivalent to the original expression with corrections for the muons and jets as described.

At the trigger level only the raw \met\ estimate is available, without any of the corrections.  At the event selection and later in the analyses employing kinematic fits, the fully correctly \met\ is used.

\subsection{$b$-quark Jets}
\label{sec:bjet}

Most jets are initiated by a parent light quark ($uds$) or gluon ($g$). Thus the  identification of at least one of the $b$-jets in \ttbar\ events significantly reduces contributions from background processes.  The $b$-jet identification algorithms employed by the Tevatron experiments generally exploit two main features of $b$-hadron decay - their long lifetime and their semi-leptonic decays.  While $c$-hadrons also have some of these same features, on average their lifetimes are shorter and they contain fewer electrons and muons than $b$-hadrons so that the algorithms employed typically achieve $b$-identification efficiencies a factor of 2-4 larger than the corresponding $c$-identification efficiencies, which is sufficient to reduce charm quark related backgrounds to acceptable levels for the \ttbar\ analyses described here.

The branching fraction for a $b$-hadron into $e$ or $\mu$ plus anything is about 20\% each (including sequential $b\ra c\ra eX$ or $\mu X$), which is about twice as large as that for $c$-hadrons, while $udsg$-jets have a negligible $e / \mu$ content.  The energy of these leptons from $b$-decay is a few GeV on average, so much ``softer'' than those from $W$- or \Zboson\ decays.  Consequently these algorithms are often called ``Soft Lepton Tagging'' algorithms.   Additionally, since they are embedded in jets, these leptons are less isolated.  The identification of these electrons and muons uses the same reconstruction algorithms as those discussed in Sec.~\ref{sec:emu}.  To further suppress backgrounds from hadrons mimicking electron or muon signatures these algorithms demand the leptons to have a transverse momentum greater than about 3 GeV/c and make stricter matching requirements between the track and EM calorimeter cluster (electrons) or track segment in the muon chamber (muons).  In addition the EM shower shape is exploited to suppress backgrounds for the electron case.  The full set of identification criteria is usually combined in a single variable using a relative likelihood function, a global $\chi^2$ or other similar multivariate techniques.  Data control samples of $J/\psi\ra\mu^{+}\mu^{-}$, $\gamma\ra e^{+}e^{-}$, and $Z\ra\mu^{+}\mu^{-}$, $e^{+}e^{-}$ are used to measure the identification efficiencies.  Typical efficiencies achieved are 80-90\% for muons in $b$-hadron jets above the $p_{T}$ threshold and fiducial to the muon chambers.   The efficiencies for electrons in $b$-hadron jets is about a factor of two lower because the EM tower often includes contributions from nearby hadrons in the jet thus degrading the shower shape criteria.  The degree to which this is a problem will partly depend on the segmentation of the EM calorimeter.   For both algorithms the rate at which hadrons are mis-identified is about 0.5\% per track and varies by about $\pm 20\%$ (relative) as a function of track $p_{T}$, track $\eta$, or hadron species (ie. $\pi^{\pm}$ vs $K^{\pm}$ vs $p$/$\overline{p}$).  These fake rates are determined using samples of      $\Lambda\ra p\pi^{-}$, $D^{+*}\ra D^{0}\pi^{+}\ra K^{-}\pi^{+}\pi^{+}$, and $K_{s}^{0}\ra\pi^{+}\pi^{-}$ decays. Including the semi-leptonic branching fractions and the track multiplicities in $udsg$-jets these algorithms in combination achieve an efficiency of $\sim 25\%$ per fiducial $b$-jet in \ttbar\ events and misidentify a $udsg$-jet at a rate of $\sim 0.02-0.03$ each.  These algorithms and their performance are described in more detail in~\cite{CDFSLTm} and~\cite{CDFSLTe}.

The most important $b$-jet identification algorithms exploit lifetime information by using tracks originating from the $b$-hadron decay and reconstructing a secondary vertex significantly displaced from the primary interaction vertex.  These algorithms exploit almost all $b$-decay topologies and thus make use of nearly the full branching fraction. They work on a per jet basis and begin by associating to each jet a set of tracks using the $\eta-\phi$ separation between the track and the jet axis.  Only tracks within a given radius (e.g. $R < 0.4$), above a given $p_{T}$ threshold (e.g. $p_{T} > 1$~GeV/c), and whose $z$ coordinate at the distance of closest approach to the beam line is consistent with having come from the primary interaction vertex of the event are used to form the candidate secondary vertices.  These tracks are vetoed if they form a good vertex and yield an invariant mass consistent with a $K^{0}_{s}$ or $\Lambda$ when paired with any other track in the same jet.  They are also vetoed if they are identified as originating from a photon conversion, $\gamma\ra e^{+}e^{-}$, in the detector material using dedicated algorithms.  At hadron colliders ``build-up'' algorithms are preferred to ``tear-down'' algorithms because non-$b$-decay tracks often out number tracks from $b$-decay owing to contributions from the under lying event and hadronization process.  The build-up algorithms begin by taking the surviving tracks pair wise and making a seed vertex, often times beginning with those tracks having the largest impact parameters relative to the primary interaction vertex of the event. Additional tracks are then added to these seed vertices.  The $p_{T}$ and impact parameter requirements for these additional tracks may be relaxed relative to the requirements made on the tracks forming the seed vertices.   Only vertices with good fit $\chi^2$ are kept and tracks making large contributions to the $\chi^2$ are removed.   Surviving vertices are then subject to additional selection criteria optimized to suppress the rate of fake tags from $udsg$-jets while maintaining high efficiency for $b$-jets.  In particular the decay length, which is the magnitude of the vector connecting the primary interaction vertex of the event to the candidate secondary vertex is required to fall within the beam pipe (e.g. $\left|\vec{L}\right| < 1.5$~cm at CDF and $\left|\vec{L}\right| < 2.6$~cm at \dzero) in order to eliminate interactions in the detector material, the angle between $\vec{L}$ and the jet momentum must be less than $\pi/2$, and the decay length significance must exceed some minimum (e.g. $\left|\vec{L}\right| / \sigma_{L} > 7$).   
There can be multiple seed vertices per jet, which are ordered by vertex quality.  Each seed vertex is then tried in turn until a good vertex is found.
The algorithms may exploit additional properties of the candidate secondary vertex, like the number of tracks included in the vertex or the invariant mass of the tracks included in the vertex, to further discriminate $b$-decay vertices from backgrounds.  The efficiency of these algorithms is determined from the data using control samples of $b$-jets collected with triggers which identify the soft electrons or muons from semi-leptonic $b$-decays.  These samples thus yield the efficiency for semi-leptonic $b$-decays and tend to have a fairly soft $E_{T}^{\mrm{jet}}$ spectrum.  Monte Carlo simulation is used to extrapolate the efficiencies to generic $b$-decays and higher jet energies.  The Monte Carlo and detector simulation are tuned to accurately reproduce the behavior of the algorithm as measured in the data control sample.  Systematic uncertainties account for discrepancies between the data and the Monte Carlo modeling and for the uncertainties associated with the extrapolation to higher jet energies.  The rate at which $udsg$-jets are mis-identified by these algorithms is measured in the data using a sample of generic QCD jets collected using only calorimetric trigger information.  To get an estimate of the mis-identification rate the experiments exploit the fact that the observed decay length distribution for $udsg$-jets is nominally symmetric about zero with a width dominated by resolution effects.  The negative decay lengths correspond to those jets for which the angle between $\vec{L}$ and the jet momentum is larger than $\pi/2$ and are dominated by contributions from $udsg$-jets.  The rate of negative decay lengths is thus used to parameterize the mis-identification rate for $udsg$-jets reconstructed with positive decay lengths.
The mis-identification rate is parameterized as a function of several variables, such as the number of tracks in the jet, the jet energy, the jet pseudo-rapidity, the number of additional interactions in the events, etc.  Small corrections are made to account for the $b$-jet and $c$-jet contributions to the negative decay length sample and for contributions to the positive decay length rate from residual photon conversions in the material.  
These algorithms are tuned to several operating points.  The most popular operating point is labeled ``Tight'' and achieves an efficiency of $\sim 50\%$ per fiducial $b$-jet in \ttbar\ events and mis-identify $udsg$-jets at a rate of $<0.01$ each.  The efficiency has a total uncertainty of about $7\%$ (relative) dominated by systematic contributions from uncertainties in the sample composition of the data control sample, the Monte Carlo modeling, and the extrapolation in $E_{T}^{\mrm{jet}}$.  The mis-identification rates have a total uncertainty of $10-20\%$ (relative) dominated by systematic contributions from uncertainties in the sample composition, corrections accounting for trigger biases, and corrections for the $b$-jet and $\gamma\ra e^{+}e^{-}$ contributions.
These algorithms and their performance are described in more detail in~\cite{CDFBtagXSttbar} and~\cite{DzBtagging}.

Other algorithms exploiting the long lifetime of $b$-decays are sometimes employed such at ``jet probability'' algorithms, or simple ``counting'' (displaced tracks) algorithms.  In general these offer additional efficiency at the cost of higher $udsg$-jet mis-identification rates.

Often multiple algorithms are employed and then combined using various
multivariate techniques to yield another 10-20\% (relative) increase in the efficiency for identifying $b$-jets for the same $udsg$ misidentification rate.  An example is described in detail in~\cite{DzBtagging}.

\subsection{Kinematic Fits}
\label{sec:kfit}

In many analyses kinematic fits are employed to fully reconstruct the \ttbar\ kinematics and improve the resolution on the reconstructed momenta of the final state partons from the top-quark decays.  The fits constrain the kinematics of each event assuming the \ttbar\ hypothesis and using the decay relevant for the final state in question.  In the $had$ channel the six highest $E_T$ jets are assumed to correspond to the quarks from the top-quark and \Wboson\ decays.  In the $ljt$ channel the four highest $E_T$ jets are assumed to correspond to the quarks from the top-quark decays and the hadronically decaying $W$ boson, while the charged lepton is assumed to come from the leptonically decaying $W$ boson.  In the $dil$ channel the two highest $E_T$ jets are assumed to correspond to the $b$-quarks jets from the top-quark decays and the two (oppositely) charged leptons are assumed to come from the two leptonically decaying $W$ bosons.  In the $ljt$ and $dil$ channels the measured \met\ is used to constrain the $x$ and $y$ components of the neutrino momenta.  The measured four-vectors for the jets are corrected using the jet energy corrections described in Sec.~\ref{sec:jets} and the four-vectors of the leptons incorporate the calibrations described in Sec.~\ref{sec:emu}.  The calibration studies also yield estimates of the corresponding resolutions.   The kinematic fit proceeds via a $\chi^2$ minimization.  As an example, the $\chi^2$ function employed in the $ljt$ channel, \ttljt , is
\begin{eqnarray}
  \chi^{2} & = & \sum_{i=\ell,\:{\rm jets}} \frac{(p_{T}^{i,{\rm fit}}-p_{T}^{i,{\rm meas}})^2}{\sigma^2_i}
  + \sum_{j=x,y} \frac{(p_{j}^{{\rm uc,\:fit}}-p_{j}^{{\rm uc,\:meas}})^2}{\sigma^2_{j,{\rm uc}}} \\
 & & + \frac{(m_{\ell\nu} -\mw)^2}{\Gamma_W^2} 
  + \frac{(m_{jj} -\mw)^2}{\Gamma_W^2}
  + \frac{(m_{b\ell\nu} -\mt)^2}{\Gamma_t^2} 
  + \frac{(m_{bjj} -\mt)^2}{\Gamma_t^2}.
\label{eq:chisq}
\end{eqnarray}
Similar expressions are used for the $had$ and $dil$ channels except that the $m_{\ell\nu}$ and $m_{b\ell\nu}$ terms are replaced by $m_{jj}$ and $m_{bjj}$ terms or vice versa, respectively.  The first term constrains the $p_T$ of the lepton and jets to their measured values within their assigned resolutions, $\sigma_i$. The second term makes the same constraints on the $x$ and $y$ components of the unclustered energy in the event.  The \met\ is thus derived using these first two terms as discussed in Sec.~\ref{sec:met}.  The last four terms constrain the reconstructed invariant mass of some combination of final state partons to the pole mass\footnote{In perturbative field theory the pole mass of a particle is defined as the pole of its renormalized propagator.} of the object from which they are assumed to have originated.  For example, $m_{jj}$ is the invariant mass of the two jets assumed to have originated from the hadronically decaying $W$ boson and so it is constrained to equal \mw\ within some tolerance, taken to be the natural width of the $W$, $\Gamma_W$.  Similarly for the decay products of the leptonically decaying $W$ boson and the decay products of each top quark.  In the fit, the $p_{T}^{i,{\rm fit}}$ and $p_j^{\rm uc,\:fit}$ terms are varied to minimize the $\chi^2$.  For most analyses the \mw\ and \mt\ values are fixed to the world average measured values.  In analyses measuring the top-quark mass, discussed in Sec.~\ref{sec:mass}, the \mt\ term is left a free parameter, \mtreco , and corresponds to the reconstructed top-quark mass most consistent with the kinematics of that particular event.

In order to perform the minimization it is necessary to assign each jet as having originated from a particular final state quark.  Since the flavors of the parent quarks are unknown, all possible combinations are tried and the combination yielding the lowest fit $\chi^2$ is used.  This procedure yields the correct jet-parton assignment 65-85\% of the time depending on the details of the selection criteria and decay channel.
This ambiguity in the jet-parton assignments gives rise to a combinatoric background from \ttbar\ events for which the incorrect combination is used.  The combinatoric background significantly dilutes the resolution with which the \ttbar\ kinematics can be reconstructed.   The number of possible jet-parton assignments for an event with $N_j$ jets is $N_{\rm comb} = N_j !$.  Since the $\chi^2$ does not distinguish between a swap in the quark assignments for the jets assumed to have come from the hadronic \Wboson\ decay, the effective number of jet-parton combinations is reduced by a factor of two for each $W\to jj$ decay in the final state.  The number of combinations considered can be further reduced by rejecting combinations for which a jet identified as a $b$-jet is assigned as one of the light-quark jets from the hadronic \Wboson\ decay.  For example, in a $ljt$ event reconstructed with 4 jets in the final state, there are $\frac{4 !}{2}=12$ jet-parton assignments considered; this is reduced to six different assignments if one of the jets is identified as a $b$-jet and only two possible assignments if two jets are identified as $b$-jets.  The combinatoric background is less of a problem in the $dil$ channel and is severe in the $had$ channel.

The presence of neutrinos in the final state introduces a further ambiguity.   In the $ljt$ channel the $x$ and $y$ components are constrained using the estimated \met, but the $z$ component is unconstrained.  In general, each jet-parton assignment yields two different solutions for $p_{z}^{\nu}$ and both are tried.  This doubles the number of $\chi^2$ minimizations performed for each $ljt$ event.   For the $dil$ channel the kinematics is actually under-constrained due to the presence of two neutrinos from each of the leptonic \Wboson\ decays.  
In this case it is necessary to make an additional kinematic assumption in order to obtain a reconstructed invariant mass for each event.  An integration over the undetermined additional kinematic variable (e.g. the azimuthal angles of the escaping neutrinos) is made and a reconstructed mass is calculated for each assumed value utilizing the kinematic fit.  A weight is assigned to each of these possibilities by comparing the observed kinematics with those expected for the assumptions being made (e.g. the direction of the measured \met ).  The weighted distribution of reconstructed masses is then used to determined an \mtreco\ estimate for each event.  Specific examples of this methodology are discussed in Sec.~\ref{sec:mass}.

\subsection{Event Selection Criteria}
\label{sec:evtsel}

After the full reconstruction software has been run, the following event selection criteria are employed to identify a sample of candidate \ttbar\ $dil$, $ljt$, and $had$ events.  The three samples are statistically independent by construction.

The selection of $dil$ events begins by requiring a pair of high energy leptons, $e^{+}e^{-}$, $\mu^{+}\mu^{-}$, or $e^{\pm}\mu^{\mp}$, each with $E_T$ exceeding a value around $20$~GeV and $\left|\eta\right| < 2 (1)$ for electrons (muons).  The events are collected using the inclusive lepton triggers described in Sec.~\ref{sec:trig} and reconstructed using the algorithms described above.  Several lepton selection criteria categories are usually employed ranging from ``tight'', which yield a very pure sample of leptons, to ``loose'', which can significantly increase the acceptance but with hadron mis-identification rates that are significantly higher.  The analyses usually treat the various lepton-lepton categories (e.g. tight-tight, tight-loose, etc) separately since their signal-to-background ratio are significantly different. The leptons are required to be isolated. Backgrounds from $Z$ events are removed by vetoing events for which either of the leptons, when combined with any other track in the event with $E_{T}>10$~GeV, yields an invariant mass consistent with the mass of the $Z$ boson.   The events are also required to have a large missing transverse energy, $\met > 20-25$~GeV, and at least two jets with $E_{T}^{\mrm{jet}} > 15-20$~GeV (after correcting to the particle level) and $\left|\eta\right|<2.0-2.5$.  To reduce contributions from backgrounds for which the \met\ arises from the mis-measurement of a jet or lepton energy, the angle between the \met\ vector and any jet or lepton in the event is required to be larger than some minimum (e.g. $\Delta\phi(\met-\mrm{jet}) > 20^{\circ}$).  To remove backgrounds from $b\ra\ell c X\ra \ell\ell q X$ decays the lepton-lepton invariant mass is required to be larger than $5\:\:\gevcc$.  If an $e^{+}e^{-}$ event is consistent with having originated from a photon conversion in material, or a $\mu^{+}\mu^{-}$ event is consistent with a through going cosmic ray event, they are vetoed. Events with 0 or 1 reconstructed jet are used as background dominated control regions to verify the accuracy of the background estimates.   The total acceptance times branching fraction achieved is typically around $0.8\%$ with a purity of around $70-75\%$.  If a requirement that at least one of the jets is identified as a $b$-jet the acceptance times efficiency falls to about $0.5\%$ and the purity rises to about $90-95\%$.

The selection of $ljt$ events begins by requiring a single high energy lepton, typically with $E_{T} > 20$~GeV, collected using the inclusive electron and muon triggers described in Sec.~\ref{sec:trig}. After employing the reconstruction algorithms described above, ``tight'' lepton selection criteria are used to identify the primary lepton in each event, which is required to be isolated.  Events in which another isolated lepton is identified with $E_{T}>10$~GeV are vetoed in order to remove dilepton backgrounds.  Dedicated algorithms are used to remove $e^\pm$ events consistent with having originated from photon conversion in the material. The events are also required to have large missing transverse energy, $\met > 20-25$~GeV, and at least three jets with $E_{T}^{\mrm{jet}} > 15-20$~GeV (after correcting to the particle level) and $\left|\eta\right|<2.0-2.5$.  Events with 1 or 2 reconstructed jets are used as background dominated control regions to verify the accuracy of background estimates.  The total acceptance times branching fraction achieved is typically around $7-8\%$ with a purity of around $15-20\%$.   To improve the purity, most \ttbar\ analyses using the $ljt$ final state require that at least one of the jets in the event be identified as a $b$-jet.  Doing so yields an acceptance times branching fraction of around $4\%$ with a purity of about $60\%$.

The selection of $had$ events begins with a data sample collected using the multi-jet trigger described in Sec.~\ref{sec:trig}.  After utilizing the reconstruction algorithms described above, events with an identified high energy lepton are removed.  Since no missing transverse energy is expected in the all-hadronic final state, the observed \met\ is required to be small relative to the uncertainty with which it is measured, $\met / \sqrt{\sigma_{\met}} < 3$.  Finally, the events are required to have 6-8 reconstructed jets, each with $E_{T}^{\mrm{jet}}>15$~GeV (after correcting to the particle level) and $\left|\eta\right|<2$.   Events with 4-5 jets are used as background dominated control regions in which to verify the accuracy of the background estimates.  The total acceptance times branching fraction is about $13\%$ with a purity of about $0.3\%$.  To improve the purity, $b$-jet identification algorithms are employed, as well as multivariate discriminates that use kinematic and event shape information to suppress the background relative to \ttbar\ events.  After these additional selection requirements are made, the total acceptance times branching fraction falls to approximately $3\%$ and the purity rises to about $9\%$.

All the analyses described below begin by using the samples obtained after employing these event selections.  Often, depending on the analysis, additional criteria are used to further improve the \ttbar\ purity of the sample.  A variety of methodologies are employed to account for the remaining background contributions as described in the next section.

% ======================================================================
\section{Background Processes}
\label{sec:bgd}
% ======================================================================

After all selection criteria, background events still remain and must be accounted for in order to extract the \ttbar\ physics parameters of interest.  A variety of background processes contribute to each \ttbar\ final state and fall into two basic categories: physics backgrounds and instrumental backgrounds.   Physics backgrounds are those processes that share the same final state as the \ttbar\ signal sample we are aiming to isolate.  For example, the $\ppbar\ra W+b\overline{b}q\overline{q}$ process, with the $W$ decaying to $e\nu$ or $\mu\nu$, is a physics background to the \ttljt\ sample.  Instrumental backgrounds are those processes which mimic the \ttbar\ final state of interest due to an instrumental effect resulting in a mis-identification of some of the final state objects.  For example, the $\ppbar\ra W+b\overline{b}q\overline{q}$ process, with the $W$ decaying to $e\nu$ or $\mu\nu$, is an instrumental background to the \ttdil\ sample when one of the jets is mis-identified as a high energy lepton.  Although the mis-identification rates are typically very small, $< 1\%$, the instrumental backgrounds can still significantly contribute to the final selected samples due to the very large production cross sections for the relevant QCD and \Wpjets\ processes.  In general, the acceptances for physics backgrounds are estimated using Monte Carlo simulations, while the instrumental backgrounds are estimated using data control samples.  For background processes making small contributions to the final sample, or with well understood theoretical cross sections, the normalization is taken from theory calculations.  For all other cases the normalization is taken from the data using background dominated control samples and then extrapolated to the final signal sample.  A brief review of the background processes that contribute to the various \ttbar\ final state event samples and how they are estimated is provided here.

The generic QCD process $\ppbar\ra \mrm{jets}$ has a production cross section that is about 9 orders of magnitude larger than the \ttbar\ production cross section.  The jets produced pre-dominantly originate from $uds$-quarks or gluons, although $b$-quark jets are produced in a few percent of these events.  These generic QCD processes are the dominant background for the \tthad\ final state with the instrumental background originating from the mis-identification of a $udsg$-jet as a $b$-quark jet being the most important.   For the \ttljt\ final state, the QCD background is sometimes referred to as the ``non-W'' background and accounts for roughly $1/5$ of the total background.  It is an instrumental background and is a consequence of the mis-identification of a jet as an isolated high energy lepton and the mis-measurement of the \met\ which makes it fall into the selected sample.   If $b$-jet identification is required, a further mis-identification of one of the $udsg$-jets is also necessary in order for the event to survive all selection criteria.  All these mis-identification rates are taken from data control samples and applied successively in order to estimate the contribution of this background to the final selected sample.  There are large uncertainties associated with this background which arise from the relatively low statistics of surviving events in the control samples once the identification criteria are applied (ie. because the mis-identification rates are quite small) and from systematic uncertainties accounting for kinematic differences between the control samples and the signal sample.   For the \ttdil\ final state this background is negligible since two jets would need to be mis-identified as an isolated high energy lepton and the \met\ would have to be mis-measured.

The \Wpjets\ processes have production cross sections about three orders of magnitude greater than the \ttbar\ process and are an important background to the \ttljt\ and \ttdil\ final states.  When the $W$ decays leptonically, there is a genuine high energy lepton and genuine \met\ in the event.  For the $ljt$ final state, prior to making a $b$-jet identification requirement, these processes appear as physics background.  Accordingly their acceptance is estimated from Monte Carlo, however their normalization is taken from the data since there are large theoretical uncertainties associated with the predicted production cross-section.  These uncertainties arise because complete calculations of the $W+3\:\mrm{jets}$ and $W+4\:\mrm{jets}$ cross sections, including heavy flavor contributions, are unavailable and current estimates rely on a mixture of partial calculations at lower orders and parton shower Monte Carlo models to extrapolate to larger jet multiplicities.  When $b$-jet identification is required, this background is separated into two pieces: the $W+$~light flavor ($W+lf$) contributions, where the additional jets all originate from $uds$-quarks or gluons, and $W+$~heavy flavor ($W+hf$) contributions, where at least one of the additional jets originates from a $c$- or $b$-quark.  The $W+lf$ contributions is then an instrumental background since one of the $udsg$-jets would need to be mis-identified as a $b$-quark jet.  As mentioned above the mis-identification rate of the $b$-jet identification algorithm is taken from data control samples to estimate the contribution from this instrumental background, sometimes referred to as the ``mis-tag'' contribution.  The $W+hf$ contribution to the background uses Monte Carlo samples to estimate the fraction of surviving events which contain $c$- or $b$-quark jets and the efficiency of the $b$-jet identification algorithm to determined the number of events surviving the full selection requirements.  There are large uncertainties associated with the heavy flavor fractions determined from the Monte Carlo, which are cross checked in data control samples.  The \Wpjets\ processes are the dominant backgrounds to the \ttljt\ final states.   For the \ttdil\ final states these processes are an instrumental background since one of the jets must be mis-identified as a high energy isolated lepton in order to survive the full selection criteria.  This background is estimated from the data starting with the single-lepton+\met\ sample and then applying lepton mis-identification rates determined from data control samples to estimate the number of ``lepton fakes'' which populate the final dilepton sample.   

A related background, \Zpjets , has a production cross section roughly a factor of ten smaller than the \Wpjets\ background.  It contributes to the \ttljt\ sample only if the \zg\ decays to $e^{+}e^{-}$ or $\mu^{+}\mu^{-}$ and if one of the leptons escapes undetected to give rise to fake \met .  This latter effect is dominated by the limited geometric acceptance of the detector and is thus estimated using Monte Carlo simulation and found to be a negligible contribution to any of the samples.  For the $ljt$ sample the only significant contributions arises from \zg\ decays to $\tau^{+}\tau^{-}$ with the subsequent decay of one of the $\tau$ leptons to $\tau\ra e\nu_{\tau}\nu_{e}$ or 
$\tau\ra \mu\nu_{\tau}\nu_{\mu}$ in order to generate a lepton and \met\ to survive the selection criteria.  Once the relevant branching fractions are included, this turns out to be a small background and is estimated from the Monte Carlo.  If $b$-jet identification is required, the relevant efficiency and mis-identification rates are taken from data control samples and applied to the surviving Monte Carlo events.   For the $dil$ sample this is potentially an important background since \zg\ decays to $e^{+}e^{-}$ or $\mu^{+}\mu^{-}$ yield two high energy isolated leptons.  However to survive the full selection criteria the \met\ must be mis-measured.  To further supress this background analyses raise the \met\ requirement for events with a di-lepton mass consistent with \mz. This is an instrumental background whose residual contribution to the final sample is estimated from the data using clean samples of identified $Z\ra e^{+}e^{-}$ or $\mu^{+}\mu^{-}$ to parameterize the fake \met\ distribution and extrapolate to the \ttbar\ signal region.  There are relatively large uncertainties associated with this background arising associated with the extrapolation from the low jet multiplicity region which dominates the data control sample, to the $>=2$~jets region of the signal region.

The diboson processes $\ppbar\ra VV$ ($V=W,\:Z$, or $\gamma$) have production cross sections within a factor of two of the \ttbar\ cross section.  When the $W$ and/or $Z/\gamma$ bosons decay leptonically these processes make moderate ($\sim 10\%$) contributions to the $ljt$ samples and large contributions (up to $\sim 30\%$) to the $dil$ samples .  These are physics backgrounds whose estimates are taken from Monte Carlo and normalized to theoretical predictions calculated to next-to-leading order.  If $b$-jet identification is required, the relevant efficiency and mis-identification rates are taken from data control samples and applied to the Monte Carlo.

The production of a single top quark via the electroweak interaction, $\ppbar\ra tq$, has a production cross section about a factor of two smaller than the \ttbar\ cross section and makes a small contribution to the \ttljt\ final state.  It is a physics background and is estimated using Monte Carlo samples normalized to the theory predicted cross section.

% ======================================================================
\section{Simulation Samples}
\label{sec:sim}
% ======================================================================

A variety of Monte Carlo event generators are employed to estimate acceptances, efficiencies, and kinematics for \ttbar\ and background events.  A summary of which generators are used to model each physics process is given here.  In several cases a comparison of multiple generators is used to assess systematic uncertainties.  For the analyses described in this review, the Monte Carlo generators are always combined with simulation software that models the detector response in detail.  The resulting simulated samples are then treated just like the recorded \ppbar\ collision data, using the same reconstruction software and particle identification algorithms described in Sec.~\ref{sec:reco}.  For all samples the decay of tau leptons is handled using the \tauola\ package~\cite{tauola} and the decay of $c$- and $b$-quark hadrons is handled using the \evtgen\ package~\cite{evtgen}.  The parton shower and fragmentation processes required to simulate the evolution of a parent quark or gluon into a final state jet are modeled using dedicated sub-routines in either the \pythia\ \cite{pythia} or \herwig\ \cite{herwig} Monte Carlo software packages.  The parton shower is modeled using DGLAP evolution~\cite{dglap} while the fragmentation process is simulated using a string model in \pythia\ and a cluster model in \herwig.  Initial and final state QED radiation effects are included for all initial and final state charged particles.  

The CDF and \dzero\ collaborations each employ several Monte Carlo generators to model \ttbar\ events.  The \pythia\ and \herwig\ programs are general purpose event generators widely used throughout high energy physics.  They model a wide variety of interactions and processes using leading order calculations (LO).  Although the programs include only LO (e.g. $\ppbar\ra\ttbar$), the parton shower and QED radiation modeling can yield more complicated event topologies (e.g. $\ppbar\ra\ttbar +\mrm{jets}$ or $\ttbar + \gamma$).  However, the resulting cross sections and kinematics of these higher order final states usually carry large uncertainties since their modeling can be quite sensitive to arbitrary choices of control parameters offered by each generator.  Reasonable choices of these parameters are made by making detailed comparisons between the data and Monte Carlo samples in dedicated studies using data samples of inclusive jet and minimum bias events.   For most of all the \ttbar\ analyses described in this review, the higher order contributions affect the physics parameters of interest at a level that is negligible compared to the sensitivity of the present measurements.   Consequently the \pythia\ program is often used to model the \ttbar\ events at the Tevatron.   A more rigorous approach expected to yield improved cross section and kinematic modeling of more complicated final states is offered by the \madevent\ \cite{madevt} and \alpgen\ \cite{alpgen} programs.  These programs explicitly calculate the lowest order contributions to a given process (e.g. $\ppbar\ra\ttbar + 1\:\mrm{parton}$) and then use \pythia\ or \herwig\ to model the parton shower and fragmentation processes.  To model an inclusive sample, as observed in the data, requires the summing of several samples.  For example, to model a $\ttbar+ \geq 0~\mrm{jets}$ sample requires an \alpgen\ simulation of $\ttbar + 0j$ plus $\ttbar +1j$ plus $\ttbar + \geq 2j$ samples.   Since the parton shower can generate additional final state jets in each sample, care must be taken to avoid double-counting.  Common choices to remove this overlap include those described in~\cite{matching}.   Several of the analyses described in this review employ a series of \alpgen + \pythia\ or \madevent +\pythia\ samples to simulate the inclusive \ttbar\ signal samples.  
The \mcnlo\ \cite{mcnlo} program is a full next-to-leading order generator used to simulate \ttbar\ samples employed for systematic studies.

The \Wpjets\ and \Zpjets\ samples are generated using \alpgen\ with the parton shower and fragmentation modeling provided by \pythia.  Sometimes the \pythia\ generator is used by itself to model the fully inclusive sample.  In all cases the normalization is taken from the data using selection criteria expected to yield a sample of events dominated by these processes.

The diboson processes, $VV$ ($V=W,\:Z$, or $\gamma$), are modeled either using \alpgen + \pythia\ or \pythia\ by itself.  The samples are normalized to the theory predicted cross sections calculated to next-to-leading order.  

Electroweak processes which result in the production of a single top quark are generated using either \singletop\ \cite{singletop} or \madevent\ with the parton shower and fragmentation modeling provided by \pythia.  The samples are normalized using the next-to-leading order theory predicted cross section.  The s-channel and t-channel processes are generated independently.

Generic QCD processes are modeled using \pythia\ or \herwig\ or \alpgen + \pythia.   In general the predictions for the QCD background contributions to the final selected sample are taken from data control samples.  However, several parameters important for \ttbar\ analyses are taken from dedicated studies performed using multi-jet samples dominated by genuine QCD production processes.  These studies often require some inputs from these QCD Monte Carlo samples.  For example, the efficiency of the $b$-jet identification algorithms is determined using a di-jet sample and the ratio of the data determined efficiency to the Monte Carlo predicted efficiency is used to correct the Monte Carlo predicted efficiency in \ttbar\ samples.  Another example, di-jet samples are used to compare the heavy-flavor contribution between data and Monte Carlo in order to derive a correction for the Monte Carlo predicted ratio $W+hf$/$W+\mrm{jets}$.  The QCD samples used in these studies must be simulated in a manner consistent with that used to model the \ttbar\ and other background Monte Carlo samples so that the corrections derived can be used in the final selected samples with confidence.

% ======================================================================
\section{Sources of Systematic Uncertainty}
\label{sec:sys}
% ======================================================================

A variety of systematic uncertainties are evaluated for each of the analyses described in this review.  They fall into three general categories: uncertainties associated with the background modeling, uncertainties associated with the signal modeling, and uncertainties associated with the methodology employed to extract the physics parameter(s) of interest.  The sources of uncertainty affecting the signal and background modeling and the methods employed in evaluating them, are common across all the \ttbar\ analyses and are summarized here.

\subsection{Systematic Uncertainties Associated with Background Processes}
\label{sec:bdgsys}
The systematic uncertainties affecting the background modeling include instrumental effects related with the various mis-identification rates required to promote a background physics process into the signal region, theory related uncertainties such as variations of the factorization and renormalization scales, k-factors related to NLO corrections, and differences in parton shower and fragmentation modeling, as well as uncertainties in normalization arising from uncertainties in lepton, $b$-quark jet, etc. identification efficiencies.  Several sources of uncertainty affect multiple background processes and their effects are accounted for in the analyses in a correlated manner.

The mis-identification rate for hadrons faking an electron or muon signature and for $udsg$-jets faking a $b$-jet are derived from data control samples.  The mis-identification rates are parameterized by an empirically determined functional form, which includes the dependence of the rate on relevant kinematic variables (e.g. $E_{T}^{\mrm{jet}}$ or $\left|\eta\right|$).  The validity of the parameterization is tested using statistically independent data control samples.  Differences between the predicted rate, as determined using the parameterization, and the observed rate are used to assign systematic uncertainties on the mis-identification rates themselves.   These uncertainties are then propagated to the number of predicted background events for those background processes for which the relevant mis-identification is required in order for that process to mimic a \ttbar\ signal event.   Since these mis-identification rates depend on the kinematics of the events, they also introduce a systematic uncertainty on the shape of the resulting kinematic distributions of background events surviving the full event selection criteria.  These ``shape'' systematics are usually evaluated by re-weighting the relevant kinematic distributions using the mis-identification parameterizations varied within their uncertainties.

The estimate for the \Zpjets\ background in the $dil$ channel is normalized using $Z\ra\ell\ell$ events identified in the data.  These events predominantly have small \met\ and low jet multiplicity.  The observed \met\ distribution is parameterized and extrapolated into the \ttbar\ signal region.   The uncertainties on this extrapolation are dominated by the limited number of events with large \met\ and the degree to which additional jets in the event affect this estimate.  These uncertainties are added in quadrature and propagated to the relevant background estimate.  Since the data events are selected requiring $m_{\ell\ell}$ be consistent with \mz , an additional correction is derived from MC samples to estimate the number of \zg\ background events outside the \mz\ region.  Comparisons of different generators are used to assess a systematic uncertainty.

The estimate for QCD background processes is data driven, taken from event samples collected using inclusive jet triggers.  The QCD processes are a significant source of background in the $ljt$ and $had$ channels.  As discussed in Sec.~\ref{sec:bgd} they are instrumental backgrounds since they require the mis-identification of multiple objects in order to survive the relevant selection criteria.  The rejection factors for each source of mis-identification are usually assessed individually and their product is taken to estimate the QCD background yield.  This methodology assumes the rejection factors are uncorrelated with each other.  Studies are performed to test the validity of these assumptions and assign systematics uncertainties.
%
%%% SHOULD I ADD MORE HEREÉ A common strategy is to apply nearly the full %%%set of %selection criteria except to loosen or reverse a few of the %%%requirements. %etc.  yields similar kinematics to final sample.  can then %%%tighten some of %the requirements and compare observed rejection to %%%that predicted using %alternative (higher statistics) control samples.  %%%%Differences assigned as %syst uncert.

When generating the \Wpjets\ samples there is an ambiguity associated with the choice of factorization and renormalization scales, represented in the literature as $\mu_F$, $\mu_R$, and/or $Q^{2}$, at which to evaluate the relevant matrix element.  Some typical choices for \Wpjets\ processes are $Q^{2}=p_{T,W}^{2}$, $(m_{W}^{2}+p_{T,W}^{2})$, or $(m_{W}^{2}+\sum_{i} p_{T,i}^{2})$, where the sum runs over all the outgoing partons at a given vertex.  The nominal background estimates are estimated using the last choice.   A full set of additional Monte Carlo samples are generated using other alternative choices of $Q^{2}$ and the background predictions and kinematic distributions are re-evaluated.   Differences are assigned as a systematic uncertainty affecting the \Wpjets\ background processes.  It should be noted that this ambiguity affects all the generated samples.  However in the other samples it produces negligibly small systematic uncertainties.

All the background processes are generated with at least two different Monte Carlo generators.   For those Monte Carlo samples which reasonably model the observed data distributions, the background yields and kinematic distributions are evaluated in full.   Differences between the nominal Monte Carlo sample and any viable alternative sample are assigned as systematic uncertainties.   This systematic uncertainty primarily accounts for differences in parton shower and fragmentation modeling.

Most of the Monte Carlo samples employed in the analyses described here are based on leading order theory calculations.  The inclusion of higher order corrections affect the predicted production cross section for the various background processes.   These effects are accounted for in the analysis either by normalizing the Monte Carlo yields to next-to-leading order calculations, or by normalizing to the data itself.   Some residual higher-order effects still remain.   Most importantly is the $W+hf$ fraction of the \Wpjets\ background.  The fraction is taken from the \alpgen\ generator, but scaled to the heavy flavor fraction measured in the $W+1$~jet sample.  There is a large uncertainty associated with extrapolating that measured fraction to the $W+\geq 2$~jets regions used in the \ttbar\ analyses.   The uncertainty is evaluated by comparing the measured heavy flavor fraction derived using different methodologies as well as comparing the result derived using the $W+1$~jet sample to that derived using the $W+2$~jet for the $ljt$ channel.  The differences are assigned as systematic uncertainties and propagated to the predicted number of $W+hf$ events contributing to the final background.

The electron and muon identification efficiencies, the relevant trigger efficiencies, and the $b$-quark jet identification efficiencies all have associated statistical and systematic uncertainties.  These uncertainties are propagated through each of the affected background processes by varying each in turn within their uncertainties.  These variations result in uncertainties on the number of contributing events for the affected background processes.   For those efficiencies that have a kinematic dependence, their effect on resulting kinematic distributions is also evaluated by re-weighting the affected distributions using parameterizations that bracket the uncertainties of the relevant kinematic dependencies.

The uncertainty on the jet energy scale is propagated by varying the parameterized jet energy corrections within its uncertainty and re-evaluating the predicted number of background events and the resulting kinematic distributions. This uncertainty affects all the background processes.

The total uncertainty on the predicted background events is the quadrature sum of these various sources.  When a particular source of uncertainty affects multiple background sources, the resulting correlation is accounted for by adding the resulting uncertainties linearly and then adding the resultant sub-total quadratically with the rest of the systematic uncertainties.

\subsection{Systematic Uncertainties Associated with \ttbar\ Modeling}
\label{sec:sigsys}
There are systematic uncertainties affecting the modeling of \ttbar\ production and decay: variations of initial and final state radiation, variations in the parton distribution functions, variations in the parton shower and fragmentation modeling, variations in the modeling of color reconnection effects in final state interactions between the $t$ and $\overline{t}$ decay products, and uncertainties associated with the trigger, lepton identification, and $b$-jet identification efficiencies.  Some of these sources of uncertainty also affect background processes and their effects are accounted for in the analyses in a correlated manner.

Uncertainties in the modeling of initial state radiation (ISR) and final state radiation (FSR) are evaluated by generating \ttbar\ Monte Carlo samples with their ISR and/or FSR parameters varied within their uncertainties as determined using $\ppbar\ra \zg\ra\mu^{+}\mu^{-}$ samples.  
Like \ttbar\ production at the Tevatron, this process is dominated by $q\overline{q}$ annihilation.  The $p_T^{\mu^{+}\mu^{-}}$ and the $N_{jet}$ distributions are used to constrain the parameters in the Monte Carlo affecting the ISR modeling.   The constraints are derived in bins of $m_{\mu\mu}^2$ and extrapolated to $m_{\mu\mu}^2 = 4m_{t}^2$.   Since in the Monte Carlo generators the ISR and FSR processes are both modeled using DGLAP evolution~\cite{dglap}, the same variations are used to assign systematics for FSR.  Samples are produced varying ISR only, FSR only, and ISR/FSR simultaneously.   The sample resulting in the largest difference relative to the nominal sample is used to assign the associated systematic uncertainty.

The parton distribution functions result from multi-dimensional fits to dozens of measurements made by a variety of different experiments.  The CTEQ~\cite{cteq} and MRST~\cite{mrst} collaborations provide eigenvectors representing 
$\pm 1\sigma$ variations of $20$ uncorrelated parameters affecting the resulting PDFs.  The difference associated with each variation is evaluated by reweighing the \ttbar\ events in the nominal Monte Carlo sample using each eigenvector in turn.  The $+ 1\sigma$ and $- 1\sigma$ variations are quadratically summed separately.   In addition, samples generated with different $\alpha_s$ and different PDF fits are also generated and differences included in the associated PDF systematic uncertainty.  

The signal \ttbar\ samples are generated with several different Monte Carlo generators.   For those Monte Carlo samples which reasonably model the observed data distributions, the \ttbar\ yields and kinematic distributions are evaluated in full.   Differences between the Monte Carlo samples are assigned as a systematic uncertainty that primarily accounts for differences in parton shower and fragmentation modeling but also includes possible effects from differences in the modeling of the
\ttbar\ $p_T$ spectrum, \ttbar\ spin correlations, and final state radiation. 

Since top quarks and $W$ bosons decay quickly relative to the timescale associated with the parton shower and fragmentation processes (i.e. $1/\Gamma_{t}$, $1/\Gamma_{W}\: <<\: 1/\Lambda_{QCD}$) it is possible that the products from the {\it{different}} top-quark decays could interact with each other via color reconnections.  By default, all Monte Carlo generators ignore such reconnections since they are non-existent or irrelevant for most processes.   These color reconnection effects were first investigated at LEP2 (e.g.~\cite{LEP2CR}).  Since Tevatron's \ppbar\ initial state carries the color charge, the situation and its modeling is significantly more complicated that at LEP, which used an $e^+ e^-$ initial state.  Recently, Monte Carlo models providing an adequate description of Tevatron data and including color reconnection effects have become available.   The associated systematic uncertainty is evaluated by generating \ttbar\ samples with these various color reconnection models enabled and comparing to the nominal Monte Carlo sample.   Difference are assigned as systematic uncertainties.

The uncertainties associated with the electron and muon identification efficiencies, the relevant trigger efficiencies, the $b$-quark jet identification efficiencies, and the jet energy scale uncertainties are all evaluated in the manner described in the sub-section above.  These uncertainties also affect some or all of the background processes.  

The total uncertainty on the predicted \ttbar\ yields and resulting kinematic shapes is the quadrature sum of these various sources.  Correlations with the background processes are included. 

% ======================================================================
\section{Measurements of Top-Quark Properties}
\label{sec:topprop}
% ======================================================================

% ======================================================================
\subsection{Production Properties}
\label{sec:production}
% ======================================================================

% ======================================================================
\subsubsection{\ttbar\ Cross Section}
% ======================================================================

At hadron colliders top quarks are predominantly produced in \ttbar\ pairs via the strong interaction.  Measuring the \ttbar\ production cross section is interesting for several reasons.  First it allows a
precise test of the predictions from perturbative QCD. Computations of the $\ppbar \to \ttbar$ total rate are available at NLO \cite{Nadolsky:2008zw}. Several approximate NNLO computations, where the dominant (next to) next to leading logarithm terms are resummed to reduce sensitivity to the renormalization and factorization scales, are also available~\cite{Ahrens:2010zv,Langenfeld:2009wd,Cacciari:2008zb,Kidonakis:2008mu}.
For a top-quark mass of 172~\gevcc , the latest calculations predict a cross section of about $\sigma(\ppbar\ra\ttbar )=7.6$~pb with a relative precision of about $\pm 8$~\%.  Comparing the measurements with these predictions is important since various BSM theories postulate new particles that couple preferentially to the top quark and thus predict a higher \ttbar\ production rate. A typical example is the production of new resonant particles that decay into \ttbar\ \cite{Frederix:2007gi}.

It is also particularly important to measure the production cross section in different top-quark decay channels since new physics contributions can affect the various \ttbar\ final states, differently.  Examples of such models are Two Higgs Doublet Models among which are SUSY models~\cite{Chung:2005a,Grossman:1994jb} predicting the existence of charged Higgs bosons, $H^{\pm}$.
If such a charged Higgs boson is light enough ($m_{H^+}<m_{t} - m_b$), the decay $t \to H^+ b$ can compete with the SM decay $t \to W b$; the measured \ttbar\ production rate will then differ among the various final states due to contributions from charged Higgs decays.

To make precise measurements of the \ttbar\ production cross section requires a good understanding of the reconstruction and identification efficiencies as well as a careful evaluation of the background processes that mimic the \ttbar\ signal.   Because of this, the cross section analysis also then serves as a foundation on which most the rest of the \ttbar\ analyses are built. 
The \ttbar\ cross section measurement is performed by evaluating the following formula:
\begin{equation}
  \sigma_{\ttbar} = \frac{N_{o}-N_{b}}
                         {\epsilon_{\ttbar} {\cal B} \int {\cal L} dt }
\label{eq:xsection}
\end{equation}
where $N_{o}$ is the observed number of events after selection, $N_{b}$ is the estimated mean number of background events,
$\epsilon_{\ttbar}$ is the signal efficiency evaluated using \ttbar\ Monte Carlo, ${\cal B}$ is the relevant final-state-dependent \ttbar\ branching fraction,  and $\int {\cal L} dt $ is the integrated luminosity for the particular set of triggers used for the measurement. It is worth noting that since $\epsilon_{\ttbar}$ is increasing with the top-quark mass the measured cross section is quoted at a given \mt .
The numerator in Eq.~\ref{eq:xsection} can be evaluated either using event counting after applying the final selection criteria or by fitting a discriminant variable that separates signal and background to estimate the relative contributions of each to the selected event sample.
The use of $b$-tagging is helpful to discriminate \ttbar\ signal from background. If $b$-tagging is employed, $N_{o}-N_{b}$ is usually evaluated by event counting since events with $b$-jets have higher signal purity.

% ======================================================================
\subsubsection*{\ttbar\ Cross Section in the Lepton+jets Channel}
% ======================================================================
In the $ljt$ channel one of the $W$ bosons decays to yield a high energy electron or muon plus a neutrino, while the other decays into two quarks. The final state thus contains at least four jets in the event, two of which are $b$-jets, plus a high energy electron or muon and large \met .
This final state has the advantage of a reasonable branching fraction while achieving a good signal-to-background ratio.  The event selection begins as described in Sec.~\ref{sec:evtsel}.  For the cross section analyses additional selection criteria are often applied.  For example, the total transverse energy of the event, $H_{T} = \sum_{l} E_{T}^{\ell} + \sum_{\mrm{jet}} E_{T}^{\mrm{jet}} + \met $, where the sums are over all high energy leptons and jets, respectively, is required to be above some minimum.  At this point, further enrichment of the selected sample in \ttbar\ events is achieved by applying either a multivariate discriminant using topological information or using $b$-jet identification.  The topological information in the multivariate discriminate usually excludes any information explicitly related to $b$-jet identification and thus yields a cross section measurement that makes no assumption about the flavor of the quark in the $t\ra Wq$ decay.  Experimentally the final event samples are split according to lepton flavor ($e$ or $\mu$), jet multiplicity (3 or $\ge$ 4 jets) and (if used) number of identified $b$-jets (0, 1 or $\ge$ 2) and treated separately. Events with $<3$ jets are used as background dominated control samples.

The main background processes that can mimic a \ttbar\ event were discussed in Sec.~\ref{sec:bgd}.  The most important contributions for the $ljt$ final state are, in decreasing order of importance, \Wpjets\ production, QCD multi-jet processes, \Zpjets\, diboson processes and electroweak single top-quark production.  The \Zpjets\, diboson, and single top backgrounds are evaluated using MC normalized to (N)NLO  theoretical cross sections.  As no reliable estimation of QCD multi-jet processes can be obtained by MC, this background is estimated using data.
No reliable NLO cross section calculation exists to estimate the \Wpjets\ cross section at the large jet multiplicities relevant for \ttbar\ analyses.  Therefore the normalization is derived from data and the MC used to estimate the \Wpjets\ kinematic distributions.

The QCD background is estimated from data. For example, \dzero\ uses a matrix method~\cite{Abazov:2006ka} that consists of solving a 2D system of linear equations:
\begin{eqnarray}
  N_{loose} & = & N_{QCD} + N_{other} \nonumber \\
  N_{tight} & = & \epsilon_{QCD} N_{QCD} + \epsilon_{other} N_{other}
\end{eqnarray}
where $N_{loose}$ is the number of events in a data sample surviving loose lepton selection criteria, while $N_{tight}$ is the number of events in a subsample surviving tight lepton isolation requirements. $N_{QCD}$ is the number of QCD events in the loose sample and $N_{other}$ is the number of events that do not come from QCD events (dominated by \Wpjets\ and \ttbar\ events). The efficiency for the loose lepton from a \Wboson\ decay to survive the tight lepton selection criteria is $\epsilon_{other}$, which is estimated using \Wpjets\ and \ttbar\ MC corrected for any data-MC difference. Similarly, $\epsilon_{QCD}$ is the expected fraction of QCD events in the loose lepton sample that are expected to survive the tight lepton selection criteria. It is measured in a data sample with low \met\ which is dominated by falsely isolated leptons after subtracting real isolated lepton contributions.  Solving this linear system yields an estimate for $N_{QCD}$.
%
%%%\begin{equation}
%%%  N_{QCD} = \frac{\epsilon_{othr} N_{loose} - N_{tight}}
%%%                 {\epsilon_{othr} - \epsilon_{QCD}}.
%%%\end{equation}
%
CDF uses a slightly different method to obtain an estimate of the QCD background contributions.  A binned likelihood fit to the shape of the \met\ distribution is used to determine the relative contributions of QCD and \Wpjets\ processes. The \met\ distribution for \Wpjets\ processes is taken from the \alpgen\ MC, while the distribution for QCD processes is formed from data events that fail at least two lepton identification criteria. The fraction of QCD events in the high \met\ region is then extrapolated from a fit to the low \met\ region.

The total amount of \Wpjets\ contribution is normalized to the number of events in data minus the estimated number of \ttbar, QCD, and electroweak events.  Among the \Wpjets\ events the contribution from $W+hf$ is found to be underestimated in the MC simulation. An additional scaling determined from data is applied to the $W+hf$ MC samples.
The $W+hf$ content is measured in data events with $\leq 2$ jets using a neural network (NN) trained to separate $W+lf$ from $W+hf$. By counting the number of events in the part of the NN distribution dominated by $W+hf$ or performing a fit to the full NN distribution using templates for $udsg$-jets, $c$-jets, and $b$-jets, the ratio $W+hf$/$W+lf$ is determined after unfolding the relevant identification efficiencies.  This can be compared with the ratio predicted in MC and a correction applied to the MC if necessary.   In practice the CDF and \dzero\ experiments find that the data-to-MC ratio for this heavy flavor fraction varies between $1.0-1.6$ depending on the method employed and the event samples used.  Their analyses typically use a value of around $1.3$ with systematic uncertainties assigned that cover the full spread.  This correction factor is then applied to all the jet multiplicity bins. \dzero\ is using an additional scale factor for the $W+c$ samples based on next leading order computations.

%Measurements in the $ljt$ channel are performed using either purely %topological information (ie. no $b$-jet identification is employed) or %requiring that at least 1 jet in the event is identified as a $b$-jet.  
%The typical signal over background ratio requiring four jets in the %lepton-plus-jets channel having applied the above selection 
%is 2/3 using a topological section and 4/1 requiring that exactly 1 jet %be identified as a $b$-quark jets.

The topological cross section measurements use the output of a multivariate discriminant, a neural network (NN) for CDF or 
a boosted decision tree (BDT) for \dzero , to further discriminate \ttbar\ signal from the background.  These discriminants exploit differences in the kinematic properties of signal and background.
Indeed because of the large top-quark mass, \ttbar\ events are on average more energetic, more central, and more isotropic than $W$+jets and QCD events. Hence several kinematic distributions that are well modeled by MC simulation and that have a good separation power between signal and background are used as input to the multivariate discriminant such as $H_T$, the aplanarity, the sphericity (defined below), the lepton-\met -jets invariant mass, the minimum angle between two jets, the minimum invariant mass of any two jets, etc.
Multivariate output templates for the \ttbar\ signal and for the QCD, \Wpjets\ and diboson backgrounds are then formed and used 
in a likelihood fit of the observed NN or BDT data distribution to determine the fraction of \ttbar\ events in the selected sample.  This fraction can then be used to determine the \ttbar\ cross section.
\dzero\ uses a nuisance parameter method~\cite{Sinervo:2003wm} to simultaneously extract the result and constrain the corresponding systematic uncertainties. 

\begin{figure}
\centering
  \epsfxsize=6cm\epsfbox{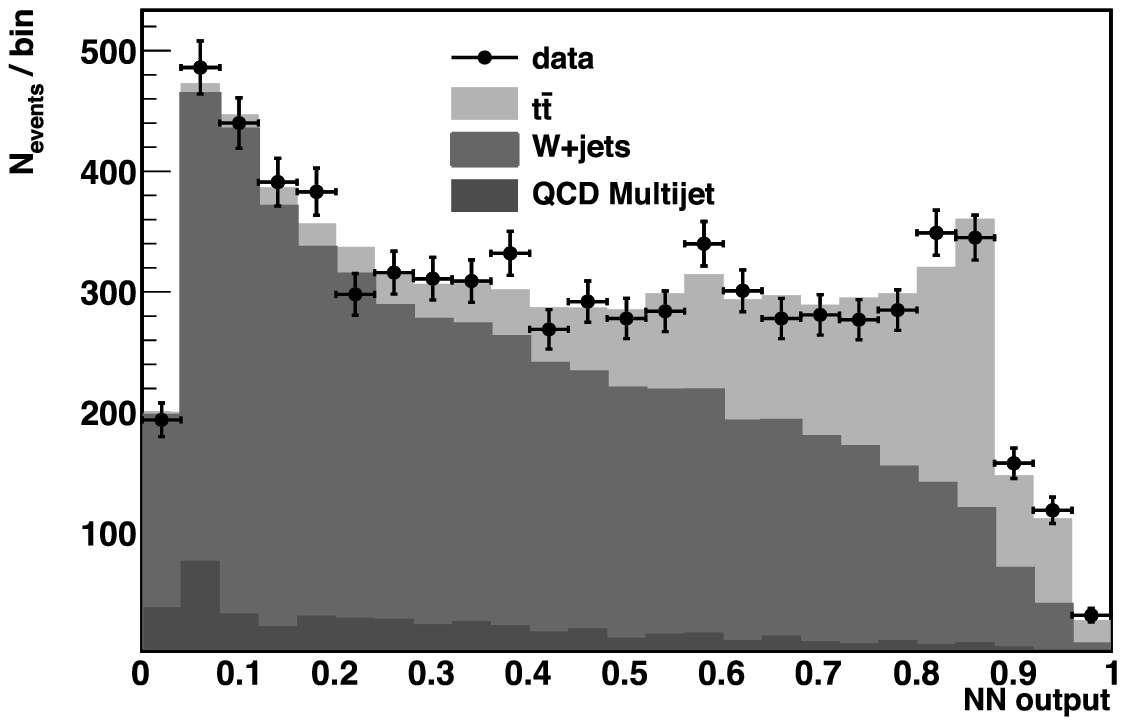}\hspace{0.5cm}
  \epsfxsize=5.6cm\epsfbox{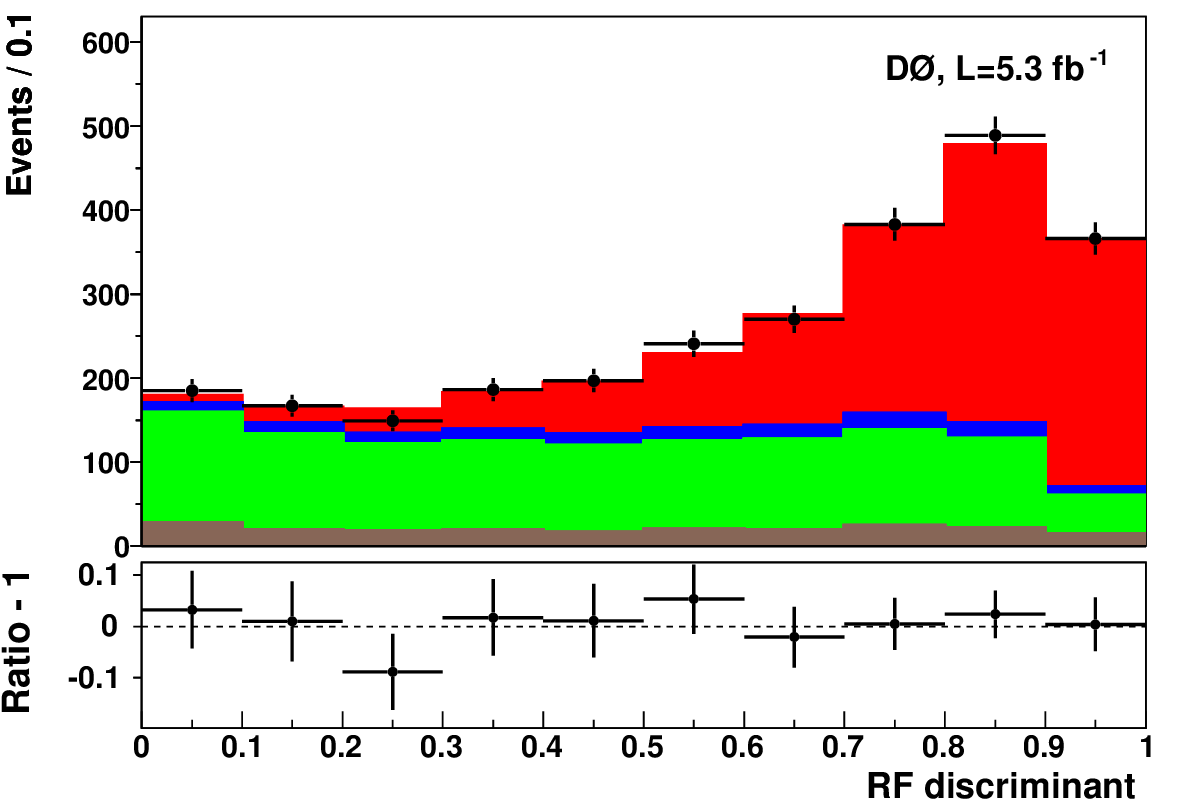}\hspace{0.5cm}
\caption{Output of the NN used by CDF (left) and the BDT used by \dzero\ 
  (right) to determine the \ttbar\ cross section in the $ljt$ channel 
  using topological information~\cite{Aaltonen:2010ic, Abazov:2010zz}.
\label{fig:xsecljetstopo}}
\end{figure}

Utilizing only topological information and a data set of 4.6~\fb\ CDF measures~\cite{Aaltonen:2010ic} 
$\sigma_{\ttbar} = 7.71 \pm 0.37 \ (\stat) \pm 0.36 \ (\syst) \pm 0.45 \ (\lumi)$~pb
while \dzero\ uses 5.3~\fb\ to measure~\cite{Abazov:2010zz} 
$\sigma_{\ttbar} = 7.68^{+0.71}_{-0.64} \ (\statsyst)$~pb, both for a 172.5~\gevcc\ top-quark mass.
These measurements are systematics limited.  Apart from the uncertainty on the luminosity, the dominant sources of systematic uncertainty are the uncertainties on the jet energy scale, jet energy resolution, and signal modeling. The precision of these measurements is around 9~\%. 
Figure~\ref{fig:xsecljetstopo} shows the distributions for the CDF NN and \dzero\ BDT discriminants used in these topological analyses.

Aside from using pure topological information, the data sample can alternatively be enriched  in \ttbar\ events by identifying $b$-jets using one of the methods described in Sec.~\ref{sec:bjet}. In that case, apart from the $W+hf$ background estimated as described above, the contribution from $W+lf$ events when a $udsg$-jet is mis-identified as a $b$-jet has to be determined. Since this mis-identification rate is difficult to model using MC, it is determined using a data sample dominated by QCD events and parametrized as a function of the relevant kinematic variables (e.g. the jet $E_T$ and $\eta$). This probability is then applied to MC $W+lf$ samples to estimate the number of mis-identified $W+lf$ events contributing to the final $b$-tagged sample.  A maximum likelihood fit is performed on the data, using the predicted background as an input, to extract the \ttbar\ production cross section.  Often times the likelihood includes nuisance parameters that allow, for example, the background normalization and $b$-jet identification efficiency, to vary within their uncertainties.  The likelihood is constructed to account for the correlations between the background and signal acceptances for each of the nuisance parameters.

\begin{figure}
\centering
  \epsfxsize=6cm\epsfbox{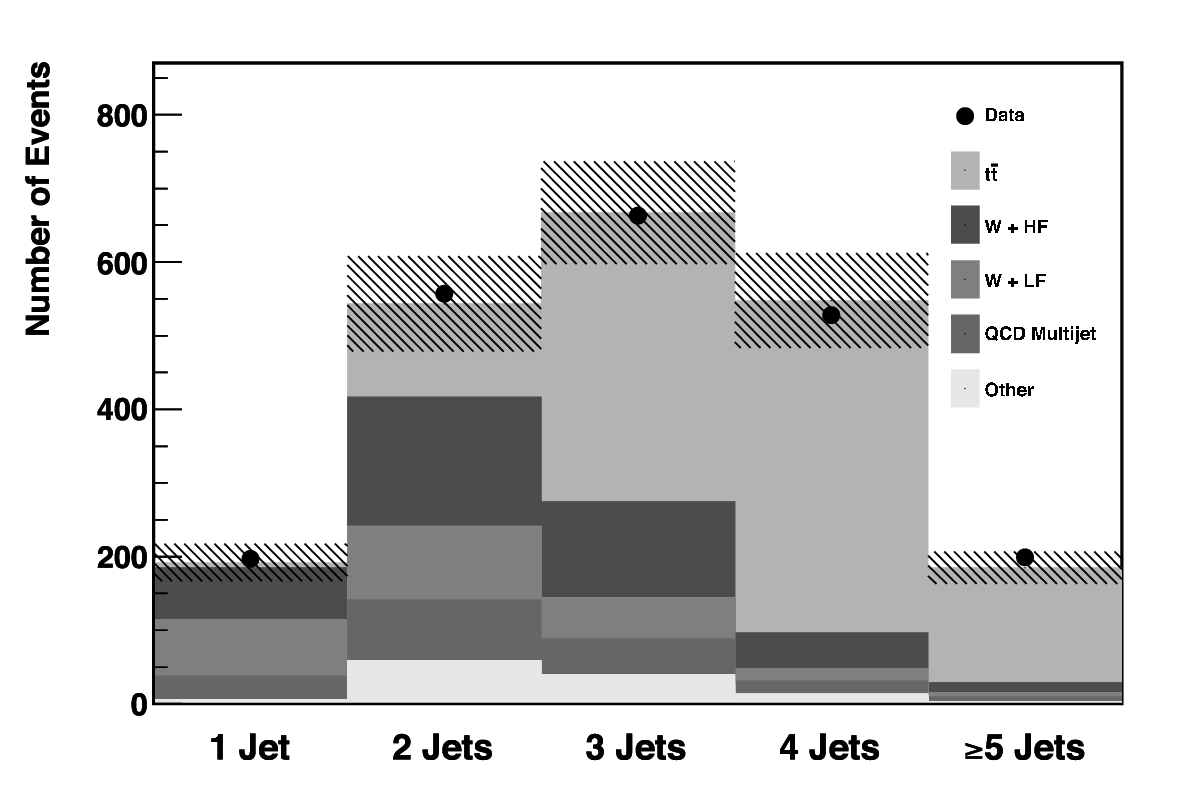}\hspace{0.5cm}
  \epsfxsize=6.3cm\epsfbox{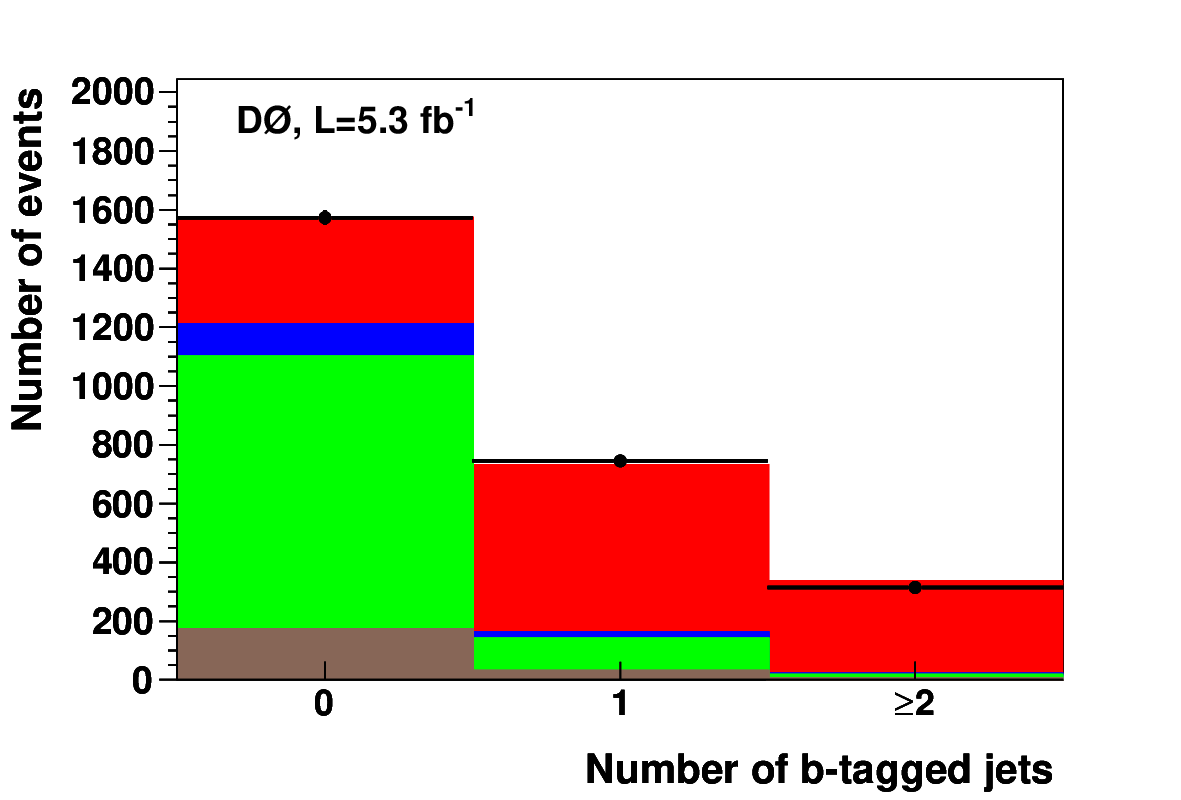}\hspace{0.5cm}
\caption{The predicted and observed number of events as a function of 
   jet multiplicity for CDF (left) and as a function of the number of identified $b$-jets for \dzero\ (right) both from a \ttbar\ cross 
   section analysis in the $ljt$ channel using $b$-jet identification~\cite{Aaltonen:2010ic, Abazov:2010zz}.
\label{fig:xsecljetsbtag}}
\end{figure}

Utilizing $b$-jet identification algorithms that exploit the long lifetime of $b$-hadrons CDF uses 4.3~\fb\ of data to measure~\cite{Aaltonen:2010ic} $\sigma_{\ttbar} = 7.22 \pm 0.35 \ (\stat) \ \pm 0.56 \ (\syst) \pm 0.44 \ (\lumi)$~pb
while \dzero\ uses 5.3~\fb\ to measure~\cite{Abazov:2010zz}
$\sigma_{\ttbar} = 8.13^{+1.02}_{-0.90} \ (\statsyst)$~pb, both for a 172.5~\gevcc\ top-quark mass.
The dominant sources of systematic uncertainty come from the luminosity determination, the estimation of the $W+hf$ background and the uncertainty on the efficiency for identifying $b$-jets. \dzero\ combines the topological and $b$-jet identification methodologies to obtain $\sigma_{\ttbar} = 7.78^{+0.48}_{-0.67}\ (\statsyst)$~pb~\cite{Abazov:2010zz}.
Using $b$-jet identification algorithms that identify leptons from semi-leptonic $b$-hadron decays, CDF measures cross sections consistent with those above but with larger uncertainties due to the lower efficiency and purity achieved by these soft-lepton algorithms~\cite{CDFSLTm, CDFSLTe}.
Figure~\ref{fig:xsecljetsbtag} shows distributions from the CDF and \dzero\ analyses that use lifetime information to identify $b$-quark jets.

The largest systematic uncertainty for both the topological and $b$-tag based measurements is the uncertainty on the luminosity.
It originates from the uncertainty on the luminosity counter acceptance and from the uncertainty on the knowledge of the \ppbar\ inelastic cross section and totals about $6\%$.  It can be effectively removed by measuring the \ttbar\ cross section relative to a cross section whose theoretical predictions are known to much better than $6\%$.  A good choice at the Tevatron is to normalize to the inclusive $\zg \to \ell \ell$
cross section.  These events can be identified using the same trigger as the \ttbar\ sample, offer reasonably large statistics (1-2 orders of magnitude more than the \ttbar\ sample), and can be reconstructed with high purity ($>95\%$).    The expression for determining the \ttbar\ production cross section is then
\begin{equation}
  \sigma_{\ttbar} = \left( \frac{\sigma_{\ttbar}}{\sigma_{\zg \to 
  \ell \ell}} \right)_{measured} (\sigma_{\zg \to \ell \ell})_{theoretical} .
\label{eq:ratiozttbar}
\end{equation}
Hence the luminosity uncertainty is traded with the theoretical uncertainty on the \zg\ cross section.
The inclusive $\zg \to \ell \ell$ cross section is measured using consistent trigger requirements and lepton identification with 
the \ttbar\ cross section. The $\zg \to \ell \ell$ cross section times branching ratio measured by CDF~\cite{Aaltonen:2010ic}
in the dilepton invariant mass range of 66\--116~\gevcc\ is 
$\sigma_{\zg \to \ell \ell} = 247.8 \pm 0.8 \ (\stat) \pm 4.4 \ (\syst) \pm 14.6 \ (\lumi)$~pb.
To extract $\sigma_{\ttbar}$ from (\ref{eq:ratiozttbar}) a theoretical cross section of $(\sigma_{\zg \to \ell \ell})_{theoretical}= 251.3 \pm 5.0$~pb~\cite{Abulencia:2005ix} is used and all
relevant systematic uncertainty correlations between the $\ttbar$ and \zg\ cross sections are included~\cite{Aaltonen:2010ic} to yield
$\sigma_{\ttbar} = 7.82 \pm 0.38 \ (\stat) \pm 0.37 \ (\syst) \pm 0.15 \ (\theo)$~pb for the CDF topological selection and
$\sigma_{\ttbar} = 7.32 \pm 0.36 \ (\stat) \pm 0.59 \ (\syst) \pm 0.14 \ (\theo)$~pb for the CDF $b$-tagging selection, both used a $4.6\:\mrm{fb}^{-1}$ data set and assumed a 172.5~\gevcc\ top-quark mass.
These two measurements are combined to give~\cite{Aaltonen:2010ic}: 
$\sigma_{\ttbar} = 7.70 \pm 0.52$~pb for a 172.5~\gevcc\ top-quark mass. The precision of the combined result is now 6.8\%, which is more precise than the latest theory predictions~\cite{Kidonakis:2008mu,Langenfeld:2009wd,Ahrens:2010zv}.

% ======================================================================
\subsubsection*{\ttbar\ Cross Section in the Dilepton Channel}
% ======================================================================
In the $dil$ channel both $W$ bosons decay to leptons. The event then contains two high $E_T$ isolated leptons, large \met , and two $b$-jets.  This channel is often sub-divided further according to the explicit lepton identification criteria employed.  When both leptons are reconstructed as an electron or muon the sample is dominated by $t \ra W b\ra e\nu b$ and $t \ra Wb \ra\mu\nu b$ decays ($\ell\ell$ channel).  The acceptance can be significantly increased, particularly for $Wb \ra\tau\nu\ra h\nu\nu$ decays, by applying very loose criteria for one of the lepton legs ($\ell \mrm{track}$). Measurements have also been performed in the final state with a $\tau$ lepton decaying into hadrons explicitly identified from the decay of one $W$ boson and an accompanying electron or muon from the other \Wboson\ decay ($\ell \tau$ channel).  In general the $dil$ channel has the advantage of a good signal to background ratio even without using $b$-tagging but suffers from a smaller branching fraction than the $ljt$ or $had$ channels.

In the $dil$ channel, the main source of background comes from the production of electroweak bosons that decay to charged leptons.
For the $\ell\ell$ channel, it arises from Drell-Yan processes, $\zg \to \ell^+ \ell^-$, and diboson processes when
the bosons decays lead to at least two leptons in the final state. These backgrounds are reduced by requiring $\geq 2$ jets and large \met .
In the $\ell\tau$ and $\ell$track channels, the main background comes from processes with jets mis-identified as an electron, $\tau$, or isolated track, or from muons from semi-leptonic $b$-quark decay. This occurs mainly in \Wpjets\ or QCD events.  The diboson backgrounds are evaluated using MC normalized to (N)NLO theory cross sections.  The contribution of \Zpjets\ backgrounds to $dil$ events in which the leptons are of different flavors (e.g. $e\mu$ events) is also evaluated using MC normalized to NLO theory cross sections.  The contribution of \Zpjets\ to $dil$ events in which the two leptons are the same flavor ($ee$ or $\mu\mu$), \Wpjets\ normalization, and the QCD backgrounds are evaluated using data control samples.  The jet-to-lepton fake rates are computed in a background dominated sample orthogonal to the signal sample
(QCD di-jets, $\gamma+jets$, same sign dilepton samples).  By applying $b$-jet identification a very pure \ttbar\ sample can be identified.  For these analyses the background after $b$-tagging is estimated in the same manner employed for the $ljt$ channel.

As it benefits from a favorable signal to background ratio, the \ttbar\ dilepton selection, at least in the case of two well identified leptons,
relies on a small number of simple requirements.  
Selecting \ttbar\ events decaying to dileptons requires first an inclusive lepton trigger as described in Sec.~\ref{sec:reco}, two high $E_T$ isolated electrons or muons 
(typically with $E_{T} > 20$~GeV) or one isolated electron or muon and one isolated high \pt\ track and generally
at least two jets with $E_{T} > 20$~GeV. In the $\ell \tau$ channel, a set of neural networks~\cite{Galea:2006mi} developed to separate hadronically decaying $\tau$ leptons and jets is used to identify a $\tau$ lepton in the event. 
Large \met\ is further required (typically $\met > 25$~to~$35$~GeV). The \met\ threshold is raised if the dilepton invariant mass is in the range of \mz .
Other topological cuts like a cut on $H_T$ sometimes replace the \met\ requirement particularly for the $e\mu$ channel since it suffers from the Drell-Yan process only through double leptonic $\tau$ decays of $\zg \to \tau^+ \tau^-$.
In the $\ell\tau$ and $\ell$track channels, to reduce the contribution from events with fake \met\ due to misconstruction, requirements are made on the lepton-\met\ and jet-\met\ opening angle.
In the $\ell\tau$ and $\ell$track channels,  additional requirements are made to further suppress the QCD background and $b$-jet identification is often employed.  In the $\ell\ell$ channel, the signal-to-background ratio after all selection criteria is about 3:1 prior to and $15:1$ after requiring at least one $b$-jet in the event.  The analyses typically use simple event counting and minimize a likelihood function that includes nuisance parameters constraining background contributions and other systematic uncertainties to determine $\sigma_{\ttbar}$.

In the $\ell$track channel, using 1.1~\fb\ of data and for $\mt = 175$~\gevcc\, CDF measures~\cite{Aaltonen:2009ve} 
$\sigmattbar = 8.3 \pm 1.3 \ (\stat) \pm 0.8 \ (\syst) \pm 0.5 \ (\lumi)$~pb without $b$-tagging and 
$\sigmattbar = 10.5^{+1.4}_{-1.3} \ (\stat) ^{+0.8}_{-0.7} \ (\syst) \pm 0.6 \ (\lumi)$~pb using $b$-tagging.
Combining these two measurements gives
$\sigmattbar = 9.6 \pm 1.2 \ (\stat) ^{+0.6}_{-0.5} \ (\syst) \pm 0.6 \ (\lumi)$~pb.
In this channel the main systematic uncertainties come from the uncertainty in the background determination for the result
without $b$-tagging and from the uncertainty on the $b$-jet identification uncertainty in the second case. 

\begin{figure}
\centering
\epsfxsize=6cm\epsfbox{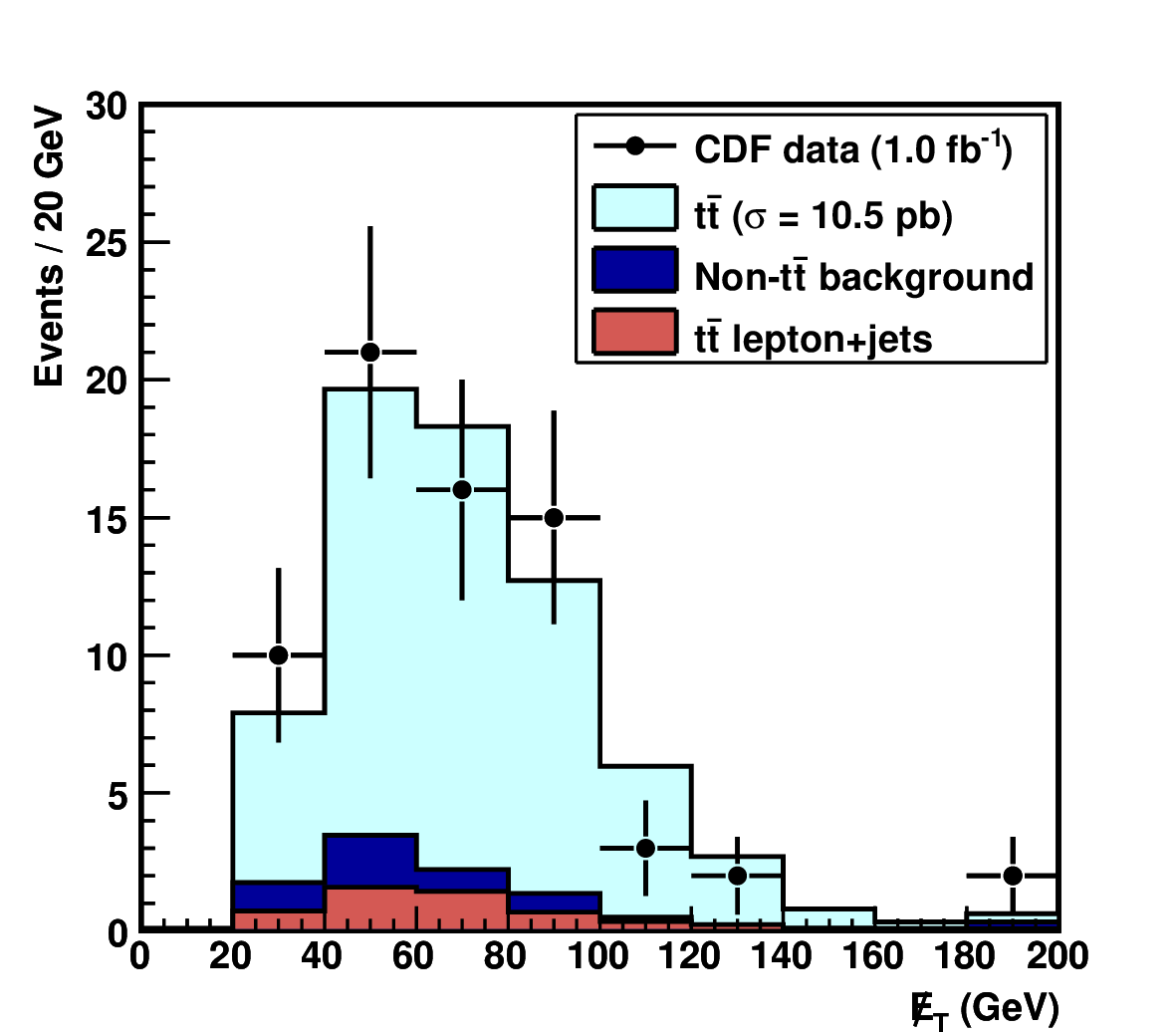}\hspace{0.5cm}
\epsfxsize=5.cm\epsfbox{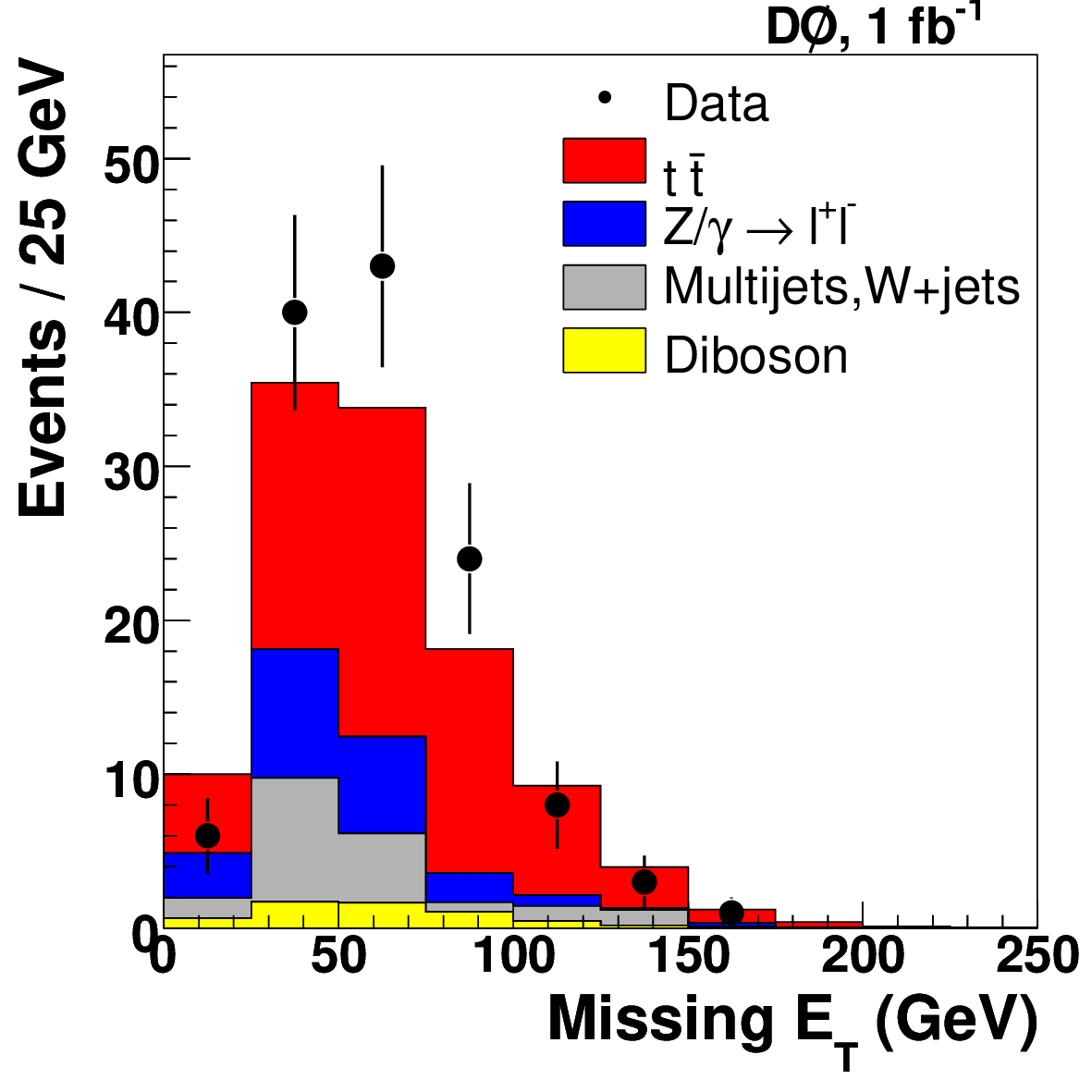}\hspace{0.5cm}
\caption{The observed \met\ distribution compared to the SM expectations for the \ttbar\ cross section measurement using CDF $dil$ events (left) and using \dzero\ $\ell\ell$ and $\ell\tau$ events combined (right)~\cite{Aaltonen:2009ve, Abazov:2009si}.
\label{fig:xsecdilep}}
\end{figure}

Using 1~\fb\ of data, \dzero\ measures in the $\ell \ell$ channel without $b$-tagging for a 170~\gevcc\ top-quark mass~~\cite{Abazov:2009si}
$\sigmattbar = 7.5 ^{+1.2}_{-1.1}  \ (\stat) ^{+0.7}_{-0.6} \ (\syst) ^{+0.7}_{-0.5} \ (\lumi) $~pb 
and in the $\ell \tau$ channel using $b$-tagging and treating \ttbar\ events containing a $\tau$ lepton that is mimicked by a jet as background
$\sigmattbar = 7.6^{+4.9}_{-4.3}  \ (\stat) ^{+3.5}_{-3.4} \ (\syst) ^{+1.4}_{-0.9} \ (\lumi) $~pb.
In the $\ell\ell$ channel, the main sources of systematic uncertainties come from the signal modeling and the jet energy scale while in the
$\ell \tau$ channel, these come from the signal modeling and the background estimation.
Figure~\ref{fig:xsecdilep} shows the predicted and observed \met\ distribution in both CDF and \dzero\ from these \ttbar\ dilepton cross section measurements.

Both collaborations have provided preliminary updates of the \ttbar\ cross section in the $dil$ channel using more than 5~\fb.
The latest measurement from CDF in the $\ell \ell$ channel for a 172.5~\gevcc\ top-quark mass using 5.1~\fb\ of data is
$\sigmattbar = 7.40 \pm 0.58 \ (\stat) \pm 0.63 \ (\syst) \pm 0.45 \ (\lumi) $~pb without $b$-tagging and
$\sigmattbar = 7.25 \pm 0.66 \ (\stat) \pm 0.47 \ (\syst) \pm 0.44 \ (\lumi) $~pb using $b$-tagging.
The latest measurement from \dzero\ in the $\ell \ell$ channel without $b$-tagging for a 172.5~\gevcc\ top-quark mass using 5.3~\fb\ of data is
$\sigmattbar = 8.4 \pm 0.5 \ (\stat) ^{+0.9}_{-0.8} \ (\syst) \ ^{+0.7}_{-0.6} \ (\lumi)$~pb.
Both of these measurements are now limited by the systematic uncertainties.

% ======================================================================
\subsubsection*{\ttbar\ Cross Section in the All Hadronic Channel}
% ======================================================================
In the $had$ channel, both $W$ bosons decay hadronically into two quarks.  The final state consists then of at least six jets among which two of them are $b$-quark jets.  This channel has the advantage of a large branching fraction and no undetectable final state particles but 
is challenged by a QCD multi-jet background that dominates the signal by several orders of magnitude in cross section. The use of $b$-tagging in this channel is therefore essential. The purity of the sample can be further increased by using kinematical and topological characteristics of the \ttbar\ events.

The trigger and event selection is described in Sec.~\ref{sec:reco}. After additionally requiring that two jets are identified as $b$-jets 
the signal-to-background ratio is around 1:2.  To further separate the all hadronic \ttbar\ signal from the QCD background multivariate approaches are used.  CDF combines several kinematical jet or event variables into a neural network while \dzero\ uses a relative likelihood method.  The choice of the input variables to these multivariate discriminants is based on the signal-to-background separation
power and agreement between the data and the background MC model as evaluated in a background dominated control region.
The input variables mainly measure the amount of energy in the event, and the topology of the event.  For example, the event centrality, aplanarity, and sphericity are used.
The centrality is defined as the scalar sum of the jet $E_T^{\mrm{jet}}$ divided by the sum of the jet energies.
The event aplanarity and sphericity are defined through the normalized momentum tensor ${\cal M}$~\cite{Brandt:1978zm}:
$ {\cal M}_{i j} = \sum_\alpha p^\alpha_i p^\alpha_j / \sum_\alpha | \vec{p}^\alpha |^2 $
where $\vec{p}^\alpha$ is the momentum vector of a reconstructed jet $\alpha$ and $i,j=1,2,3$ denote the three 
spatial components of the jet momenta. The three eigenvalues of this tensor are noted in decreasing order 
$\lambda_1, \lambda_2$ and $\lambda_3$ with $\lambda_1 + \lambda_2 + \lambda_3=1$.
The aplanarity is defined as ${\cal A} = \frac{3}{2} \lambda_3$ and reflects the isotropy of the event.
The sphericity is defined as ${\cal S} = \frac{3}{2} ( \lambda_2 + \lambda_3)$.
Furthermore di-jet or tri-jet invariant masses are also used. 
Additional variables like the $\eta$ and $\phi$ moments of a jet~\cite{Aaltonen:2010pe} can be added as they show a good 
discrimination between quark-initiated and gluon-initiated jets, the former being the dominant case for \ttbar\ events while the latter is the dominant case for QCD processes.

The QCD background estimate in the $had$ channel is taken from the data using events from the lower jet multiplicity bins. To measure the amount of background remaining after $b$-tagging, CDF first evaluates the probability of tagging a $udsg$-jet in a sample with exactly four jets, which is QCD dominated. This rate is parametrized in terms of jet $E_T$, the number of tracks associated to the jet, and the number of primary vertices reconstructed in the event. This tag rate is then used to predict the number of tagged events in the signal region taking into account the possible correlation among the jets. The accuracy of this background model is validated by comparing the data with the estimated background in an intermediate NN region ($0.75 < NN < 0.85$) correcting for the small \ttbar\ contamination that lies in this region.  \dzero\ creates a background sample from data by attaching low-$E_T$ jets selected from events with six or more jets to events with four or five jets ensuring a compatible phase-space configuration between the two sources of jets. This is achieved by requiring some amount of
\met\ in the four or five jets events with a low $\met/H_T$ ratio to reduce effects from mis-reconstruction or from events containing hard neutrinos.  This model of background is validated by comparing the five jets events in data with the constructed four+one jet background model.
The background in the final sample used for the cross section measurement is constructed with four+two jet events mixed with
five+one jet events. The difference between these two constructions is taken as systematic uncertainty.

\begin{figure}
\centering
  \epsfxsize=6cm\epsfbox{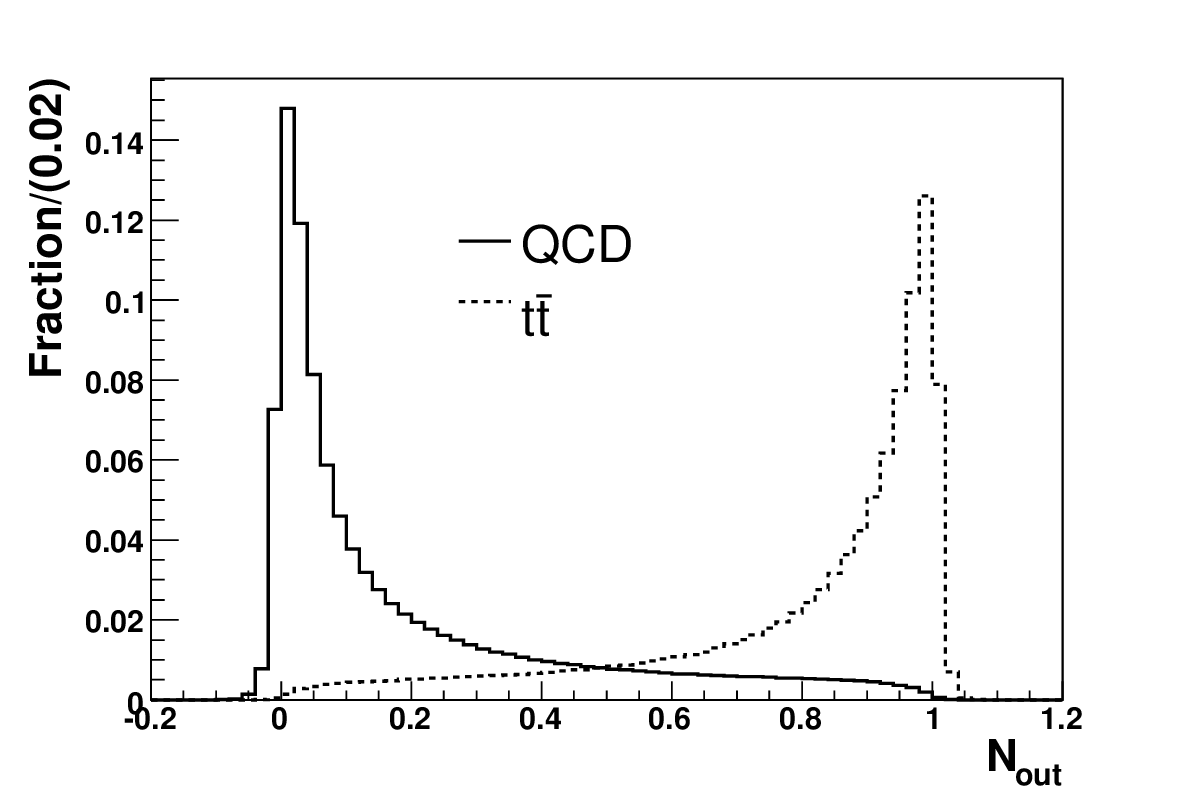}\hspace{0.5cm}
  \epsfxsize=4.5cm\epsfbox{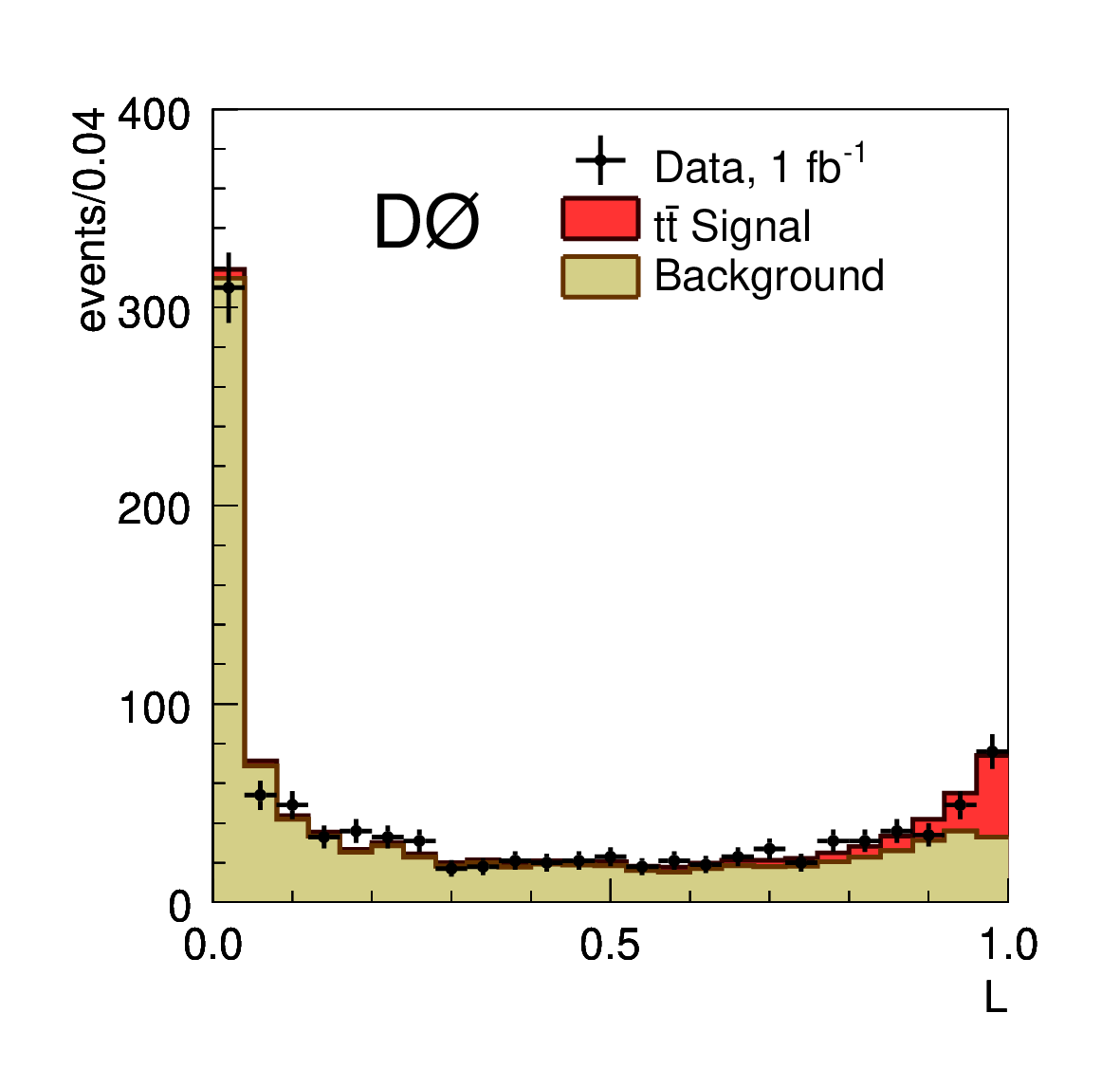}\hspace{0.5cm}
\caption{The NN output distribution for QCD data (solid) and \ttbar\ MC 
  events (dashed) from CDF (left). A comparison of the likelihood 
  distribution between data and MC using the signal purity fit from the 
  data from \dzero\ (right).  Fits to these distributions are used to 
  measure the \ttbar\ cross section in the all hadronic channel~\cite{Aaltonen:2010pe,Abazov:2009ss}.
\label{fig:xsecallhad}}
\end{figure}

Using 2.9~\fb, CDF measures~\cite{Aaltonen:2010pe}
$\sigmattbar = 7.2 \pm 0.5 \ (\stat) \pm 1.0 \ (\syst) \pm 0.4 \ (\lumi)$~pb while \dzero\ uses 1~\fb\ to measure~\cite{Abazov:2009ss}
$\sigmattbar = 6.9 \pm 1.3 \ (\stat) \pm 1.4 \ (\syst) \pm 0.4 \ (\lumi)$~pb, both for $\mt = 175$~\gevcc .
The uncertainty on the background modeling, on the jet energy scale, and on the $b$-jet identification efficiency dominate the systematic uncertainty on these measurements.
Figure~\ref{fig:xsecallhad} shows the distributions of the multivariate discriminants employed by CDF and \dzero\ in their \ttbar\ cross section measurements in the $had$ channel.

 \dzero\ also performed a \ttbar\ cross section measurement in the $\tau$+jets final state with the $\tau$ lepton decaying hadronically. 
The $\tau$ lepton identification criteria are similar to these used in the $\ell\tau$ channel described above. 
A neural network is built to separate \ttbar\ signal from the main background coming from QCD events.
The cross section is extracted from a fit to the entire output discriminant distribution. 
Using 1~\fb\ of data, \dzero\ measures~\cite{Abazov:2010pa} $\sigmattbar = 6.9 \pm 1.2 \ (\stat) ^{+0.8}_{-0.7} \ (\syst) \pm 0.4 \ (\lumi)$~pb for a 170~\gevcc\ top-quark mass.

% ======================================================================
\subsubsection*{\ttbar\  Cross Section Combination and Ratio of Cross Sections}
% ======================================================================
CDF and \dzero\ measure the \ttbar\ cross section in almost all possible decay channels assuming the SM branching fractions.  A comparison of the results is sensitive to potential BSM contributions that affect the various decay channel in ratios differing from the SM, while a combination of the results yields an improved precision on the \ttbar\ cross section.

Within uncertainties all the measured \ttbar\ cross sections in the different final states agree with each other and are thus combined.  To simplify the combination all channels are constructed to be statistically independent.  To compute the combined cross section, \dzero\ defines a joint likelihood function multiplying the 
Poisson probability for the different channels~\cite{Abazov:2009ae}. Additional Poisson terms are added to constrain the QCD background in the relevant channels. Each systematic uncertainty is included in the likelihood function as a nuisance parameter. Correlations between channels are taken into account by using the same nuisance parameter. 
Combining the $ljt$, the $\ell \ell$ and the $\ell \tau$ channels measured using 1~\fb, \dzero\ finds~\cite{Abazov:2009ae} $\sigmattbar = 8.18^{+0.98}_{-0.87}$~pb for $\mt = 170$~\gevcc .
CDF performs a combination of its latest preliminary measurements, using up to $5.7$~\fb\ of data in the $dil$, $had$, topological $ljt$, and $b$-tag $ljt$ channels, by forming a best linear unbiased estimate~\cite{Lyons:1988rp,Valassi:2003mu} and taking into account the statistical and systematic correlations.  For $\mt = 172.5$~\gevcc\ the preliminary combination yields  $7.46^{+0.66}_{-0.80}$~pb.  Both of these combinations are in agreement with the theoretical computation~\cite{Langenfeld:2009wd}. By the end of the Tevatron run, a precision of about 6\% could be achieved.
%
%%%Figures~\ref{fig:xsecsummary} show a summary of the latest \ttbar\ cross %%%section measurements performed by CDF and \dzero.
%%%\begin{figure}
%%%\centering
%%%  \epsfxsize=6cm\epsfbox{cdf_summary}\hspace{0.5cm}
%%%  \epsfxsize=5cm\epsfbox{d0_summary}\hspace{0.5cm}
%%%\caption{Summary of the latest \ttbar\ cross section measurements 
%%%  performed by CDF (left) and \dzero\ (right).
%%%\label{fig:xsecsummary}}
%%%\end{figure}

Calculating the ratio of \sigmattbar\ measured in different final states or with different numbers of $b$-tagged jets probes for the presence of non-SM decays of the top quark.  Some BSM theories which might give rise to such effects were discussed in Sec.~\ref{sec:topsearch}. 
An example of such an approach is the measurement of the ratio of top-quark branching fractions, $R_b = \frac{ {\cal B} ( t \to W b)}{ {\cal B} (t \to W q)}$, 
which can be expressed in terms of CKM matrix elements.  
%
%%%By combining this with measurements of the single top-quark production %%%cross section direct measurements of $\left| V_{tq} \right|$ can be %%%obtained without invoking any assumptions about the number of quark %%%families or the unitarity of the CKM matrix.  
%
\dzero\ performed a simultaneous determination of $R_b$ and \sigmattbar~\cite{Abazov:2008yn}.
In addition to applying $b$-tagging, a topological discriminant is used in the $\ge 4$~jets and 0~$b$-tag subsamples to further
constrain the number of \ttbar\ events. Using 0.9~\fb, the result is: $R_b = 0.97^{+0.09}_{-0.08} \ (\statsyst)$ for
a 175~\gevcc\ top-quark mass in agreement with the SM expectation. The main uncertainty comes from the limited statistics while
the systematic uncertainty is dominated by the uncertainty on the $b$-jet identification efficiency.

Another example of BSM top-quark decays is $t \to H^{+}b$~\cite{Chung:2005a,Guasch:1995rn}.  
A large $H^{+}\ra\tau^{+}\nu$ branching fraction would result in a larger fraction of \ttbar\ events showing up in the $\ell \tau$ channel and fewer events in other channels than expected by the SM.  On the other hand, a
lepto-phobic charged Higgs would lead to fewer events in the $\ell\ell$ channel compared to $ljt$.  The ratios of the \ttbar\ cross sections as measured in the various decay channels was determined by \dzero\ after taking correlations into account.  The ratio of the $\ell\ell$ to $ljt$ channel was measured as~\cite{Abazov:2009ae} $R^{\ell \ell / ljt}_{\sigma} = 0.86^{+0.19}_{-0.17}$ while the ratio of the $\ell\tau$ determined cross section relative to all the others was measured as 
$R^{\ell \tau/ \ell \ell + ljt}_{\sigma} = 0.97^{+0.32}_{-0.29}$, both of which are compatible with the SM expectation.  These were used to set limits on tau-onic and lepto-phobic charged Higgs decays as well as on the parameter space of the MSSM~\cite{Abazov:2009zh}.
CDF has also reinterpreted its cross section results to set limits on the charged Higgs production~\cite{Abulencia:2005jd} and assuming five possible Higgs decay modes: 
$t \to W^+ b$, $t \to H^+ b \to \bar{\tau} \nu b$, $t \to H^+b \to c \bar{s} b$, $t \to H^+ b \to t^* \bar{b} b$ and 
$t \to H^+ b$ where $H^+ \to W^+ h$ and $h \to b \bar{b}$. Limits are set in the plane $(H^+ ,\tan \beta)$  for several MSSM scenarios. If no assumption is made on the charged Higgs decay, an upper limit
on ${\cal B}(t \to H^{+}b)$ of 0.91 is set at 95~\% confidence level.

% ======================================================================
\subsubsection*{\ttbar\ Differential Cross Section}
% ======================================================================
The large statistics \ttbar\ samples now available at the Tevatron enable a more thorough study of \ttbar\ production.  In addition to providing good statistics, the $ljt$ channel also offers good resolution on reconstructed quantities, and good sample purity.  Thus, the most constraining measurements of top-quark properties and searches for BSM physics in the top-quark sample predominantly come from the $ljt$ channel.  Both CDF and \dzero\ have used fully reconstructed $ljt$ events to measure differential cross sections.  Several BSM theories like technicolor, topcolor, or models with extra dimensions can distort the \ttbar\ invariant mass spectrum or the transverse momentum distribution of the top quarks compared to the SM expectations~\cite{Hill:1993hs,Frederix:2007gi}.  Measurements of differential cross sections also test the perturbative QCD heavy-quark predictions.

CDF measures the \ttbar\ differential cross section with respect to the \ttbar\ invariant mass, $d\sigma / d\mttbar$~\cite{Aaltonen:2009iz} using the $ljt$ channel with $b$-jet identification.
The event selection and background estimate is described above.
To reduce the systematic uncertainty introduced by the jet energy corrections, which affects the measured \mttbar\ spectrum, the jet energy scale is constrained {\it{in situ}} by comparing the measured 
invariant mass of the hadronically-decaying $W$ boson to the world average $W$-boson mass. The reconstructed background subtracted \mttbar\ distribution is corrected to the parton level using a regularized unfolding method~\cite{Hocker:1995kb}.
%to prevent statistical fluctuations in the inversion of the matrix that %links the true binned distribution with the measured one.
Apart from the statistical uncertainty, the largest systematic uncertainty arises from uncertainties in the parton distribution functions since the tail of the \mttbar\ distribution is very sensitive to the large-$x$ contributions to the PDF.  The resulting differential cross section shows no evidence of BSM physics.

\begin{figure}
\centering
  \epsfxsize=6cm\epsfbox{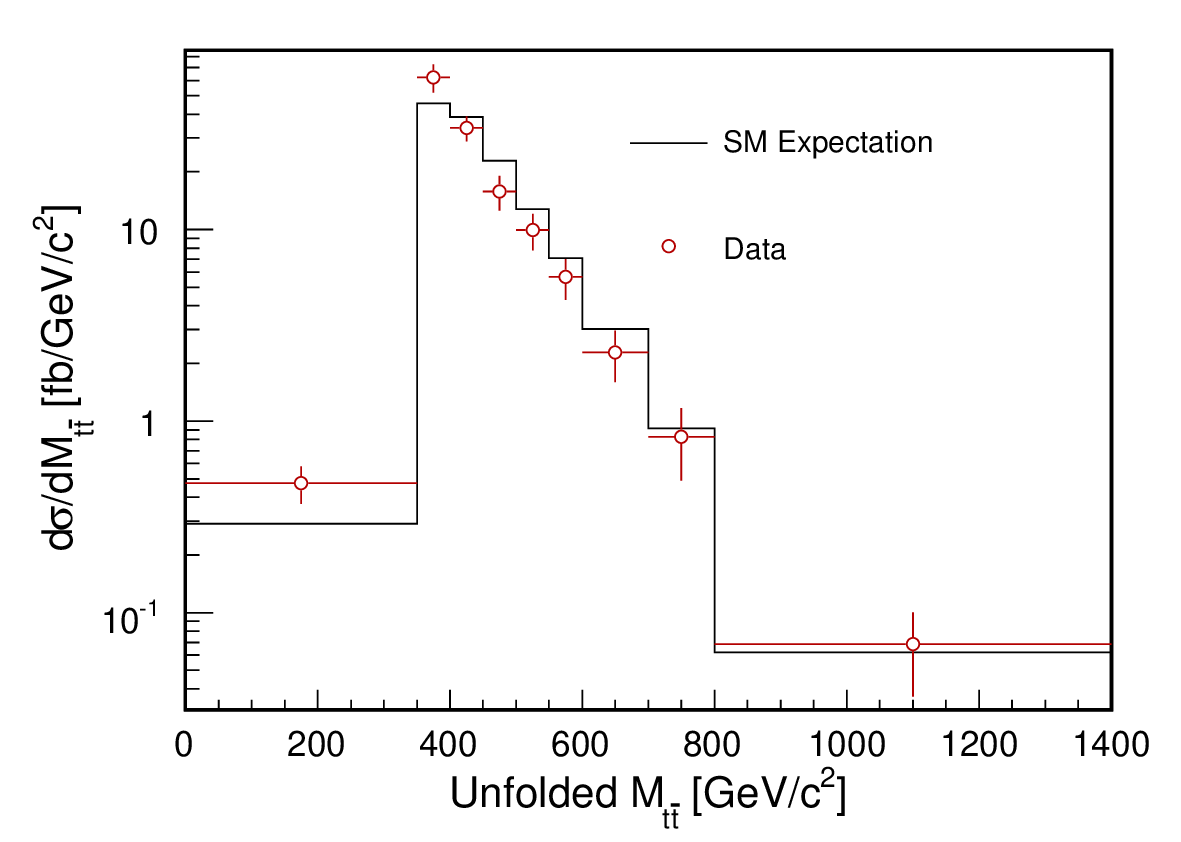}\hspace{0.5cm}
  \epsfxsize=4.5cm\epsfbox{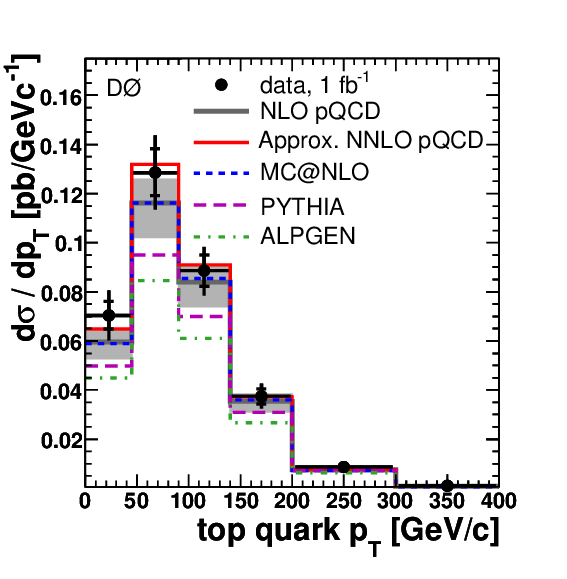}\hspace{0.5cm}
\caption{ The measured $d\sigma/d \mttbar$ from CDF (left) compared to 
  the SM expectation.  The measured 
  $d\sigma/d\pt$ from \dzero\ (right) compared to QCD calculations 
  and various MC generators~\cite{Aaltonen:2009iz, Abazov:2010js}.
\label{fig:diffxsec}}
\end{figure}

\dzero\ measures the inclusive differential cross section for $\ppbar \to \ttbar + X$ as a function of the top-quark \pt , $d\sigma / dp_{T}^{t}$~\cite{Abazov:2010js} using the $ljt$ channel selected as described above.  A kinematic fit, which constrains the $W$-boson and top-quark masses to fixed values, is performed for all possible jet-parton assignments and the combination yielding the minimum fit chi-squared is used to associate the leptons and jets to the top quark.  The reconstructed $p_{T}^{t}$ spectrum is corrected to the parton level. 
The uncertainties on the shape of the $p_{T}^{t}$ spectrum are dominated by statistics while the total systematic uncertainty integrated over 
$p_{T}^{t}$ is around 10\%. The NLO or NNLO perturbative QCD calculations agree with the measured cross section in normalization and shape while the results from the {\alpgen+\pythia} and {\pythia}
generators describe the shape of the data distribution but not its normalization.
Figure~\ref{fig:diffxsec} shows the measured differential \ttbar\ cross sections with respect to \mttbar\ or the top-quark \pt\ measured by CDF and \dzero.

% ======================================================================
\subsubsection{Forward-Backward Charge Asymmetry}
% ======================================================================

At lowest order in QCD, the SM predicts that the top-quark pair production in \ppbar\ interactions is charge symmetric.
However radiative corrections involving either virtual or real gluon emissions lead to a small charge asymmetry~\cite{Kuhn:1998kw,Kuhn:1998jr}.  This asymmetry has its origin in radiative corrections to $q\overline{q}$ fusion from interference of the box diagram with the Born diagram while the contribution to \ttbar\ production from gluon fusion remains charge symmetric.  The real emission corrections tend to push the top quark backward, opposite to the proton beam direction, while the virtual corrections push it forward, in the direction of the proton beam~\cite{Bowen:2005ap}. This asymmetry is predicted to be around 5\%~\cite{Dittmaier:2007wz,Bowen:2005ap,Almeida:2008ug} while the measured asymmetry depends strongly on the region of phase space being probed.
Due to charge conjugation symmetry, the charge asymmetry can also be interpreted as a top-quark forward-backward asymmetry.
Some extensions of the SM~\cite{Ferrario:2009ns}, such as $Z' \to \ttbar$ decays~\cite{Malkawi:1996fs},   t-channel $W' $ exchange~\cite{Cheung:2009ch}, a new color octet resonance~\cite{Antunano:2007da, Ferrario:2008wm} or extra-dimension models~\cite{Djouadi:2009nb} predict different asymmetries.
Both CDF~\cite{Aaltonen:2008hc} and \dzero~\cite{Abazov:2007qb} measure the forward-backward charge asymmetry using $ljt$ events with $b$-jet identification.
%%%$A_{FB} = 0.050 \pm 0.015$ 

The asymmetry $A_{fb}$ is frame dependent and can be defined either in the \ppbar\ or \ttbar\ rest frame.   The \ttbar\ rest frame is more difficult to reconstruct but the asymmetry in the \ppbar\ rest frame is predicted to be about $30$~\% smaller~\cite{Antunano:2007da}.  In the \ppbar\ rest frame the asymmetry can be written as:
\begin{equation}
  A^{\ppbar}_{fb} = \frac{N_t(\cos \theta > 0) - N_t(\cos \theta < 0)}  {N_t(\cos \theta > 0) + N_t(\cos \theta < 0)}
\end{equation}
where $N_t(x)$ is the number of top quarks satisfying requirement $x$, $\cos \theta = - Q_\ell \cdot \cos\alpha_p$, $\alpha_p$ is the polar angle between the top quark with the hadronic \Wboson\ decay and the proton beam, and 
$Q_\ell$ is the lepton charge from the leptonic decaying $W$ boson. 
In the \ttbar\ rest frame the asymmetry can be defined similarly except that $\theta^*$ is the production angle of the top quark in the \ttbar\ rest frame. This angle is related to the rapidity, $y$, of the $t$ and $\bar{t}$ in the \ppbar\ frame by
\begin{equation}
  \Delta y = y_t - y_{\bar{t}} = 2 \tanh^{-1} \left( \frac{\cos \theta^*}{ \sqrt{1 + \frac{4 m^2_t}{s - 4 m^2_t}}} \right)
\end{equation}
where $s$ is the square of the center-of-mass energy. So the
\ttbar\ rapidity difference in the \ppbar\ rest frame can be used to measure the production angle $\theta^*$ in the \ttbar\ rest frame.
As both $\Delta y$ and $\cos \theta^*$ have the same sign, the asymmetry can be written
\begin{equation}
  A^{\ttbar}_{fb} = \frac{N_t(\Delta y > 0) - N_t(\Delta y < 0)}{N_t(\Delta y > 0) + N_t(\Delta y < 0)}.
\label{eq:afb_y}
\end{equation}

\begin{figure}
\centering
  \epsfxsize=3.7cm\epsfbox{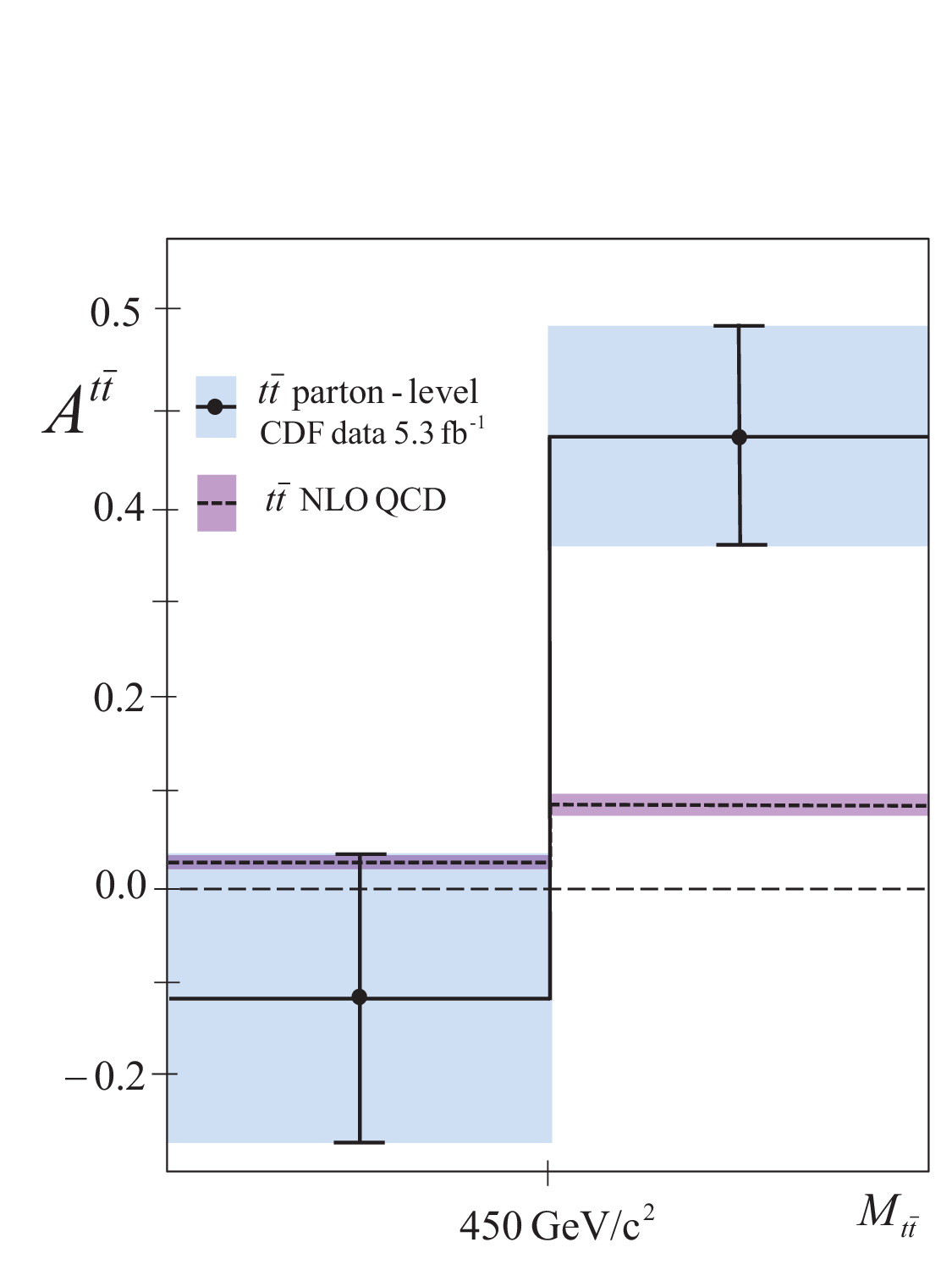}\hspace{0.5cm}
  \epsfxsize=6cm\epsfbox{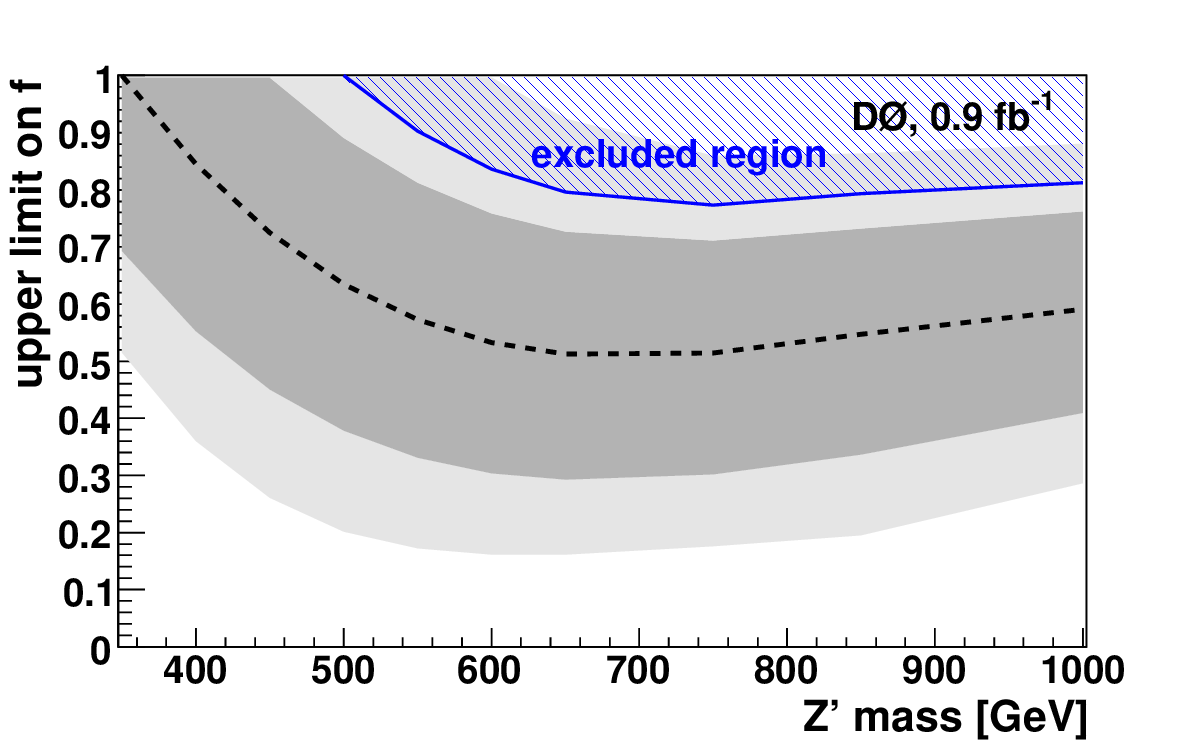}\hspace{0.5cm}
\caption{The CDF determined $A_{fb}$ in the \ttbar\ frame, corrected to the parton level, compared to NLO predictions in bins of the \ttbar\ invariant mass (left). The $95\%$ confidence level
  limits on the fraction of \ttbar\ produced via a $Z' $ resonance as a 
  function of the $Z' $ mass (right) obtained from the \dzero\ $A^
  {\ttbar \ obs}_{fb}$ measurement~\cite{CdfAfb53,Abazov:2007qb}. 
\label{fig:afb}}
\end{figure}

CDF measures the asymmetry in both frames using fits to the $\cos \theta$ and to the $\Delta y$ data distributions after background subtraction.  The background fraction is constrained in these fits using the estimates from the cross section analysis described above.  
The \ttbar\ events are reconstructed using a kinematic fit that constrains the \Wboson\ mass and the $m_t - m_{\overline{t}}$ difference.  The combination of jet-parton assignments which minimizes the fit $\chi^2$ is used in the analysis.   The results are corrected to the parton level using a matrix inversion technique.  Using 1.9~\fb\ of data CDF measures~\cite{Aaltonen:2008hc}
$A^{\ppbar}_{fb} = 0.17 \pm 0.07 (\stat) \pm 0.04 (\syst)$ and 
$A^{\ttbar}_{fb} = 0.24 \pm 0.13 (\stat) \pm 0.04 (\syst)$.
These results agree within 2 standard deviations with the SM expectation.   Using the same methodology CDF has a result using $5.3\:\mrm{fb}^{-1}$.  The resulting forward backward asymmetry in the \ttbar\ rest frame is $A^{\ttbar}_{fb}=0.158\pm 0.075\:(\mrm{stat}+\mrm{syst})$, consistent with the SM NLO prediction within two standard deviations.  As shown in Figure~\ref{fig:afb}, they additionally observe that the asymmetry is rising as a function of the \ttbar\ invariant mass, $m_{\ttbar}$, and determine $A^{\ttbar}_{fb}(m_{\ttbar} > 450\gevcc )=0.475\pm0.114\:(\mrm{stat}+\mrm{syst})$~\cite{CdfAfb53}, which is about 3.4 standard deviations away from the NLO prediction of $0.088\pm0.013$ obtained using the MCFM program~\cite{MCFM}.  A preliminary CDF analysis using $5.1\:\mrm{fb}^{-1}$ of data in the $dil$ channel measures a total asymmetry consistent with that observed in $ljt$ events and which also grows with $m_{\ttbar}$~\cite{CdfAfbDil51}.

\dzero\ measures the asymmetry in the \ttbar\ frame and performs a simultaneous fit for the asymmetry and the background fraction using the $\Delta y$ distribution and a likelihood discriminant designed to separate \ttbar\ from \Wpjets\ processes.  In an effort to remain independent of any potential BSM contributions, they do not correct for reconstruction effects but provide a parameterization of the resolution.  The \dzero\ measured asymmetry corrected for background contributions from $0.9$~\fb\ of data is~\cite{Abazov:2007qb}
$A^{\ttbar \ obs}_{fb} = 0.12 \pm 0.08 (\stat) \pm 0.01 (\syst)$.
\dzero\ derives from this limits on the mass of a leptophobic $Z'$ decaying to \ttbar\ as shown Fig.~\ref{fig:afb}.  The result has been recently updated using the same analysis techniques and $4.3\:\mrm{fb}^{-1}$ of data to obtain $A^{\ttbar\ obs}_{fb}=0.08 \pm 0.04 (\stat) \pm 0.01 (\syst)$~\cite{DzAfb43}.  Both these results are consistent within two standard deviations of the SM expectation, which is estimated using a NLO QCD MC generator, applying the event selection criteria, and determining the resulting asymmetry.

These analyses will remain interesting as the data sets increase since the 
Tevatron's \ppbar\ initial state is a $CP$ eigenstate and offers increased sensitivity to \ttbar\ $A_{fb}$ effects relative to the LHC, a proton-proton collider.

% ======================================================================
\subsubsection{Fraction of gluon-gluon Fusion Produced \ttbar }
% ======================================================================

At the Tevatron, a variety of production mechanisms can yield a pair of heavy quarks in the final state.   Distinguishing between the various initial state mechanisms is difficult since information is lost as the heavy quarks hadronize and form bound states.  However, as discussed in Sec.~\ref{sec:proddecay}, top quarks decay before they hadronize and the initial state spin information survives to the final state.  This can be used together with other discriminating variables to determine the fraction of \ttbar\ pairs produced via gluon-gluon fusion.
At the Tevatron, the SM predicts that $gg\ra \ttbar$ accounts for about 15\% of \ttbar\ production, with the remainder coming from $\qqbar \ra \ttbar$ processes.  This fraction varies by about $\pm5\%$ (absolute) due to uncertainties in the gluon PDFs.   The measured $gg\ra\ttbar$ contribution can be affected by BSM particles decaying to
 \ttbar\ \cite{Hill:1993hs,Lane:1995gw,Zhang:1999qy}.

CDF uses the $ljt$ channel with $b$-jet identification to measure the relative fraction of gluon-gluon produced \ttbar\ pairs using two different methodologies.  The first method exploits the fact that gluons are more likely to radiate additional gluons so that gluon initiated processes result in a higher track multiplicity on average than quark initiated processes.  The observed low-$p_T$ track multiplicity distribution is fit to a contribution from background, from $\qqbar\ra\ttbar$, and from $gg\ra\ttbar$ processes.   The distributions used in the fit are taken from MC samples calibrated using data samples known to be gluon-rich or gluon-depleted (\Wpjets\ and QCD di-jet events).  Corrections are made for the number of high energy jets in the event, for background contributions, and for acceptance effects.   Using this first method CDF measures $\sigma(gg \to \ttbar) / \sigma(\ppbar \to \ttbar) = 0.07 \pm 0.14 \ (\stat) \pm 0.07 \ (\syst)$ which corresponds to an upper limit of 0.33 at 95~\% confidence level~\cite{Aaltonen:2007kq}.  

The second method uses kinematic information to discriminate between the two production mechanisms.   A kinematic fit is used to determined the best jet-parton assignment to use.  This allows a full reconstruction of the final state and of the top quarks and their kinematics.   A variety of variables are combined using a NN, trained using \ttbar\ MC samples.   The variables include several top-quark production angles, and several that are sensitive to \ttbar\ spin correlations as suggested in~\cite{Mahlon:1997uc,Parke:1996pr}.  A fit including background, $gg\ra\ttbar$, and $\qqbar\ra\ttbar$ contributions, is performed to the observed NN distribution.  Using this second method CDF employs a Feldman-Cousins prescription~\cite{Feldman:1997qc} to set an upper limit of $\sigma(gg \to \ttbar) / \sigma(\ppbar \to \ttbar)<0.61$ at 95~\% confidence level~\cite{Abulencia:2008su}. The dominant systematic uncertainties on this measurement come from the background shape and composition and from the difference between LO and NLO predictions.

The two measurements are combined to yield
$\sigma(gg \to \ttbar) / \sigma(\ppbar \to \ttbar) = 0.07^{+0.15}_{-0.07}$~\cite{Abulencia:2008su} consistent with the SM expectation.  

% ======================================================================
\subsection{Decay Properties}
\label{sec:decay}
% ======================================================================

\subsubsection{W-Boson Polarization Fractions}
\label{sec:WHel}

Since the top quark decays before forming a bound state, the $t\ra W^{+}b$ process provides an opportunity to directly probe the $tWb$ interaction vertex.  In particular, measurements of the polarization fractions of the resulting $W$ bosons can be used to test for BSM $tWb$ interactions.   In the SM the fraction of $W^+$ bosons from top-quark decay produced with longitudinal polarization is about $f_0 = 0.70$, with left-handed polarization is about $f_- = 0.30$, and with right-handed polarization is about  $f_+ = 0.0004$.  Contributions from BSM physics can give rise to anomalous couplings at the $tWb$ vertex that change these polarization fractions~\cite{Kane:1991bg, AguilarWHel}.  Some examples include the SM extensions discussed in~\cite{Bernreuther:2003xj, Nam:2002rq}.  Measurements of the \Wboson\ polarization fractions can be used to set constrains on the vector and tensor form factors, $f^R_1$, $f^L_1$, $f^R_2$, and $f^L_2$.  These terms appear in the general form of the Lagrangian describing the $tWb$ vertex.  In the SM only $f^L_1$ has a non-zero value.

The \Wboson\ polarization affects several kinematic variables, which can be used to measure $(f_0 , f_+ )$.  The most commonly used is $\theta^*$, the angle between the $W$-boson momentum direction in the rest frame of the top quark and the direction of the down-type fermion decay product of the $W$ boson (charged lepton or $d-$, or $s$-quark) in the rest frame of the $W$ boson.  The \Wboson\ differential decay rate can be written in terms of $\cos \theta^*$ as:
\begin{equation}
\frac{1}{\Gamma} \frac{d\Gamma}{d \cos \theta^*} = \frac{3}{8} f_- (1 - \cos \theta^*)^2 + \frac{3}{4} f_0 (1 - \cos^2 \theta^*) + \frac{3}{8} f_+ ( 1 + \cos \theta^*)^2.
\end{equation}
Estimating $\cos\theta^*$ requires each \ttbar\ event be fully reconstructed using, for example, a kinematic fit.
The charged lepton $E_T$ is also sometimes used since it is correlated to the \Wboson\ polarization.  It produces a less precise measurement, but is insensitive to uncertainties in the jet energy scale.  Both CDF and \dzero\ have recently used $\cos \theta^*$ to measure the \Wboson\ polarization fractions with 2.7~\fb and 1~\fb of data, respectively.

CDF uses the matrix element method described in Sec.~\ref{sec:mass} to determine the polarization fractions in the $ljt$ channel with $b$-jet identification.   In this analysis the LO \ttbar\ differential cross section is expressed in terms of $\cos\theta^*$ and the fractions $(f_0 , f_+ )$ while the top mass is fixed at 175~\gevcc .  The fit to the data is performed under three different scenarios, a) a model independent fit to both $f_0$ and $f_+$ simultaneously, b) a fit to determine $f_+$ with $f_0$ fixed to its SM value, and c) a fit to determine $f_0$ with $f_+$ fixed to zero (its SM value given the current level of precision).  Corrections for background contributions and corrections for the variation of the event selection acceptance as a function of $(f_0 , f_+ )$ are included.
The model independent fit yields~\cite{Aaltonen:2010WHel}
$f_0 =  0.88 \pm 0.11 \ (\stat) \pm 0.06 \ (\syst) $ and 
$f_+ = -0.15 \pm 0.07 \ (\stat) \pm 0.06 \ (\syst)$ with a correlation coefficient of $-0.59$.  These measurements are statistics limited.  Fixing $f_0$ to its SM value yields $f_+ = -0.01 \pm 0.02 \ (\stat) \pm 0.05 \ (\syst)$ while fixing $f_+$ to zero yields $f_0 = 0.70 \pm 0.07 \ (\stat) \pm 0.04 (\syst)$.  The dominant systematic uncertainties arise from uncertainties in the background modeling and from uncertainties in ISR/FSR and parton shower modeling in \ttbar\ events.

\dzero\ uses a kinematic fit to fully reconstruct the event and calculate $\cos\theta^*$ in the $ljt$ and $dil$ channels.
In the $ljt$ channel events are reconstructed using a kinematic fit with fixed \Wboson\ and top-quark masses.  In the $dil$ channel, given a top-quark mass, a four-fold ambiguity exists for each set of jet-parton assignments considered.  In the analysis all real solutions are considered.  The resulting reconstructed $\cos \theta^*$ distributions are compared with templates derived from \ttbar\ MC with different \Wboson\ polarization fractions and taking into account background and reconstruction effects.
For a top-quark mass of 172.5~\gevcc\ the measured \Wboson\ polarization fractions determined from a model independent fit are~\cite{Abazov:2007ve}
$f_0 = 0.43 \pm 0.17 \ (\stat) \pm 0.10 \ (\syst)$ and $f_+=0.12 \pm 0.09 \ (\stat) \pm 0.05 \ (\syst)$. These measurements are statistics limited.  The main systematic uncertainties come from the background or \ttbar\ modeling and from the MC template statistics. 
%
%
%\begin{figure}
%\centering
%  \epsfxsize=5cm\epsfbox{cdf_whel}\hspace{0.5cm}
%  \epsfxsize=6cm\epsfbox{d0_whel}\hspace{0.5cm}
%\caption{(left) Unfolded $\cos \theta^*$ distribution in CDF~\cite
%{Aaltonen:2008ei}.
%(right) MC and data $\cos \theta^*$ distribution in the dilepton %events at \dzero~\cite{Abazov:2007ve} The dashed line shows the SM %expectation.
%\label{fig:whel}}
%\end{figure}

\dzero\ combines the measured \Wboson\ polarization fractions with the measurement of the electroweak single top production 
cross section~\cite{Abazov:2006gd,Abazov:2008kt} to set limits on the $f^L_2$, $f^R_1$, and $f^R_2$ BSM form factors~\cite{Abazov:2009ky}.

% ======================================================================
\subsubsection{Decays via Flavor Changing Neutral Currents}
\label{sec:fcnc}
% ======================================================================

In the SM quarks can change flavor only via \Wboson\ interactions, so that at tree level decays of the type, $t\to qZ$ or $t\to q\gamma$, where $q=u,\: c$, are forbidden.  Thus the rates for these decays are suppressed because they can only occur through quantum loops involving the $W$ boson. Because the top quark is so massive, the quantum loop corrections arising from new particle loops are much larger than is the case for the light quarks.  This makes the top quark a potentially great place to look for evidence of BSM Flavor Changing Neutral Currents (FCNC) decays.  For example, the SM predicts ${\cal B}(t\to qZ )\approx {\cal O}(10^{-14})$~\cite{AguilarSaavedra:2004wm}, while new physics models like 2HDM, supersymmetry, and topcolor-assisted technicolor predict branching fractions as large as ${\cal O}(10^{-4})$~\cite{Larios:2006pb}.  An observation of a branching fraction significantly larger than predicted by the SM would be unambiguous evidence of BSM physics.

CDF has searched for the FCNC decay $t \to qZ$ using \ttbar\ events when the top quarks decay as $\ttbar\ra qZ qZ$ or $\ttbar\ra qZ bW$ and one of the $Z$ bosons decays into a $e^+ e^-$ or $\mu^+ \mu^-$ pair~\cite{Aaltonen:2008aaa}, while the other boson in the event decays hadronically.  The experimental signature is thus very similar to the $ljt$ channel except that the final state neutrino is replaced by a charged lepton.  The dominant background process is the production of \Zpjets . Smaller background contributions come from the SM \ttbar, $WZ$, and $ZZ$ processes. The event sample is divided into sub-samples depending on whether or not any of the jets have been identified as $b$-quark jets.  A kinematic fit is employed similar to the one described in Sec.~\ref{sec:kfit} except that one or both of \mw\ constraints are replaced by \mz\ constraints depending on which final state hypothesis is being tested.  To separate the FCNC signal from the background, requirements are made on the transverse mass of the $Z$ and the four leading jets, the $E_T^{\mrm{jet}}$ of the four leading jets, and on the kinematic fit $\chi^2$.  A binned likelihood fit to the $\chi^2$ distribution is simultaneously performed for the sub-samples with and without an identified $b$-jet and the fraction of FCNC decays in the data is extracted.   In the fit the normalization for the dominant \Zpjets\ background is allowed to vary while the \ttbar\ cross section is fixed to the observed \ttljt\ cross section.  The systematic uncertainties affecting the signal and background shapes and normalizations are taken into account using the technique of~\cite{Read:1999kh}.  The dominant systematic uncertainty arises from the jet energy corrections and in uncertainties related to the fraction of \Zpjets\ events that have identified $b$-jets.  A Feldman-Cousins procedure~\cite{Feldman:1997qc} 
is employed to set the limit. Using 1.9~\fb, CDF data is consistent with the SM expectation and an upper limit of ${\cal B}(t \to qZ^0 )<3.7 \%$ is established at 95~\% confidence level.
A similar analysis has been performed by \dzero\ searching for $\ttbar\ra qZ qZ$ or $\ttbar\ra qZ bW$ but with a leptonic decay of the $Z$ or $W$ bosons.
Using 4.1~\fb\ in the trilepton final state, \dzero\ does not observe  any sign of anomalous coupling and set a limit of ${\cal B}(t \to qZ^0 )<3.2 \%$ at 
95\% confidence level~\cite{Abazov:2011qf}.
Similar searches can also be performed using the electroweak production of single top quarks~\cite{Abazov:2010qk}.

%
%\begin{figure}
%\centering
%  \epsfxsize=6cm\epsfbox{cdf_fcnc}\hspace{0.5cm}
%\caption{FCNC\label{fig:fcnc}}
%\end{figure}

% ======================================================================
\subsubsection{Decays to Charged Higgs Bosons}
\label{sec:chhiggs}
% ======================================================================

Many theories beyond the SM propose an extension of the Higgs sector to multiple Higgs doublets~\cite{Grossman:1994jb}. The simplest extension of the SM Higgs sector is the two Higgs-doublet model (2HDM)~\cite{Gunion:1989we}. In the 2HDMs, the Higgs sector consists of three neutral Higgs bosons $(h,H,A)$ and two charged ones $(H^\pm)$. Several types of 2HDMs exist and differ in their strategies to avoid large flavor changing neutral current effects. In type I 2HDM, only one of the Higgs doublets couple to fermions.  In type II, one doublet couples to the up-type quarks and neutrinos and the other to down-type quarks and charged leptons.
The minimal supersymmetric standard model (MSSM) is an example of type II 2HDM. In models of type III, both doublets couple to fermions and FCNCs are suppressed through other mechanisms. An observation of a charged Higgs boson would be unambiguous proof of a BSM Higgs sector.

At the Tevatron direct production of a single charged Higgs boson can occur by $q\bar{q}^\prime$ annihilation. Searches for $q\bar{q}^\prime\to H^{+}\to tb$ have been performed by \dzero\ assuming a charged Higgs boson mass larger than the top-quark mass~\cite{Abazov:2008rn}. 
This channel has the same signature as the electroweak production of single top quarks. Direct production of $H^+H^-$ through the weak interaction has a rather small expected cross section, of the order of 0.1~pb~\cite{Carena:2002es} and so no such
search was performed so far.  For $m_{H^+}<m_t$ in 2HDM, the top quark can decay to $H^+ b$, hence competing with the SM decay to $W^+ b$.  As seen in section~\ref{sec:production}, the ratio of \ttbar\ cross sections as measured in different decay channels is thus sensitive to BSM top-quark decays.  The measured \ttbar\ cross sections have been used to set limits in the plane $(m_{H^+},\tan \beta)$ where $\tan \beta$ is the ratio of vacuum expectation values of the two Higgs doublets~\cite{Abazov:2009zh}.
At small $\tan \beta$ ($\lesssim 1$), the top branching fraction into charged Higgs, ${\cal B}(t \to H^+ b)$, can be large. 
In that regime, the charged Higgs boson predominately decays to $c\bar{s}$ for low charged Higgs mass ($\lesssim 130$~GeV ) or 
to $t^*\bar{b} \to W b \bar{b}$ for large mass. At larger $\tan \beta$ ($\gtrsim 15$), ${\cal B}(t \to H^+ b)$ can also be large, but in that case the charged Higgs boson decays almost 100~\% of the time to $\tau^+ \nu$.

Using 2.2 \fb, CDF has performed a direct search for decays of top quarks into charged Higgs bosons assuming $H^+ \to c \bar{s}$ using the $ljt$ final state including $b$-jet identification.  The di-jet invariant mass distribution, which is otherwise assumed to come from the \Wboson\ decay, is used to search for evidence of such decays~\cite{Aaltonen:2009ke}.   The \ttbar\ events are reconstructed using a kinematic fitter similar to the one described in Sec.~\ref{sec:kfit} except that the \mw\ constraint for the hadronically decaying \Wboson\ candidate is dropped and templates are built for signal and background processes. The signal is modeled using $H^+$ MC where the $H^+$ decay is forced to $c \bar{s}$ and $\Gamma_{H^+}=0$ is assumed for $60\: < \: m_{H^+} \: < \: 150$~\gevcc.   A binned likelihood fit, including systematic effects, is used to exclude  ${\cal B}(t \to H^+ b)$ larger than 0.1 to 0.3 at the $95\%$ confidence level assuming ${\cal B}(H^+ \to c \bar{s})=1$.
 
% ======================================================================
\subsection{Intrinsic Properties}
\label{sec:intrinsec}
% ======================================================================

% ======================================================================
\subsubsection{Mass}
\label{sec:mass}
% ======================================================================

The top-quark mass, \mt, is a free parameter in the SM and must be experimentally determined.  A precision determination of \mt\ is important since quantum loops including top quarks contribute large corrections to theory predictions for many precision electroweak observables.  For example, the SM predicts a precise relationship between the $W$- and \Zboson\ masses,
\begin{equation}
  \left( \frac{\mw}{\mz} \right)^{2} = \left( 1 - \sin^{2}\theta_{W}\right) \left( 1 + \Delta\rho \right)
\end{equation}
where $\sin\theta_{W}$ is the weak mixing angle and $\Delta\rho$ is $0$ at tree level and non-zero once quantum loop corrections are included.  The dominant quantum corrections are quadratically dependent on the top-quark mass and logarithmically dependent on the Higgs boson mass, $\Delta\rho = f(m^{2}_{t}, \ln \mh)$.  The experimental program then consists of measuring \mw, \mz, and \mt\ as precisely as possible in order to constrain \mh\ - detailed discussions can be found in~\cite{PDG,gfitter}.  The experiments at LEP and SLAC precisely determined \mz, the Tevatron and LEP2 experiments precisely determine \mw, and the Tevatron experiments alone precisely determine \mt.  It is important 
to measure \mt\ in all the different top decay channels since BSM contributions can affect them differently~\cite{Kane:1996ny}.

For a free particle, the physical mass is usually defined as the pole of its renormalized propagator. However this definition is ambiguous when dealing with colored particles like quarks.  Thus the definition of the top-quark pole mass is intrinsically ambiguous on the order of $\Lambda_{QCD}$ due to non-perturbative QCD effects~\cite{Bigi:1994em,Beneke:1994sw,Smith:1996xz}.  For an unstable particle (like the top quark), the pole mass enters the description of the resonance through a Breit-Wigner function.  Other mass definitions exist such as the $\overline{\rm MS}$-mass using the $\overline{\rm MS}$ renormalization scheme~\cite{Bardeen:1978yd} or those discussed in~\cite{Hoang:2008xm}.   The $\overline{\rm MS}$-mass is only sensitive to short distance QCD effects and is often used to describe the mass of the light quarks when the typical energy of the process is much larger than the quark mass in question.  At the Tevatron and LHC, the top-quark mass will be directly measured through the reconstruction of its decay products and the measured mass is taken to be the pole mass. However, aside from the theoretical ambiguity mentioned above, additional ambiguities are introduced since the measurements are calibrated using MC generators that include model dependent descriptions of the parton shower and hadronization processes. These affect the experimentally reconstructed top-quark mass in a model dependent way and introduce a set of modeling related systematic uncertainties that are of the order $\Lambda_{QCD}$ or even larger~\cite{Buckley:2011ms}.

Aside from having to isolate a relatively pure sample of \ttbar\ events, there are a few experimental challenges to making a precise determination of \mt.  The most sensitive analyses fully reconstruct the \ttbar\ kinematics using a kinematic fit or a matrix element technique.  In order to do so it is necessary to assign each final state jet as having originated from a particular final state quark.  As discussed in Sec.~\ref{sec:kfit}, this gives rise to a combinatoric ambiguity that dilutes the resolution of the reconstructed \ttbar\ kinematics.   The resolution with which the \ttbar\ kinematics can be reconstructed is further diluted by the addition of ISR/FSR jets, the loss of quark jets due to limited geometric acceptance, and jet algorithm merging/splitting effects.  Moreover, large systematic uncertainties can arise from uncertainties in the jet energy corrections.  Roughly, each $1\%$ of uncertainty in the jet energy corrections gives a $1$~\gevcc\ uncertainty in the top-quark mass.  This was a limiting systematic uncertainty for early \mt\ measurements.  With larger statistics samples the jet energy corrections can be calibrated {\it{in situ}} by constraining the corresponding jet-jet invariant mass to the world average \mw\ in events containing a $W\ra qq^{\prime}$ decay.  Doing so significantly reduces the associated systematic uncertainty, which then scales with the statistics of the \ttbar\ sample itself.  Most measurements in the $ljt$ and $had$ channels now take advantage of this {\it{in situ}} calibration by performing simultaneous fits to \mt\ and the jet energy scale.  Other measurements seek to identify experimental observables that are only weakly dependent on the jet energy corrections.

For all analyses described below it is necessary to calibrate the measured \mt\ using MC pseudo-experiments (ie. ensemble testing) in order to quantify the influence of the simplifications made in the employed techniques.  Each pseudo-experiment is built from a mix of \ttbar\ signal and background events in the proportions estimated from the relevant cross section analysis. The number of events contributed by each physics process is allowed to vary according to a Poisson distribution.  Once a set of pseudo-events have been assembled, they are then treated just like the data and a determination of \mt\ is made.  Then, from an ensemble of pseudo-experiments, the relationship between the true top-quark mass (ie. the \mt\ value used when generating the \ttbar\ MC sample) and the mean measured \mt\ can be established and, if necessary, corrections can be derived - usually as a linear function of the measured value.  This is done separately for each analysis technique.  The ensemble of pseudo-experiments is also used to study the expected statistical uncertainty and the statistical behavior of each analysis technique.  Once a particular analysis has been calibrated, its performance is checked using MC samples.

Four methods are used at the Tevatron to measure the top-quark mass. They are described below.  For recent detailed reviews on top-quark mass measurements, see~\cite{Fiedler:2010sy,Wicke:2010}. 

% ======================================================================
\subsubsection*{Template Method}
\label{sec:tplate}
% ======================================================================

This traditional method begins by choosing an observable correlated with the top-quark mass.  Distributions of this observable are then constructed using MC samples generated with varying \mt\ as input.  The data distribution is then compared with these MC templates using a maximum likelihood fit.  The observable most correlated with \mt\ is the reconstructed invariant mass of the \ttbar\ decay products, \mtreco, which is the most often used and is estimated using the kinematic fits discussed in Sec.~\ref{sec:kfit}.  If a given analysis uses the hadronic \Wboson\ decay to calibrate the jet energy corrections, an observable correlated with \mw\ is used - usually the invariant mass of the associated jets, $m_{jj}$ - and the maximum likelihood fit is extended to two dimensions.  The template method is relatively simple and can be easily extended to several decay channels or sub-samples.   The observable(s) used as estimators can be chosen to minimize sensitivity to specific systematic uncertainties.  The statistical sensitivity of this method is sometimes worse than other methods discussed because it does not use the full event information, nor take advantage of event-by-event differences to weight more heavily those events with kinematics resulting in improved \mtreco\ resolution.

\begin{figure}
\centering
  \epsfxsize=6cm\epsfbox{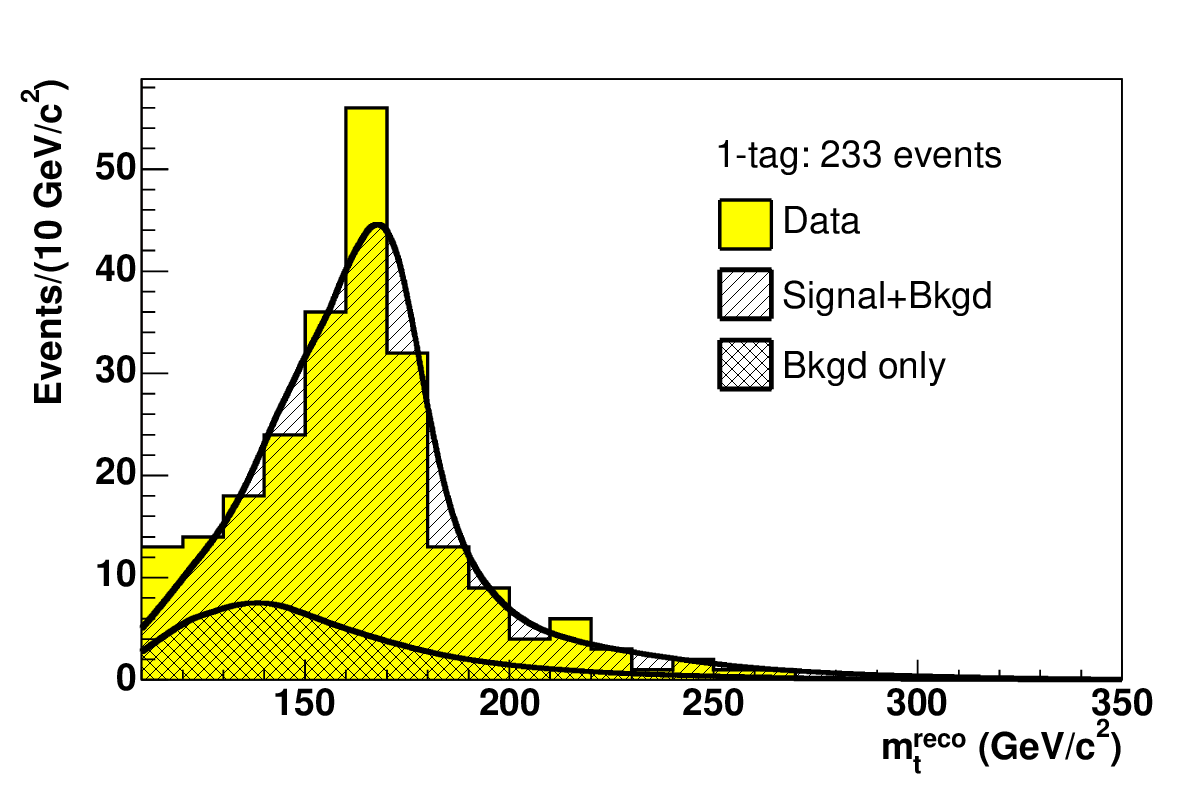}\hspace{0.5cm}
  \epsfxsize=6cm\epsfbox{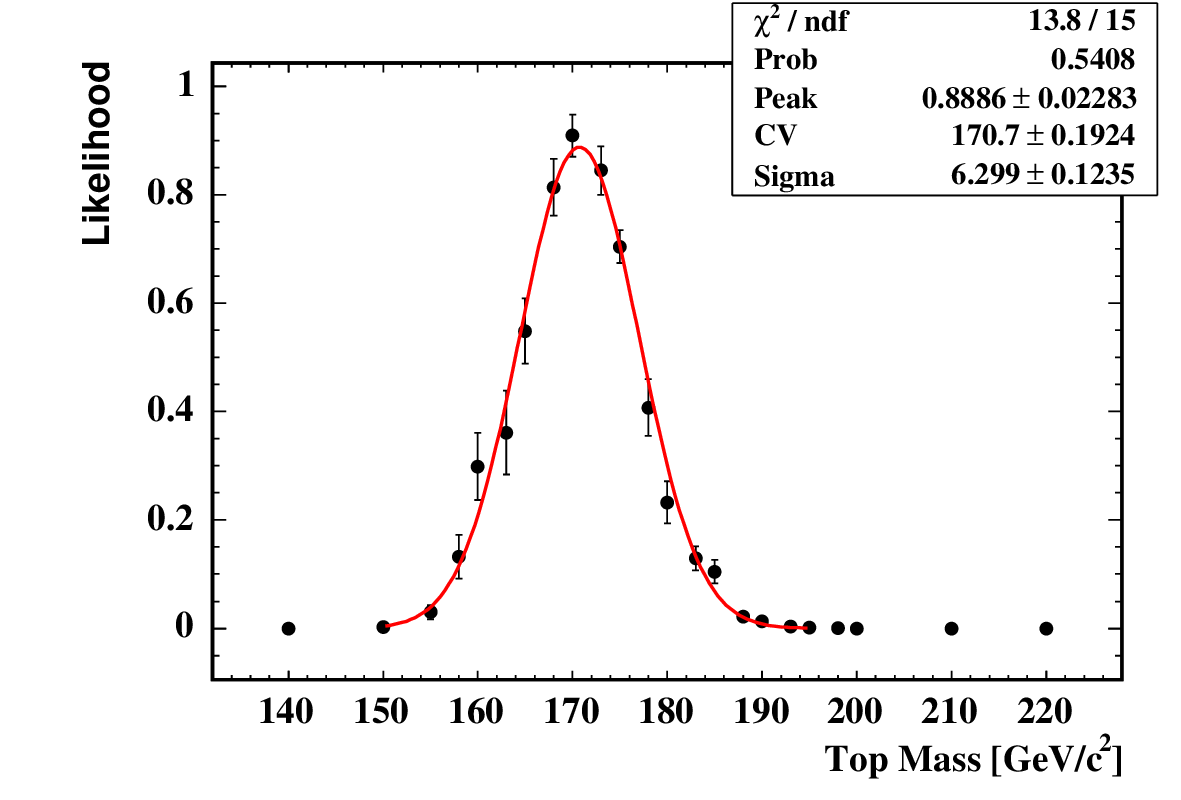}\hspace{0.5cm}
\caption{The \mtreco\ distribution (left) in the CDF $ljt$ channel with the signal and background template estimates overlaid using $\mt = 172$~\gevcc\ and the nominal jet energy corrections for events with 1 identified $b$-jet~\cite{Aaltonen:2008gj}.  The likelihood curve from the data fit (right) in the $ljt$ channel using $L_T$ and lepton $E_T$ as top-quark mass estimators~\cite{Aaltonen:2009hd}.
\label{fig:mass_template_ljets}}
\end{figure}

A detailed description of using the template method to measure \mt\ in the $ljt$ channel can be found in~\cite{CDFMtTplate2006}.  The most recent published result uses $1.9$~\fb\ of CDF data and simultaneously fits the $ljt$ and $dil$ channels using the {\it{in situ}} $W\ra qq^{\prime}$ decays to constrain the jet energy corrections and reduce the associated systematic uncertainties~\cite{Aaltonen:2008gj}.
In the $ljt$ channel a kinematic fit is used to determine $(\mtreco, m_{jj})$ for each event.  Since the \ttbar\ kinematics are under constrained in the $dil$ channel, a neutrino weighting algorithm described below is used to determine \mtreco . It is used together with $H_T$ in this particular analysis.   Two dimensional templates are constructed for each channel and a joint likelihood fit is performed to measure \mt.  Templates are built from simulated samples of \ttbar\ and background events using MC generated at discrete values of \mt.  The shape of the templates at these discrete points is described using kernel density estimators~\cite{Scott:1992,Silverman:1998} and then smoothed and interpolated using the method of~\cite{Loader:1999} to enable an estimate of the template shapes for any arbitrary value of \mt.  
The data in each channel is further divided into sub-samples of varying signal purity and \mtreco\ resolution using the number of reconstructed jets and the number of identified $b$-jets in the events.  A full set of templates are created for each sub-sample.  The two dimensional data distributions are compared to the resulting templates using an unbinned maximum likelihood fit to determine \mt, the best-fit jet energy scale correction, and their associated uncertainties.  The resulting top-quark mass is $\mt = 171.9 \pm 1.7 \ ({\rm stat + JES}) \pm 1.1 \ (\syst)$~\gevcc\ for the simultaneous fit of the $ljt$ and $dil$ samples.  The dominant systematic uncertainties arise from residual uncertainties in the jet energy corrections\footnote{The measurement treats the jet energy scale as a constant.  The residual jet energy scale systematic accounts for the known variations of these corrections as a function of jet $E_T$ and $\eta$.} and from variations of the MC generator used to model \ttbar\ events.  An example of one of the \mtreco\ distributions is shown in Fig.~\ref{fig:mass_template_ljets}.  CDF has also used the template method to determine \mt\ from a sample of $ljt$ events in which the $b$-jets are identified using one of the soft lepton tagging algorithms described in Sec.~\ref{sec:bjet}~\cite{Aaltonen:2009zi}.

CDF also employs the template method to determine \mt\ using observables with minimal sensitivity to the jet energy corrections~\cite{Aaltonen:2009hd,Aaltonen:2009zi}.  Since the transverse momenta of the top-quark decay products depend approximately linearly on \mt, the decay length of the $b$-hadron measured in the transverse plane, $L_T$, or the $E_T$ of the lepton from the \Wboson\ decays can be used to build templates.
These observables are almost completely independent of the jet energy corrections. Small dependencies remain due to uncertainties in \ttbar\ and background acceptances that arise from changes in the event selection efficiencies when the jet energies are varied.  The shape of the $L_T$ distribution is calibrated using di-jet data samples enriched in $b\overline{b}$.  The lepton $E_T$ scales are calibrated as discussed in Sec.~\ref{sec:emu}.  The two-dimensional $L_T$ and lepton $E_T$ distribution is used to determine \mt\ and the resulting likelihood curve is shown in Fig.~\ref{fig:mass_template_ljets}.
Using 1.9~\fb\ CDF measures with these observables  $\mt = 170.7 \pm 6.3 \ (\stat) \pm 2.6 \ (\syst)$~\gevcc . The measurement will clearly benefit from more statistics and has a very low systematic uncertainty coming
from jet energy calibration. However it suffers from relatively large systematic uncertainties arising from the $L_T$ and $E_T$ calibration.  The precision of these calibrations should improve as the statistics of the relevant control samples increase.

A detailed description of using the template method to measure \mt\ in the $dil$ channel can be found in~\cite{CDFMtDil2006}.  Recent measurements use essentially the same methodologies on significantly larger data sets.   As described in Sec.~\ref{sec:kfit}, since the kinematics of the $dil$ events is under constrained it is necessary to make an additional kinematic assumption in order to obtain a reconstructed invariant mass for each event.  The CDF and \dzero\ template analyses in the $dil$ channel primarily differ in the additional kinematic assumption they choose to make.

CDF's most recent template analyses in the $dil$ channel use the neutrino $\phi$ weighting method~\cite{Aaltonen:2009tza,Aaltonen:2010Mt2}, which integrates over the azimuthal angles for the two neutrinos in each event.   A $\chi^2$ function is minimized to estimate \mtreco\ for each event.  This $\chi^2$ function includes a term that constrains the measured quantities within their uncertainties to the assumed \ttbar\ kinematics. 
Templates of the \mtreco\ distributions are obtained from simulated samples of \ttbar\ signal and background events generated at discrete values of \mt.  The shape of the templates is parameterized and interpolated to enable an estimate of the template shape at any arbitrary \mt\ value.  The top-quark mass is determined using an unbinned likelihood fit to the data.  In one of these analyses~\cite{Aaltonen:2010Mt2} the fit to the \mtreco\ distribution is extended to include information from the $m_{T2}$ variable suggested in~\cite{Lester:1999, Barr:2003}.
CDF also developed a template based measurement where the \mt\ dependent SM \ttbar\ cross section is  included as an additional constraint~\cite{Aaltonen:2007jw}.

The most recent \dzero\ template analyses in the $dil$ channel use a neutrino weighting method and the so called matrix weighting method~\cite{Abazov:2009eq}.  The neutrino weighting algorithm employed integrates over the neutrino rapidities as the assumed input to the kinematic fit.  Weights are assigned by comparing the resulting neutrino momenta solutions  to the measured \met . The first two moments of the weighted distributions are used from each event to build \mtreco\ templates and extract from the data distribution the most probable top-quark mass.  The templates are constructed in two different manners, using binned probability density histograms and using probability density fit functions, which are later combined.   The matrix weighting technique integrates over assumed top-quark masses and solves for the $t$ and $\overline{t}$ momenta. Each solution is then weighted by the probability to measure 
the observed lepton energy in the top-quark rest frame given the assumed top-quark mass using the matrix element based expressions in~\cite{Dalitz:1991wa}. For each event \mtreco\ is taken from the maximum of the resulting weighted distribution and used to build templates.

Using 3.4~\fb, CDF uses a combination of the \mtreco\ and $m_{T2}$ distributions in the $dil$ channel to measure~\cite{Aaltonen:2010Mt2}: $\mt = 169.3 \pm 2.7 \ (\stat) \pm 3.2 \ (\syst)$~\gevcc.   Using 1~\fb, \dzero\ combines the results from the neutrino weighting and the matrix weighting techniques in the $dil$ channel to measure~\cite{Abazov:2009eq}
$\mt = 174.7 \pm 4.4 \ (\stat) \pm 2.0 \ (\syst)$~\gevcc.
These measurements are still limited by statistics.  The systematic uncertainty is completely dominated by the uncertainty in the jet energy corrections as no {\it{in situ}} jet energy scale calibration is possible in the $dil$ channel.

A detailed description of using the template method to measure \mt\ in the $had$ channel can be found in~\cite{Aaltonen:2010pe} in which CDF performs a simultaneous measurement of the \ttbar\ cross section and the top-quark mass using a NN based event selection and requiring at least one identified $b$-jet.  A kinematic fit is used and all possible jet-parton assignments consistent with the $b$-jet information are considered. The combination yielding the smallest fit $\chi^2$ is used to determine \mtreco , $m_{jj}^{W+}$, and $m_{jj}^{W-}$ for each event.  Templates are built from the \mtreco\ distribution and from the $m_{jj}$ distribution for a set of MC samples generated at discrete \mt\ and jet energy correction values.   These templates are fit to a Gamma plus Gaussian function, and the resulting parameters are smoothed and interpolated to enable an estimate of the template shape at any arbitrary \mt\ value.  The top-quark mass is determined from an unbinned likelihood fit to the data that includes an {\it{in situ}} constraint on the jet energy corrections using the $m_{jj}$ distributions. 
Using 2.9~\fb, CDF measures $\mt = 174.8 \pm 2.4 \ ({\rm stat+JES}) ^{+1.2}_{-1.0} \ (\syst)$~\gevcc.  The measurement is limited by statistics and the dominant source of systematic uncertainty arises from residual jet energy correction uncertainties and in method related systematics.

% ======================================================================
\subsubsection*{Matrix Element Method}
\label{sec:ME}
% ======================================================================

The matrix element method is a more sophisticated method for measuring a particle property and constructs a likelihood curve for each event by comparing the observe kinematics to those expected as a function of the particle property of interest (e.g. \mt ).  The event probability density is estimated using a leading order matrix element and integrating over the unmeasured quantities.   Detector effects are incorporated by integrating over resolution functions.  The total likelihood for a given sample of events is obtained from the product of the individual event probability densities.   Although it will not be discussed separately, it should be noted that the dynamic likelihood method is very similar to the matrix element method described here.  Of all the methods explored at the Tevatron for determining \mt, the matrix element method offers the best statistical sensitivity since it uses the full kinematic information available in each event and since it effectively gives more weight to events whose kinematics afford a more precise estimate of \mtreco\ by virtue of having a narrower event probability density.  The principal downside is that it is enormously CPU intensive, typically requiring hours of CPU per event.  This limitation renders these analyses less nimble than analyses using the template method.

The method is based on an approach suggested in~\cite{Dalitz:1998zn,Kondo:1993in} and close to the method
used for some of the \Wboson\ mass measurements at LEP~\cite{Berends:1997dm,Abreu:1997ic,Juste:1998rw}. 
%%%The method was first employed by \dzero\ \cite{Abazov:2004cs} to measure the top-quark mass.  
Currently, all the most precise \mt\ determinations use the matrix element method.  It has been applied in all the decay final states.  In the $ljt$ and $had$ channels it is extended to include an integration over the jet energy corrections, which are then constrained {\it{in situ}} using hadronic \Wboson\ decays.  A detailed description of the matrix element technique applied to measure the top-quark mass can be found in~\cite{Fiedler:2010sg}.

The event probability $P_{\rm evt}$ is built from a \ttbar\ probability ($P_{\ttbar}$) and a background probability ($P_{\rm b}$),
\begin{equation}
  P_{\rm evt}(x;\mt,\fttbar) = \fttbar \cdot P_{\ttbar}(x;\mt)+(1-\fttbar) \cdot P_{\rm b}(x) 
\end{equation}
where $x$ denotes the set of observed variables (i.e. the jet and lepton momenta) and $\fttbar$ is the \ttbar\ signal fraction in the event sample.  The signal and background probability densities are constructed by integrating over the appropriate parton-level differential cross section, $d\sigma(y)/dy$, convoluted with parton distribution functions ($k(q)$) and resolution effects.  The resolution effects are described by the transfer functions, $W (x;y)$, which give the probability of observing the set of variables $x$ given the underlying partonic quantities, $y$.  Typically the jet and lepton angles are taken to be exactly measured so that the relevant transfer functions are simply Dirac delta functions.   The transfer functions for the jet energies are parameterized as a function of parton energy and rapidity using fully simulated MC events.  Separate transfer functions are derived for $uds$, gluon, and $b$-jets.  The transfer functions for the lepton energies are usually taken to be Dirac delta functions but are sometimes treated in a manner similar to the jets. It should be noted that the transfer functions are not assumed to be Gaussian, and more sophisticated functions can be used to obtain a more accurate description of the relevant resolutions.  The \ttbar\ probability density is expressed as 
\begin{equation}
  P(x) = \frac{1}{\sigma^{\rm obs}} \sum_{\rm j-p\:comb} 
    \int \sum_{\rm flavors} \frac{d\sigma(y)}{dy} k(q_1) k(q_2) dq_1 
    dq_2 W(x;y) dy
\label{eq:me}
\end{equation}
where the first sum is over all possible jet-parton combinations and the second sum is over all relevant PDF parton flavors.  The probability density is normalized to the total observed cross section, $\sigma^{\rm obs}$, after including event selection effects.  For signal the differential cross section is taken from the leading order matrix element, ${\cal M}(\qqbar \to \ttbar \to y)$, found in~\cite{Mahlon:1997uc}, which depends on \mt, $\frac{d\sigma(y)}{dy}\ra\frac{\sigma(\mt,y)}{dy}$.  For background the differential cross section is taken from a sum of matrix elements, calculated by dedicated MC generators, and is independent of \mt.  The background probability density usually includes contributions from just the dominant background processes.  The effect of the resulting approximation on the measured \mt\ is included as part of the systematic uncertainties and is usually found to be quite small.  For analyses that constrain the jet energy corrections using hadronic \Wboson\ decays, the jet energy transfer functions, and thus the event probability densities, are additionally expressed in terms of an overall jet energy scale factor, JES, $W(x;y)\ra W(x;y,{\rm JES})$.

\begin{figure}
\centering
  \epsfxsize=6cm\epsfbox{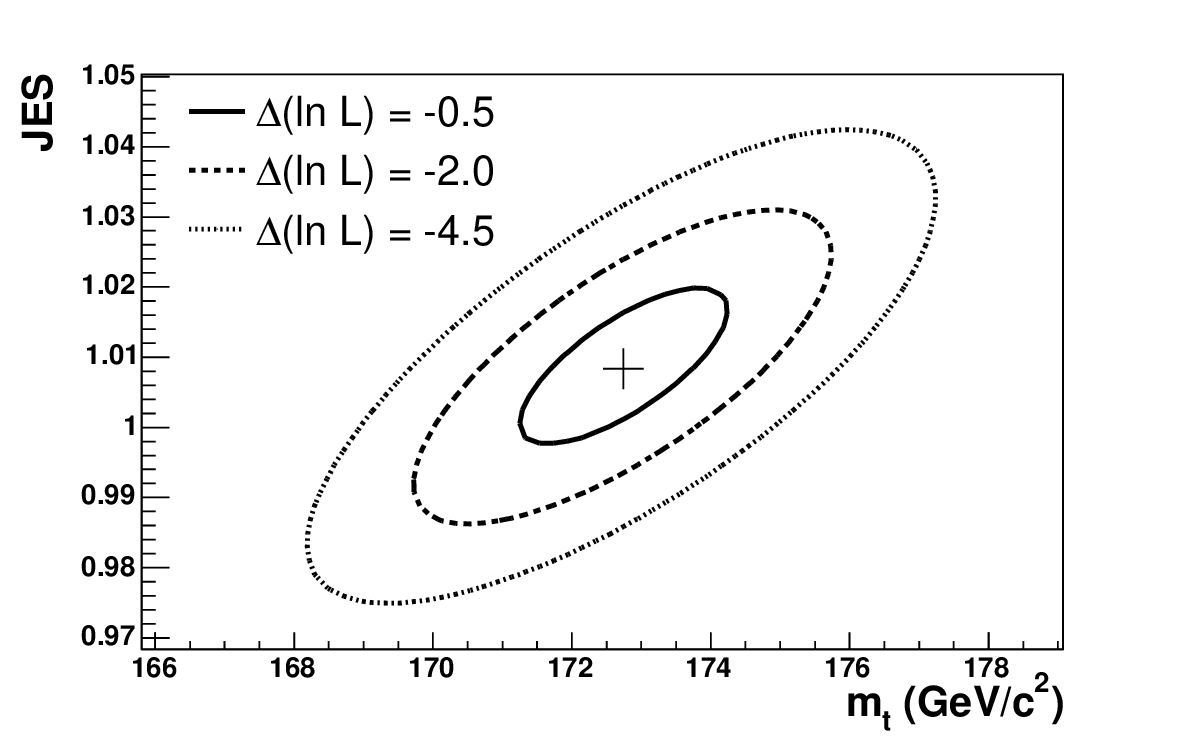}\hspace{0.5cm}
  \epsfxsize=4cm\epsfbox{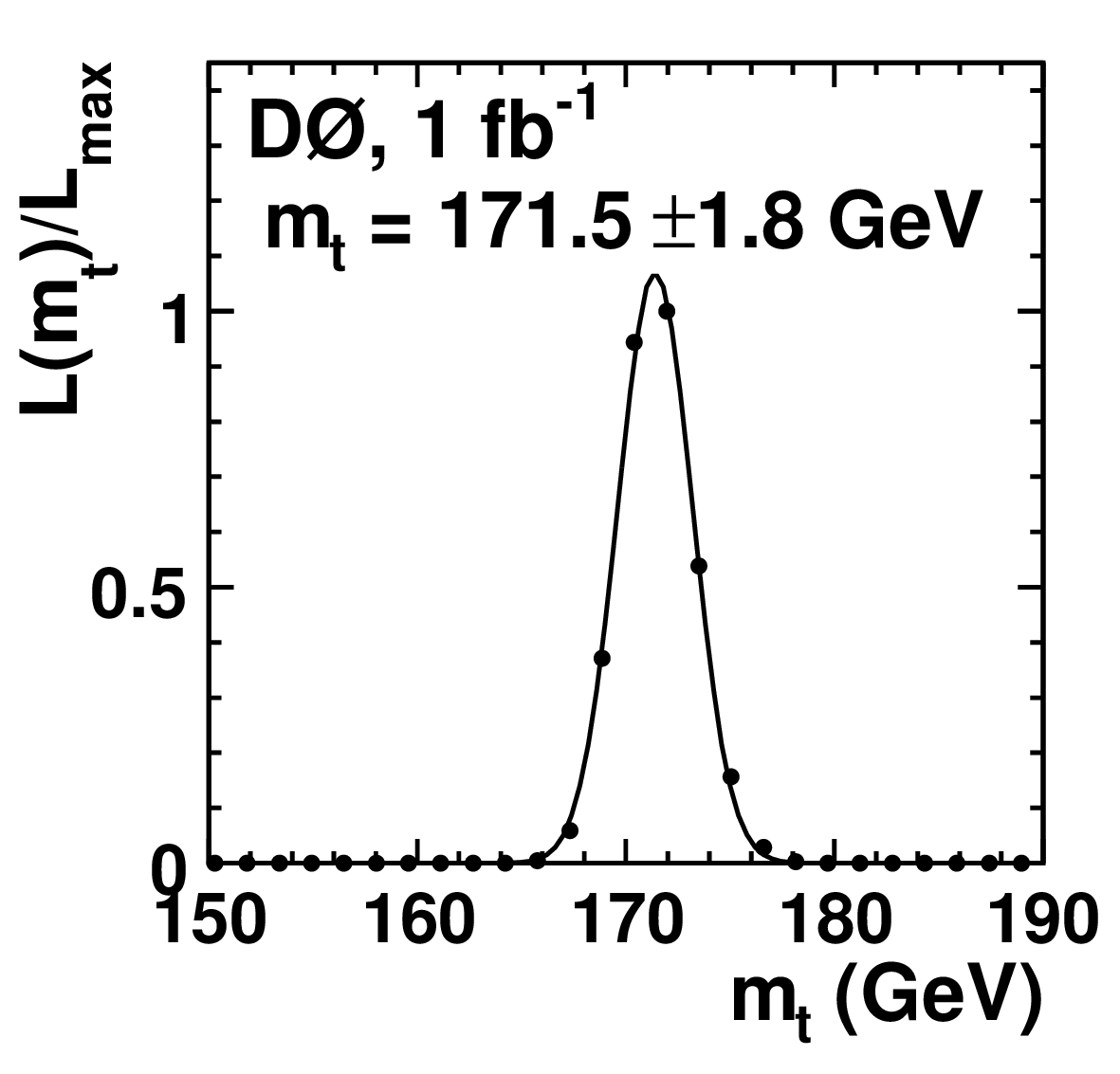}\hspace{0.5cm}
\caption{Results from \mt\ measurements in the $ljt$ channel using the matrix element method.  The 2-D likelihood (left) from CDF data and the data likelihood projected onto the \mt\ axis (right) from \dzero\ 
\cite{Aaltonen:2008mx, Abazov:2008ds}.
\label{fig:mass_me_ljets}}
\end{figure}

Both CDF and \dzero\ have recently published measurements of the top-quark mass in the $ljt$ channel using the matrix element technique.  In their most recent analysis~\cite{Aaltonen:2008mx} CDF also takes into account the finite angular resolution of the jets. 
%
%%%by replacing the Breit-Wigner terms in the matrix element with %%%``effective propagators'' estimated from MC by replacing the parton %%%direction with th direction of their corresponding reconstructed jet. 
%
In this analysis the gluon fusion diagrams are also included in the \ttbar\ matrix element and the jet-parton combinations are weighted using information from a $b$-jet identification algorithm.  A NN discriminant is used to separate \ttbar\ from background events rather than the matrix elements for the background processes.  Events with very small likelihoods are vetoed in order to remove poorly behaved \ttbar\ events arising, for example, from the presence of ISR/FSR jets.  The jet energy corrections are constrained using the $W\ra qq^{\prime}$ decays.
The method is calibrated using MC samples generated at various values of \mt\ and assuming different JES factors.  Using $1.9$~\fb\ of data CDF measures
$\mt = 172.7 \pm 1.8 \ ({\rm stat+JES}) \pm 1.2 \ (\syst)$~\gevcc\
(see Fig.~\ref{fig:mass_me_ljets}).  The dominant systematic uncertainty arises from difference between MC generators.

The \dzero\ analysis~\cite{Abazov:2008ds} includes only the $q\overline{q}\ra\ttbar$ component in the signal matrix element and calculates the background probability density using only the matrix element for the $W+4$~partons process.  The jet-parton combinations are weighted using information from a $b$-jet identification algorithm and an {\it in-situ} JES calibration is performed. The method is calibrated using MC samples generated at various values of \mt\ and assuming different JES values.  Using 1~\fb\ of data \dzero\ measures
$\mt = 171.5 \pm 1.8 \ ({\rm stat+JES}) \pm 1.1 \ (\syst)$~\gevcc\ (see Fig.~\ref{fig:mass_me_ljets}).  The dominant systematic uncertainty arises from uncertainties associated with calibrating the response of the calorimeter to $b$-quark jets.

CDF has also published a measurement using the matrix element technique in the $dil$ channel~\cite{Aaltonen:2008bd}.
In this channel, no JES {\it in-situ} calibration is possible.
In this analysis, the event selection has been optimized to get the best statistical precision on \mt\ using neuroevolution~\cite{Stanley:2002zz} to train a NN discriminant.   The signal matrix element includes only the $q\overline{q}\ra\ttbar$ component.  Background probability densities are computed using the matrix elements for the $\zg \to ee,\mu\mu$+jets, $W+\ge 3$~jets where a jet is misidentified as a lepton, and $WW$+jets processes.  Events with or without identified $b$-jets are fit separately.  Using 2~\fb, CDF measures $\mt = 171.2 \pm 2.7 \ (\stat) \pm 2.9 \ (\syst)$~\gevcc.  The dominant systematic uncertainty is due to uncertainties in the jet energy corrections.

% ======================================================================
\subsubsection*{Ideogram Method}
% ======================================================================

The ideogram method can be thought of as an approximation to the matrix element method.   A kinematic fit is used to determine \mtreco\ for each event.  A per event probability density is calculated as a function of \mt\ by calculating the probability of observing \mtreco\ assuming the true top-quark mass is \mt\ and knowing the resolution $\sigma_{\mtreco}$.   All jet-parton combinations are considered for each event and each carries a weight derived from its corresponding fit $\chi^2$.  Typically, additional weights are included that use information from $b$-jet identification algorithms.  Like the matrix element method the event probability density is built from a signal and a background piece.  The signal probability density is a convolution of a Gaussian with a Breit-Wigner while the background probability density is taken from MC simulation.  The total likelihood for a given sample of events is obtained from the product of the individual event probability densities.  A JES constraint can be incorporated by repeating the kinematic fits for different assumptions on the jet energy corrections scale factor.  Like the matrix element method, the ideogram method offers improved statistical sensitivity.  In terms of sophistication, CPU budget, and statistical sensitivity the ideogram method typically falls between the template and matrix element methods.  It should be noted that since it requires a full reconstruction of the \ttbar\ kinematics, this method can only be used in the $ljt$ and $had$ final states.

\dzero\ has performed a measurement using this technique in the $ljt$ channel~\cite{Abazov:2007rk}.  The event probability density is factorized into the product of two separate probabilities for signal and background each.  The first probability depends on \mt\ and JES and provides the necessary information to constrain  the jet energy corrections and to measure the top-quark mass.  The second probability depends on the output of a multivariate discriminant designed to separate \ttbar\ from background processes.  The variables used in the discriminant are chosen and the discriminant itself is constructed to be uncorrelated with \mt\ and JES.  This second probability helps constrain the observed \ttbar\ fraction in the event sample and to de-weight events that have kinematics more consistent with having originated from background processes.   MC simulation is used to account for contributions from wrong jet-parton assignments and background events.  Using 0.43~\fb, \dzero\ measures $\mt =173.7 \pm 4.4 \ ({\rm stat+JES})^{+2.1}_{-2.0} \ (\syst)$~\gevcc.  The dominant systematic uncertainty arises from uncertainties associated with calibrating the response of the calorimeter to $b$-quark jets and from uncertainties associated with modeling the \ttbar\ signal.

The method was used by CDF to produced a result in the $had$ channel using $310$~\pb~\cite{Aaltonen:2006xc}.  The result was limited by the uncertainties in the jet energy corrections.  As described above subsequent \mt\ determinations in the $had$ channel exploit $W\ra qq^{\prime}$ decays to reduce the jet energy correction systematic and also include significantly larger data sets.

% ======================================================================
\subsubsection*{Using the \ttbar\ Cross Section}
\label{sec:MtFromXS}
% ======================================================================

As mentioned above, there is some ambiguity associated with the theoretical interpretation of the \mt\ parameter measured by the above techniques.   Assuming SM production and decay, an estimate of the top-quark mass can also be made by comparing the measured production cross section to theory predictions.  While less precise, this method has the advantage that the \mt\ parameter in the predictions is theoretically well defined. 

\begin{figure}
\centering
\epsfxsize=6cm\epsfbox{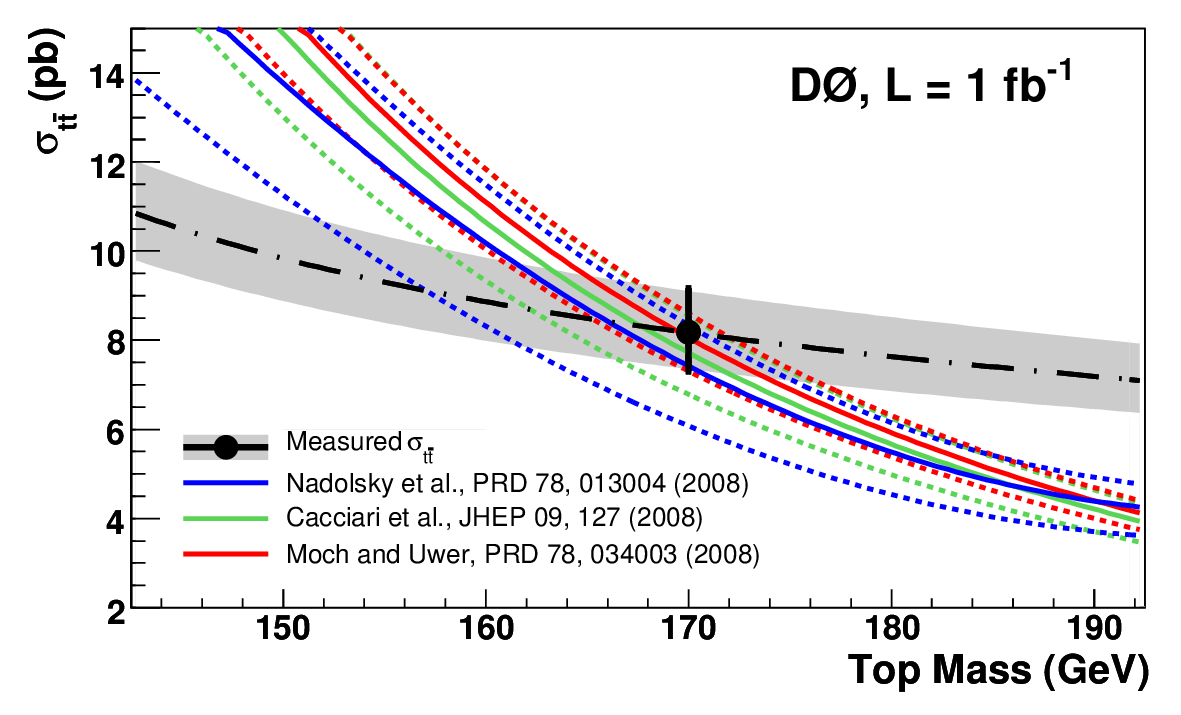}\hspace{0.5cm}
\caption{ \dzero\ measurement of the top-quark mass determined by comparing the measured and the SM predicted \ttbar\ cross section as a function of \mt\ \cite{Abazov:2009ae}.
\label{fig:xsvsmt}}
\end{figure}

\dzero\ has performed this measurement~\cite{Abazov:2009ae}. The signal acceptance is estimated as a function of \mt\ using MC samples generated at various \mt\ values.  The resulting acceptances are smoothed and the production cross section is then measured as a function of \mt\ as shown in Fig.~\ref{fig:xsvsmt}.  The \mt\ dependence of the acceptance arises from the event selection lepton and jet $E_T$ requirements.  The measured cross section is compared to several NLO predictions~\cite{Nadolsky:2008zw,Beenakker:1988bq,Cacciari:2008zb,Moch:2008qy} that define \mt\ as the pole mass.  A normalized likelihood function is formed as a function of \mt\ by comparing the theory prediction with the measured cross section at each assumed top-quark mass.  The uncertainty in the theory arising from variations in PDFs and in the renormalization and factorization scales and the uncertainty in the measured cross section are accounted for in the comparison.
Using~\cite{Moch:2008qy} and the combined $ljt+dil$ measured cross section from 1~\fb\ of data, \dzero\ measures
$\mt = 169.1^{+5.9}_{-5.2}$~\gevcc. This result is in agreement with the direct top-quark mass measurements.  
Using the same theoretical predictions but an expected experimental uncertainty on the \ttbar\ cross section of 6\% at the end of the Tevatron run, a top-quark mass uncertainty of around 3.3~\gevcc\ could be achieved with this method. 
%It should be noted that since the uncertainty on \mt\ in this method scales with the uncertainty on the cross section as $\Delta\mt / \mt \approx 5\cdot\Delta\sigma_{\ttbar} / \sigma_{\ttbar}$, it will be difficult to obtain a determination of the top-quark mass with a precision comparable to that obtained using the techniques described above.
%
It should be noted that an extraction of the running mass in the $\overline{\rm MS}$ scheme have also been performed~\cite{Langenfeld:2009wd}.

% ======================================================================
\subsubsection*{Combination}
\label{sec:MtCombo}
% ======================================================================

The world average top-quark mass is obtained from a combination of CDF and \dzero\ most precise \mt\ measurements using up to $5.6$~\pb\ of data.  Five measurements using data taken in an earlier run (1990-1995) and six measurements using data from the ongoing run (2001-) are used.
Work has been performed between the two collaborations to standardize the assessment of systematic uncertainties. 
For the combination a detailed breakdown of the various sources of uncertainty has been established in order to properly account for correlations among the various measurements.  The combination uses the method described in~\cite{Lyons:1988rp}.  The preliminary combined value for the top-quark mass is $\mt = 173.32 \pm 0.56 \ (\stat) \pm 0.89 \ (\syst)$~\gevcc\, \cite{:1900yx} which
corresponds to a total uncertainty of 1.06~\gevcc\ and a relative precision of 0.61~\%. The combination has a $\chi^2$
of 6.1 for 10 degrees-of-freedom indicating good consistency among the measurements across different experiments and different decay channels.  The most recent CDF and \dzero\ measurements in the $ljt$ channel using the matrix element method carry the largest
weights in the combination followed by the CDF template measurement in the $had$ channel.

In the future, the statistical and {\it{in-situ}} JES related uncertainties will continue to decrease with the statistics of the data samples employed.  The systematic uncertainty has roughly equal contributions from residual jet energy correction uncertainties and uncertainties associated with modeling the \ttbar\ signal from variations in MC generators and color reconnection effects. Work is in progress to better understand these sources of systematic uncertainty with the aim of reducing them. A total top-quark mass uncertainty below 1~\gevcc\ is potentially achievable in the near future.
%
%\begin{figure}
%\centering
%\epsfxsize=6cm\epsfbox{tev_mt}\hspace{0.5cm}
%\caption{Summary of the top-quark mass measurements by the CDF and
%\dzero\ collaborations 
%combined to obtain the world average top-quark mass~\cite{:1900yx}.
%\label{fig:mtcombi}}
%\end{figure}

% ======================================================================
\subsubsection{Mass Difference}
% ======================================================================

The triple product of the discrete symmetries of charge conjugation (C), parity conjugation ( P), and time reversal (T) is an exact symmetry of any local Lorentz-invariant quantum field theory and requires that the particle and anti-particle masses be identical. 
However, CPT can be violated in some non-local new physics models~\cite{Colladay:1996iz}.
Tests of CPT have been performed in many sectors with no evidence of CPT violation~\cite{PDG}.
Measuring a $q-\overline{q}$ mass difference is usually difficult since hadronization introduces QCD binding and non-perturbative showering effects.   However, since it decays before hadronizing the top quark provides a unique opportunity to measure a $q-\overline{q}$
mass difference free from these QCD effects~\cite{Cembranos:2006hj}.

\dzero\ performed the measurement of the $t-\overline{t}$ mass difference in the $ljt$ final state using 1~\fb\ of data~\cite{Abazov:2009xq}. This analysis uses the matrix element technique as in~\cite{Abazov:2008ds} to form probability densities for each event as a function of the mass of the $t$ and the $\overline{t}$ and 
the \ttbar\ signal fraction.  SM \ttbar\ production and decay is assumed and the lepton charge is used to determine whether the $\ell\nu b$ invariant mass should be assigned as $\mt$ or $m_{\overline{t}}$.
Many of the systematic uncertainties relevant for the top-quark mass measurements cancel here. The result is  $\mt - m_{\bar{t}} = 3.8 \pm 3.4 \ (\stat) \pm 1.2 \ (\syst)$~\gevcc\ \cite{Abazov:2009xq},
consistent with CPT invariance. 

A new CDF measurement using the template method in the $ljt$ channel with 5.6~\fb\ of data finds
$\mt - m_{\bar{t}} = -3.3 \pm 1.4 \ (\stat) \pm 1.0 \ (\syst)$~\gevcc, also consistent with CPT invariance~\cite{CdfMtMtbar56}.

% ======================================================================
\subsubsection{Charge}
% ======================================================================
The electric charge is a fundamental property of the top quark predicted in the SM to be $+2/3$ in units of the electric charge of the electron.
An experimental determination of the top-quark electric charge serves as a check of SM consistency.  In all the \ttbar\ analyses so far discussed the assumption is made that $t\ra W^+ b$ and $\overline{t}\ra W^- \overline{b}$.  However, since none of these analyses have attempted to differentiate a $b$-jet from a $\overline{b}-jet$, it is possible that what is being observed is an exotic top-like particle that decays to $W^+ \overline{b}$ and $W^- b$.  Just such a particle can be found in~\cite{Chang:1998pt,Choudhury:2001hs}, a four generation model that postulates a $-4/3$ charged quark that decays to $W^- b$. In this model, the right-handed $b$-quark mixes with a $-1/3$ charged heavy quark $Q_1$. Its weak isospin partner $Q_4$ has charge -4/3, a mass around 172.5~\gevcc\ and would be the particle observed at the Tevatron
while the left-handed real top quark would have a mass around 270~\gevcc\ and would thus far have escaped detection.  It is worth noting that the precision electroweak data can be satisfactorily described by this model.

The electric charge of the top quark can be determined from measurements of the electromagnetic coupling strength using $\ttbar\gamma$ events or by reconstructing the electric charges of the top-quark decay products.
The production cross section for $\ppbar \to \ttbar \gamma$ is proportional to the square of the top-quark electric charge. However this measurement suffers from a large irreducible background coming from processes where the photon is radiated from the $b$-quark or the $W$ boson. Both CDF and \dzero\ use the second approach to measure the top-quark electric charge in the $ljt$ channel. A kinematic fit is used to fully reconstruct the \ttbar\ kinematics and choose particular jet-parton and $b$-jet pairings (ie. is the $b$-quark jet paired with the leptonically decaying $W$, $\ell\nu b$, or the hadronic decaying $W$, $qq^{\prime} b$).  The flavor of the $b$-jets must then be specified by determining whether the jet originated from a $b$-quark or a $\bar{b}$-quark.  The sum of the electric charges of the charged lepton and the appropriate electric charge of the $b$-jet for the $\ell\nu b-{\rm jet}$ pairing is used as an estimate of the top-quark charge.

CDF requires that at least one jet is identified as a $b$-jet using an algorithm that exploits the long lifetime of $b$-hadrons.  A kinematic fit is employed and the jet-parton and $b$-jet pairings are taken from the permutation yielding the lowest fit chi-squared.  This corresponds to the correct assignment about 76\% of the time.  Soft lepton taggers identify the flavor of the $b$-jets with a 69\% purity.   The purity of the $b$-jet flavor determination is calibrated using $b\bar{b}$ data and MC events.  Using 2.7~\fb\ the CDF data is consistent with the SM expectation~\cite{Aaltonen:2010js} and excludes the exotic -4/3 charged quark interpretation at 95\% confidence level.

\dzero\ requires that at least two jets are identified as $b$-jets and also uses a kinematic fit to reconstruct the \ttbar\ kinematics and choose the jet-parton and $b$-jet pairings.   Using the permutation with the lowest fit chi-squared yields the correct assignments about 84\% of the time.  A jet-charge algorithm is used to identify the flavor of the $b$-jet.  The optimized estimator of the jet charge is $q_{jet} = ( \sum_i q_i \pt_i^{0.6})/(\sum_i \pt_i^{0.6})$ where 
$i$ runs over all tracks associated with the jet in question. The expected $q_{jet}$ distributions for $b$, $\bar{b}$, c and $\bar{c}$ jets are derived from di-jet data samples enhanced in heavy flavor. 
The charge of the leptonic decaying top quark is reconstructed as the sum of lepton charge and the charge of the $b$-jet 
assigned to this top quark while the charge of the hadronic decaying top quark is computed as the sum of the second $b$-jet minus the lepton charge.  Using 0.4~\fb\ the \dzero\ data is consistent with the SM expectation~\cite{Abazov:2006vd} and excludes the exotic -4/3 charged quark interpretation at the 92\% confidence level.
%
%\begin{figure}
%\centering
%  \epsfxsize=6cm\epsfbox{cdf_charge}\hspace{0.5cm}
%  \epsfxsize=6cm\epsfbox{d0_charge}\hspace{0.5cm}
%\caption{(left) SM and exotic model p-values for the normalized %difference between the number of SM and exotic events at 
%CDF~\cite{Aaltonen:2010js} (right) Measured values for the top-quark 
%charge compared to the SM and the tested exotic model at \dzero~\cite
%{Abazov:2006vd}.
%\label{fig:charge}}
%\end{figure}

% ======================================================================
\subsubsection{Width}
% ======================================================================

The total width of the top quark is precisely determined in the SM once \mt\ is specified.  Because \mt\ is large, so is the top-quark width, $\Gamma_t$, and consequently its lifetime, $\sim 1/\Gamma_{t}$, is extremely short, which precludes measuring the width using reconstructed displaced vertices from $t\ra Wb$ decays.
A measurement of $\Gamma_t$ can also be determined from the width of the \mtreco\ distribution.  In the SM at NLO, neglecting terms of order $m^2_b/\mt^2$, $\alpha^2_s$, and $\alpha^2_s/\pi m^2_W/\mt^2$, the top-quark total width is~\cite{Jezabek:1988iv}:
 \begin{equation}
 \Gamma_t =  \Gamma^0_t  \left( 1 - \frac{m^2_W}{\mt^2} \right)^2 \left( 1 + 2 \frac{m^2_W}{\mt^2} \right) 
 \left[ 1 - \frac{2 \alpha_s}{3 \pi} \left( \frac{2 \pi^2}{3} - \frac{5}{2} \right) \right]
 \end{equation}
where $\Gamma^0_t=\frac{G_F \mt^3}{8 \pi \sqrt{2}}|V_{tb}|^2$ with $G_F$ being the Fermi coupling constant. This yields approximately to $\Gamma_t=1.4$~\gevcc\ for a 172~\gevcc\ top-quark mass.  Deviations from the SM expectation could signal the precense of BSM physics contributions arising, for example, from $t \to H^+b$ decays, anomalous $tWb$ form factors, enhanced rates of $t \to dW^+$ or $t \to sW^+$ decays, or the existence of a fourth-generation $b'$-quark.

\begin{figure}
\centering
  \epsfxsize=4.5cm\epsfbox{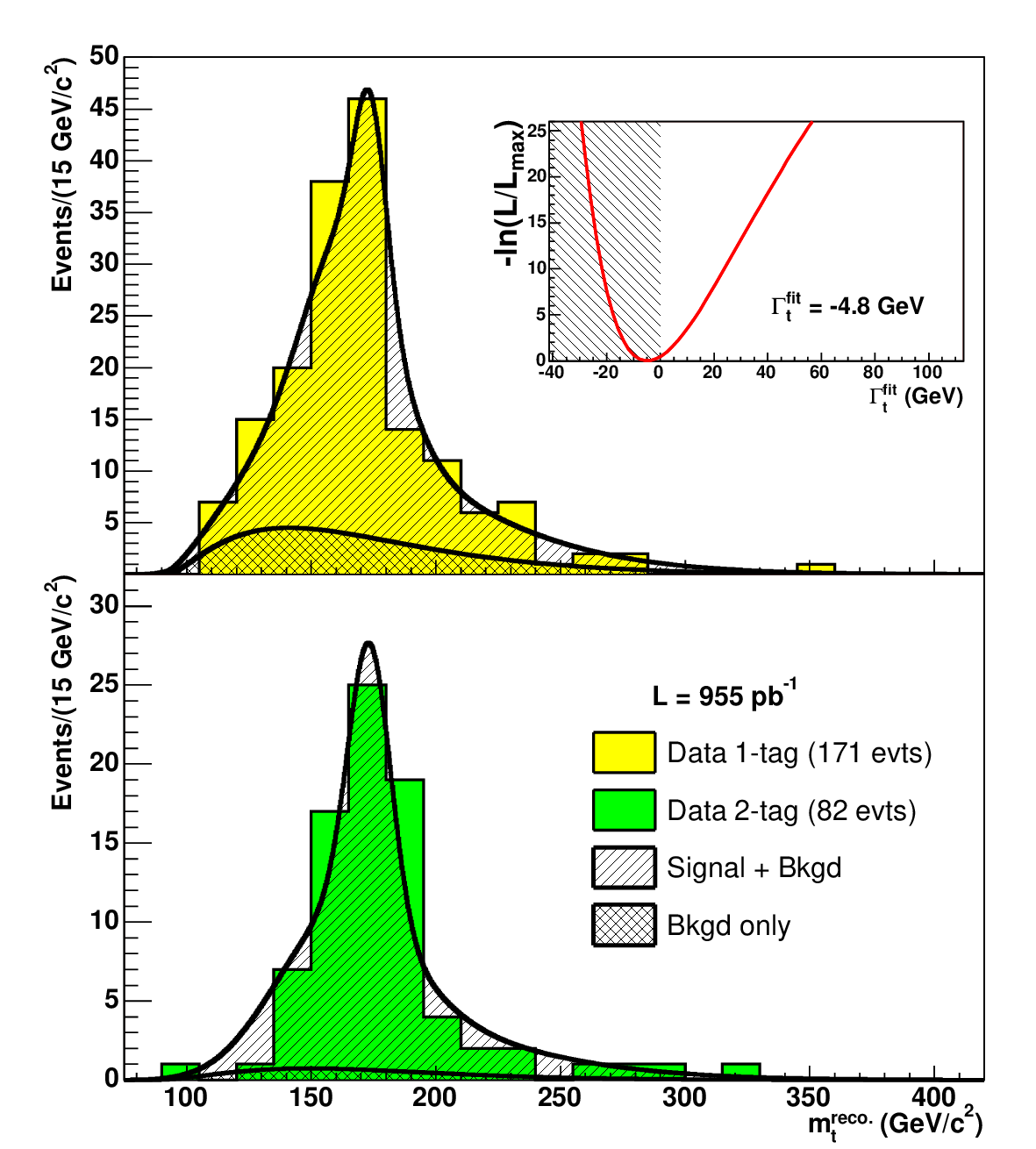}\hspace{0.5cm}
  \epsfxsize=5cm\epsfbox{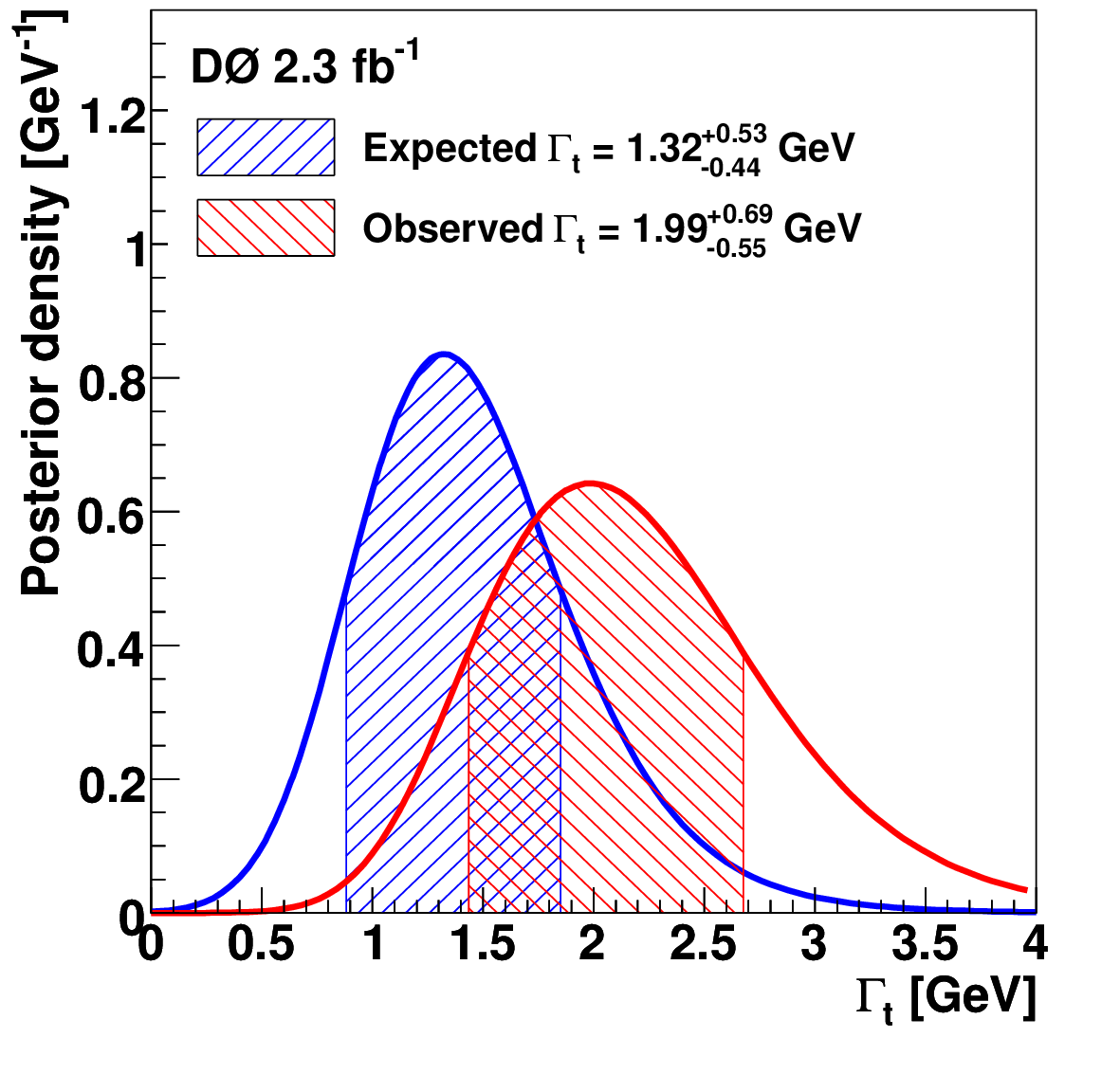}\hspace{0.5cm}
\caption{The \mtreco\ distribution (left) for CDF events with identified $b$-jets overlaid with signal and background distributions and the posterior probability density (right) for the expected and measured $\Gamma_t$ determination from \dzero~\cite{Aaltonen:2008ir, Abazov:2010tm}.
\label{fig:width}}
\end{figure}

CDF uses 1~\fb\ of data in the $ljt$ channel with at least one identified $b$-jets and employs a kinematic fit to obtain the \mtreco\ distribution. The shape of the \mtreco\ distribution is sensitive to the top-quark width and a template method is employed to measure $\Gamma_t$.  The CDF data is shown in Fig.~\ref{fig:width}.  A confidence interval based on the method of~\cite{Feldman:1997qc} is used to determine $\Gamma_t < 13.1$~\gevcc\ at 95\% confidence level.  The dominant systematic uncertainties arise from uncertainties in the jet energy corrections and in modeling the jet energy resolutions~\cite{Aaltonen:2008ir}.

Since the SM top-quark width is far smaller than the experimental resolution on \mtreco\ it is difficult to reach high precision using the direct measurement technique described above. However it is also possible to extract $\Gamma_t$ by determining its partial width $\Gamma(t \to Wb)$ using the measured cross section for the electroweak production of single top quarks and the measured top-quark branching fraction ${\cal B}(t \to Wb)$, $\Gamma_t = \Gamma(t \to Wb)/{\cal B}(t \to Wb)$
~\cite{Yuan:1993ck,Carlson:1995ck,Yuan:1995cm}. This method assumes that the couplings at the $tWb$ production and decay vertices are the same, that the production via flavor changing neutral currents is negligible, and that $|V_{td}|,\: |V_{ts}| << |V_{tb}|$.  The ${\cal B}(t \to Wb)$ is determined from the $R_b$ measurement discussed in Sec.~\ref{sec:production} assuming ${\cal B}(t \to Wq)=1$.  Electroweak single top production can occur via an $s$-channel or a $t$-channel exchange of a $W$ boson. If the $s+t$-channel cross section is used to extract $\Gamma_t$, then the SM ratio ($s$/$t$) is assumed.  To minimize the model dependence of this method, the dominant $t$-channel cross section can be used instead.

\dzero\ uses the 1~\fb\ $R_b$ measurement~\cite{Abazov:2008yn} and a $2.3$~\fb\ $t$-channel single top-quark production cross section measurement~\cite{Abazov:2009pa} to determine the top-quark width using this method. A NLO QCD calculation is used to determine ${\cal B}(t\to Wb)$ from the $t$-channel cross section and the correlated systematic uncertainties have been taken into account.  Using these inputs the top-quark width~\cite{Abazov:2010tm} is measured to be $\Gamma_t = 1.99^{+0.69}_{-0.55}$~\gevcc\ 
%%%which is equivalent to a lifetime of $\tau_t = (3.3^{+1.3}_{-0.9}) %%%%%%%\times 10^{-25}$~s 
%
in agreement with the SM expectation (cf. Fig.~\ref{fig:width}).  This top-quark width measurement can be used to extract a limit on $|V_{tb'}|$, an element of a four generation CKM matrix characterizing the coupling strength of the top quark with a hypothetical fourth-generation $b'$-quark. Assuming $m_{b'}> \mt - m_W$ and the unitarity of the extended CKM matrix, this measurement yields $|V_{tb'}|<0.63$ at 95\% confidence level. 

% ======================================================================
\subsubsection{Spin Correlations}
\label{sec:spincorrelation}
% ======================================================================

At the Tevatron the spins of the $t$ and $\bar{t}$ at production are correlated~\cite{Barger:1988jj, Mahlon:1995zn}.   The top quarks decay on a time scale that is short relative to the hadronization and spin-flip time scale and, hence, the spin correlations are preserved and affect the angular distributions of the top-quark decay products~\cite{Falk:1993rf}. 
If we note $\theta$ the angle between the spin quantization axis and the direction of flight of the $W$-boson decay particle in the rest frame of its parent top quark,
calculations show that the angular correlations are strongest for the charged leptons ($\theta_{\ell}$) or the down-type quarks ($\theta_{d}$) from the decay of the $W$ bosons~\cite{Mahlon:2010gw}.  Experimentally, the two dimensional distribution $(\cos\theta_{\ell},\:\cos\theta_{d})$ or $(\cos\theta_{\ell^+},\:\cos\theta_{\ell^-})$ is used to determine the spin-correlation coefficient, $C$,
\begin{equation}
  \frac{1}{\sigma} \frac{d \sigma}{d \cos \theta_1 d \cos \theta_2} =   
  \frac{1}{4} ( 1 - C \cos \theta_1 \cos \theta_2)
\label{eq:dsigmaspin}
\end{equation}
where $\sigma$ denotes the total \ttbar\ cross section in the relevant decay channel, and the indices $1$ and $2$ refer to the top-quark and antitop-quark decay products respectively.  This relationship has been calculated to order $\alpha_{s}^{3}$~\cite{Bernreuther:2004jv}.  The angles are measured relative to a given axis.  Three common choices are a) the beam basis, which uses the direction of one of the colliding hadrons as the reference axis, b) the helicity basis, which uses the $t$ flight direction as the reference axis, and c) the off-diagonal basis, which uses a reference axis chosen so that the contribution from like-spin $t-\bar{t}$ pairs vanishes at leading order in perturbation theory~\cite{Parke:1996pr}.  The measured coefficient changes as a function of \mttbar\ and depends on the fraction of \ttbar\ pairs produced at threshold and the fraction produced by gluon-fusion.  In the absence of selection or acceptance requirements $C$ can be computed in terms of expectation values~\cite{Bernreuther:2004jv} $C = - 9 < \cos \theta_1 \cos \theta_2 >$, which vary with the choice of basis employed as shown in Table~\ref{tab:spinC}.
\begin{table}
\caption{NLO computation for the spin-correlation coefficient, $C$, discussed in Sec.~\ref{sec:spincorrelation}~\cite{Bernreuther:2004jv} } 
\label{tab:spinC}
\begin{center}
\begin{tabular}{|c|c|c|}
\hline
basis  & dilepton channel & lepton+jets channel \\ \hline
helicity & -0.352 & -0.168 \\
beam   & 0.777   & 0.370 \\
off-diagonal & 0.782 & 0.372 \\ 
\hline
\end{tabular}

\end{center}
\end{table}

The observation of $t-\bar{t}$ spin correlations would confirm that the top quark is a spin $1/2$ particle and that it decays before strong interaction processes affect the $t$ and $\bar{t}$ spins.  Several BSM models could affect the measured coefficient.  The presence of a techni-$\eta$ in two scale technicolor~\cite{Eichten:1994nc} would enhance the unlike spin contributions. 
A new vector particle associated with topcolor~\cite{Hill:1993hs} would affect the spin correlations by changing the fraction of gluon-fusion produced \ttbar\ pairs.  On the decay side, BSM contributions from $t\to H^+ b$ decays, for example, would affect the correlations between the top-quark spin and its decay products and therefore modify the observed angular correlations~\cite{Mahlon:1995zn}.

Both CDF and \dzero\ have performed preliminary measurements of the spin correlation in the $dil$ channel using 2.8~\fb\ and 5.4~\fb\ of data, respectively.  To fully reconstruct the \ttbar\ kinematics CDF uses a likelihood method to determine the best solution for the unknown neutrino momenta using the observed lepton, jet, and \met\ energies and taking into account their relevant resolutions, while \dzero\ uses the neutrino weighting technique described above in Sec.~\ref{sec:mass}.  After reconstructing the \ttbar\ final state, two dimensional $(\cos\theta_{\ell^-},\:\cos\theta_{\ell^+})$ templates are constructed.  The signal templates are built by reweighing \pythia\ \ttbar\ samples using Eq.~\ref{eq:dsigmaspin}.  A maximum likelihood fit to the data, including systematic uncertainties via nuisance parameters, gives
$C = 0.32^{+0.55}_{-0.78}$ for CDF in the off-diagonal basis~\cite{CdfttSpinDil28} and $C=0.10^{+0.45}_{-0.45}$ for \dzero\ in the beam basis~\cite{DzttSpinDil54}.

\begin{figure}
\centering
\epsfxsize=6cm\epsfbox{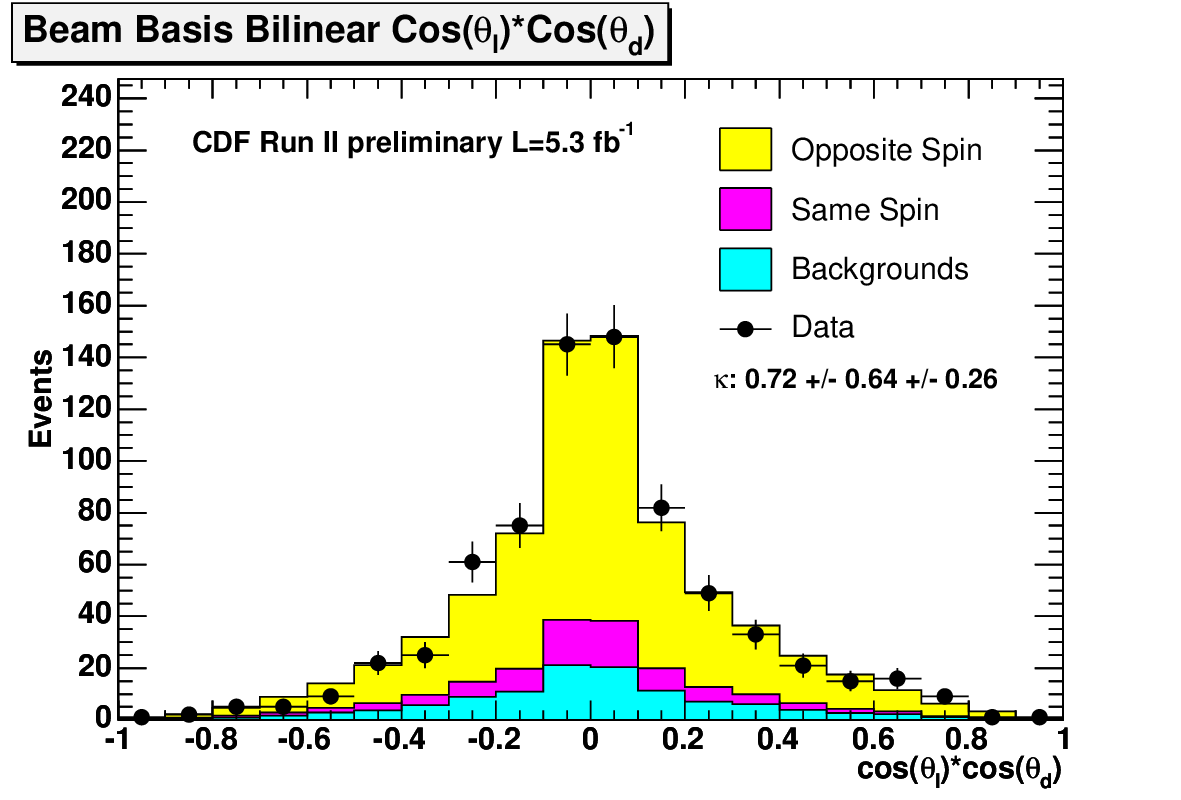}\hspace{0.5cm}
\epsfxsize=6cm\epsfbox{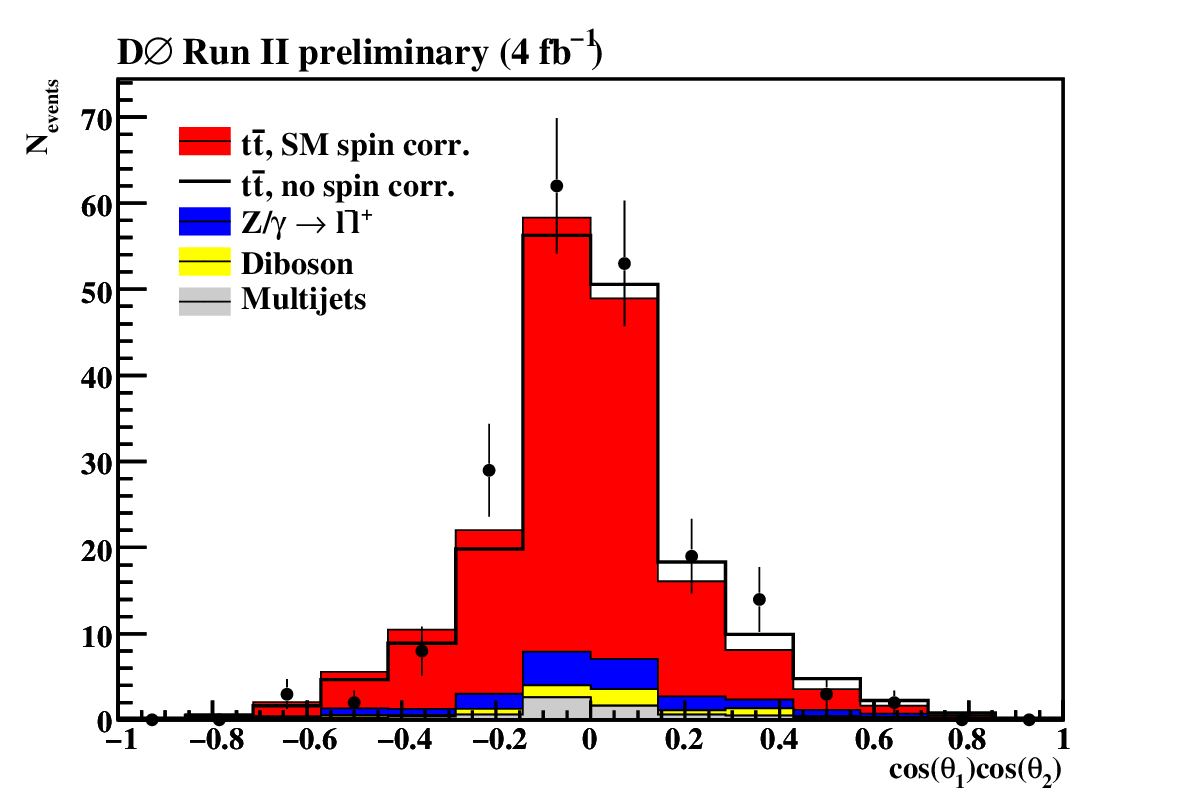}\hspace{0.5cm}
\caption{The CDF $\cos \theta_{\ell}\cdot\cos \theta_{d}$ distribution (left) in $ljt$ events using the beam basis.  The \dzero\ $cos \theta_{\ell^-}\cdot\cos\theta_{\ell^+}$ distribution (right) in $dil$ events using the beam basis.
\label{fig:spin}}
\end{figure}

CDF has also performed a measurement of the \ttbar\ spin correlation in the helicity basis using the $ljt$ channel in 4.3~\fb\ of data.  The difficulty here lies in determining which jet comes from the hadronization of the down-type quark. It is assumed to be the jet closest to the $b$-jet in the \Wboson\ rest frame. This prescription gives the right assignment about 60~\% of the time~\cite{Mahlon:1995zn}.  A template fit to the two dimension distribution $(\cos\theta_{\ell},\:\cos\theta_{d})$ is used to determine 
$C = 0.60 \pm 0.50\ (\stat) \pm 0.16\ (\syst)$~\cite{CdfttSpinLjt43}.  Using the same methodology CDF updated the analysis to use both the helicity and beam bases and to include 5.3~\fb\ of data.  The preliminary results are
$C = 0.48 \pm 0.48 \ (\stat) \pm 0.22 \ (\syst)$ in the helicity basis and $C = 0.72 \pm 0.64 \ (\stat) \pm 0.26 \ (\syst)$ in the beam basis~\cite{CdfttSpinLjt53}.  One dimensional distributions of $cos\theta_{\ell}\cdot\cos\theta_{d,\ell}$ from the \dzero\ and CDF analyses are shown in Fig.~\ref{fig:spin}.

All these results are completely dominated by the statistical uncertainty and will benefit from including the full Tevatron statistics in the future. This measurement will remain interesting at the Tevatron 
since a different spin correlation will be produced at the LHC where \ttbar\ production is dominated by gluon-gluon fusion rather than $q\bar{q}$ annihilation.

% ======================================================================
\section{Direct Searches for New Physics in the Top-Quark Sector}
\label{sec:topsearch}
% ======================================================================

% ======================================================================
\subsection{Search for \ttbar\ Resonances}
% ======================================================================

Many SM extensions predict massive new particles that can decay to a \ttbar\ pair.  For example, the scalar color-singlets predicted in two-Higgs doublets models~\cite{Dicus:1994bm},  the vector color-singlet $Z^{\prime}$ particles postulated in extended gauge theories~\cite{Leike:1998wr}, the $KK_{\rm gluon}$ and $KK_Z$ states of extra dimension models~\cite{Lillie:2007yh}, and the axigluons, colorons, and vector color-octet particles that appear in top-color models~\cite{Sehgal:1987wi, Hill:1993hs}, all couple to \ttbar\ pairs and would affect the \mttbar\ distribution.   In particular, resonances whose natural width is small relative to the experimental resolution would appear as an anomalous peak.  Both CDF and \dzero\ have searched for such peaks using the $ljt$ channel.  In all such searches so far performed at the Tevatron, the natural width of the resonance, generically denoted $Z^{\prime}$ since $Z^{\prime}$ models are often used as benchmark models, is taken to be $\Gamma_{Z'} = 0.012 m_{Z'}$, well below the typical experimental resolution achieved of $0.05-0.10 \times\mttbar$.  The resulting upper limits derived from these analyses are, in general, valid for all widths up to $\Gamma_{Z'} \approx 0.05 m_{Z'}$ unless otherwise specified.

Different approaches were used to reconstruct \mttbar\ for each event. CDF uses~\cite{Aaltonen:2007dz} a matrix element technique similar to the one developed for the top-quark mass measurement as described in Sec.~\ref{sec:mass}. The mean of the event probability density distribution is used to determine the reconstructed \mttbar\ value for that particular event. Templates are then formed for a variety of postulated $Z^\prime$ resonances and including background contributions from \ttbar, \Wpjets, and QCD processes.  A binned likelihood fit to the data is performed letting the $Z^\prime$ production cross section vary in the fit.  Using 0.68~\fb, the data is found to be consistent with SM expectations and CDF excludes a leptophobic top-color resonance candidate with a mass less than 725~\gevcc\ \cite{Aaltonen:2007dz} at the $95\%$ confidence level.  A complementary analysis uses a slightly different event selection and a kinematic fit to obtain the reconstructed \mttbar\ value for each event.  With a $0.96$~\fb\ data set this CDF analysis excludes a leptophobic $Z^\prime$ with a mass less than 720~\gevcc\ \cite{Aaltonen:2007dia} at the $95\%$ confidence level.  The production cross section for any narrow $Z^\prime$-like new particle decaying to \ttbar\ is determined to be less that 0.64~pb at the $95\%$ confidence level for all $m_{Z^\prime} > 600$~\gevcc.  All the above limits include the effects of systematic uncertainties on the background normalizations, the signal efficiencies, and the shapes of the signal and background \mttbar\ distributions.

\begin{figure}
\centering
  \epsfxsize=4.5cm\epsfbox{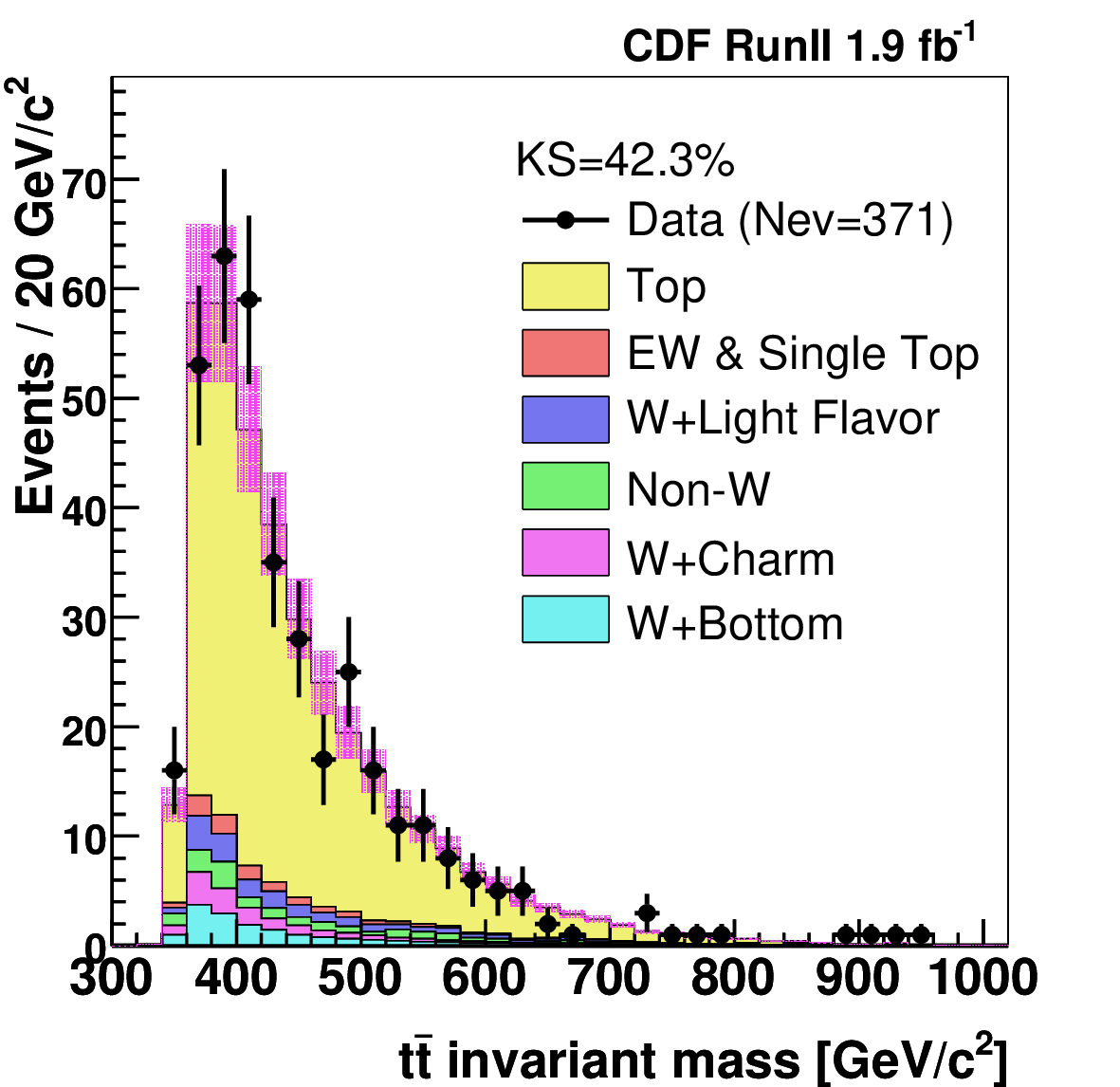}\hspace{0.5cm}
  \epsfxsize=6cm\epsfbox{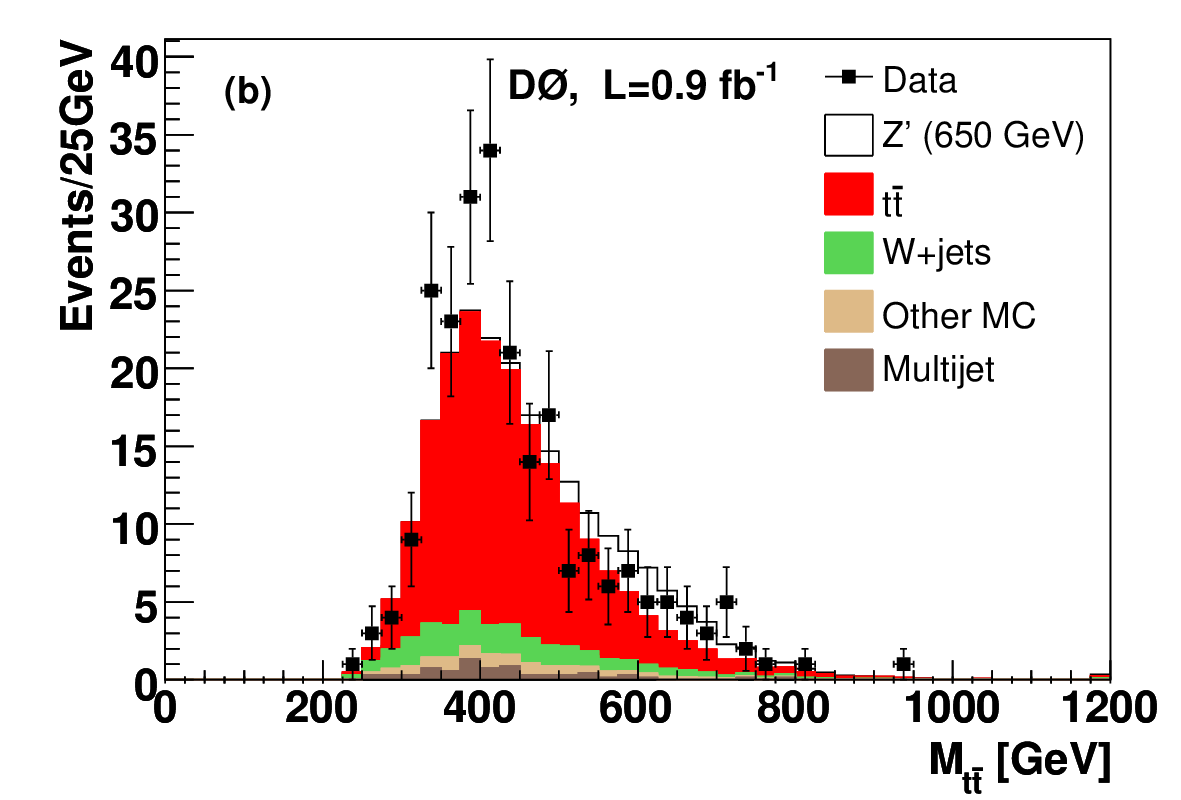}\hspace{0.5cm}
\caption{\label{fig:ttbar_res}
A comparison of the observed \mttbar\ distribution (points) to that expected from SM sources (stacked histograms) for CDF (left) and 
\dzero\ (right)~\cite{Aaltonen:2009tx, Abazov:2008ny}. 
}
\end{figure}

\dzero\ has performed a similar search in the $ljt$ channel requiring at least one identified $b$-jet in each event.  A kinematic fit is employed to determine the reconstructed \mttbar\ value for each event and a fit to the data is performed using a binned likelihood based on the template method.  No significant deviation from the SM expectation is observed, as shown in Fig.~\ref{fig:ttbar_res}, and a Bayesian approach is used to set 95~\% confidence level limits on $\sigma_{X} \cdot {\cal B}(X \to \ttbar)$ for different values of $m_X$. Systematic uncertainties that affect both the normalization and the shape of the signal and background \mttbar\ distributions are taken into account. Using 0.9 \fb\, \dzero\ excludes a lepto-phobic $Z^\prime$ with a mass less than 700~\gevcc\ at 95~\% confidence level~\cite{Abazov:2008ny}.

CDF has also searched for a massive vector color-octet particle, $G$, such as a massive gluon that preferentially couples to top quarks~\cite{Aaltonen:2009tx}.  Neglecting the gluon-$G$ coupling, the strength of the massive gluon coupling to any quark is $\lambda_{q} \alpha_s$, where $\alpha_s$ is the SM strong coupling constant. In this analysis the mass and width of the particle $G$ are denoted $M$ and $\Gamma$.
If only SM top-quark decays are assumed, the \ttbar\ production through an intermediate $G$ interferes with the SM \ttbar\ production. After selecting $ljt$ events the \mttbar\ spectrum is reconstructed using a dynamical likelihood method similar to the matrix element methods described in Sec.~\ref{sec:mass}. A standard unbinned likelihood technique is then applied to fit the data. If the mass $M$ and $\Gamma/M$ are fixed, the signal \mttbar\ distribution depends only on the assumed coupling strength $\lambda_t$.  Taking into account systematic effects,
limits on $\lambda_t$ as a function of $M$ for $\Gamma/M=0.1$ and $\Gamma/M=0.5$ are determined.  Using 1.9~\fb\, the data, shown in Fig.~\ref{fig:ttbar_res}, are consistent with SM expectations and CDF excludes the presence of massive gluons
with couplings to the top quark larger than 0.5 for $400\: < \: M \: < \: 800$~\gevcc\ and $0.05 \: < \: \Gamma/M \: < \: 0.5$ at the 95\% confidence level.

% ======================================================================
\subsection{Search for Top-quark like $t'$-Quarks}
\label{sec:tprime}
% ======================================================================

A number of BSM models predict a fourth generation of massive fermions~\cite{Frampton:1999xi,Holdom:2009rf}.  
The precise measurement of the \Zboson\ invisible width excludes the existence of a fourth neutrino family with mass less than half the \Zboson\ mass but a new neutral lepton with a mass larger than that is still compatible with electroweak data.
An extra chiral fermion family appears in grand-unified theories (GUTs)~\cite{Ross:1985ai} or $N=2$ SUSY models~\cite{He:2001tp}.
Often a small mass difference between the fourth generation $t^\prime$ and its down-type partner $b^\prime$ is preferred so that for $m_{t^\prime} > m_{b^\prime} + \mw$ and moderate mixing between $t^\prime$ and the SM light down-type quarks $q$, the decay mode $t^\prime \to W q$ dominates.

In case the fourth generation fermions are vector-like, experimental constraints are even weaker.  Such heavy exotic vector-like quarks are predicted by little Higgs models~\cite{Han:2003gf,Dobrescu:1997nm} 
or models with large extra-dimensions.   If not too heavy, such particles
can be pair produced at the Tevatron and can decay to $Wq$ with a substantial branching fraction.  Some models predict the production of a single vector-like heavy $t^\prime$ in association with a SM top quark~\cite{Dobrescu:2009vz}.

Using 0.76 \fb, CDF has searched for the pair production of massive up-type $t^\prime$ quark/anti-quark pairs that each decay to $Wq$ using the $ljt$ channel~\cite{Aaltonen:2008nf}.  The analysis employs a kinematic fit to determine $m_{t^\prime}^{\rm reco}$ for each event and uses the $H_T$ distribution to further discriminate a possible signal from SM processes. A two dimensional binned likelihood fit of the background and signal to the observed $(H_T , m_{t^\prime}^{\rm reco})$ distribution
in data is performed to set limits on the $t^\prime$ mass. Systematic uncertainties are treated as nuisance parameters and include variations in signal and background normalizations and shape.
While still statistically limited systematic uncertainties are dominated by the uncertainty on the jet energy calibration. In the fit the \ttbar\ production cross 
section is constrained to its SM value. Assuming ${\cal B}(t^\prime \to Wq)=1$, CDF excludes a $t^\prime$ mass below 256~\gevcc\ at 95~\% confidence level.

CDF and \dzero\ have provided preliminary updates for this $t'$ search with more luminosity using similar analysis techniques.
Using 5.6 \fb, CDF excludes $t^\prime$ with masses below 358~\gevcc, while \dzero\ using 4.3 \fb\ excludes $t^\prime$ masses below 296~\gevcc\ 
both at 95~\% confidence level~\cite{CdfTprime56, DzTprime43}.
%
%\begin{figure}
%\centering
%  \epsfxsize=6cm\epsfbox{cdf_tprime}\hspace{0.5cm}
%\caption{$t'$\label{fig:tprime}}
%\end{figure}

% ======================================================================
\subsection{Search for Scalar Top Quarks}
\label{sec:stop}
% ======================================================================

In supersymmetric models~\cite{Nilles:1983ge,Haber:1984rc,Martin:1997ns} each of the SM fermions has a scalar partner. In particular the left- and right-handed top quarks each have a scalar super-partner: $\st_R$ and $\st_L$. The mass eigenstates $\st_1$ and $\st_2$ of these two scalar top quarks are rotated with respect to $\st_R$ and $\st_L$. Due to the large top-quark mass, this mixing can be substantial and $\st_1$ and $\st_2$ can have a large mass difference. Hence, the lightest scalar top quark can be the lightest scalar quark and even lighter than the top quark itself. In the MSSM the scalar top quarks are mainly produced in $\st_1 \bar{\st_1}$ pairs via the strong interaction.  In general the production cross section is relatively small.  For example, when $m_{\st}=\mt$ the $\st_1 \bar{\st_1}$ pair production cross section is about 10 times smaller than the \ttbar\ production cross section.
The \st\ decay differs according to the parameters of the model and is particularly dependent on the masses of the super-particles involved. If kinematically possible, it can decay via the two body processes
$\st \to t \tchi^0_1$, $\st \to b \tchi^+_1$, or $\st \to c \tchi^0_1$, where $\tchi^0_1$ is the lightest neutralino and $\tchi^+_1$ the lightest chargino. Three-body decays are also possible, $\st \to W^+ b \tchi^0_1$, $\st \to b \tilde{\ell}^+ \nu_\ell$, or $\st \to b \ell^+ \tilde{\nu}_\ell$, where $\tilde{\ell}$ and $\tilde{\nu}_\ell$ are the super-partners of a given lepton $\ell$ and neutrino $\nu_\ell$.
For scalar top quarks lighter than the top quark the decay channel $\st \to \tchi^+_1 b \to ( \tchi^0_1 W^+ )b $ can dominate.
In this decay channel the $\st_1 \bar{\st_1}$ pair signature is similar to the \ttbar\ one apart from the presence of the neutralinos, whose experimental signature mimics that of a neutrino and are mostly produced back-to-back.  

\dzero\ searched for some admixture of $\st_1 \bar{\st_1}$ pairs in the \ttbar\ final state using the $ljt$ channel~\cite{Abazov:2009ps}.
In this analysis the neutralino mass has been fixed to a value close to the experimental LEP limit, $m_{\tchi^0_1}=50$~\gevcc, while
the \st\ mass has been varied from 130 to 190~\gevcc\ and the chargino mass from 90 to 150~\gevcc. Events with at least three jets and 
at least one identified $b$-jet have been used. For events with at least four jets, a kinematic reconstruction to the \ttbar\ hypothesis is performed.  As the \ttbar\ signature is very similar to the \st\ signal, a multivariate likelihood discriminant is employed to separate the two processes.  Separate discriminants have been trained for the three-jet and the four-jet events. Variables based on the two jet or three jet invariant masses, the reconstructed top-quark mass, the jet $E_T$, the scalar sum of the jet $E_T$, and the angular separation between the jets are used as inputs to the discriminants. As no excess over the SM expectation is observed, a binned likelihood over the discriminant distributions is constructed using Bayesian statistics to extract the upper limits on the $\st_1 \bar{\st_1}$  pair production cross section taking systematic uncertainties into account. Using 0.9~\fb, \dzero\ obtains limits that are a factor of 2 to 13 times larger than the MSSM theory prediction and agree with the expected limits within uncertainties~\cite{Abazov:2009ps}.

When both scalar top quarks decay as $\st \to b l \tilde{\nu}_l$ the experimental signature is similar to the \ttbar\ $dil$ signature.  Searches exploiting this decay channels have been performed by CDF~\cite{Aaltonen:2010uf} and \dzero~\cite{Abazov:2007im}.  A similar experimental signature is obtained if the scalar top quark is assumed to decay as
$\st\ra \tchi^+_{1}b \ra \tchi^0_1 \ell^+ \nu b$ as is done in~\cite{Aaltonen:2010zzz}.  In all three of these analyses the selection criteria are optimized to suppress the \ttbar\ contributions.  The data are consistent with the SM expectation and the resulting limits significantly extend the LEP2 exclusions when $m_{\st} >> m_{\tilde{\nu}}, m_{\tchi^0_1}$.

% ======================================================================
\section{Prospects and Conclusions}
\label{sec:conclusion}
% ======================================================================

Owing to its large mass, the top quark was discovered relatively recently and potentially offers a unique window into Beyond-the-Standard-Model physics.  Using data samples with ${\cal O}(10^3)$ candidate \ttbar\ events, the CDF and \dzero\ experiments at the Fermilab Tevatron have pioneered numerous analyses to exploit these unique events.   They have instituted a broad physics program that tests the Standard Model descriptions of the \ttbar\ production mechanisms, the top-quark decay widths, and the intrinsic properties of the top quark such as its electric charge and mass.   They have also probed the \ttbar\ event samples for direct evidence of new particles that might mimic the experimental signature of the top quark, such as the scalar top quarks of Supersymmetry or the fourth generation $t^\prime$-quark of Grand Unified Theories.  A few of the measurements are systematics limited and have reached precisions comparable to those associated with the relevant theory predictions, such as the measured production cross section and the measured top-quark mass.  The data provide strong evidence that the top quark carries electric charge $+2/3$ in units of the electron electric charge.   Measurements of the polarization fractions of $W$ bosons from top-quark decay and of \ttbar\ spin correlations are consistent with the Standard Model expectation of a spin $1/2$ particle with a $V-A$ coupling to the charged weak current.  The decay widths of the top quark are also consistent with the Standard Model expectations, being dominated by $t\to W^+ b$ and with no evidence of anomalous flavor-changing-neutral-current or charged higgs decays.  The results of these many measurements when compared across the various \ttbar\ final states yield consistent results - again as expected by the Standard Model.   In general, the detailed exploration of the \ttbar\ sample so far performed by the Tevatron experiments reveals no significant discrepancies with the Standard Model.    Except for a few cases, most all the measurements are limited by their statistical uncertainty.   
So the precise study of the top quark will persevere as the Tevatron is continuing to deliver collision data.
%
%%%focusing on the unique features offered by a \ttbar\ production 
%%%dominated by $q\overline{q}$ annihilation process.
%
The LHC experiments expect to collect \ttbar\ samples that are a few orders of magnitude larger in the coming several years.  With these sorts of statistics most all the measurements should achieve levels of precision that more thoroughly probe the Standard Model expectations and more stringently restrict the parameter space of a wide variety of new physics models.   Even for those measurements already systematics limited at the Tevatron, the LHC experiments may be able to exploit the larger statistics to identify a sub-set of events less affected by the systematics in question or to employ analysis techniques which trade statistical uncertainties for systematic uncertainties to achieve an improvement in the total uncertainty.  
The LHC is poised to build upon the foundation laid by the Tevatron experiments, and use the physics of the top quark to more fully elucidate the workings of nature at its most fundamental level.

% ============================================================================
% --- ACKNOWLEDGEMENTS
% ============================================================================

\section{Acknowledgements}
\label{sec:ack}

We gratefully acknowledge Marc Besan\c{c}on and Yvonne Peters for their 
comments and suggestions.

% ============================================================================
% --- BIBLIOGRAPHY
% ============================================================================
% Auto generated using bibtex and file RMPTop.bib
\bibliography{RMPTop}

%Merlin.mbs v4.21 2009-07-09.
\begin{thebibliography}{248}%
\makeatletter
\providecommand \@ifxundefined [1]{%
 \ifx #1\undefined \expandafter \@firstoftwo
 \else \expandafter \@secondoftwo
\fi
}%
\providecommand \@ifnum [1]{%
 \ifnum #1\expandafter \@firstoftwo
 \else \expandafter \@secondoftwo
\fi
}%
\providecommand \natexlab [1]{#1}%
\providecommand \enquote [1]{``#1''}%
\providecommand \bibnamefont  [1]{#1}%
\providecommand \bibfnamefont [1]{#1}%
\providecommand \citenamefont [1]{#1}%
\providecommand\href[0]{\@sanitize\@href}%
\providecommand\@href[1]{\endgroup\@@startlink{#1}\endgroup\@@href}%
\providecommand\@@href[1]{#1\@@endlink}%
\providecommand \@sanitize [0]{\begingroup\catcode`\&12\catcode`\#12\relax}%
\@ifxundefined \pdfoutput {\@firstoftwo}{%
 \@ifnum{\z@=\pdfoutput}{\@firstoftwo}{\@secondoftwo}%
}{%
 \providecommand\@@startlink[1]{\leavevmode\special{html:<a href="#1">}}%
 \providecommand\@@endlink[0]{\special{html:</a>}}%
}{%
 \providecommand\@@startlink[1]{%
  \leavevmode
  \pdfstartlink
   attr{/Border[0 0 1 ]/H/I/C[0 1 1]}%
   user{/Subtype/Link/A<</Type/Action/S/URI/URI(#1)>>}%
  \relax
 }%
 \providecommand\@@endlink[0]{\pdfendlink}%
}%
\providecommand \url  [0]{\begingroup\@sanitize \@url }%
\providecommand \@url [1]{\endgroup\@href {#1}{\urlprefix}}%
\providecommand \urlprefix [0]{URL }%
\providecommand \Eprint[0]{\href }%
\@ifxundefined \urlstyle {%
  \providecommand \doi [1]{doi:\discretionary{}{}{}#1}%
}{%
  \providecommand \doi [0]{doi:\discretionary{}{}{}\begingroup
  \urlstyle{rm}\Url }%
}%
\providecommand \doibase [0]{http://dx.doi.org/}%
\providecommand \Doi[1]{\href{\doibase#1}}%
\providecommand \bibAnnote [3]{%
  \BibitemShut{#1}%
  \begin{quotation}\noindent
    \textsc{Key:}\ #2\\\textsc{Annotation:}\ #3%
  \end{quotation}%
}%
\providecommand \bibAnnoteFile [2]{%
  \IfFileExists{#2}{\bibAnnote {#1} {#2} {\input{#2}}}{}%
}%
\providecommand \typeout [0]{\immediate \write \m@ne }%
\providecommand \selectlanguage [0]{\@gobble}%
\providecommand \bibinfo [0]{\@secondoftwo}%
\providecommand \bibfield [0]{\@secondoftwo}%
\providecommand \translation [1]{[#1]}%
\providecommand \BibitemOpen[0]{}%
\providecommand \bibitemStop [0]{}%
\providecommand \bibitemNoStop [0]{.\EOS\space}%
\providecommand \EOS [0]{\spacefactor3000\relax}%
\providecommand \BibitemShut [1]{\csname bibitem#1\endcsname}%
%</preamble>
\bibitem[{\citenamefont{Aaltonen}\
  \emph{et~al.}(2010{\natexlab{a}})\citenamefont{Aaltonen}, \citenamefont{V.M.}
  \emph{et~al.}}]{:1900yx}%
  \BibitemOpen
  \bibfield{author}{%
  \bibinfo {author} {\bibnamefont{Aaltonen}, \bibfnamefont{T.}}, \bibinfo
  {author} {\bibfnamefont{A.}~\bibnamefont{V.M.}}, \emph{et~al.} (\bibinfo
  {collaboration} {Tevatron Electroweak Working Group})}%
  , \bibinfo {year} {2010}{\natexlab{a}}\
  \Eprint{http://arxiv.org/abs/1007.3178}{arXiv:1007.3178 [hep-ex]}%
  \bibAnnoteFile{NoStop}{:1900yx}%
%%CITATION = 1007.3178;%%
\bibitem[{\citenamefont{Aaltonen}\ \emph{et~al.}(2007)\citenamefont{Aaltonen}
  \emph{et~al.}}]{Aaltonen:2006xc}%
  \BibitemOpen
  \bibfield{author}{%
  \bibinfo {author} {\bibnamefont{Aaltonen}, \bibfnamefont{T.}}, \emph{et~al.}
  (\bibinfo {collaboration} {CDF})}%
  , \bibinfo {year} {2007},\ \bibfield{journal}{%
  \Doi{10.1103/PhysRevLett.98.142001}{\bibinfo {journal} {Phys. Rev. Lett.}}\
  }%
  \textbf{\bibinfo {volume} {98}},\ \bibinfo {pages} {142001},\
  \Eprint{http://arxiv.org/abs/hep-ex/0612026}{arXiv:hep-ex/0612026}%
  \bibAnnoteFile{NoStop}{Aaltonen:2006xc}%
%%CITATION = HEP-EX/0612026;%%
\bibitem[{\citenamefont{Aaltonen}\
  \emph{et~al.}(2008{\natexlab{a}})\citenamefont{Aaltonen}
  \emph{et~al.}}]{Aaltonen:2007jw}%
  \BibitemOpen
  \bibfield{author}{%
  \bibinfo {author} {\bibnamefont{Aaltonen}, \bibfnamefont{T.}}, \emph{et~al.}
  (\bibinfo {collaboration} {CDF})}%
  , \bibinfo {year} {2008}{\natexlab{a}},\ \bibfield{journal}{%
  \Doi{10.1103/PhysRevLett.100.062005}{\bibinfo {journal} {Phys. Rev. Lett.}}\
  }%
  \textbf{\bibinfo {volume} {100}},\ \bibinfo {pages} {062005},\
  \Eprint{http://arxiv.org/abs/0710.4037}{arXiv:0710.4037 [hep-ex]}%
  \bibAnnoteFile{NoStop}{Aaltonen:2007jw}%
%%CITATION = 0710.4037;%%
\bibitem[{\citenamefont{Aaltonen}\
  \emph{et~al.}(2008{\natexlab{b}})\citenamefont{Aaltonen}
  \emph{et~al.}}]{Aaltonen:2007kq}%
  \BibitemOpen
  \bibfield{author}{%
  \bibinfo {author} {\bibnamefont{Aaltonen}, \bibfnamefont{T.}}, \emph{et~al.}
  (\bibinfo {collaboration} {CDF})}%
  , \bibinfo {year} {2008}{\natexlab{b}},\ \bibfield{journal}{%
  \Doi{10.1103/PhysRevD.78.111101}{\bibinfo {journal} {Phys. Rev.}}\ }%
  \textbf{\bibinfo {volume} {D78}},\ \bibinfo {pages} {111101},\
  \Eprint{http://arxiv.org/abs/0712.3273}{arXiv:0712.3273 [hep-ex]}%
  \bibAnnoteFile{NoStop}{Aaltonen:2007kq}%
%%CITATION = 0712.3273;%%
\bibitem[{\citenamefont{Aaltonen}\
  \emph{et~al.}(2008{\natexlab{c}})\citenamefont{Aaltonen}
  \emph{et~al.}}]{CDFMw}%
  \BibitemOpen
  \bibfield{author}{%
  \bibinfo {author} {\bibnamefont{Aaltonen}, \bibfnamefont{T.}}, \emph{et~al.}
  (\bibinfo {collaboration} {CDF})}%
  , \bibinfo {year} {2008}{\natexlab{c}},\ \bibfield{journal}{%
  \Doi{10.1103/PhysRevD.77.112001}{\bibinfo {journal} {Phys. Rev.}}\ }%
  \textbf{\bibinfo {volume} {D77}},\ \bibinfo {pages} {112001},\
  \Eprint{http://arxiv.org/abs/0708.3642}{arXiv:0708.3642 [hep-ex]}%
  \bibAnnoteFile{NoStop}{CDFMw}%
\bibitem[{\citenamefont{Aaltonen}\
  \emph{et~al.}(2008{\natexlab{d}})\citenamefont{Aaltonen}
  \emph{et~al.}}]{Aaltonen:2008hc}%
  \BibitemOpen
  \bibfield{author}{%
  \bibinfo {author} {\bibnamefont{Aaltonen}, \bibfnamefont{T.}}, \emph{et~al.}
  (\bibinfo {collaboration} {CDF})}%
  , \bibinfo {year} {2008}{\natexlab{d}},\ \bibfield{journal}{%
  \Doi{10.1103/PhysRevLett.101.202001}{\bibinfo {journal} {Phys. Rev. Lett.}}\
  }%
  \textbf{\bibinfo {volume} {101}},\ \bibinfo {pages} {202001},\
  \Eprint{http://arxiv.org/abs/0806.2472}{arXiv:0806.2472 [hep-ex]}%
  \bibAnnoteFile{NoStop}{Aaltonen:2008hc}%
%%CITATION = 0806.2472;%%
\bibitem[{\citenamefont{Aaltonen}\
  \emph{et~al.}(2008{\natexlab{e}})\citenamefont{Aaltonen}
  \emph{et~al.}}]{Aaltonen:2007dia}%
  \BibitemOpen
  \bibfield{author}{%
  \bibinfo {author} {\bibnamefont{Aaltonen}, \bibfnamefont{T.}}, \emph{et~al.}
  (\bibinfo {collaboration} {CDF})}%
  , \bibinfo {year} {2008}{\natexlab{e}},\ \bibfield{journal}{%
  \Doi{10.1103/PhysRevD.77.051102}{\bibinfo {journal} {Phys. Rev.}}\ }%
  \textbf{\bibinfo {volume} {D77}},\ \bibinfo {pages} {051102},\
  \Eprint{http://arxiv.org/abs/0710.5335}{arXiv:0710.5335 [hep-ex]}%
  \bibAnnoteFile{NoStop}{Aaltonen:2007dia}%
%%CITATION = 0710.5335;%%
\bibitem[{\citenamefont{Aaltonen}\
  \emph{et~al.}(2008{\natexlab{f}})\citenamefont{Aaltonen}
  \emph{et~al.}}]{craigspaper}%
  \BibitemOpen
  \bibfield{author}{%
  \bibinfo {author} {\bibnamefont{Aaltonen}, \bibfnamefont{T.}}, \emph{et~al.}
  (\bibinfo {collaboration} {CDF})}%
  , \bibinfo {year} {2008}{\natexlab{f}},\ \bibfield{journal}{%
  \Doi{10.1103/PhysRevD.78.052006}{\bibinfo {journal} {Phys. Rev.}}\ }%
  \textbf{\bibinfo {volume} {D 78}},\ \bibinfo {pages} {052006},\
  \Eprint{http://arxiv.org/abs/0807.2204}{arXiv:0807.2204 [hep-ex]}%
  \bibAnnoteFile{NoStop}{craigspaper}%
\bibitem[{\citenamefont{Aaltonen}\
  \emph{et~al.}(2008{\natexlab{g}})\citenamefont{Aaltonen}
  \emph{et~al.}}]{Aaltonen:2008nf}%
  \BibitemOpen
  \bibfield{author}{%
  \bibinfo {author} {\bibnamefont{Aaltonen}, \bibfnamefont{T.}}, \emph{et~al.}
  (\bibinfo {collaboration} {CDF})}%
  , \bibinfo {year} {2008}{\natexlab{g}},\ \bibfield{journal}{%
  \Doi{10.1103/PhysRevLett.100.161803}{\bibinfo {journal} {Phys. Rev. Lett.}}\
  }%
  \textbf{\bibinfo {volume} {100}},\ \bibinfo {pages} {161803},\
  \Eprint{http://arxiv.org/abs/0801.3877}{arXiv:0801.3877 [hep-ex]}%
  \bibAnnoteFile{NoStop}{Aaltonen:2008nf}%
%%CITATION = 0801.3877;%%
\bibitem[{\citenamefont{Aaltonen}\
  \emph{et~al.}(2008{\natexlab{h}})\citenamefont{Aaltonen}
  \emph{et~al.}}]{Aaltonen:2007dz}%
  \BibitemOpen
  \bibfield{author}{%
  \bibinfo {author} {\bibnamefont{Aaltonen}, \bibfnamefont{T.}}, \emph{et~al.}
  (\bibinfo {collaboration} {CDF})}%
  , \bibinfo {year} {2008}{\natexlab{h}},\ \bibfield{journal}{%
  \Doi{10.1103/PhysRevLett.100.231801}{\bibinfo {journal} {Phys. Rev. Lett.}}\
  }%
  \textbf{\bibinfo {volume} {100}},\ \bibinfo {pages} {231801},\
  \Eprint{http://arxiv.org/abs/0709.0705}{arXiv:0709.0705 [hep-ex]}%
  \bibAnnoteFile{NoStop}{Aaltonen:2007dz}%
%%CITATION = 0709.0705;%%
\bibitem[{\citenamefont{Aaltonen}\
  \emph{et~al.}(2008{\natexlab{i}})\citenamefont{Aaltonen}
  \emph{et~al.}}]{Aaltonen:2008aaa}%
  \BibitemOpen
  \bibfield{author}{%
  \bibinfo {author} {\bibnamefont{Aaltonen}, \bibfnamefont{T.}}, \emph{et~al.}
  (\bibinfo {collaboration} {CDF})}%
  , \bibinfo {year} {2008}{\natexlab{i}},\ \bibfield{journal}{%
  \Doi{10.1103/PhysRevLett.101.192002}{\bibinfo {journal} {Phys. Rev. Lett.}}\
  }%
  \textbf{\bibinfo {volume} {101}},\ \bibinfo {pages} {192002},\
  \Eprint{http://arxiv.org/abs/0805.2109}{arXiv:0805.2109 [hep-ex]}%
  \bibAnnoteFile{NoStop}{Aaltonen:2008aaa}%
%%CITATION = 0805.2109;%%
\bibitem[{\citenamefont{Aaltonen}\
  \emph{et~al.}(2009{\natexlab{a}})\citenamefont{Aaltonen}
  \emph{et~al.}}]{Aaltonen:2009ve}%
  \BibitemOpen
  \bibfield{author}{%
  \bibinfo {author} {\bibnamefont{Aaltonen}, \bibfnamefont{T.}}, \emph{et~al.}
  (\bibinfo {collaboration} {CDF})}%
  , \bibinfo {year} {2009}{\natexlab{a}},\ \bibfield{journal}{%
  \Doi{10.1103/PhysRevD.79.112007}{\bibinfo {journal} {Phys. Rev.}}\ }%
  \textbf{\bibinfo {volume} {D79}},\ \bibinfo {pages} {112007},\
  \Eprint{http://arxiv.org/abs/0903.5263}{arXiv:0903.5263 [hep-ex]}%
  \bibAnnoteFile{NoStop}{Aaltonen:2009ve}%
%%CITATION = 0903.5263;%%
\bibitem[{\citenamefont{Aaltonen}\
  \emph{et~al.}(2009{\natexlab{b}})\citenamefont{Aaltonen}
  \emph{et~al.}}]{Aaltonen:2008ir}%
  \BibitemOpen
  \bibfield{author}{%
  \bibinfo {author} {\bibnamefont{Aaltonen}, \bibfnamefont{T.}}, \emph{et~al.}
  (\bibinfo {collaboration} {CDF})}%
  , \bibinfo {year} {2009}{\natexlab{b}},\ \bibfield{journal}{%
  \Doi{10.1103/PhysRevLett.102.042001}{\bibinfo {journal} {Phys. Rev. Lett.}}\
  }%
  \textbf{\bibinfo {volume} {102}},\ \bibinfo {pages} {042001},\
  \Eprint{http://arxiv.org/abs/0808.2167}{arXiv:0808.2167 [hep-ex]}%
  \bibAnnoteFile{NoStop}{Aaltonen:2008ir}%
%%CITATION = 0808.2167;%%
\bibitem[{\citenamefont{Aaltonen}\
  \emph{et~al.}(2009{\natexlab{c}})\citenamefont{Aaltonen}
  \emph{et~al.}}]{Aaltonen:2009iz}%
  \BibitemOpen
  \bibfield{author}{%
  \bibinfo {author} {\bibnamefont{Aaltonen}, \bibfnamefont{T.}}, \emph{et~al.}
  (\bibinfo {collaboration} {CDF})}%
  , \bibinfo {year} {2009}{\natexlab{c}},\ \bibfield{journal}{%
  \Doi{10.1103/PhysRevLett.102.222003}{\bibinfo {journal} {Phys. Rev. Lett.}}\
  }%
  \textbf{\bibinfo {volume} {102}},\ \bibinfo {pages} {222003},\
  \Eprint{http://arxiv.org/abs/0903.2850}{arXiv:0903.2850 [hep-ex]}%
  \bibAnnoteFile{NoStop}{Aaltonen:2009iz}%
%%CITATION = 0903.2850;%%
\bibitem[{\citenamefont{Aaltonen}\
  \emph{et~al.}(2009{\natexlab{d}})\citenamefont{Aaltonen}
  \emph{et~al.}}]{Aaltonen:2009st}%
  \BibitemOpen
  \bibfield{author}{%
  \bibinfo {author} {\bibnamefont{Aaltonen}, \bibfnamefont{T.}}, \emph{et~al.}
  (\bibinfo {collaboration} {CDF})}%
  , \bibinfo {year} {2009}{\natexlab{d}},\ \bibfield{journal}{%
  \Doi{10.1103/PhysRevLett.103.092002}{\bibinfo {journal} {Phys. Rev. Lett.}}\
  }%
  \textbf{\bibinfo {volume} {103}},\ \bibinfo {pages} {092002},\
  \Eprint{http://arxiv.org/abs/0903.0885}{arXiv:0903.0885 [hep-ex]}%
  \bibAnnoteFile{NoStop}{Aaltonen:2009st}%
%%CITATION = 0903.0885;%%
\bibitem[{\citenamefont{Aaltonen}\
  \emph{et~al.}(2009{\natexlab{e}})\citenamefont{Aaltonen}
  \emph{et~al.}}]{Aaltonen:2008gj}%
  \BibitemOpen
  \bibfield{author}{%
  \bibinfo {author} {\bibnamefont{Aaltonen}, \bibfnamefont{T.}}, \emph{et~al.}
  (\bibinfo {collaboration} {CDF})}%
  , \bibinfo {year} {2009}{\natexlab{e}},\ \bibfield{journal}{%
  \Doi{10.1103/PhysRevD.79.092005}{\bibinfo {journal} {Phys. Rev.}}\ }%
  \textbf{\bibinfo {volume} {D79}},\ \bibinfo {pages} {092005},\
  \Eprint{http://arxiv.org/abs/0809.4808}{arXiv:0809.4808 [hep-ex]}%
  \bibAnnoteFile{NoStop}{Aaltonen:2008gj}%
%%CITATION = 0809.4808;%%
\bibitem[{\citenamefont{Aaltonen}\
  \emph{et~al.}(2009{\natexlab{f}})\citenamefont{Aaltonen}
  \emph{et~al.}}]{CdfttSpinDil28}%
  \BibitemOpen
  \bibfield{author}{%
  \bibinfo {author} {\bibnamefont{Aaltonen}, \bibfnamefont{T.}}, \emph{et~al.}
  (\bibinfo {collaboration} {CDF})}%
  , \bibinfo {year} {2009}{\natexlab{f}}\ \bibinfo {note} {preliminary result
  available in CDF public note 9824.}%
  \bibAnnoteFile{Stop}{CdfttSpinDil28}%
\bibitem[{\citenamefont{Aaltonen}\
  \emph{et~al.}(2009{\natexlab{g}})\citenamefont{Aaltonen}
  \emph{et~al.}}]{Abulencia:2008su}%
  \BibitemOpen
  \bibfield{author}{%
  \bibinfo {author} {\bibnamefont{Aaltonen}, \bibfnamefont{T.}}, \emph{et~al.}
  (\bibinfo {collaboration} {CDF})}%
  , \bibinfo {year} {2009}{\natexlab{g}},\ \bibfield{journal}{%
  \Doi{10.1103/PhysRevD.79.031101}{\bibinfo {journal} {Phys. Rev.}}\ }%
  \textbf{\bibinfo {volume} {D79}},\ \bibinfo {pages} {031101},\
  \Eprint{http://arxiv.org/abs/0807.4262}{arXiv:0807.4262 [hep-ex]}%
  \bibAnnoteFile{NoStop}{Abulencia:2008su}%
%%CITATION = 0807.4262;%%
\bibitem[{\citenamefont{Aaltonen}\
  \emph{et~al.}(2009{\natexlab{h}})\citenamefont{Aaltonen}
  \emph{et~al.}}]{Aaltonen:2009tza}%
  \BibitemOpen
  \bibfield{author}{%
  \bibinfo {author} {\bibnamefont{Aaltonen}, \bibfnamefont{T.}}, \emph{et~al.}
  (\bibinfo {collaboration} {CDF})}%
  , \bibinfo {year} {2009}{\natexlab{h}},\ \bibfield{journal}{%
  \Doi{10.1103/PhysRevD.79.072005}{\bibinfo {journal} {Phys. Rev.}}\ }%
  \textbf{\bibinfo {volume} {D79}},\ \bibinfo {pages} {072005},\
  \Eprint{http://arxiv.org/abs/0901.3773}{arXiv:0901.3773 [hep-ex]}%
  \bibAnnoteFile{NoStop}{Aaltonen:2009tza}%
%%CITATION = 0901.3773;%%
\bibitem[{\citenamefont{Aaltonen}\
  \emph{et~al.}(2009{\natexlab{i}})\citenamefont{Aaltonen}
  \emph{et~al.}}]{Aaltonen:2009zi}%
  \BibitemOpen
  \bibfield{author}{%
  \bibinfo {author} {\bibnamefont{Aaltonen}, \bibfnamefont{T.}}, \emph{et~al.}
  (\bibinfo {collaboration} {CDF})}%
  , \bibinfo {year} {2009}{\natexlab{i}},\ \bibfield{journal}{%
  \Doi{10.1103/PhysRevD.80.051104}{\bibinfo {journal} {Phys. Rev.}}\ }%
  \textbf{\bibinfo {volume} {D80}},\ \bibinfo {pages} {051104},\
  \Eprint{http://arxiv.org/abs/0906.5371}{arXiv:0906.5371 [hep-ex]}%
  \bibAnnoteFile{NoStop}{Aaltonen:2009zi}%
%%CITATION = 0906.5371;%%
\bibitem[{\citenamefont{Aaltonen}\
  \emph{et~al.}(2009{\natexlab{j}})\citenamefont{Aaltonen}
  \emph{et~al.}}]{Aaltonen:2008bd}%
  \BibitemOpen
  \bibfield{author}{%
  \bibinfo {author} {\bibnamefont{Aaltonen}, \bibfnamefont{T.}}, \emph{et~al.}
  (\bibinfo {collaboration} {CDF})}%
  , \bibinfo {year} {2009}{\natexlab{j}},\ \bibfield{journal}{%
  \Doi{10.1103/PhysRevLett.102.152001}{\bibinfo {journal} {Phys. Rev. Lett.}}\
  }%
  \textbf{\bibinfo {volume} {102}},\ \bibinfo {pages} {152001},\
  \Eprint{http://arxiv.org/abs/0807.4652}{arXiv:0807.4652 [hep-ex]}%
  \bibAnnoteFile{NoStop}{Aaltonen:2008bd}%
%%CITATION = 0807.4652;%%
\bibitem[{\citenamefont{Aaltonen}\
  \emph{et~al.}(2009{\natexlab{k}})\citenamefont{Aaltonen}
  \emph{et~al.}}]{Aaltonen:2009ke}%
  \BibitemOpen
  \bibfield{author}{%
  \bibinfo {author} {\bibnamefont{Aaltonen}, \bibfnamefont{T.}}, \emph{et~al.}
  (\bibinfo {collaboration} {CDF})}%
  , \bibinfo {year} {2009}{\natexlab{k}},\ \bibfield{journal}{%
  \Doi{10.1103/PhysRevLett.103.101803}{\bibinfo {journal} {Phys. Rev. Lett.}}\
  }%
  \textbf{\bibinfo {volume} {103}},\ \bibinfo {pages} {101803},\
  \Eprint{http://arxiv.org/abs/0907.1269}{arXiv:0907.1269 [hep-ex]}%
  \bibAnnoteFile{NoStop}{Aaltonen:2009ke}%
%%CITATION = 0907.1269;%%
\bibitem[{\citenamefont{Aaltonen}\
  \emph{et~al.}(2009{\natexlab{l}})\citenamefont{Aaltonen}
  \emph{et~al.}}]{Aaltonen:2008mx}%
  \BibitemOpen
  \bibfield{author}{%
  \bibinfo {author} {\bibnamefont{Aaltonen}, \bibfnamefont{T.}}, \emph{et~al.}
  (\bibinfo {collaboration} {CDF})}%
  , \bibinfo {year} {2009}{\natexlab{l}},\ \bibfield{journal}{%
  \Doi{10.1103/PhysRevD.79.072001}{\bibinfo {journal} {Phys. Rev.}}\ }%
  \textbf{\bibinfo {volume} {D79}},\ \bibinfo {pages} {072001},\
  \Eprint{http://arxiv.org/abs/0812.4469}{arXiv:0812.4469 [hep-ex]}%
  \bibAnnoteFile{NoStop}{Aaltonen:2008mx}%
%%CITATION = 0812.4469;%%
\bibitem[{\citenamefont{Aaltonen}\
  \emph{et~al.}(2010{\natexlab{b}})\citenamefont{Aaltonen}
  \emph{et~al.}}]{Aaltonen:2010js}%
  \BibitemOpen
  \bibfield{author}{%
  \bibinfo {author} {\bibnamefont{Aaltonen}, \bibfnamefont{T.}}, \emph{et~al.}
  (\bibinfo {collaboration} {CDF})}%
  , \bibinfo {year} {2010}{\natexlab{b}},\ \bibfield{journal}{%
  \bibinfo {journal} {Phys. Rev. Lett.}\ }%
  \textbf{\bibinfo {volume} {105}},\ \bibinfo {pages} {101801},\
  \Eprint{http://arxiv.org/abs/1006.4597}{arXiv:1006.4597 [hep-ex]}%
  \bibAnnoteFile{NoStop}{Aaltonen:2010js}%
%%CITATION = 1006.4597;%%
\bibitem[{\citenamefont{Aaltonen}\
  \emph{et~al.}(2010{\natexlab{c}})\citenamefont{Aaltonen}
  \emph{et~al.}}]{Aaltonen:2010ic}%
  \BibitemOpen
  \bibfield{author}{%
  \bibinfo {author} {\bibnamefont{Aaltonen}, \bibfnamefont{T.}}, \emph{et~al.}
  (\bibinfo {collaboration} {CDF})}%
  , \bibinfo {year} {2010}{\natexlab{c}},\ \bibfield{journal}{%
  \Doi{10.1103/PhysRevLett.105.012001}{\bibinfo {journal} {Phys. Rev. Lett.}}\
  }%
  \textbf{\bibinfo {volume} {105}},\ \bibinfo {pages} {012001},\
  \Eprint{http://arxiv.org/abs/1004.3224}{arXiv:1004.3224 [hep-ex]}%
  \bibAnnoteFile{NoStop}{Aaltonen:2010ic}%
%%CITATION = 1004.3224;%%
\bibitem[{\citenamefont{Aaltonen}\
  \emph{et~al.}(2010{\natexlab{d}})\citenamefont{Aaltonen}
  \emph{et~al.}}]{CdfttSpinLjt53}%
  \BibitemOpen
  \bibfield{author}{%
  \bibinfo {author} {\bibnamefont{Aaltonen}, \bibfnamefont{T.}}, \emph{et~al.}
  (\bibinfo {collaboration} {CDF})}%
  , \bibinfo {year} {2010}{\natexlab{d}}\ \bibinfo {note} {preliminary result
  available in CDF public note 10211.}%
  \bibAnnoteFile{Stop}{CdfttSpinLjt53}%
\bibitem[{\citenamefont{Aaltonen}\
  \emph{et~al.}(2010{\natexlab{e}})\citenamefont{Aaltonen}
  \emph{et~al.}}]{Aaltonen:2010pe}%
  \BibitemOpen
  \bibfield{author}{%
  \bibinfo {author} {\bibnamefont{Aaltonen}, \bibfnamefont{T.}}, \emph{et~al.}
  (\bibinfo {collaboration} {CDF})}%
  , \bibinfo {year} {2010}{\natexlab{e}},\ \bibfield{journal}{%
  \Doi{10.1103/PhysRevD.81.052011}{\bibinfo {journal} {Phys. Rev.}}\ }%
  \textbf{\bibinfo {volume} {D81}},\ \bibinfo {pages} {052011},\
  \Eprint{http://arxiv.org/abs/1002.0365}{arXiv:1002.0365 [hep-ex]}%
  \bibAnnoteFile{NoStop}{Aaltonen:2010pe}%
%%CITATION = 1002.0365;%%
\bibitem[{\citenamefont{Aaltonen}\
  \emph{et~al.}(2010{\natexlab{f}})\citenamefont{Aaltonen}
  \emph{et~al.}}]{CDFSLTe}%
  \BibitemOpen
  \bibfield{author}{%
  \bibinfo {author} {\bibnamefont{Aaltonen}, \bibfnamefont{T.}}, \emph{et~al.}
  (\bibinfo {collaboration} {CDF})}%
  , \bibinfo {year} {2010}{\natexlab{f}},\ \bibfield{journal}{%
  \Doi{10.1103/PhysRevD.81.092002}{\bibinfo {journal} {Phys. Rev.}}\ }%
  \textbf{\bibinfo {volume} {D81}},\ \bibinfo {pages} {092002},\
  \Eprint{http://arxiv.org/abs/1002.3783}{arXiv:1002.3783 [hep-ex]}%
  \bibAnnoteFile{NoStop}{CDFSLTe}%
\bibitem[{\citenamefont{Aaltonen}\
  \emph{et~al.}(2010{\natexlab{g}})\citenamefont{Aaltonen}
  \emph{et~al.}}]{Aaltonen:2010WHel}%
  \BibitemOpen
  \bibfield{author}{%
  \bibinfo {author} {\bibnamefont{Aaltonen}, \bibfnamefont{T.}}, \emph{et~al.}
  (\bibinfo {collaboration} {CDF})}%
  , \bibinfo {year} {2010}{\natexlab{g}},\ \bibfield{journal}{%
  \bibinfo {journal} {Phys. Rev. Lett.}\ }%
  \textbf{\bibinfo {volume} {105}},\ \bibinfo {pages} {042002},\
  \Eprint{http://arxiv.org/abs/1003.0224}{arXiv:1003.0224 [hep-ex]}%
  \bibAnnoteFile{NoStop}{Aaltonen:2010WHel}%
%%CITATION = 1003.0224;%%
\bibitem[{\citenamefont{Aaltonen}\
  \emph{et~al.}(2010{\natexlab{h}})\citenamefont{Aaltonen}
  \emph{et~al.}}]{Aaltonen:2009hd}%
  \BibitemOpen
  \bibfield{author}{%
  \bibinfo {author} {\bibnamefont{Aaltonen}, \bibfnamefont{T.}}, \emph{et~al.}
  (\bibinfo {collaboration} {CDF})}%
  , \bibinfo {year} {2010}{\natexlab{h}},\ \bibfield{journal}{%
  \Doi{10.1103/PhysRevD.81.032002}{\bibinfo {journal} {Phys. Rev.}}\ }%
  \textbf{\bibinfo {volume} {D81}},\ \bibinfo {pages} {032002},\
  \Eprint{http://arxiv.org/abs/0910.0969}{arXiv:0910.0969 [hep-ex]}%
  \bibAnnoteFile{NoStop}{Aaltonen:2009hd}%
%%CITATION = 0910.0969;%%
\bibitem[{\citenamefont{Aaltonen}\
  \emph{et~al.}(2010{\natexlab{i}})\citenamefont{Aaltonen}
  \emph{et~al.}}]{Aaltonen:2010st}%
  \BibitemOpen
  \bibfield{author}{%
  \bibinfo {author} {\bibnamefont{Aaltonen}, \bibfnamefont{T.}}, \emph{et~al.}
  (\bibinfo {collaboration} {CDF})}%
  , \bibinfo {year} {2010}{\natexlab{i}},\ \bibfield{journal}{%
  \Doi{10.1103/PhysRevD.82.112005}{\bibinfo {journal} {Phys. Rev.}}\ }%
  \textbf{\bibinfo {volume} {D82}},\ \bibinfo {pages} {112005},\
  \Eprint{http://arxiv.org/abs/1004.1181}{arXiv:1004.1181 [hep-ex]}%
  \bibAnnoteFile{NoStop}{Aaltonen:2010st}%
%%CITATION = 1004.1181;%%
\bibitem[{\citenamefont{Aaltonen}\
  \emph{et~al.}(2010{\natexlab{j}})\citenamefont{Aaltonen}
  \emph{et~al.}}]{Aaltonen:2009tx}%
  \BibitemOpen
  \bibfield{author}{%
  \bibinfo {author} {\bibnamefont{Aaltonen}, \bibfnamefont{T.}}, \emph{et~al.}
  (\bibinfo {collaboration} {CDF})}%
  , \bibinfo {year} {2010}{\natexlab{j}},\ \bibfield{journal}{%
  \Doi{10.1016/j.physletb.2010.06.036}{\bibinfo {journal} {Phys. Lett.}}\ }%
  \textbf{\bibinfo {volume} {B691}},\ \bibinfo {pages} {183},\
  \Eprint{http://arxiv.org/abs/0911.3112}{arXiv:0911.3112 [hep-ex]}%
  \bibAnnoteFile{NoStop}{Aaltonen:2009tx}%
%%CITATION = 0911.3112;%%
\bibitem[{\citenamefont{Aaltonen}\
  \emph{et~al.}(2010{\natexlab{k}})\citenamefont{Aaltonen}
  \emph{et~al.}}]{Aaltonen:2010zzz}%
  \BibitemOpen
  \bibfield{author}{%
  \bibinfo {author} {\bibnamefont{Aaltonen}, \bibfnamefont{T.}}, \emph{et~al.}
  (\bibinfo {collaboration} {CDF})}%
  , \bibinfo {year} {2010}{\natexlab{k}},\ \bibfield{journal}{%
  \Doi{10.1103/PhysRevLett.104.251801}{\bibinfo {journal} {Phys. Rev. Lett.}}\
  }%
  \textbf{\bibinfo {volume} {104}},\ \bibinfo {pages} {251801},\
  \Eprint{http://arxiv.org/abs/0912.1308}{arXiv:0912.1308 [hep-ex]}%
  \bibAnnoteFile{NoStop}{Aaltonen:2010zzz}%
%%CITATION = 0912.1308;%%
\bibitem[{\citenamefont{Aaltonen}\
  \emph{et~al.}(2010{\natexlab{l}})\citenamefont{Aaltonen}
  \emph{et~al.}}]{Aaltonen:2010uf}%
  \BibitemOpen
  \bibfield{author}{%
  \bibinfo {author} {\bibnamefont{Aaltonen}, \bibfnamefont{T.}}, \emph{et~al.}
  (\bibinfo {collaboration} {CDF})}%
  , \bibinfo {year} {2010}{\natexlab{l}},\ \bibfield{journal}{%
  \Doi{10.1103/PhysRevD.82.092001}{\bibinfo {journal} {Phys. Rev.}}\ }%
  \textbf{\bibinfo {volume} {D82}},\ \bibinfo {pages} {092001},\
  \Eprint{http://arxiv.org/abs/1009.0266}{arXiv:1009.0266 [hep-ex]}%
  \bibAnnoteFile{NoStop}{Aaltonen:2010uf}%
%%CITATION = 1009.0266;%%
\bibitem[{\citenamefont{Aaltonen}\
  \emph{et~al.}(2010{\natexlab{m}})\citenamefont{Aaltonen}
  \emph{et~al.}}]{Aaltonen:2010Mt2}%
  \BibitemOpen
  \bibfield{author}{%
  \bibinfo {author} {\bibnamefont{Aaltonen}, \bibfnamefont{T.}}, \emph{et~al.}
  (\bibinfo {collaboration} {CDF})}%
  , \bibinfo {year} {2010}{\natexlab{m}},\ \bibfield{journal}{%
  \Doi{10.1103/PhysRevD.81.031102}{\bibinfo {journal} {Phys. Rev.}}\ }%
  \textbf{\bibinfo {volume} {D81}},\ \bibinfo {pages} {031102},\
  \Eprint{http://arxiv.org/abs/0911.2956}{arXiv:0911.2956 [hep-ex]}%
  \bibAnnoteFile{NoStop}{Aaltonen:2010Mt2}%
%%CITATION = 0911.2956;%%
\bibitem[{\citenamefont{Aaltonen}\
  \emph{et~al.}(2011{\natexlab{a}})\citenamefont{Aaltonen}
  \emph{et~al.}}]{CdfAfb53}%
  \BibitemOpen
  \bibfield{author}{%
  \bibinfo {author} {\bibnamefont{Aaltonen}, \bibfnamefont{T.}}, \emph{et~al.}
  (\bibinfo {collaboration} {CDF})}%
  , \bibinfo {year} {2011}{\natexlab{a}}\ \bibinfo {note} {submitted to Phys.
  Rev. D},\ \Eprint{http://arxiv.org/abs/1101.0034}{arXiv:1101.0034 [hep-ex]}%
  \bibAnnoteFile{NoStop}{CdfAfb53}%
%%CITATION = 1101.0034;%%
\bibitem[{\citenamefont{Aaltonen}\
  \emph{et~al.}(2011{\natexlab{b}})\citenamefont{Aaltonen}
  \emph{et~al.}}]{CdfttSpinLjt43}%
  \BibitemOpen
  \bibfield{author}{%
  \bibinfo {author} {\bibnamefont{Aaltonen}, \bibfnamefont{T.}}, \emph{et~al.}
  (\bibinfo {collaboration} {CDF})}%
  , \bibinfo {year} {2011}{\natexlab{b}},\ \bibfield{journal}{%
  \Doi{10.1103/PhysRevD.83.031104}{\bibinfo {journal} {Phys. Rev.}}\ }%
  \textbf{\bibinfo {volume} {D83}},\ \bibinfo {pages} {031104},\
  \Eprint{http://arxiv.org/abs/1012.3093}{arXiv:1012.3093 [hep-ex]}%
  \bibAnnoteFile{NoStop}{CdfttSpinLjt43}%
%%CITATION = 1012.3093;%%
\bibitem[{\citenamefont{Aaltonen}\
  \emph{et~al.}(2011{\natexlab{c}})\citenamefont{Aaltonen}
  \emph{et~al.}}]{CdfAfbDil51}%
  \BibitemOpen
  \bibfield{author}{%
  \bibinfo {author} {\bibnamefont{Aaltonen}, \bibfnamefont{T.}}, \emph{et~al.}
  (\bibinfo {collaboration} {CDF})}%
  , \bibinfo {year} {2011}{\natexlab{c}}\ \bibinfo {note} {preliminary result
  available in CDF public note 10436.}%
  \bibAnnoteFile{Stop}{CdfAfbDil51}%
\bibitem[{\citenamefont{Aaltonen}\
  \emph{et~al.}(2011{\natexlab{d}})\citenamefont{Aaltonen}
  \emph{et~al.}}]{CdfMtMtbar56}%
  \BibitemOpen
  \bibfield{author}{%
  \bibinfo {author} {\bibnamefont{Aaltonen}, \bibfnamefont{T.}}, \emph{et~al.}
  (\bibinfo {collaboration} {CDF})}%
  , \bibinfo {year} {2011}{\natexlab{d}}\ \bibinfo {note} {accepted by Phys.
  Rev. Lett.},\ \Eprint{http://arxiv.org/abs/1103.2782}{arXiv:1103.2782
  [hep-ex]}%
  \bibAnnoteFile{NoStop}{CdfMtMtbar56}%
\bibitem[{\citenamefont{Aaltonen}\
  \emph{et~al.}(2011{\natexlab{e}})\citenamefont{Aaltonen}
  \emph{et~al.}}]{CdfTprime56}%
  \BibitemOpen
  \bibfield{author}{%
  \bibinfo {author} {\bibnamefont{Aaltonen}, \bibfnamefont{T.}}, \emph{et~al.}
  (\bibinfo {collaboration} {CDF})}%
  , \bibinfo {year} {2011}{\natexlab{e}}\ \bibinfo {note} {preliminary result
  available in CDF public note 10395.}%
  \bibAnnoteFile{Stop}{CdfTprime56}%
\bibitem[{\citenamefont{Abachi}\ \emph{et~al.}(1995)\citenamefont{Abachi}
  \emph{et~al.}}]{Top1995d}%
  \BibitemOpen
  \bibfield{author}{%
  \bibinfo {author} {\bibnamefont{Abachi}, \bibfnamefont{S.}}, \emph{et~al.}
  (\bibinfo {collaboration} {D0})}%
  , \bibinfo {year} {1995},\ \bibfield{journal}{%
  \Doi{10.1103/PhysRevLett.74.2632}{\bibinfo {journal} {Phys. Rev. Lett.}}\ }%
  \textbf{\bibinfo {volume} {74}},\ \bibinfo {pages} {2632},\
  \Eprint{http://arxiv.org/abs/9503003}{arXiv:9503003 [hep-ex]}%
  \bibAnnoteFile{NoStop}{Top1995d}%
\bibitem[{\citenamefont{Abazov}\
  \emph{et~al.}(2006{\natexlab{a}})\citenamefont{Abazov}
  \emph{et~al.}}]{dzero}%
  \BibitemOpen
  \bibfield{author}{%
  \bibinfo {author} {\bibnamefont{Abazov}, \bibfnamefont{V.}}, \emph{et~al.}
  (\bibinfo {collaboration} {D0})}%
  , \bibinfo {year} {2006}{\natexlab{a}},\ \bibfield{journal}{%
  \Doi{10.1016/j.nima.2006.05.248}{\bibinfo {journal} {Nucl. Instrum. Meth.}}\
  }%
  \textbf{\bibinfo {volume} {A565}},\ \bibinfo {pages} {463},\
  \Eprint{http://arxiv.org/abs/0507191}{arXiv:0507191 [physics]}%
  \bibAnnoteFile{NoStop}{dzero}%
\bibitem[{\citenamefont{Abazov}\
  \emph{et~al.}(2008{\natexlab{a}})\citenamefont{Abazov}
  \emph{et~al.}}]{dzjes}%
  \BibitemOpen
  \bibfield{author}{%
  \bibinfo {author} {\bibnamefont{Abazov}, \bibfnamefont{V.}}, \emph{et~al.}
  (\bibinfo {collaboration} {D0 Collaboration})}%
  , \bibinfo {year} {2008}{\natexlab{a}},\ \bibfield{journal}{%
  \Doi{10.1103/PhysRevLett.101.062001}{\bibinfo {journal} {Phys.Rev.Lett.}}\ }%
  \textbf{\bibinfo {volume} {101}},\ \bibinfo {pages} {062001},\
  \Eprint{http://arxiv.org/abs/0802.2400}{arXiv:0802.2400 [hep-ex]}%
  \bibAnnoteFile{NoStop}{dzjes}%
\bibitem[{\citenamefont{Abazov}\
  \emph{et~al.}(2009{\natexlab{a}})\citenamefont{Abazov} \emph{et~al.}}]{DzMw}%
  \BibitemOpen
  \bibfield{author}{%
  \bibinfo {author} {\bibnamefont{Abazov}, \bibfnamefont{V.}}, \emph{et~al.}
  (\bibinfo {collaboration} {D0})}%
  , \bibinfo {year} {2009}{\natexlab{a}},\ \bibfield{journal}{%
  \Doi{10.1103/PhysRevLett.103.141801}{\bibinfo {journal} {Phys. Rev. Lett.}}\
  }%
  \textbf{\bibinfo {volume} {103}},\ \bibinfo {pages} {141801},\
  \Eprint{http://arxiv.org/abs/0908.0766}{arXiv:0908.0766 [hep-ex]}%
  \bibAnnoteFile{NoStop}{DzMw}%
\bibitem[{\citenamefont{Abazov}\
  \emph{et~al.}(2009{\natexlab{b}})\citenamefont{Abazov}
  \emph{et~al.}}]{Abazov:2009st}%
  \BibitemOpen
  \bibfield{author}{%
  \bibinfo {author} {\bibnamefont{Abazov}, \bibfnamefont{V.}}, \emph{et~al.}
  (\bibinfo {collaboration} {D0})}%
  , \bibinfo {year} {2009}{\natexlab{b}},\ \bibfield{journal}{%
  \Doi{10.1103/PhysRevLett.103.092001}{\bibinfo {journal} {Phys. Rev. Lett.}}\
  }%
  \textbf{\bibinfo {volume} {103}},\ \bibinfo {pages} {092001},\
  \Eprint{http://arxiv.org/abs/0903.0850}{arXiv:0903.0850 [hep-ex]}%
  \bibAnnoteFile{NoStop}{Abazov:2009st}%
%%CITATION = 0903.0850;%%
\bibitem[{\citenamefont{Abazov}\
  \emph{et~al.}(2010{\natexlab{a}})\citenamefont{Abazov}
  \emph{et~al.}}]{DzTprime43}%
  \BibitemOpen
  \bibfield{author}{%
  \bibinfo {author} {\bibnamefont{Abazov}, \bibfnamefont{V.}}, \emph{et~al.}
  (\bibinfo {collaboration} {D0})}%
  , \bibinfo {year} {2010}{\natexlab{a}}\ \bibinfo {note} {preliminary result
  available in \dzero\ conference note 5892.}%
  \bibAnnoteFile{Stop}{DzTprime43}%
\bibitem[{\citenamefont{Abazov}\
  \emph{et~al.}(2011{\natexlab{a}})\citenamefont{Abazov}
  \emph{et~al.}}]{DzttSpinDil54}%
  \BibitemOpen
  \bibfield{author}{%
  \bibinfo {author} {\bibnamefont{Abazov}, \bibfnamefont{V.}}, \emph{et~al.}
  (\bibinfo {collaboration} {D0})}%
  , \bibinfo {year} {2011}{\natexlab{a}}\ \bibinfo {note} {submitted to Phys.
  Lett. B.},\ \Eprint{http://arxiv.org/abs/1103.1871}{arXiv:1103.1871
  [hep-ex]}%
  \bibAnnoteFile{NoStop}{DzttSpinDil54}%
%%CITATION = 1103.1871;%%
\bibitem[{\citenamefont{Abazov}\
  \emph{et~al.}(2011{\natexlab{b}})\citenamefont{Abazov}
  \emph{et~al.}}]{DzAfb43}%
  \BibitemOpen
  \bibfield{author}{%
  \bibinfo {author} {\bibnamefont{Abazov}, \bibfnamefont{V.}}, \emph{et~al.}
  (\bibinfo {collaboration} {D0})}%
  , \bibinfo {year} {2011}{\natexlab{b}}\ \bibinfo {note} {preliminary result
  available in \dzero\ conference note 6062.}%
  \bibAnnoteFile{Stop}{DzAfb43}%
\bibitem[{\citenamefont{Abazov}\
  \emph{et~al.}(2006{\natexlab{b}})\citenamefont{Abazov}
  \emph{et~al.}}]{Abazov:2006ka}%
  \BibitemOpen
  \bibfield{author}{%
  \bibinfo {author} {\bibnamefont{Abazov}, \bibfnamefont{V.~M.}}, \emph{et~al.}
  (\bibinfo {collaboration} {D0})}%
  , \bibinfo {year} {2006}{\natexlab{b}},\ \bibfield{journal}{%
  \Doi{10.1103/PhysRevD.74.112004}{\bibinfo {journal} {Phys. Rev.}}\ }%
  \textbf{\bibinfo {volume} {D74}},\ \bibinfo {pages} {112004},\
  \Eprint{http://arxiv.org/abs/hep-ex/0611002}{arXiv:hep-ex/0611002}%
  \bibAnnoteFile{NoStop}{Abazov:2006ka}%
%%CITATION = HEP-EX/0611002;%%
\bibitem[{\citenamefont{Abazov}\
  \emph{et~al.}(2007{\natexlab{a}})\citenamefont{Abazov}
  \emph{et~al.}}]{Abazov:2006gd}%
  \BibitemOpen
  \bibfield{author}{%
  \bibinfo {author} {\bibnamefont{Abazov}, \bibfnamefont{V.~M.}}, \emph{et~al.}
  (\bibinfo {collaboration} {D0})}%
  , \bibinfo {year} {2007}{\natexlab{a}},\ \bibfield{journal}{%
  \Doi{10.1103/PhysRevLett.98.181802}{\bibinfo {journal} {Phys. Rev. Lett.}}\
  }%
  \textbf{\bibinfo {volume} {98}},\ \bibinfo {pages} {181802},\
  \Eprint{http://arxiv.org/abs/hep-ex/0612052}{arXiv:hep-ex/0612052}%
  \bibAnnoteFile{NoStop}{Abazov:2006gd}%
%%CITATION = HEP-EX/0612052;%%
\bibitem[{\citenamefont{Abazov}\
  \emph{et~al.}(2007{\natexlab{b}})\citenamefont{Abazov}
  \emph{et~al.}}]{Abazov:2006vd}%
  \BibitemOpen
  \bibfield{author}{%
  \bibinfo {author} {\bibnamefont{Abazov}, \bibfnamefont{V.~M.}}, \emph{et~al.}
  (\bibinfo {collaboration} {D0})}%
  , \bibinfo {year} {2007}{\natexlab{b}},\ \bibfield{journal}{%
  \Doi{10.1103/PhysRevLett.98.041801}{\bibinfo {journal} {Phys. Rev. Lett.}}\
  }%
  \textbf{\bibinfo {volume} {98}},\ \bibinfo {pages} {041801},\
  \Eprint{http://arxiv.org/abs/hep-ex/0608044}{arXiv:hep-ex/0608044}%
  \bibAnnoteFile{NoStop}{Abazov:2006vd}%
%%CITATION = HEP-EX/0608044;%%
\bibitem[{\citenamefont{Abazov}\
  \emph{et~al.}(2007{\natexlab{c}})\citenamefont{Abazov}
  \emph{et~al.}}]{Abazov:2007rk}%
  \BibitemOpen
  \bibfield{author}{%
  \bibinfo {author} {\bibnamefont{Abazov}, \bibfnamefont{V.~M.}}, \emph{et~al.}
  (\bibinfo {collaboration} {D0})}%
  , \bibinfo {year} {2007}{\natexlab{c}},\ \bibfield{journal}{%
  \Doi{10.1103/PhysRevD.75.092001}{\bibinfo {journal} {Phys. Rev.}}\ }%
  \textbf{\bibinfo {volume} {D75}},\ \bibinfo {pages} {092001},\
  \Eprint{http://arxiv.org/abs/hep-ex/0702018}{arXiv:hep-ex/0702018}%
  \bibAnnoteFile{NoStop}{Abazov:2007rk}%
%%CITATION = HEP-EX/0702018;%%
\bibitem[{\citenamefont{Abazov}\
  \emph{et~al.}(2008{\natexlab{b}})\citenamefont{Abazov}
  \emph{et~al.}}]{Abazov:2008kt}%
  \BibitemOpen
  \bibfield{author}{%
  \bibinfo {author} {\bibnamefont{Abazov}, \bibfnamefont{V.~M.}}, \emph{et~al.}
  (\bibinfo {collaboration} {D0})}%
  , \bibinfo {year} {2008}{\natexlab{b}},\ \bibfield{journal}{%
  \Doi{10.1103/PhysRevD.78.012005}{\bibinfo {journal} {Phys. Rev.}}\ }%
  \textbf{\bibinfo {volume} {D78}},\ \bibinfo {pages} {012005},\
  \Eprint{http://arxiv.org/abs/0803.0739}{arXiv:0803.0739 [hep-ex]}%
  \bibAnnoteFile{NoStop}{Abazov:2008kt}%
%%CITATION = 0803.0739;%%
\bibitem[{\citenamefont{Abazov}\
  \emph{et~al.}(2008{\natexlab{c}})\citenamefont{Abazov}
  \emph{et~al.}}]{Abazov:2007qb}%
  \BibitemOpen
  \bibfield{author}{%
  \bibinfo {author} {\bibnamefont{Abazov}, \bibfnamefont{V.~M.}}, \emph{et~al.}
  (\bibinfo {collaboration} {D0})}%
  , \bibinfo {year} {2008}{\natexlab{c}},\ \bibfield{journal}{%
  \Doi{10.1103/PhysRevLett.100.142002}{\bibinfo {journal} {Phys. Rev. Lett.}}\
  }%
  \textbf{\bibinfo {volume} {100}},\ \bibinfo {pages} {142002},\
  \Eprint{http://arxiv.org/abs/0712.0851}{arXiv:0712.0851 [hep-ex]}%
  \bibAnnoteFile{NoStop}{Abazov:2007qb}%
%%CITATION = 0712.0851;%%
\bibitem[{\citenamefont{Abazov}\
  \emph{et~al.}(2008{\natexlab{d}})\citenamefont{Abazov}
  \emph{et~al.}}]{Abazov:2007ve}%
  \BibitemOpen
  \bibfield{author}{%
  \bibinfo {author} {\bibnamefont{Abazov}, \bibfnamefont{V.~M.}}, \emph{et~al.}
  (\bibinfo {collaboration} {D0})}%
  , \bibinfo {year} {2008}{\natexlab{d}},\ \bibfield{journal}{%
  \Doi{10.1103/PhysRevLett.100.062004}{\bibinfo {journal} {Phys. Rev. Lett.}}\
  }%
  \textbf{\bibinfo {volume} {100}},\ \bibinfo {pages} {062004},\
  \Eprint{http://arxiv.org/abs/0711.0032}{arXiv:0711.0032 [hep-ex]}%
  \bibAnnoteFile{NoStop}{Abazov:2007ve}%
%%CITATION = 0711.0032;%%
\bibitem[{\citenamefont{Abazov}\
  \emph{et~al.}(2008{\natexlab{e}})\citenamefont{Abazov}
  \emph{et~al.}}]{Abazov:2008ds}%
  \BibitemOpen
  \bibfield{author}{%
  \bibinfo {author} {\bibnamefont{Abazov}, \bibfnamefont{V.~M.}}, \emph{et~al.}
  (\bibinfo {collaboration} {D0})}%
  , \bibinfo {year} {2008}{\natexlab{e}},\ \bibfield{journal}{%
  \Doi{10.1103/PhysRevLett.101.182001}{\bibinfo {journal} {Phys. Rev. Lett.}}\
  }%
  \textbf{\bibinfo {volume} {101}},\ \bibinfo {pages} {182001},\
  \Eprint{http://arxiv.org/abs/0807.2141}{arXiv:0807.2141 [hep-ex]}%
  \bibAnnoteFile{NoStop}{Abazov:2008ds}%
%%CITATION = 0807.2141;%%
\bibitem[{\citenamefont{Abazov}\
  \emph{et~al.}(2008{\natexlab{f}})\citenamefont{Abazov}
  \emph{et~al.}}]{Abazov:2008ny}%
  \BibitemOpen
  \bibfield{author}{%
  \bibinfo {author} {\bibnamefont{Abazov}, \bibfnamefont{V.~M.}}, \emph{et~al.}
  (\bibinfo {collaboration} {D0})}%
  , \bibinfo {year} {2008}{\natexlab{f}},\ \bibfield{journal}{%
  \Doi{10.1016/j.physletb.2008.08.027}{\bibinfo {journal} {Phys. Lett.}}\ }%
  \textbf{\bibinfo {volume} {B668}},\ \bibinfo {pages} {98},\
  \Eprint{http://arxiv.org/abs/0804.3664}{arXiv:0804.3664 [hep-ex]}%
  \bibAnnoteFile{NoStop}{Abazov:2008ny}%
%%CITATION = 0804.3664;%%
\bibitem[{\citenamefont{Abazov}\
  \emph{et~al.}(2008{\natexlab{g}})\citenamefont{Abazov}
  \emph{et~al.}}]{Abazov:2007im}%
  \BibitemOpen
  \bibfield{author}{%
  \bibinfo {author} {\bibnamefont{Abazov}, \bibfnamefont{V.~M.}}, \emph{et~al.}
  (\bibinfo {collaboration} {D0})}%
  , \bibinfo {year} {2008}{\natexlab{g}},\ \bibfield{journal}{%
  \Doi{10.1016/j.physletb.2007.11.086}{\bibinfo {journal} {Phys. Lett.}}\ }%
  \textbf{\bibinfo {volume} {B659}},\ \bibinfo {pages} {500},\
  \Eprint{http://arxiv.org/abs/0707.2864}{arXiv:0707.2864 [hep-ex]}%
  \bibAnnoteFile{NoStop}{Abazov:2007im}%
%%CITATION = 0707.2864;%%
\bibitem[{\citenamefont{Abazov}\
  \emph{et~al.}(2008{\natexlab{h}})\citenamefont{Abazov}
  \emph{et~al.}}]{Abazov:2008yn}%
  \BibitemOpen
  \bibfield{author}{%
  \bibinfo {author} {\bibnamefont{Abazov}, \bibfnamefont{V.~M.}}, \emph{et~al.}
  (\bibinfo {collaboration} {D0})}%
  , \bibinfo {year} {2008}{\natexlab{h}},\ \bibfield{journal}{%
  \Doi{10.1103/PhysRevLett.100.192003}{\bibinfo {journal} {Phys. Rev. Lett.}}\
  }%
  \textbf{\bibinfo {volume} {100}},\ \bibinfo {pages} {192003},\
  \Eprint{http://arxiv.org/abs/0801.1326}{arXiv:0801.1326 [hep-ex]}%
  \bibAnnoteFile{NoStop}{Abazov:2008yn}%
%%CITATION = 0801.1326;%%
\bibitem[{\citenamefont{Abazov}\
  \emph{et~al.}(2009{\natexlab{c}})\citenamefont{Abazov}
  \emph{et~al.}}]{Abazov:2009ae}%
  \BibitemOpen
  \bibfield{author}{%
  \bibinfo {author} {\bibnamefont{Abazov}, \bibfnamefont{V.~M.}}, \emph{et~al.}
  (\bibinfo {collaboration} {D0})}%
  , \bibinfo {year} {2009}{\natexlab{c}},\ \bibfield{journal}{%
  \Doi{10.1103/PhysRevD.80.071102}{\bibinfo {journal} {Phys. Rev.}}\ }%
  \textbf{\bibinfo {volume} {D80}},\ \bibinfo {pages} {071102},\
  \Eprint{http://arxiv.org/abs/0903.5525}{arXiv:0903.5525 [hep-ex]}%
  \bibAnnoteFile{NoStop}{Abazov:2009ae}%
%%CITATION = 0903.5525;%%
\bibitem[{\citenamefont{Abazov}\
  \emph{et~al.}(2009{\natexlab{d}})\citenamefont{Abazov}
  \emph{et~al.}}]{Abazov:2009xq}%
  \BibitemOpen
  \bibfield{author}{%
  \bibinfo {author} {\bibnamefont{Abazov}, \bibfnamefont{V.~M.}}, \emph{et~al.}
  (\bibinfo {collaboration} {D0})}%
  , \bibinfo {year} {2009}{\natexlab{d}},\ \bibfield{journal}{%
  \Doi{10.1103/PhysRevLett.103.132001}{\bibinfo {journal} {Phys. Rev. Lett.}}\
  }%
  \textbf{\bibinfo {volume} {103}},\ \bibinfo {pages} {132001},\
  \Eprint{http://arxiv.org/abs/0906.1172}{arXiv:0906.1172 [hep-ex]}%
  \bibAnnoteFile{NoStop}{Abazov:2009xq}%
%%CITATION = 0906.1172;%%
\bibitem[{\citenamefont{Abazov}\
  \emph{et~al.}(2009{\natexlab{e}})\citenamefont{Abazov}
  \emph{et~al.}}]{Abazov:2009eq}%
  \BibitemOpen
  \bibfield{author}{%
  \bibinfo {author} {\bibnamefont{Abazov}, \bibfnamefont{V.~M.}}, \emph{et~al.}
  (\bibinfo {collaboration} {D0})}%
  , \bibinfo {year} {2009}{\natexlab{e}},\ \bibfield{journal}{%
  \Doi{10.1103/PhysRevD.80.092006}{\bibinfo {journal} {Phys. Rev.}}\ }%
  \textbf{\bibinfo {volume} {D80}},\ \bibinfo {pages} {092006},\
  \Eprint{http://arxiv.org/abs/0904.3195}{arXiv:0904.3195 [hep-ex]}%
  \bibAnnoteFile{NoStop}{Abazov:2009eq}%
%%CITATION = 0904.3195;%%
\bibitem[{\citenamefont{Abazov}\
  \emph{et~al.}(2009{\natexlab{f}})\citenamefont{Abazov}
  \emph{et~al.}}]{Abazov:2009si}%
  \BibitemOpen
  \bibfield{author}{%
  \bibinfo {author} {\bibnamefont{Abazov}, \bibfnamefont{V.~M.}}, \emph{et~al.}
  (\bibinfo {collaboration} {D0})}%
  , \bibinfo {year} {2009}{\natexlab{f}},\ \bibfield{journal}{%
  \Doi{10.1016/j.physletb.2009.07.032}{\bibinfo {journal} {Phys. Lett.}}\ }%
  \textbf{\bibinfo {volume} {B679}},\ \bibinfo {pages} {177},\
  \Eprint{http://arxiv.org/abs/0901.2137}{arXiv:0901.2137 [hep-ex]}%
  \bibAnnoteFile{NoStop}{Abazov:2009si}%
%%CITATION = 0901.2137;%%
\bibitem[{\citenamefont{Abazov}\
  \emph{et~al.}(2009{\natexlab{g}})\citenamefont{Abazov}
  \emph{et~al.}}]{Abazov:2009ps}%
  \BibitemOpen
  \bibfield{author}{%
  \bibinfo {author} {\bibnamefont{Abazov}, \bibfnamefont{V.~M.}}, \emph{et~al.}
  (\bibinfo {collaboration} {D0})}%
  , \bibinfo {year} {2009}{\natexlab{g}},\ \bibfield{journal}{%
  \Doi{10.1016/j.physletb.2009.02.027}{\bibinfo {journal} {Phys. Lett.}}\ }%
  \textbf{\bibinfo {volume} {B674}},\ \bibinfo {pages} {4},\
  \Eprint{http://arxiv.org/abs/0901.1063}{arXiv:0901.1063 [hep-ex]}%
  \bibAnnoteFile{NoStop}{Abazov:2009ps}%
%%CITATION = 0901.1063;%%
\bibitem[{\citenamefont{Abazov}\
  \emph{et~al.}(2009{\natexlab{h}})\citenamefont{Abazov}
  \emph{et~al.}}]{Abazov:2009ky}%
  \BibitemOpen
  \bibfield{author}{%
  \bibinfo {author} {\bibnamefont{Abazov}, \bibfnamefont{V.~M.}}, \emph{et~al.}
  (\bibinfo {collaboration} {D0})}%
  , \bibinfo {year} {2009}{\natexlab{h}},\ \bibfield{journal}{%
  \Doi{10.1103/PhysRevLett.102.092002}{\bibinfo {journal} {Phys. Rev. Lett.}}\
  }%
  \textbf{\bibinfo {volume} {102}},\ \bibinfo {pages} {092002},\
  \Eprint{http://arxiv.org/abs/0901.0151}{arXiv:0901.0151 [hep-ex]}%
  \bibAnnoteFile{NoStop}{Abazov:2009ky}%
%%CITATION = 0901.0151;%%
\bibitem[{\citenamefont{Abazov}\
  \emph{et~al.}(2009{\natexlab{i}})\citenamefont{Abazov}
  \emph{et~al.}}]{Abazov:2008rn}%
  \BibitemOpen
  \bibfield{author}{%
  \bibinfo {author} {\bibnamefont{Abazov}, \bibfnamefont{V.~M.}}, \emph{et~al.}
  (\bibinfo {collaboration} {D0})}%
  , \bibinfo {year} {2009}{\natexlab{i}},\ \bibfield{journal}{%
  \Doi{10.1103/PhysRevLett.102.191802}{\bibinfo {journal} {Phys. Rev. Lett.}}\
  }%
  \textbf{\bibinfo {volume} {102}},\ \bibinfo {pages} {191802},\
  \Eprint{http://arxiv.org/abs/0807.0859}{arXiv:0807.0859 [hep-ex]}%
  \bibAnnoteFile{NoStop}{Abazov:2008rn}%
%%CITATION = 0807.0859;%%
\bibitem[{\citenamefont{Abazov}\
  \emph{et~al.}(2009{\natexlab{j}})\citenamefont{Abazov}
  \emph{et~al.}}]{Abazov:2009zh}%
  \BibitemOpen
  \bibfield{author}{%
  \bibinfo {author} {\bibnamefont{Abazov}, \bibfnamefont{V.~M.}}, \emph{et~al.}
  (\bibinfo {collaboration} {D0})}%
  , \bibinfo {year} {2009}{\natexlab{j}},\ \bibfield{journal}{%
  \Doi{10.1016/j.physletb.2009.11.016}{\bibinfo {journal} {Phys. Lett.}}\ }%
  \textbf{\bibinfo {volume} {B682}},\ \bibinfo {pages} {278},\
  \Eprint{http://arxiv.org/abs/0908.1811}{arXiv:0908.1811 [hep-ex]}%
  \bibAnnoteFile{NoStop}{Abazov:2009zh}%
%%CITATION = 0908.1811;%%
\bibitem[{\citenamefont{Abazov}\
  \emph{et~al.}(2010{\natexlab{b}})\citenamefont{Abazov}
  \emph{et~al.}}]{DzBtagging}%
  \BibitemOpen
  \bibfield{author}{%
  \bibinfo {author} {\bibnamefont{Abazov}, \bibfnamefont{V.~M.}}, \emph{et~al.}
  (\bibinfo {collaboration} {D0})}%
  , \bibinfo {year} {2010}{\natexlab{b}},\ \bibfield{journal}{%
  \Doi{10.1016/j.nima.2010.03.118}{\bibinfo {journal} {Nucl. Instrum. Meth.}}\
  }%
  \textbf{\bibinfo {volume} {A620}},\ \bibinfo {pages} {490},\
  \Eprint{http://arxiv.org/abs/1002.4224}{arXiv:1002.4224 [hep-ex]}%
  \bibAnnoteFile{NoStop}{DzBtagging}%
%%CITATION = 1002.4224;%%
\bibitem[{\citenamefont{Abazov}\
  \emph{et~al.}(2010{\natexlab{c}})\citenamefont{Abazov}
  \emph{et~al.}}]{Abazov:2010js}%
  \BibitemOpen
  \bibfield{author}{%
  \bibinfo {author} {\bibnamefont{Abazov}, \bibfnamefont{V.~M.}}, \emph{et~al.}
  (\bibinfo {collaboration} {D0})}%
  , \bibinfo {year} {2010}{\natexlab{c}},\ \bibfield{journal}{%
  \Doi{10.1016/j.physletb.2010.09.011}{\bibinfo {journal} {Phys. Lett.}}\ }%
  \textbf{\bibinfo {volume} {B693}},\ \bibinfo {pages} {515},\
  \Eprint{http://arxiv.org/abs/1001.1900}{arXiv:1001.1900 [hep-ex]}%
  \bibAnnoteFile{NoStop}{Abazov:2010js}%
%%CITATION = 1001.1900;%%
\bibitem[{\citenamefont{Abazov}\
  \emph{et~al.}(2010{\natexlab{d}})\citenamefont{Abazov}
  \emph{et~al.}}]{Abazov:2010pa}%
  \BibitemOpen
  \bibfield{author}{%
  \bibinfo {author} {\bibnamefont{Abazov}, \bibfnamefont{V.~M.}}, \emph{et~al.}
  (\bibinfo {collaboration} {D0})}%
  , \bibinfo {year} {2010}{\natexlab{d}},\ \bibfield{journal}{%
  \Doi{10.1103/PhysRevD.82.071102}{\bibinfo {journal} {Phys. Rev.}}\ }%
  \textbf{\bibinfo {volume} {D82}},\ \bibinfo {pages} {071102},\
  \Eprint{http://arxiv.org/abs/1008.4284}{arXiv:1008.4284 [hep-ex]}%
  \bibAnnoteFile{NoStop}{Abazov:2010pa}%
%%CITATION = 1008.4284;%%
\bibitem[{\citenamefont{Abazov}\
  \emph{et~al.}(2010{\natexlab{e}})\citenamefont{Abazov}
  \emph{et~al.}}]{Abazov:2009pa}%
  \BibitemOpen
  \bibfield{author}{%
  \bibinfo {author} {\bibnamefont{Abazov}, \bibfnamefont{V.~M.}}, \emph{et~al.}
  (\bibinfo {collaboration} {D0})}%
  , \bibinfo {year} {2010}{\natexlab{e}},\ \bibfield{journal}{%
  \Doi{10.1016/j.physletb.2009.11.038}{\bibinfo {journal} {Phys. Lett.}}\ }%
  \textbf{\bibinfo {volume} {B682}},\ \bibinfo {pages} {363},\
  \Eprint{http://arxiv.org/abs/0907.4259}{arXiv:0907.4259 [hep-ex]}%
  \bibAnnoteFile{NoStop}{Abazov:2009pa}%
%%CITATION = 0907.4259;%%
\bibitem[{\citenamefont{Abazov}\
  \emph{et~al.}(2010{\natexlab{f}})\citenamefont{Abazov}
  \emph{et~al.}}]{Abazov:2009ss}%
  \BibitemOpen
  \bibfield{author}{%
  \bibinfo {author} {\bibnamefont{Abazov}, \bibfnamefont{V.~M.}}, \emph{et~al.}
  (\bibinfo {collaboration} {D0})}%
  , \bibinfo {year} {2010}{\natexlab{f}},\ \bibfield{journal}{%
  \Doi{10.1103/PhysRevD.82.032002}{\bibinfo {journal} {Phys. Rev.}}\ }%
  \textbf{\bibinfo {volume} {D82}},\ \bibinfo {pages} {032002},\
  \Eprint{http://arxiv.org/abs/0911.4286}{arXiv:0911.4286 [hep-ex]}%
  \bibAnnoteFile{NoStop}{Abazov:2009ss}%
%%CITATION = 0911.4286;%%
\bibitem[{\citenamefont{Abazov}\
  \emph{et~al.}(2010{\natexlab{g}})\citenamefont{Abazov}
  \emph{et~al.}}]{Abazov:2010qk}%
  \BibitemOpen
  \bibfield{author}{%
  \bibinfo {author} {\bibnamefont{Abazov}, \bibfnamefont{V.~M.}}, \emph{et~al.}
  (\bibinfo {collaboration} {D0})}%
  , \bibinfo {year} {2010}{\natexlab{g}},\ \bibfield{journal}{%
  \Doi{10.1016/j.physletb.2010.08.011}{\bibinfo {journal} {Phys. Lett.}}\ }%
  \textbf{\bibinfo {volume} {B693}},\ \bibinfo {pages} {81},\
  \Eprint{http://arxiv.org/abs/1006.3575}{arXiv:1006.3575 [hep-ex]}%
  \bibAnnoteFile{NoStop}{Abazov:2010qk}%
%%CITATION = 1006.3575;%%
\bibitem[{\citenamefont{Abazov}\
  \emph{et~al.}(2011{\natexlab{c}})\citenamefont{Abazov}
  \emph{et~al.}}]{Abazov:2010tm}%
  \BibitemOpen
  \bibfield{author}{%
  \bibinfo {author} {\bibnamefont{Abazov}, \bibfnamefont{V.~M.}}, \emph{et~al.}
  (\bibinfo {collaboration} {D0 Collaboration})}%
  , \bibinfo {year} {2011}{\natexlab{c}},\ \bibfield{journal}{%
  \Doi{10.1103/PhysRevLett.106.022001}{\bibinfo {journal} {Phys.Rev.Lett.}}\ }%
  \textbf{\bibinfo {volume} {106}},\ \bibinfo {pages} {022001},\
  \Eprint{http://arxiv.org/abs/1009.5686}{arXiv:1009.5686 [hep-ex]}%
  \bibAnnoteFile{NoStop}{Abazov:2010tm}%
\bibitem[{\citenamefont{Abazov}\
  \emph{et~al.}(2011{\natexlab{d}})\citenamefont{Abazov}
  \emph{et~al.}}]{Abazov:2010zz}%
  \BibitemOpen
  \bibfield{author}{%
  \bibinfo {author} {\bibnamefont{Abazov}, \bibfnamefont{V.~M.}}, \emph{et~al.}
  (\bibinfo {collaboration} {D0})}%
  , \bibinfo {year} {2011}{\natexlab{d}}\ \bibinfo {note} {submitted to Phys.
  Rev. D},\ \Eprint{http://arxiv.org/abs/1101.0124}{arXiv:1101.0124 [hep-ex]}%
  \bibAnnoteFile{NoStop}{Abazov:2010zz}%
%%CITATION = 1101.0124;%%
\bibitem[{\citenamefont{Abazov}\
  \emph{et~al.}(2011{\natexlab{e}})\citenamefont{Abazov}
  \emph{et~al.}}]{Abazov:2011qf}%
  \BibitemOpen
  \bibfield{author}{%
  \bibinfo {author} {\bibnamefont{Abazov}, \bibfnamefont{V.~M.}}, \emph{et~al.}
  (\bibinfo {collaboration} {D0})}%
  , \bibinfo {year} {2011}{\natexlab{e}}\
  \Eprint{http://arxiv.org/abs/1103.4574}{arXiv:1103.4574 [hep-ex]}%
  \bibAnnoteFile{NoStop}{Abazov:2011qf}%
%%CITATION = 1103.4574;%%
\bibitem[{\citenamefont{Abbiendi}\ \emph{et~al.}(2006)\citenamefont{Abbiendi}
  \emph{et~al.}}]{LEP2CR}%
  \BibitemOpen
  \bibfield{author}{%
  \bibinfo {author} {\bibnamefont{Abbiendi}, \bibfnamefont{G.}},
  \emph{et~al.}}%
  , \bibinfo {year} {2006},\ \bibfield{journal}{%
  \Doi{10.1140/epjc/s2005-02439-x}{\bibinfo {journal} {Eur. Phys. J}}\ }%
  \textbf{\bibinfo {volume} {C45}},\ \bibinfo {pages} {291},\
  \Eprint{http://arxiv.org/abs/0508062}{0508062 [hep-ex]}%
  \bibAnnoteFile{NoStop}{LEP2CR}%
\bibitem[{\citenamefont{Abe}\ \emph{et~al.}(1994)\citenamefont{Abe}
  \emph{et~al.}}]{Top1994}%
  \BibitemOpen
  \bibfield{author}{%
  \bibinfo {author} {\bibnamefont{Abe}, \bibfnamefont{F.}}, \emph{et~al.}
  (\bibinfo {collaboration} {CDF})}%
  , \bibinfo {year} {1994},\ \bibfield{journal}{%
  \Doi{10.1103/PhysRevLett.73.225}{\bibinfo {journal} {Phys. Rev. Lett.}}\ }%
  \textbf{\bibinfo {volume} {73}},\ \bibinfo {pages} {225},\
  \Eprint{http://arxiv.org/abs/9405005}{arXiv:9405005 [hep-ex]}%
  \bibAnnoteFile{NoStop}{Top1994}%
\bibitem[{\citenamefont{Abe}\ \emph{et~al.}(1995)\citenamefont{Abe}
  \emph{et~al.}}]{Top1995c}%
  \BibitemOpen
  \bibfield{author}{%
  \bibinfo {author} {\bibnamefont{Abe}, \bibfnamefont{F.}}, \emph{et~al.}
  (\bibinfo {collaboration} {CDF})}%
  , \bibinfo {year} {1995},\ \bibfield{journal}{%
  \Doi{10.1103/PhysRevLett.74.2626}{\bibinfo {journal} {Phys. Rev. Lett.}}\ }%
  \textbf{\bibinfo {volume} {74}},\ \bibinfo {pages} {2626},\
  \Eprint{http://arxiv.org/abs/9503002}{arXiv:9503002 [hep-ex]}%
  \bibAnnoteFile{NoStop}{Top1995c}%
\bibitem[{\citenamefont{Abreu}\ \emph{et~al.}(1998)\citenamefont{Abreu}
  \emph{et~al.}}]{Abreu:1997ic}%
  \BibitemOpen
  \bibfield{author}{%
  \bibinfo {author} {\bibnamefont{Abreu}, \bibfnamefont{P.}}, \emph{et~al.}
  (\bibinfo {collaboration} {DELPHI})}%
  , \bibinfo {year} {1998},\ \bibfield{journal}{%
  \Doi{10.1007/s100520050163}{\bibinfo {journal} {Eur. Phys. J.}}\ }%
  \textbf{\bibinfo {volume} {C2}},\ \bibinfo {pages} {581}%
  \bibAnnoteFile{NoStop}{Abreu:1997ic}%
%%CITATION = EPHJA,C2,581;%%
\bibitem[{\citenamefont{Abulencia}\
  \emph{et~al.}(2006{\natexlab{a}})\citenamefont{Abulencia}
  \emph{et~al.}}]{CDFMtDil2006}%
  \BibitemOpen
  \bibfield{author}{%
  \bibinfo {author} {\bibnamefont{Abulencia}, \bibfnamefont{A.}}, \emph{et~al.}
  (\bibinfo {collaboration} {CDF})}%
  , \bibinfo {year} {2006}{\natexlab{a}},\ \bibfield{journal}{%
  \Doi{10.1103/PhysRevD.73.112006}{\bibinfo {journal} {Phys. Rev.}}\ }%
  \textbf{\bibinfo {volume} {D73}},\ \bibinfo {pages} {112006},\
  \Eprint{http://arxiv.org/abs/hep-ex/0602008}{arXiv:hep-ex/0602008 [hep-ex]}%
  \bibAnnoteFile{NoStop}{CDFMtDil2006}%
%%CITATION = hep-ex/0602008;%%
\bibitem[{\citenamefont{Abulencia}\
  \emph{et~al.}(2006{\natexlab{b}})\citenamefont{Abulencia}
  \emph{et~al.}}]{Abulencia:2005jd}%
  \BibitemOpen
  \bibfield{author}{%
  \bibinfo {author} {\bibnamefont{Abulencia}, \bibfnamefont{A.}}, \emph{et~al.}
  (\bibinfo {collaboration} {CDF})}%
  , \bibinfo {year} {2006}{\natexlab{b}},\ \bibfield{journal}{%
  \Doi{10.1103/PhysRevLett.96.042003}{\bibinfo {journal} {Phys. Rev. Lett.}}\
  }%
  \textbf{\bibinfo {volume} {96}},\ \bibinfo {pages} {042003},\
  \Eprint{http://arxiv.org/abs/hep-ex/0510065}{arXiv:hep-ex/0510065}%
  \bibAnnoteFile{NoStop}{Abulencia:2005jd}%
%%CITATION = HEP-EX/0510065;%%
\bibitem[{\citenamefont{Abulencia}\
  \emph{et~al.}(2006{\natexlab{c}})\citenamefont{Abulencia}
  \emph{et~al.}}]{CDFMtTplate2006}%
  \BibitemOpen
  \bibfield{author}{%
  \bibinfo {author} {\bibnamefont{Abulencia}, \bibfnamefont{A.}}, \emph{et~al.}
  (\bibinfo {collaboration} {CDF})}%
  , \bibinfo {year} {2006}{\natexlab{c}},\ \bibfield{journal}{%
  \Doi{10.1103/PhysRevD.73.032003}{\bibinfo {journal} {Phys. Rev.}}\ }%
  \textbf{\bibinfo {volume} {D73}},\ \bibinfo {pages} {032003},\
  \Eprint{http://arxiv.org/abs/hep-ex/0510048}{arXiv:hep-ex/0510048 [hep-ex]}%
  \bibAnnoteFile{NoStop}{CDFMtTplate2006}%
%%CITATION = hep-ex/0510048;%%
\bibitem[{\citenamefont{Abulencia}\
  \emph{et~al.}(2007{\natexlab{a}})\citenamefont{Abulencia}
  \emph{et~al.}}]{Abulencia:2005ix}%
  \BibitemOpen
  \bibfield{author}{%
  \bibinfo {author} {\bibnamefont{Abulencia}, \bibfnamefont{A.}}, \emph{et~al.}
  (\bibinfo {collaboration} {CDF Collaboration})}%
  , \bibinfo {year} {2007}{\natexlab{a}},\ \bibfield{journal}{%
  \Doi{10.1088/0954-3899/34/12/001}{\bibinfo {journal} {J.Phys.G}}\ }%
  \textbf{\bibinfo {volume} {G34}},\ \bibinfo {pages} {2457},\
  \Eprint{http://arxiv.org/abs/hep-ex/0508029}{arXiv:hep-ex/0508029 [hep-ex]}%
  \bibAnnoteFile{NoStop}{Abulencia:2005ix}%
\bibitem[{\citenamefont{Abulencia}\
  \emph{et~al.}(2007{\natexlab{b}})\citenamefont{Abulencia}
  \emph{et~al.}}]{CDFWZxsec}%
  \BibitemOpen
  \bibfield{author}{%
  \bibinfo {author} {\bibnamefont{Abulencia}, \bibfnamefont{A.}}, \emph{et~al.}
  (\bibinfo {collaboration} {CDF})}%
  , \bibinfo {year} {2007}{\natexlab{b}},\ \bibfield{journal}{%
  \Doi{10.1088/0954-3899/34/12/001}{\bibinfo {journal} {J.Phys.}}\ }%
  \textbf{\bibinfo {volume} {G34}},\ \bibinfo {pages} {2457},\
  \Eprint{http://arxiv.org/abs/0508029}{arXiv:0508029 [hep-ex]}%
  \bibAnnoteFile{NoStop}{CDFWZxsec}%
\bibitem[{\citenamefont{Acosta}\
  \emph{et~al.}(2005{\natexlab{a}})\citenamefont{Acosta}
  \emph{et~al.}}]{CDFSLTm}%
  \BibitemOpen
  \bibfield{author}{%
  \bibinfo {author} {\bibnamefont{Acosta}, \bibfnamefont{D.}}, \emph{et~al.}
  (\bibinfo {collaboration} {CDF})}%
  , \bibinfo {year} {2005}{\natexlab{a}},\ \bibfield{journal}{%
  \Doi{10.1103/PhysRevD.72.032002}{\bibinfo {journal} {Phys. Rev.}}\ }%
  \textbf{\bibinfo {volume} {D72}},\ \bibinfo {pages} {032002},\
  \Eprint{http://arxiv.org/abs/0506001}{arXiv:0506001 [hep-ex]}%
  \bibAnnoteFile{NoStop}{CDFSLTm}%
\bibitem[{\citenamefont{Acosta}\
  \emph{et~al.}(2005{\natexlab{b}})\citenamefont{Acosta}
  \emph{et~al.}}]{CDFBtagXSttbar}%
  \BibitemOpen
  \bibfield{author}{%
  \bibinfo {author} {\bibnamefont{Acosta}, \bibfnamefont{D.}}, \emph{et~al.}
  (\bibinfo {collaboration} {CDF})}%
  , \bibinfo {year} {2005}{\natexlab{b}},\ \bibfield{journal}{%
  \Doi{10.1103/PhysRevD.71.052003}{\bibinfo {journal} {Phys. Rev.}}\ }%
  \textbf{\bibinfo {volume} {D71}},\ \bibinfo {pages} {052003},\
  \Eprint{http://arxiv.org/abs/0410041}{arXiv:0410041 [hep-ex]}%
  \bibAnnoteFile{NoStop}{CDFBtagXSttbar}%
\bibitem[{\citenamefont{Agashe}\ \emph{et~al.}(2003)\citenamefont{Agashe},
  \citenamefont{Delgado}, \citenamefont{May},\ and\
  \citenamefont{Sundrum}}]{Agashe:2003zs}%
  \BibitemOpen
  \bibfield{author}{%
  \bibinfo {author} {\bibnamefont{Agashe}, \bibfnamefont{K.}}, \bibinfo
  {author} {\bibfnamefont{A.}~\bibnamefont{Delgado}}, \bibinfo {author}
  {\bibfnamefont{M.~J.}\ \bibnamefont{May}},\ and\ \bibinfo {author}
  {\bibfnamefont{R.}~\bibnamefont{Sundrum}}}%
  , \bibinfo {year} {2003},\ \bibfield{journal}{%
  \bibinfo {journal} {JHEP}\ }%
  \textbf{\bibinfo {volume} {08}},\ \bibinfo {pages} {050},\
  \Eprint{http://arxiv.org/abs/hep-ph/0308036}{arXiv:hep-ph/0308036}%
  \bibAnnoteFile{NoStop}{Agashe:2003zs}%
%%CITATION = HEP-PH/0308036;%%
\bibitem[{\citenamefont{Aguilar-Saavedra}(2004)}]{AguilarSaavedra:2004wm}%
  \BibitemOpen
  \bibfield{author}{%
  \bibinfo {author} {\bibnamefont{Aguilar-Saavedra}, \bibfnamefont{J.~A.}}}%
  , \bibinfo {year} {2004},\ \bibfield{journal}{%
  \bibinfo {journal} {Acta Phys. Polon.}\ }%
  \textbf{\bibinfo {volume} {B35}},\ \bibinfo {pages} {2695},\
  \Eprint{http://arxiv.org/abs/hep-ph/0409342}{arXiv:hep-ph/0409342}%
  \bibAnnoteFile{NoStop}{AguilarSaavedra:2004wm}%
%%CITATION = HEP-PH/0409342;%%
\bibitem[{\citenamefont{Aguilar-Saavedra}\
  \emph{et~al.}(2007)\citenamefont{Aguilar-Saavedra}
  \emph{et~al.}}]{AguilarWHel}%
  \BibitemOpen
  \bibfield{author}{%
  \bibinfo {author} {\bibnamefont{Aguilar-Saavedra}, \bibfnamefont{J.~A.}},
  \emph{et~al.}}%
  , \bibinfo {year} {2007},\ \bibfield{journal}{%
  \bibinfo {journal} {Eur. Phys. J}\ }%
  \textbf{\bibinfo {volume} {C50}},\ \bibinfo {pages} {519},\
  \Eprint{http://arxiv.org/abs/hep-ph/0605190}{arXiv:hep-ph/0605190}%
  \bibAnnoteFile{NoStop}{AguilarWHel}%
%%CITATION = HEP-PH/0605190;%%
\bibitem[{\citenamefont{Ahrens}\ \emph{et~al.}(2010)\citenamefont{Ahrens},
  \citenamefont{Ferroglia}, \citenamefont{Neubert}, \citenamefont{Pecjak},\
  and\ \citenamefont{Yang}}]{Ahrens:2010zv}%
  \BibitemOpen
  \bibfield{author}{%
  \bibinfo {author} {\bibnamefont{Ahrens}, \bibfnamefont{V.}}, \bibinfo
  {author} {\bibfnamefont{A.}~\bibnamefont{Ferroglia}}, \bibinfo {author}
  {\bibfnamefont{M.}~\bibnamefont{Neubert}}, \bibinfo {author}
  {\bibfnamefont{B.~D.}\ \bibnamefont{Pecjak}},\ and\ \bibinfo {author}
  {\bibfnamefont{L.~L.}\ \bibnamefont{Yang}}}%
  , \bibinfo {year} {2010}\
  \Eprint{http://arxiv.org/abs/1003.5827}{arXiv:1003.5827 [hep-ph]}%
  \bibAnnoteFile{NoStop}{Ahrens:2010zv}%
%%CITATION = 1003.5827;%%
\bibitem[{\citenamefont{Almeida}\ \emph{et~al.}(2008)\citenamefont{Almeida},
  \citenamefont{Sterman},\ and\ \citenamefont{Vogelsang}}]{Almeida:2008ug}%
  \BibitemOpen
  \bibfield{author}{%
  \bibinfo {author} {\bibnamefont{Almeida}, \bibfnamefont{L.~G.}}, \bibinfo
  {author} {\bibfnamefont{G.}~\bibnamefont{Sterman}},\ and\ \bibinfo {author}
  {\bibfnamefont{W.}~\bibnamefont{Vogelsang}}}%
  , \bibinfo {year} {2008},\ \bibfield{journal}{%
  \Doi{10.1103/PhysRevD.78.014008}{\bibinfo {journal} {Phys. Rev.}}\ }%
  \textbf{\bibinfo {volume} {D78}},\ \bibinfo {pages} {014008},\
  \Eprint{http://arxiv.org/abs/0805.1885}{arXiv:0805.1885 [hep-ph]}%
  \bibAnnoteFile{NoStop}{Almeida:2008ug}%
%%CITATION = 0805.1885;%%
\bibitem[{\citenamefont{Alwall}\ \emph{et~al.}(2007)\citenamefont{Alwall}
  \emph{et~al.}}]{madevt}%
  \BibitemOpen
  \bibfield{author}{%
  \bibinfo {author} {\bibnamefont{Alwall}, \bibfnamefont{J.}}, \emph{et~al.}}%
  , \bibinfo {year} {2007},\ \bibfield{journal}{%
  \Doi{10.1088/1126-6708/2007/09/028}{\bibinfo {journal} {JHEP}}\ }%
  \textbf{\bibinfo {volume} {0709}},\ \bibinfo {pages} {028},\
  \Eprint{http://arxiv.org/abs/0706.2334}{arXiv:0706.2334 [hep-ph]}%
  \bibAnnoteFile{NoStop}{madevt}%
\bibitem[{\citenamefont{Alwall}\ \emph{et~al.}(2008)\citenamefont{Alwall}
  \emph{et~al.}}]{matching}%
  \BibitemOpen
  \bibfield{author}{%
  \bibinfo {author} {\bibnamefont{Alwall}, \bibfnamefont{J.}}, \emph{et~al.}}%
  , \bibinfo {year} {2008},\ \bibfield{journal}{%
  \Doi{10.1140/epjc/s10052-007-0490-5}{\bibinfo {journal} {Eur. Phys. J.}}\ }%
  \textbf{\bibinfo {volume} {C53}},\ \bibinfo {pages} {473},\
  \Eprint{http://arxiv.org/abs/0706.2569}{arXiv:0706.2569 [hep-ph]}%
  \bibAnnoteFile{NoStop}{matching}%
\bibitem[{\citenamefont{Antoniadis}\
  \emph{et~al.}(1994)\citenamefont{Antoniadis}, \citenamefont{Benakli},\ and\
  \citenamefont{Quiros}}]{Antoniadis:1994yi}%
  \BibitemOpen
  \bibfield{author}{%
  \bibinfo {author} {\bibnamefont{Antoniadis}, \bibfnamefont{I.}}, \bibinfo
  {author} {\bibfnamefont{K.}~\bibnamefont{Benakli}},\ and\ \bibinfo {author}
  {\bibfnamefont{M.}~\bibnamefont{Quiros}}}%
  , \bibinfo {year} {1994},\ \bibfield{journal}{%
  \Doi{10.1016/0370-2693(94)91058-8}{\bibinfo {journal} {Phys. Lett.}}\ }%
  \textbf{\bibinfo {volume} {B331}},\ \bibinfo {pages} {313},\
  \Eprint{http://arxiv.org/abs/hep-ph/9403290}{arXiv:hep-ph/9403290}%
  \bibAnnoteFile{NoStop}{Antoniadis:1994yi}%
%%CITATION = HEP-PH/9403290;%%
\bibitem[{\citenamefont{Antoniadis}\
  \emph{et~al.}(1999)\citenamefont{Antoniadis}, \citenamefont{Benakli},\ and\
  \citenamefont{Quiros}}]{Antoniadis:1999bq}%
  \BibitemOpen
  \bibfield{author}{%
  \bibinfo {author} {\bibnamefont{Antoniadis}, \bibfnamefont{I.}}, \bibinfo
  {author} {\bibfnamefont{K.}~\bibnamefont{Benakli}},\ and\ \bibinfo {author}
  {\bibfnamefont{M.}~\bibnamefont{Quiros}}}%
  , \bibinfo {year} {1999},\ \bibfield{journal}{%
  \Doi{10.1016/S0370-2693(99)00764-9}{\bibinfo {journal} {Phys. Lett.}}\ }%
  \textbf{\bibinfo {volume} {B460}},\ \bibinfo {pages} {176},\
  \Eprint{http://arxiv.org/abs/hep-ph/9905311}{arXiv:hep-ph/9905311}%
  \bibAnnoteFile{NoStop}{Antoniadis:1999bq}%
%%CITATION = HEP-PH/9905311;%%
\bibitem[{\citenamefont{Antunano}\ \emph{et~al.}(2008)\citenamefont{Antunano},
  \citenamefont{Kuhn},\ and\ \citenamefont{Rodrigo}}]{Antunano:2007da}%
  \BibitemOpen
  \bibfield{author}{%
  \bibinfo {author} {\bibnamefont{Antunano}, \bibfnamefont{O.}}, \bibinfo
  {author} {\bibfnamefont{J.~H.}\ \bibnamefont{Kuhn}},\ and\ \bibinfo {author}
  {\bibfnamefont{G.}~\bibnamefont{Rodrigo}}}%
  , \bibinfo {year} {2008},\ \bibfield{journal}{%
  \Doi{10.1103/PhysRevD.77.014003}{\bibinfo {journal} {Phys. Rev.}}\ }%
  \textbf{\bibinfo {volume} {D77}},\ \bibinfo {pages} {014003},\
  \Eprint{http://arxiv.org/abs/0709.1652}{arXiv:0709.1652 [hep-ph]}%
  \bibAnnoteFile{NoStop}{Antunano:2007da}%
%%CITATION = 0709.1652;%%
\bibitem[{\citenamefont{Appelquist}\
  \emph{et~al.}(2001)\citenamefont{Appelquist}, \citenamefont{Cheng},\ and\
  \citenamefont{Dobrescu}}]{Appelquist:2000nn}%
  \BibitemOpen
  \bibfield{author}{%
  \bibinfo {author} {\bibnamefont{Appelquist}, \bibfnamefont{T.}}, \bibinfo
  {author} {\bibfnamefont{H.-C.}\ \bibnamefont{Cheng}},\ and\ \bibinfo {author}
  {\bibfnamefont{B.~A.}\ \bibnamefont{Dobrescu}}}%
  , \bibinfo {year} {2001},\ \bibfield{journal}{%
  \Doi{10.1103/PhysRevD.64.035002}{\bibinfo {journal} {Phys. Rev.}}\ }%
  \textbf{\bibinfo {volume} {D64}},\ \bibinfo {pages} {035002},\
  \Eprint{http://arxiv.org/abs/hep-ph/0012100}{arXiv:hep-ph/0012100}%
  \bibAnnoteFile{NoStop}{Appelquist:2000nn}%
%%CITATION = HEP-PH/0012100;%%
\bibitem[{\citenamefont{Aubert}\ \emph{et~al.}(1974)\citenamefont{Aubert}
  \emph{et~al.}}]{JPsi1974ST}%
  \BibitemOpen
  \bibfield{author}{%
  \bibinfo {author} {\bibnamefont{Aubert}, \bibfnamefont{J.}}, \emph{et~al.}}%
  , \bibinfo {year} {1974},\ \bibfield{journal}{%
  \Doi{10.1103/PhysRevLett.33.1404}{\bibinfo {journal} {Phys. Rev. Lett.}}\ }%
  \textbf{\bibinfo {volume} {33}},\ \bibinfo {pages} {1404}%
  \bibAnnoteFile{NoStop}{JPsi1974ST}%
\bibitem[{\citenamefont{Augustin}\ \emph{et~al.}(1974)\citenamefont{Augustin}
  \emph{et~al.}}]{JPsi1974BR}%
  \BibitemOpen
  \bibfield{author}{%
  \bibinfo {author} {\bibnamefont{Augustin}, \bibfnamefont{J.}},
  \emph{et~al.}}%
  , \bibinfo {year} {1974},\ \bibfield{journal}{%
  \Doi{10.1103/PhysRevLett.33.1406}{\bibinfo {journal} {Phys. Rev. Lett.}}\ }%
  \textbf{\bibinfo {volume} {33}},\ \bibinfo {pages} {1406}%
  \bibAnnoteFile{NoStop}{JPsi1974BR}%
\bibitem[{\citenamefont{Bardeen}\ \emph{et~al.}(1978)\citenamefont{Bardeen},
  \citenamefont{Buras}, \citenamefont{Duke},\ and\
  \citenamefont{Muta}}]{Bardeen:1978yd}%
  \BibitemOpen
  \bibfield{author}{%
  \bibinfo {author} {\bibnamefont{Bardeen}, \bibfnamefont{W.~A.}}, \bibinfo
  {author} {\bibfnamefont{A.~J.}\ \bibnamefont{Buras}}, \bibinfo {author}
  {\bibfnamefont{D.~W.}\ \bibnamefont{Duke}},\ and\ \bibinfo {author}
  {\bibfnamefont{T.}~\bibnamefont{Muta}}}%
  , \bibinfo {year} {1978},\ \bibfield{journal}{%
  \Doi{10.1103/PhysRevD.18.3998}{\bibinfo {journal} {Phys. Rev.}}\ }%
  \textbf{\bibinfo {volume} {D18}},\ \bibinfo {pages} {3998}%
  \bibAnnoteFile{NoStop}{Bardeen:1978yd}%
%%CITATION = PHRVA,D18,3998;%%
\bibitem[{\citenamefont{Barger}\ \emph{et~al.}(1989)\citenamefont{Barger},
  \citenamefont{Ohnemus},\ and\ \citenamefont{Phillips}}]{Barger:1988jj}%
  \BibitemOpen
  \bibfield{author}{%
  \bibinfo {author} {\bibnamefont{Barger}, \bibfnamefont{V.~D.}}, \bibinfo
  {author} {\bibfnamefont{J.}~\bibnamefont{Ohnemus}},\ and\ \bibinfo {author}
  {\bibfnamefont{R.~J.~N.}\ \bibnamefont{Phillips}}}%
  , \bibinfo {year} {1989},\ \bibfield{journal}{%
  \Doi{10.1142/S0217751X89000297}{\bibinfo {journal} {Int. J. Mod. Phys.}}\ }%
  \textbf{\bibinfo {volume} {A4}},\ \bibinfo {pages} {617}%
  \bibAnnoteFile{NoStop}{Barger:1988jj}%
%%CITATION = IMPAE,A4,617;%%
\bibitem[{\citenamefont{Barr}\ \emph{et~al.}(2003)\citenamefont{Barr},
  \citenamefont{Lester},\ and\ \citenamefont{Stephens}}]{Barr:2003}%
  \BibitemOpen
  \bibfield{author}{%
  \bibinfo {author} {\bibnamefont{Barr}, \bibfnamefont{A.}}, \bibinfo {author}
  {\bibfnamefont{C.}~\bibnamefont{Lester}},\ and\ \bibinfo {author}
  {\bibfnamefont{P.}~\bibnamefont{Stephens}}}%
  , \bibinfo {year} {2003},\ \bibfield{journal}{%
  \Doi{10.1088/0954-3899/29/10/304}{\bibinfo {journal} {J. Phys.}}\ }%
  \textbf{\bibinfo {volume} {G29}},\ \bibinfo {pages} {2343},\
  \Eprint{http://arxiv.org/abs/0304226}{arXiv:0304226 [hep-ph]}%
  \bibAnnoteFile{NoStop}{Barr:2003}%
%%CITATION = hep-ph/0304226;%%
\bibitem[{\citenamefont{Beenakker}\
  \emph{et~al.}(1989)\citenamefont{Beenakker}, \citenamefont{Kuijf},
  \citenamefont{van Neerven},\ and\ \citenamefont{Smith}}]{Beenakker:1988bq}%
  \BibitemOpen
  \bibfield{author}{%
  \bibinfo {author} {\bibnamefont{Beenakker}, \bibfnamefont{W.}}, \bibinfo
  {author} {\bibfnamefont{H.}~\bibnamefont{Kuijf}}, \bibinfo {author}
  {\bibfnamefont{W.~L.}\ \bibnamefont{van Neerven}},\ and\ \bibinfo {author}
  {\bibfnamefont{J.}~\bibnamefont{Smith}}}%
  , \bibinfo {year} {1989},\ \bibfield{journal}{%
  \Doi{10.1103/PhysRevD.40.54}{\bibinfo {journal} {Phys. Rev.}}\ }%
  \textbf{\bibinfo {volume} {D40}},\ \bibinfo {pages} {54}%
  \bibAnnoteFile{NoStop}{Beenakker:1988bq}%
%%CITATION = PHRVA,D40,54;%%
\bibitem[{\citenamefont{Beneke}\ and\
  \citenamefont{Braun}(1994)}]{Beneke:1994sw}%
  \BibitemOpen
  \bibfield{author}{%
  \bibinfo {author} {\bibnamefont{Beneke}, \bibfnamefont{M.}},\ and\ \bibinfo
  {author} {\bibfnamefont{V.~M.}\ \bibnamefont{Braun}}}%
  , \bibinfo {year} {1994},\ \bibfield{journal}{%
  \Doi{10.1016/0550-3213(94)90314-X}{\bibinfo {journal} {Nucl. Phys.}}\ }%
  \textbf{\bibinfo {volume} {B426}},\ \bibinfo {pages} {301},\
  \Eprint{http://arxiv.org/abs/hep-ph/9402364}{arXiv:hep-ph/9402364}%
  \bibAnnoteFile{NoStop}{Beneke:1994sw}%
%%CITATION = HEP-PH/9402364;%%
\bibitem[{\citenamefont{Berends}\ \emph{et~al.}(1998)\citenamefont{Berends},
  \citenamefont{Papadopoulos},\ and\ \citenamefont{Pittau}}]{Berends:1997dm}%
  \BibitemOpen
  \bibfield{author}{%
  \bibinfo {author} {\bibnamefont{Berends}, \bibfnamefont{F.~A.}}, \bibinfo
  {author} {\bibfnamefont{C.~G.}\ \bibnamefont{Papadopoulos}},\ and\ \bibinfo
  {author} {\bibfnamefont{R.}~\bibnamefont{Pittau}}}%
  , \bibinfo {year} {1998},\ \bibfield{journal}{%
  \Doi{10.1016/S0370-2693(97)01374-9}{\bibinfo {journal} {Phys. Lett.}}\ }%
  \textbf{\bibinfo {volume} {B417}},\ \bibinfo {pages} {385},\
  \Eprint{http://arxiv.org/abs/hep-ph/9709257}{arXiv:hep-ph/9709257}%
  \bibAnnoteFile{NoStop}{Berends:1997dm}%
%%CITATION = HEP-PH/9709257;%%
\bibitem[{\citenamefont{Bernreuther}\
  \emph{et~al.}(2004{\natexlab{a}})\citenamefont{Bernreuther},
  \citenamefont{Brandenburg}, \citenamefont{Si},\ and\
  \citenamefont{Uwer}}]{Bernreuther:2004jv}%
  \BibitemOpen
  \bibfield{author}{%
  \bibinfo {author} {\bibnamefont{Bernreuther}, \bibfnamefont{W.}}, \bibinfo
  {author} {\bibfnamefont{A.}~\bibnamefont{Brandenburg}}, \bibinfo {author}
  {\bibfnamefont{Z.~G.}\ \bibnamefont{Si}},\ and\ \bibinfo {author}
  {\bibfnamefont{P.}~\bibnamefont{Uwer}}}%
  , \bibinfo {year} {2004}{\natexlab{a}},\ \bibfield{journal}{%
  \Doi{10.1016/j.nuclphysb.2004.04.019}{\bibinfo {journal} {Nucl. Phys.}}\ }%
  \textbf{\bibinfo {volume} {B690}},\ \bibinfo {pages} {81},\
  \Eprint{http://arxiv.org/abs/hep-ph/0403035}{arXiv:hep-ph/0403035}%
  \bibAnnoteFile{NoStop}{Bernreuther:2004jv}%
%%CITATION = HEP-PH/0403035;%%
\bibitem[{\citenamefont{Bernreuther}\
  \emph{et~al.}(2004{\natexlab{b}})\citenamefont{Bernreuther},
  \citenamefont{Fuecker},\ and\ \citenamefont{Umeda}}]{Bernreuther:2003xj}%
  \BibitemOpen
  \bibfield{author}{%
  \bibinfo {author} {\bibnamefont{Bernreuther}, \bibfnamefont{W.}}, \bibinfo
  {author} {\bibfnamefont{M.}~\bibnamefont{Fuecker}},\ and\ \bibinfo {author}
  {\bibfnamefont{Y.}~\bibnamefont{Umeda}}}%
  , \bibinfo {year} {2004}{\natexlab{b}},\ \bibfield{journal}{%
  \Doi{10.1016/j.physletb.2003.12.010}{\bibinfo {journal} {Phys. Lett.}}\ }%
  \textbf{\bibinfo {volume} {B582}},\ \bibinfo {pages} {32},\
  \Eprint{http://arxiv.org/abs/hep-ph/0308296}{arXiv:hep-ph/0308296}%
  \bibAnnoteFile{NoStop}{Bernreuther:2003xj}%
%%CITATION = HEP-PH/0308296;%%
\bibitem[{\citenamefont{Bhatti}\ \emph{et~al.}(2006)\citenamefont{Bhatti}
  \emph{et~al.}}]{cdfjes}%
  \BibitemOpen
  \bibfield{author}{%
  \bibinfo {author} {\bibnamefont{Bhatti}, \bibfnamefont{A.}}, \emph{et~al.}
  (\bibinfo {collaboration} {CDF})}%
  , \bibinfo {year} {2006},\ \bibfield{journal}{%
  \Doi{10.1016/j.nima.2006.05.269}{\bibinfo {journal} {Nucl. Instrum. Meth.}}\
  }%
  \textbf{\bibinfo {volume} {A566}},\ \bibinfo {pages} {375},\
  \Eprint{http://arxiv.org/abs/0510047}{arXiv:0510047 [hep-ex]}%
  \bibAnnoteFile{NoStop}{cdfjes}%
\bibitem[{\citenamefont{Bigi}\ \emph{et~al.}(1994)\citenamefont{Bigi},
  \citenamefont{Shifman}, \citenamefont{Uraltsev},\ and\
  \citenamefont{Vainshtein}}]{Bigi:1994em}%
  \BibitemOpen
  \bibfield{author}{%
  \bibinfo {author} {\bibnamefont{Bigi}, \bibfnamefont{I.~I.~Y.}}, \bibinfo
  {author} {\bibfnamefont{M.~A.}\ \bibnamefont{Shifman}}, \bibinfo {author}
  {\bibfnamefont{N.~G.}\ \bibnamefont{Uraltsev}},\ and\ \bibinfo {author}
  {\bibfnamefont{A.~I.}\ \bibnamefont{Vainshtein}}}%
  , \bibinfo {year} {1994},\ \bibfield{journal}{%
  \Doi{10.1103/PhysRevD.50.2234}{\bibinfo {journal} {Phys. Rev.}}\ }%
  \textbf{\bibinfo {volume} {D50}},\ \bibinfo {pages} {2234},\
  \Eprint{http://arxiv.org/abs/hep-ph/9402360}{arXiv:hep-ph/9402360}%
  \bibAnnoteFile{NoStop}{Bigi:1994em}%
%%CITATION = HEP-PH/9402360;%%
\bibitem[{\citenamefont{Blaire}\ \emph{et~al.}(1996)\citenamefont{Blaire}
  \emph{et~al.}}]{cdf}%
  \BibitemOpen
  \bibfield{author}{%
  \bibinfo {author} {\bibnamefont{Blaire}, \bibfnamefont{R.}}, \emph{et~al.}
  (\bibinfo {collaboration} {CDF})}%
  , \bibinfo {year} {1996},\ \enquote{\bibinfo {title} {{The CDF II Detector:
  Technical Design Report}},}\ \bibinfo {note} {FERMILAB-Pub-96/390-E}%
  \bibAnnoteFile{NoStop}{cdf}%
\bibitem[{\citenamefont{Boos}\ \emph{et~al.}(2006)\citenamefont{Boos},
  \citenamefont{Bunichev}, \citenamefont{Dudko}, \citenamefont{Savrin},\ and\
  \citenamefont{Sherstnev}}]{singletop}%
  \BibitemOpen
  \bibfield{author}{%
  \bibinfo {author} {\bibnamefont{Boos}, \bibfnamefont{E.~E.}}, \bibinfo
  {author} {\bibfnamefont{V.~E.}\ \bibnamefont{Bunichev}}, \bibinfo {author}
  {\bibfnamefont{L.~V.}\ \bibnamefont{Dudko}}, \bibinfo {author}
  {\bibfnamefont{V.~I.}\ \bibnamefont{Savrin}},\ and\ \bibinfo {author}
  {\bibfnamefont{A.~V.}\ \bibnamefont{Sherstnev}}}%
  , \bibinfo {year} {2006},\ \bibfield{journal}{%
  \Doi{10.1134/S1063778806080084}{\bibinfo {journal} {Phys. Atom. Nucl.}}\ }%
  \textbf{\bibinfo {volume} {69}},\ \bibinfo {pages} {1317}%
  \bibAnnoteFile{NoStop}{singletop}%
%%CITATION = PANUE,69,1317;%%
\bibitem[{\citenamefont{Bowen}\ \emph{et~al.}(2006)\citenamefont{Bowen},
  \citenamefont{Ellis},\ and\ \citenamefont{Rainwater}}]{Bowen:2005ap}%
  \BibitemOpen
  \bibfield{author}{%
  \bibinfo {author} {\bibnamefont{Bowen}, \bibfnamefont{M.~T.}}, \bibinfo
  {author} {\bibfnamefont{S.~D.}\ \bibnamefont{Ellis}},\ and\ \bibinfo {author}
  {\bibfnamefont{D.}~\bibnamefont{Rainwater}}}%
  , \bibinfo {year} {2006},\ \bibfield{journal}{%
  \Doi{10.1103/PhysRevD.73.014008}{\bibinfo {journal} {Phys. Rev.}}\ }%
  \textbf{\bibinfo {volume} {D73}},\ \bibinfo {pages} {014008},\
  \Eprint{http://arxiv.org/abs/hep-ph/0509267}{arXiv:hep-ph/0509267}%
  \bibAnnoteFile{NoStop}{Bowen:2005ap}%
%%CITATION = HEP-PH/0509267;%%
\bibitem[{\citenamefont{Brandt}\ and\
  \citenamefont{Dahmen}(1979)}]{Brandt:1978zm}%
  \BibitemOpen
  \bibfield{author}{%
  \bibinfo {author} {\bibnamefont{Brandt}, \bibfnamefont{S.}},\ and\ \bibinfo
  {author} {\bibfnamefont{H.~D.}\ \bibnamefont{Dahmen}}}%
  , \bibinfo {year} {1979},\ \bibfield{journal}{%
  \Doi{10.1007/BF01450381}{\bibinfo {journal} {Zeit. Phys.}}\ }%
  \textbf{\bibinfo {volume} {C1}},\ \bibinfo {pages} {61}%
  \bibAnnoteFile{NoStop}{Brandt:1978zm}%
%%CITATION = ZEPYA,C1,61;%%
\bibitem[{\citenamefont{Buckley}\ \emph{et~al.}(2011)\citenamefont{Buckley}
  \emph{et~al.}}]{Buckley:2011ms}%
  \BibitemOpen
  \bibfield{author}{%
  \bibinfo {author} {\bibnamefont{Buckley}, \bibfnamefont{A.}}, \emph{et~al.}}%
  , \bibinfo {year} {2011}\
  \Eprint{http://arxiv.org/abs/1101.2599}{arXiv:1101.2599 [hep-ph]}%
  \bibAnnoteFile{NoStop}{Buckley:2011ms}%
%%CITATION = 1101.2599;%%
\bibitem[{\citenamefont{Cacciari}\ \emph{et~al.}(2008)\citenamefont{Cacciari},
  \citenamefont{Frixione}, \citenamefont{Mangano}, \citenamefont{Nason},\ and\
  \citenamefont{Ridolfi}}]{Cacciari:2008zb}%
  \BibitemOpen
  \bibfield{author}{%
  \bibinfo {author} {\bibnamefont{Cacciari}, \bibfnamefont{M.}}, \bibinfo
  {author} {\bibfnamefont{S.}~\bibnamefont{Frixione}}, \bibinfo {author}
  {\bibfnamefont{M.~L.}\ \bibnamefont{Mangano}}, \bibinfo {author}
  {\bibfnamefont{P.}~\bibnamefont{Nason}},\ and\ \bibinfo {author}
  {\bibfnamefont{G.}~\bibnamefont{Ridolfi}}}%
  , \bibinfo {year} {2008},\ \bibfield{journal}{%
  \Doi{10.1088/1126-6708/2008/09/127}{\bibinfo {journal} {JHEP}}\ }%
  \textbf{\bibinfo {volume} {09}},\ \bibinfo {pages} {127},\
  \Eprint{http://arxiv.org/abs/0804.2800}{arXiv:0804.2800 [hep-ph]}%
  \bibAnnoteFile{NoStop}{Cacciari:2008zb}%
%%CITATION = 0804.2800;%%
\bibitem[{\citenamefont{Campbell}\ and\ \citenamefont{Ellis}(1999)}]{MCFM}%
  \BibitemOpen
  \bibfield{author}{%
  \bibinfo {author} {\bibnamefont{Campbell}, \bibfnamefont{J.}},\ and\ \bibinfo
  {author} {\bibfnamefont{R.}~\bibnamefont{Ellis}}}%
  , \bibinfo {year} {1999},\ \bibfield{journal}{%
  \Doi{10.1103/PhysRevD.60.113006}{\bibinfo {journal} {Phys. Rev.}}\ }%
  \textbf{\bibinfo {volume} {D60}},\ \bibinfo {pages} {113006},\ \bibinfo
  {note} {also see http://mcfm.fnal.gov/},\
  \Eprint{http://arxiv.org/abs/hep-ph/9905386}{arXiv:hep-ph/9905386}%
  \bibAnnoteFile{NoStop}{MCFM}%
%%CITATION = HEP-PH/9905386;%%
\bibitem[{\citenamefont{Cao}\ \emph{et~al.}(2010)\citenamefont{Cao},
  \citenamefont{Heng}, \citenamefont{Wu},\ and\
  \citenamefont{Yang}}]{Cao:2009uz}%
  \BibitemOpen
  \bibfield{author}{%
  \bibinfo {author} {\bibnamefont{Cao}, \bibfnamefont{J.}}, \bibinfo {author}
  {\bibfnamefont{Z.}~\bibnamefont{Heng}}, \bibinfo {author}
  {\bibfnamefont{L.}~\bibnamefont{Wu}},\ and\ \bibinfo {author}
  {\bibfnamefont{J.~M.}\ \bibnamefont{Yang}}}%
  , \bibinfo {year} {2010},\ \bibfield{journal}{%
  \Doi{10.1103/PhysRevD.81.014016}{\bibinfo {journal} {Phys. Rev.}}\ }%
  \textbf{\bibinfo {volume} {D81}},\ \bibinfo {pages} {014016},\
  \Eprint{http://arxiv.org/abs/0912.1447}{arXiv:0912.1447 [hep-ph]}%
  \bibAnnoteFile{NoStop}{Cao:2009uz}%
%%CITATION = 0912.1447;%%
\bibitem[{\citenamefont{Carena}\ and\
  \citenamefont{Haber}(2003)}]{Carena:2002es}%
  \BibitemOpen
  \bibfield{author}{%
  \bibinfo {author} {\bibnamefont{Carena}, \bibfnamefont{M.~S.}},\ and\
  \bibinfo {author} {\bibfnamefont{H.~E.}\ \bibnamefont{Haber}}}%
  , \bibinfo {year} {2003},\ \bibfield{journal}{%
  \Doi{10.1016/S0146-6410(02)00177-1}{\bibinfo {journal} {Prog. Part. Nucl.
  Phys.}}\ }%
  \textbf{\bibinfo {volume} {50}},\ \bibinfo {pages} {63},\
  \Eprint{http://arxiv.org/abs/hep-ph/0208209}{arXiv:hep-ph/0208209}%
  \bibAnnoteFile{NoStop}{Carena:2002es}%
%%CITATION = HEP-PH/0208209;%%
\bibitem[{\citenamefont{Carlson}\ and\
  \citenamefont{Yuan}(1995)}]{Carlson:1995ck}%
  \BibitemOpen
  \bibfield{author}{%
  \bibinfo {author} {\bibnamefont{Carlson}, \bibfnamefont{D.~O.}},\ and\
  \bibinfo {author} {\bibfnamefont{C.~P.}\ \bibnamefont{Yuan}}}%
  , \bibinfo {year} {1995}\
  \Eprint{http://arxiv.org/abs/hep-ph/9509208}{arXiv:hep-ph/9509208}%
  \bibAnnoteFile{NoStop}{Carlson:1995ck}%
%%CITATION = HEP-PH/9509208;%%
\bibitem[{\citenamefont{Cembranos}\
  \emph{et~al.}(2008)\citenamefont{Cembranos}, \citenamefont{Rajaraman},\ and\
  \citenamefont{Takayama}}]{Cembranos:2006hj}%
  \BibitemOpen
  \bibfield{author}{%
  \bibinfo {author} {\bibnamefont{Cembranos}, \bibfnamefont{J.~A.~R.}},
  \bibinfo {author} {\bibfnamefont{A.}~\bibnamefont{Rajaraman}},\ and\ \bibinfo
  {author} {\bibfnamefont{F.}~\bibnamefont{Takayama}}}%
  , \bibinfo {year} {2008},\ \bibfield{journal}{%
  \Doi{10.1209/0295-5075/82/21001}{\bibinfo {journal} {Europhys. Lett.}}\ }%
  \textbf{\bibinfo {volume} {82}},\ \bibinfo {pages} {21001},\
  \Eprint{http://arxiv.org/abs/hep-ph/0609244}{arXiv:hep-ph/0609244}%
  \bibAnnoteFile{NoStop}{Cembranos:2006hj}%
%%CITATION = HEP-PH/0609244;%%
\bibitem[{\citenamefont{Chang}\ \emph{et~al.}(1999)\citenamefont{Chang},
  \citenamefont{Chang},\ and\ \citenamefont{Ma}}]{Chang:1998pt}%
  \BibitemOpen
  \bibfield{author}{%
  \bibinfo {author} {\bibnamefont{Chang}, \bibfnamefont{D.}}, \bibinfo {author}
  {\bibfnamefont{W.-F.}\ \bibnamefont{Chang}},\ and\ \bibinfo {author}
  {\bibfnamefont{E.}~\bibnamefont{Ma}}}%
  , \bibinfo {year} {1999},\ \bibfield{journal}{%
  \Doi{10.1103/PhysRevD.59.091503}{\bibinfo {journal} {Phys. Rev.}}\ }%
  \textbf{\bibinfo {volume} {D59}},\ \bibinfo {pages} {091503},\
  \Eprint{http://arxiv.org/abs/hep-ph/9810531}{arXiv:hep-ph/9810531}%
  \bibAnnoteFile{NoStop}{Chang:1998pt}%
%%CITATION = HEP-PH/9810531;%%
\bibitem[{\citenamefont{Chang}\ \emph{et~al.}(2000)\citenamefont{Chang},
  \citenamefont{Hisano}, \citenamefont{Nakano}, \citenamefont{Okada},\ and\
  \citenamefont{Yamaguchi}}]{Chang:1999nh}%
  \BibitemOpen
  \bibfield{author}{%
  \bibinfo {author} {\bibnamefont{Chang}, \bibfnamefont{S.}}, \bibinfo {author}
  {\bibfnamefont{J.}~\bibnamefont{Hisano}}, \bibinfo {author}
  {\bibfnamefont{H.}~\bibnamefont{Nakano}}, \bibinfo {author}
  {\bibfnamefont{N.}~\bibnamefont{Okada}},\ and\ \bibinfo {author}
  {\bibfnamefont{M.}~\bibnamefont{Yamaguchi}}}%
  , \bibinfo {year} {2000},\ \bibfield{journal}{%
  \Doi{10.1103/PhysRevD.62.084025}{\bibinfo {journal} {Phys. Rev.}}\ }%
  \textbf{\bibinfo {volume} {D62}},\ \bibinfo {pages} {084025},\
  \Eprint{http://arxiv.org/abs/hep-ph/9912498}{arXiv:hep-ph/9912498}%
  \bibAnnoteFile{NoStop}{Chang:1999nh}%
%%CITATION = HEP-PH/9912498;%%
\bibitem[{\citenamefont{Cheng}\ \emph{et~al.}(2002)\citenamefont{Cheng},
  \citenamefont{Matchev},\ and\ \citenamefont{Schmaltz}}]{Cheng:2002ab}%
  \BibitemOpen
  \bibfield{author}{%
  \bibinfo {author} {\bibnamefont{Cheng}, \bibfnamefont{H.-C.}}, \bibinfo
  {author} {\bibfnamefont{K.~T.}\ \bibnamefont{Matchev}},\ and\ \bibinfo
  {author} {\bibfnamefont{M.}~\bibnamefont{Schmaltz}}}%
  , \bibinfo {year} {2002},\ \bibfield{journal}{%
  \Doi{10.1103/PhysRevD.66.056006}{\bibinfo {journal} {Phys. Rev.}}\ }%
  \textbf{\bibinfo {volume} {D66}},\ \bibinfo {pages} {056006},\
  \Eprint{http://arxiv.org/abs/hep-ph/0205314}{arXiv:hep-ph/0205314}%
  \bibAnnoteFile{NoStop}{Cheng:2002ab}%
%%CITATION = HEP-PH/0205314;%%
\bibitem[{\citenamefont{Cheung}\ \emph{et~al.}(2009)\citenamefont{Cheung},
  \citenamefont{Keung},\ and\ \citenamefont{Yuan}}]{Cheung:2009ch}%
  \BibitemOpen
  \bibfield{author}{%
  \bibinfo {author} {\bibnamefont{Cheung}, \bibfnamefont{K.}}, \bibinfo
  {author} {\bibfnamefont{W.-Y.}\ \bibnamefont{Keung}},\ and\ \bibinfo {author}
  {\bibfnamefont{T.-C.}\ \bibnamefont{Yuan}}}%
  , \bibinfo {year} {2009},\ \bibfield{journal}{%
  \Doi{10.1016/j.physletb.2009.11.015}{\bibinfo {journal} {Phys. Lett.}}\ }%
  \textbf{\bibinfo {volume} {B682}},\ \bibinfo {pages} {287},\
  \Eprint{http://arxiv.org/abs/0908.2589}{arXiv:0908.2589 [hep-ph]}%
  \bibAnnoteFile{NoStop}{Cheung:2009ch}%
%%CITATION = 0908.2589;%%
\bibitem[{\citenamefont{Cheung}\ and\
  \citenamefont{Landsberg}(2002)}]{Cheung:2001mq}%
  \BibitemOpen
  \bibfield{author}{%
  \bibinfo {author} {\bibnamefont{Cheung}, \bibfnamefont{K.-M.}},\ and\
  \bibinfo {author} {\bibfnamefont{G.~L.}\ \bibnamefont{Landsberg}}}%
  , \bibinfo {year} {2002},\ \bibfield{journal}{%
  \Doi{10.1103/PhysRevD.65.076003}{\bibinfo {journal} {Phys. Rev.}}\ }%
  \textbf{\bibinfo {volume} {D65}},\ \bibinfo {pages} {076003},\
  \Eprint{http://arxiv.org/abs/hep-ph/0110346}{arXiv:hep-ph/0110346}%
  \bibAnnoteFile{NoStop}{Cheung:2001mq}%
%%CITATION = HEP-PH/0110346;%%
\bibitem[{\citenamefont{Chivukula}\
  \emph{et~al.}(1996)\citenamefont{Chivukula}, \citenamefont{Cohen},\ and\
  \citenamefont{Simmons}}]{Chivukula:1996yr}%
  \BibitemOpen
  \bibfield{author}{%
  \bibinfo {author} {\bibnamefont{Chivukula}, \bibfnamefont{R.~S.}}, \bibinfo
  {author} {\bibfnamefont{A.~G.}\ \bibnamefont{Cohen}},\ and\ \bibinfo {author}
  {\bibfnamefont{E.~H.}\ \bibnamefont{Simmons}}}%
  , \bibinfo {year} {1996},\ \bibfield{journal}{%
  \Doi{10.1016/0370-2693(96)00464-9}{\bibinfo {journal} {Phys. Lett.}}\ }%
  \textbf{\bibinfo {volume} {B380}},\ \bibinfo {pages} {92},\
  \Eprint{http://arxiv.org/abs/hep-ph/9603311}{arXiv:hep-ph/9603311}%
  \bibAnnoteFile{NoStop}{Chivukula:1996yr}%
%%CITATION = HEP-PH/9603311;%%
\bibitem[{\citenamefont{Choudhury}\
  \emph{et~al.}(2007)\citenamefont{Choudhury}, \citenamefont{Godbole},
  \citenamefont{Singh},\ and\ \citenamefont{Wagh}}]{Choudhury:2007ux}%
  \BibitemOpen
  \bibfield{author}{%
  \bibinfo {author} {\bibnamefont{Choudhury}, \bibfnamefont{D.}}, \bibinfo
  {author} {\bibfnamefont{R.~M.}\ \bibnamefont{Godbole}}, \bibinfo {author}
  {\bibfnamefont{R.~K.}\ \bibnamefont{Singh}},\ and\ \bibinfo {author}
  {\bibfnamefont{K.}~\bibnamefont{Wagh}}}%
  , \bibinfo {year} {2007},\ \bibfield{journal}{%
  \Doi{10.1016/j.physletb.2007.09.057}{\bibinfo {journal} {Phys. Lett.}}\ }%
  \textbf{\bibinfo {volume} {B657}},\ \bibinfo {pages} {69},\
  \Eprint{http://arxiv.org/abs/0705.1499}{arXiv:0705.1499 [hep-ph]}%
  \bibAnnoteFile{NoStop}{Choudhury:2007ux}%
%%CITATION = 0705.1499;%%
\bibitem[{\citenamefont{Choudhury}\
  \emph{et~al.}(2002)\citenamefont{Choudhury}, \citenamefont{Tait},\ and\
  \citenamefont{Wagner}}]{Choudhury:2001hs}%
  \BibitemOpen
  \bibfield{author}{%
  \bibinfo {author} {\bibnamefont{Choudhury}, \bibfnamefont{D.}}, \bibinfo
  {author} {\bibfnamefont{T.~M.~P.}\ \bibnamefont{Tait}},\ and\ \bibinfo
  {author} {\bibfnamefont{C.~E.~M.}\ \bibnamefont{Wagner}}}%
  , \bibinfo {year} {2002},\ \bibfield{journal}{%
  \Doi{10.1103/PhysRevD.65.053002}{\bibinfo {journal} {Phys. Rev.}}\ }%
  \textbf{\bibinfo {volume} {D65}},\ \bibinfo {pages} {053002},\
  \Eprint{http://arxiv.org/abs/hep-ph/0109097}{arXiv:hep-ph/0109097}%
  \bibAnnoteFile{NoStop}{Choudhury:2001hs}%
%%CITATION = HEP-PH/0109097;%%
\bibitem[{\citenamefont{Chung}\ \emph{et~al.}(2005)\citenamefont{Chung},
  \citenamefont{Everett}, \citenamefont{Kane}, \citenamefont{King},
  \citenamefont{Lykken},\ and\ \citenamefont{Wang}}]{Chung:2005a}%
  \BibitemOpen
  \bibfield{author}{%
  \bibinfo {author} {\bibnamefont{Chung}, \bibfnamefont{D.}}, \bibinfo {author}
  {\bibfnamefont{L.}~\bibnamefont{Everett}}, \bibinfo {author}
  {\bibfnamefont{G.}~\bibnamefont{Kane}}, \bibinfo {author}
  {\bibfnamefont{S.}~\bibnamefont{King}}, \bibinfo {author}
  {\bibfnamefont{J.}~\bibnamefont{Lykken}},\ and\ \bibinfo {author}
  {\bibfnamefont{L.}~\bibnamefont{Wang}}}%
  , \bibinfo {year} {2005},\ \bibfield{journal}{%
  \bibinfo {journal} {Phys. Rept.}\ }%
  \textbf{\bibinfo {volume} {407}},\ \bibinfo {pages} {1}%
  \bibAnnoteFile{NoStop}{Chung:2005a}%
%%CITATION = PHYSREP,407,1;%%
\bibitem[{\citenamefont{Colladay}\ and\
  \citenamefont{Kostelecky}(1997)}]{Colladay:1996iz}%
  \BibitemOpen
  \bibfield{author}{%
  \bibinfo {author} {\bibnamefont{Colladay}, \bibfnamefont{D.}},\ and\ \bibinfo
  {author} {\bibfnamefont{V.~A.}\ \bibnamefont{Kostelecky}}}%
  , \bibinfo {year} {1997},\ \bibfield{journal}{%
  \Doi{10.1103/PhysRevD.55.6760}{\bibinfo {journal} {Phys. Rev.}}\ }%
  \textbf{\bibinfo {volume} {D55}},\ \bibinfo {pages} {6760},\
  \Eprint{http://arxiv.org/abs/hep-ph/9703464}{arXiv:hep-ph/9703464}%
  \bibAnnoteFile{NoStop}{Colladay:1996iz}%
%%CITATION = HEP-PH/9703464;%%
\bibitem[{\citenamefont{Contino}\ and\
  \citenamefont{Servant}(2008)}]{Contino:2008hi}%
  \BibitemOpen
  \bibfield{author}{%
  \bibinfo {author} {\bibnamefont{Contino}, \bibfnamefont{R.}},\ and\ \bibinfo
  {author} {\bibfnamefont{G.}~\bibnamefont{Servant}}}%
  , \bibinfo {year} {2008},\ \bibfield{journal}{%
  \Doi{10.1088/1126-6708/2008/06/026}{\bibinfo {journal} {JHEP}}\ }%
  \textbf{\bibinfo {volume} {06}},\ \bibinfo {pages} {026},\
  \Eprint{http://arxiv.org/abs/0801.1679}{arXiv:0801.1679 [hep-ph]}%
  \bibAnnoteFile{NoStop}{Contino:2008hi}%
%%CITATION = 0801.1679;%%
\bibitem[{\citenamefont{Corcella}\ \emph{et~al.}(2001)\citenamefont{Corcella}
  \emph{et~al.}}]{herwig}%
  \BibitemOpen
  \bibfield{author}{%
  \bibinfo {author} {\bibnamefont{Corcella}, \bibfnamefont{G.}},
  \emph{et~al.}}%
  , \bibinfo {year} {2001},\ \bibfield{journal}{%
  \Doi{10.1088/1126-6708/2001/01/010}{\bibinfo {journal} {JHEP}}\ }%
  \textbf{\bibinfo {volume} {0101}},\ \bibinfo {pages} {010},\ \bibinfo {note}
  {see also http://hepwww.rl.ac.uk/theory/seymour/herwig/},\
  \Eprint{http://arxiv.org/abs/0011363}{arXiv:0011363 [hep-ph]}%
  \bibAnnoteFile{NoStop}{herwig}%
\bibitem[{\citenamefont{Csaki}\ \emph{et~al.}(2002)\citenamefont{Csaki},
  \citenamefont{Erlich},\ and\ \citenamefont{Terning}}]{Csaki:2002gy}%
  \BibitemOpen
  \bibfield{author}{%
  \bibinfo {author} {\bibnamefont{Csaki}, \bibfnamefont{C.}}, \bibinfo {author}
  {\bibfnamefont{J.}~\bibnamefont{Erlich}},\ and\ \bibinfo {author}
  {\bibfnamefont{J.}~\bibnamefont{Terning}}}%
  , \bibinfo {year} {2002},\ \bibfield{journal}{%
  \Doi{10.1103/PhysRevD.66.064021}{\bibinfo {journal} {Phys. Rev.}}\ }%
  \textbf{\bibinfo {volume} {D66}},\ \bibinfo {pages} {064021},\
  \Eprint{http://arxiv.org/abs/hep-ph/0203034}{arXiv:hep-ph/0203034}%
  \bibAnnoteFile{NoStop}{Csaki:2002gy}%
%%CITATION = HEP-PH/0203034;%%
\bibitem[{\citenamefont{Dalitz}\ and\
  \citenamefont{Goldstein}(1992)}]{Dalitz:1991wa}%
  \BibitemOpen
  \bibfield{author}{%
  \bibinfo {author} {\bibnamefont{Dalitz}, \bibfnamefont{R.~H.}},\ and\
  \bibinfo {author} {\bibfnamefont{G.~R.}\ \bibnamefont{Goldstein}}}%
  , \bibinfo {year} {1992},\ \bibfield{journal}{%
  \Doi{10.1103/PhysRevD.45.1531}{\bibinfo {journal} {Phys. Rev.}}\ }%
  \textbf{\bibinfo {volume} {D45}},\ \bibinfo {pages} {1531}%
  \bibAnnoteFile{NoStop}{Dalitz:1991wa}%
%%CITATION = PHRVA,D45,1531;%%
\bibitem[{\citenamefont{Dalitz}\ and\
  \citenamefont{Goldstein}(1999)}]{Dalitz:1998zn}%
  \BibitemOpen
  \bibfield{author}{%
  \bibinfo {author} {\bibnamefont{Dalitz}, \bibfnamefont{R.~H.}},\ and\
  \bibinfo {author} {\bibfnamefont{G.~R.}\ \bibnamefont{Goldstein}}}%
  , \bibinfo {year} {1999},\ \bibfield{journal}{%
  \Doi{10.1098/rspa.1999.0428}{\bibinfo {journal} {Proc. Roy. Soc. Lond.}}\ }%
  \textbf{\bibinfo {volume} {A455}},\ \bibinfo {pages} {2803},\
  \Eprint{http://arxiv.org/abs/hep-ph/9802249}{arXiv:hep-ph/9802249}%
  \bibAnnoteFile{NoStop}{Dalitz:1998zn}%
%%CITATION = HEP-PH/9802249;%%
\bibitem[{\citenamefont{Davidson}\ \emph{et~al.}(2010)\citenamefont{Davidson},
  \citenamefont{Przedzinski}, \citenamefont{Richter-Was},\ and\
  \citenamefont{Was}}]{tauola}%
  \BibitemOpen
  \bibfield{author}{%
  \bibinfo {author} {\bibnamefont{Davidson}, \bibfnamefont{G., N.~Nanava}},
  \bibinfo {author} {\bibfnamefont{T.}~\bibnamefont{Przedzinski}}, \bibinfo
  {author} {\bibfnamefont{E.}~\bibnamefont{Richter-Was}},\ and\ \bibinfo
  {author} {\bibfnamefont{Z.}~\bibnamefont{Was}}}%
  , \bibinfo {year} {2010},\ \enquote{\bibinfo {title} {{Universal Interface of
  TAUOLA Technical and Physics Documentation}},}\ \bibinfo {note}
  {IFJPAN-IV-2009-10; see also http://wasm.home.cern.ch/wasm/goodies.html},\
  \Eprint{http://arxiv.org/abs/1002.0543}{arXiv:1002.0543 [hep-ph]}%
  \bibAnnoteFile{NoStop}{tauola}%
\bibitem[{\citenamefont{Davoudiasl}\
  \emph{et~al.}(2000)\citenamefont{Davoudiasl}, \citenamefont{Hewett},\ and\
  \citenamefont{Rizzo}}]{Davoudiasl:1999tf}%
  \BibitemOpen
  \bibfield{author}{%
  \bibinfo {author} {\bibnamefont{Davoudiasl}, \bibfnamefont{H.}}, \bibinfo
  {author} {\bibfnamefont{J.~L.}\ \bibnamefont{Hewett}},\ and\ \bibinfo
  {author} {\bibfnamefont{T.~G.}\ \bibnamefont{Rizzo}}}%
  , \bibinfo {year} {2000},\ \bibfield{journal}{%
  \Doi{10.1016/S0370-2693(99)01430-6}{\bibinfo {journal} {Phys. Lett.}}\ }%
  \textbf{\bibinfo {volume} {B473}},\ \bibinfo {pages} {43},\
  \Eprint{http://arxiv.org/abs/hep-ph/9911262}{arXiv:hep-ph/9911262}%
  \bibAnnoteFile{NoStop}{Davoudiasl:1999tf}%
%%CITATION = HEP-PH/9911262;%%
\bibitem[{\citenamefont{Davoudiasl}\
  \emph{et~al.}(2001)\citenamefont{Davoudiasl}, \citenamefont{Hewett},\ and\
  \citenamefont{Rizzo}}]{Davoudiasl:2000wi}%
  \BibitemOpen
  \bibfield{author}{%
  \bibinfo {author} {\bibnamefont{Davoudiasl}, \bibfnamefont{H.}}, \bibinfo
  {author} {\bibfnamefont{J.~L.}\ \bibnamefont{Hewett}},\ and\ \bibinfo
  {author} {\bibfnamefont{T.~G.}\ \bibnamefont{Rizzo}}}%
  , \bibinfo {year} {2001},\ \bibfield{journal}{%
  \Doi{10.1103/PhysRevD.63.075004}{\bibinfo {journal} {Phys. Rev.}}\ }%
  \textbf{\bibinfo {volume} {D63}},\ \bibinfo {pages} {075004},\
  \Eprint{http://arxiv.org/abs/hep-ph/0006041}{arXiv:hep-ph/0006041}%
  \bibAnnoteFile{NoStop}{Davoudiasl:2000wi}%
%%CITATION = HEP-PH/0006041;%%
\bibitem[{\citenamefont{Demina}\ and\
  \citenamefont{Thomson}(2008)}]{Demina:2008}%
  \BibitemOpen
  \bibfield{author}{%
  \bibinfo {author} {\bibnamefont{Demina}, \bibfnamefont{R.}},\ and\ \bibinfo
  {author} {\bibfnamefont{E.}~\bibnamefont{Thomson}}}%
  , \bibinfo {year} {2008},\ \bibfield{journal}{%
  \Doi{10.1146/annurev.nucl.58.110707.171224}{\bibinfo {journal}
  {Ann.Rev.Nucl.Part.Sci.}}\ }%
  \textbf{\bibinfo {volume} {58}},\ \bibinfo {pages} {125}%
  \bibAnnoteFile{NoStop}{Demina:2008}%
\bibitem[{\citenamefont{Dicus}\ \emph{et~al.}(1994)\citenamefont{Dicus},
  \citenamefont{Stange},\ and\ \citenamefont{Willenbrock}}]{Dicus:1994bm}%
  \BibitemOpen
  \bibfield{author}{%
  \bibinfo {author} {\bibnamefont{Dicus}, \bibfnamefont{D.}}, \bibinfo {author}
  {\bibfnamefont{A.}~\bibnamefont{Stange}},\ and\ \bibinfo {author}
  {\bibfnamefont{S.}~\bibnamefont{Willenbrock}}}%
  , \bibinfo {year} {1994},\ \bibfield{journal}{%
  \Doi{10.1016/0370-2693(94)91017-0}{\bibinfo {journal} {Phys. Lett.}}\ }%
  \textbf{\bibinfo {volume} {B333}},\ \bibinfo {pages} {126},\
  \Eprint{http://arxiv.org/abs/hep-ph/9404359}{arXiv:hep-ph/9404359}%
  \bibAnnoteFile{NoStop}{Dicus:1994bm}%
%%CITATION = HEP-PH/9404359;%%
\bibitem[{\citenamefont{Dienes}\ \emph{et~al.}(1998)\citenamefont{Dienes},
  \citenamefont{Dudas},\ and\ \citenamefont{Gherghetta}}]{Dienes:1998vh}%
  \BibitemOpen
  \bibfield{author}{%
  \bibinfo {author} {\bibnamefont{Dienes}, \bibfnamefont{K.~R.}}, \bibinfo
  {author} {\bibfnamefont{E.}~\bibnamefont{Dudas}},\ and\ \bibinfo {author}
  {\bibfnamefont{T.}~\bibnamefont{Gherghetta}}}%
  , \bibinfo {year} {1998},\ \bibfield{journal}{%
  \Doi{10.1016/S0370-2693(98)00977-0}{\bibinfo {journal} {Phys. Lett.}}\ }%
  \textbf{\bibinfo {volume} {B436}},\ \bibinfo {pages} {55},\
  \Eprint{http://arxiv.org/abs/hep-ph/9803466}{arXiv:hep-ph/9803466}%
  \bibAnnoteFile{NoStop}{Dienes:1998vh}%
%%CITATION = HEP-PH/9803466;%%
\bibitem[{\citenamefont{Dienes}\ \emph{et~al.}(1999)\citenamefont{Dienes},
  \citenamefont{Dudas},\ and\ \citenamefont{Gherghetta}}]{Dienes:1998vg}%
  \BibitemOpen
  \bibfield{author}{%
  \bibinfo {author} {\bibnamefont{Dienes}, \bibfnamefont{K.~R.}}, \bibinfo
  {author} {\bibfnamefont{E.}~\bibnamefont{Dudas}},\ and\ \bibinfo {author}
  {\bibfnamefont{T.}~\bibnamefont{Gherghetta}}}%
  , \bibinfo {year} {1999},\ \bibfield{journal}{%
  \Doi{10.1016/S0550-3213(98)00669-5}{\bibinfo {journal} {Nucl. Phys.}}\ }%
  \textbf{\bibinfo {volume} {B537}},\ \bibinfo {pages} {47},\
  \Eprint{http://arxiv.org/abs/hep-ph/9806292}{arXiv:hep-ph/9806292}%
  \bibAnnoteFile{NoStop}{Dienes:1998vg}%
%%CITATION = HEP-PH/9806292;%%
\bibitem[{\citenamefont{Dittmaier}\
  \emph{et~al.}(2007)\citenamefont{Dittmaier}, \citenamefont{Uwer},\ and\
  \citenamefont{Weinzierl}}]{Dittmaier:2007wz}%
  \BibitemOpen
  \bibfield{author}{%
  \bibinfo {author} {\bibnamefont{Dittmaier}, \bibfnamefont{S.}}, \bibinfo
  {author} {\bibfnamefont{P.}~\bibnamefont{Uwer}},\ and\ \bibinfo {author}
  {\bibfnamefont{S.}~\bibnamefont{Weinzierl}}}%
  , \bibinfo {year} {2007},\ \bibfield{journal}{%
  \Doi{10.1103/PhysRevLett.98.262002}{\bibinfo {journal} {Phys. Rev. Lett.}}\
  }%
  \textbf{\bibinfo {volume} {98}},\ \bibinfo {pages} {262002},\
  \Eprint{http://arxiv.org/abs/hep-ph/0703120}{arXiv:hep-ph/0703120}%
  \bibAnnoteFile{NoStop}{Dittmaier:2007wz}%
%%CITATION = HEP-PH/0703120;%%
\bibitem[{\citenamefont{Djouadi}\ \emph{et~al.}(2009)\citenamefont{Djouadi},
  \citenamefont{Moreau}, \citenamefont{Richard},\ and\
  \citenamefont{Singh}}]{Djouadi:2009nb}%
  \BibitemOpen
  \bibfield{author}{%
  \bibinfo {author} {\bibnamefont{Djouadi}, \bibfnamefont{A.}}, \bibinfo
  {author} {\bibfnamefont{G.}~\bibnamefont{Moreau}}, \bibinfo {author}
  {\bibfnamefont{F.}~\bibnamefont{Richard}},\ and\ \bibinfo {author}
  {\bibfnamefont{R.~K.}\ \bibnamefont{Singh}}}%
  , \bibinfo {year} {2009}\
  \Eprint{http://arxiv.org/abs/0906.0604}{arXiv:0906.0604 [hep-ph]}%
  \bibAnnoteFile{NoStop}{Djouadi:2009nb}%
%%CITATION = 0906.0604;%%
\bibitem[{\citenamefont{Dobrescu}\ and\
  \citenamefont{Hill}(1998)}]{Dobrescu:1997nm}%
  \BibitemOpen
  \bibfield{author}{%
  \bibinfo {author} {\bibnamefont{Dobrescu}, \bibfnamefont{B.~A.}},\ and\
  \bibinfo {author} {\bibfnamefont{C.~T.}\ \bibnamefont{Hill}}}%
  , \bibinfo {year} {1998},\ \bibfield{journal}{%
  \Doi{10.1103/PhysRevLett.81.2634}{\bibinfo {journal} {Phys. Rev. Lett.}}\ }%
  \textbf{\bibinfo {volume} {81}},\ \bibinfo {pages} {2634},\
  \Eprint{http://arxiv.org/abs/hep-ph/9712319}{arXiv:hep-ph/9712319}%
  \bibAnnoteFile{NoStop}{Dobrescu:1997nm}%
%%CITATION = HEP-PH/9712319;%%
\bibitem[{\citenamefont{Dobrescu}\ \emph{et~al.}(2009)\citenamefont{Dobrescu},
  \citenamefont{Kong},\ and\ \citenamefont{Mahbubani}}]{Dobrescu:2009vz}%
  \BibitemOpen
  \bibfield{author}{%
  \bibinfo {author} {\bibnamefont{Dobrescu}, \bibfnamefont{B.~A.}}, \bibinfo
  {author} {\bibfnamefont{K.}~\bibnamefont{Kong}},\ and\ \bibinfo {author}
  {\bibfnamefont{R.}~\bibnamefont{Mahbubani}}}%
  , \bibinfo {year} {2009},\ \bibfield{journal}{%
  \Doi{10.1088/1126-6708/2009/06/001}{\bibinfo {journal} {JHEP}}\ }%
  \textbf{\bibinfo {volume} {06}},\ \bibinfo {pages} {001},\
  \Eprint{http://arxiv.org/abs/0902.0792}{arXiv:0902.0792 [hep-ph]}%
  \bibAnnoteFile{NoStop}{Dobrescu:2009vz}%
%%CITATION = 0902.0792;%%
\bibitem[{\citenamefont{Eberhardt}\
  \emph{et~al.}(2010)\citenamefont{Eberhardt}, \citenamefont{Lenz},\ and\
  \citenamefont{Rohrwild}}]{Eberhardt:2010bm}%
  \BibitemOpen
  \bibfield{author}{%
  \bibinfo {author} {\bibnamefont{Eberhardt}, \bibfnamefont{O.}}, \bibinfo
  {author} {\bibfnamefont{A.}~\bibnamefont{Lenz}},\ and\ \bibinfo {author}
  {\bibfnamefont{J.}~\bibnamefont{Rohrwild}}}%
  , \bibinfo {year} {2010},\ \bibfield{journal}{%
  \Doi{10.1103/PhysRevD.82.095006}{\bibinfo {journal} {Phys.Rev.}}\ }%
  \textbf{\bibinfo {volume} {D82}},\ \bibinfo {pages} {095006},\
  \Eprint{http://arxiv.org/abs/1005.3505}{arXiv:1005.3505 [hep-ph]}%
  \bibAnnoteFile{NoStop}{Eberhardt:2010bm}%
\bibitem[{\citenamefont{Eichten}\ \emph{et~al.}(1986)\citenamefont{Eichten},
  \citenamefont{Hinchliffe}, \citenamefont{Lane},\ and\
  \citenamefont{Quigg}}]{Eichten:1986eq}%
  \BibitemOpen
  \bibfield{author}{%
  \bibinfo {author} {\bibnamefont{Eichten}, \bibfnamefont{E.}}, \bibinfo
  {author} {\bibfnamefont{I.}~\bibnamefont{Hinchliffe}}, \bibinfo {author}
  {\bibfnamefont{K.~D.}\ \bibnamefont{Lane}},\ and\ \bibinfo {author}
  {\bibfnamefont{C.}~\bibnamefont{Quigg}}}%
  , \bibinfo {year} {1986},\ \bibfield{journal}{%
  \Doi{10.1103/PhysRevD.34.1547}{\bibinfo {journal} {Phys. Rev.}}\ }%
  \textbf{\bibinfo {volume} {D34}},\ \bibinfo {pages} {1547}%
  \bibAnnoteFile{NoStop}{Eichten:1986eq}%
%%CITATION = PHRVA,D34,1547;%%
\bibitem[{\citenamefont{Eichten}\ and\
  \citenamefont{Lane}(1980)}]{Eichten:1979ah}%
  \BibitemOpen
  \bibfield{author}{%
  \bibinfo {author} {\bibnamefont{Eichten}, \bibfnamefont{E.}},\ and\ \bibinfo
  {author} {\bibfnamefont{K.~D.}\ \bibnamefont{Lane}}}%
  , \bibinfo {year} {1980},\ \bibfield{journal}{%
  \Doi{10.1016/0370-2693(80)90065-9}{\bibinfo {journal} {Phys. Lett.}}\ }%
  \textbf{\bibinfo {volume} {B90}},\ \bibinfo {pages} {125}%
  \bibAnnoteFile{NoStop}{Eichten:1979ah}%
%%CITATION = PHLTA,B90,125;%%
\bibitem[{\citenamefont{Eichten}\ and\
  \citenamefont{Lane}(1994)}]{Eichten:1994nc}%
  \BibitemOpen
  \bibfield{author}{%
  \bibinfo {author} {\bibnamefont{Eichten}, \bibfnamefont{E.}},\ and\ \bibinfo
  {author} {\bibfnamefont{K.~D.}\ \bibnamefont{Lane}}}%
  , \bibinfo {year} {1994},\ \bibfield{journal}{%
  \Doi{10.1016/0370-2693(94)91540-7}{\bibinfo {journal} {Phys. Lett.}}\ }%
  \textbf{\bibinfo {volume} {B327}},\ \bibinfo {pages} {129},\
  \Eprint{http://arxiv.org/abs/hep-ph/9401236}{arXiv:hep-ph/9401236}%
  \bibAnnoteFile{NoStop}{Eichten:1994nc}%
%%CITATION = HEP-PH/9401236;%%
\bibitem[{\citenamefont{Falk}\ and\ \citenamefont{Peskin}(1994)}]{Falk:1993rf}%
  \BibitemOpen
  \bibfield{author}{%
  \bibinfo {author} {\bibnamefont{Falk}, \bibfnamefont{A.~F.}},\ and\ \bibinfo
  {author} {\bibfnamefont{M.~E.}\ \bibnamefont{Peskin}}}%
  , \bibinfo {year} {1994},\ \bibfield{journal}{%
  \Doi{10.1103/PhysRevD.49.3320}{\bibinfo {journal} {Phys. Rev.}}\ }%
  \textbf{\bibinfo {volume} {D49}},\ \bibinfo {pages} {3320},\
  \Eprint{http://arxiv.org/abs/hep-ph/9308241}{arXiv:hep-ph/9308241}%
  \bibAnnoteFile{NoStop}{Falk:1993rf}%
%%CITATION = HEP-PH/9308241;%%
\bibitem[{\citenamefont{Feldman}\ and\
  \citenamefont{Cousins}(1998)}]{Feldman:1997qc}%
  \BibitemOpen
  \bibfield{author}{%
  \bibinfo {author} {\bibnamefont{Feldman}, \bibfnamefont{G.~J.}},\ and\
  \bibinfo {author} {\bibfnamefont{R.~D.}\ \bibnamefont{Cousins}}}%
  , \bibinfo {year} {1998},\ \bibfield{journal}{%
  \Doi{10.1103/PhysRevD.57.3873}{\bibinfo {journal} {Phys. Rev.}}\ }%
  \textbf{\bibinfo {volume} {D57}},\ \bibinfo {pages} {3873},\
  \Eprint{http://arxiv.org/abs/physics/9711021}{arXiv:physics/9711021}%
  \bibAnnoteFile{NoStop}{Feldman:1997qc}%
%%CITATION = PHYSICS/9711021;%%
\bibitem[{\citenamefont{Ferrario}\ and\
  \citenamefont{Rodrigo}(2008)}]{Ferrario:2008wm}%
  \BibitemOpen
  \bibfield{author}{%
  \bibinfo {author} {\bibnamefont{Ferrario}, \bibfnamefont{P.}},\ and\ \bibinfo
  {author} {\bibfnamefont{G.}~\bibnamefont{Rodrigo}}}%
  , \bibinfo {year} {2008},\ \bibfield{journal}{%
  \Doi{10.1103/PhysRevD.78.094018}{\bibinfo {journal} {Phys. Rev.}}\ }%
  \textbf{\bibinfo {volume} {D78}},\ \bibinfo {pages} {094018},\
  \Eprint{http://arxiv.org/abs/0809.3354}{arXiv:0809.3354 [hep-ph]}%
  \bibAnnoteFile{NoStop}{Ferrario:2008wm}%
%%CITATION = 0809.3354;%%
\bibitem[{\citenamefont{Ferrario}\ and\
  \citenamefont{Rodrigo}(2009)}]{Ferrario:2009ns}%
  \BibitemOpen
  \bibfield{author}{%
  \bibinfo {author} {\bibnamefont{Ferrario}, \bibfnamefont{P.}},\ and\ \bibinfo
  {author} {\bibfnamefont{G.}~\bibnamefont{Rodrigo}}}%
  , \bibinfo {year} {2009},\ \bibfield{journal}{%
  \Doi{10.1088/1742-6596/171/1/012091}{\bibinfo {journal} {J. Phys. Conf.
  Ser.}}\ }%
  \textbf{\bibinfo {volume} {171}},\ \bibinfo {pages} {012091},\
  \Eprint{http://arxiv.org/abs/0907.0096}{arXiv:0907.0096 [hep-ph]}%
  \bibAnnoteFile{NoStop}{Ferrario:2009ns}%
%%CITATION = 0907.0096;%%
\bibitem[{\citenamefont{Fiedler}(2010)}]{Fiedler:2010sy}%
  \BibitemOpen
  \bibfield{author}{%
  \bibinfo {author} {\bibnamefont{Fiedler}, \bibfnamefont{F.}}}%
  , \bibinfo {year} {2010}\ \bibinfo {note} {habilitation thesis at Munich
  University},\ \Eprint{http://arxiv.org/abs/1003.0521}{arXiv:1003.0521
  [hep-ex]}%
  \bibAnnoteFile{NoStop}{Fiedler:2010sy}%
%%CITATION = 1003.0521;%%
\bibitem[{\citenamefont{Fiedler}\ \emph{et~al.}(2010)\citenamefont{Fiedler},
  \citenamefont{Grohsjean}, \citenamefont{Haefner},\ and\
  \citenamefont{Schieferdecker}}]{Fiedler:2010sg}%
  \BibitemOpen
  \bibfield{author}{%
  \bibinfo {author} {\bibnamefont{Fiedler}, \bibfnamefont{F.}}, \bibinfo
  {author} {\bibfnamefont{A.}~\bibnamefont{Grohsjean}}, \bibinfo {author}
  {\bibfnamefont{P.}~\bibnamefont{Haefner}},\ and\ \bibinfo {author}
  {\bibfnamefont{P.}~\bibnamefont{Schieferdecker}}}%
  , \bibinfo {year} {2010}\
  \Eprint{http://arxiv.org/abs/1003.1316}{arXiv:1003.1316 [hep-ex]}%
  \bibAnnoteFile{NoStop}{Fiedler:2010sg}%
%%CITATION = 1003.1316;%%
\bibitem[{\citenamefont{Fl{\"{a}}cher}\
  \emph{et~al.}(2009)\citenamefont{Fl{\"{a}}cher} \emph{et~al.}}]{gfitter}%
  \BibitemOpen
  \bibfield{author}{%
  \bibinfo {author} {\bibnamefont{Fl{\"{a}}cher}, \bibfnamefont{H.}},
  \emph{et~al.} (\bibinfo {collaboration} {Gfitter})}%
  , \bibinfo {year} {2009},\ \bibfield{journal}{%
  \Doi{10.1140/epjc/s10052-009-0966-6}{\bibinfo {journal} {Eur. Phys. J.}}\ }%
  \textbf{\bibinfo {volume} {C60}},\ \bibinfo {pages} {593},\ \bibinfo {note}
  {see also http://gfitter.desy.de/},\
  \Eprint{http://arxiv.org/abs/0811.0009}{arXiv:0811.0009 [hep-ph]}%
  \bibAnnoteFile{NoStop}{gfitter}%
\bibitem[{\citenamefont{Frampton}\ \emph{et~al.}(2000)\citenamefont{Frampton},
  \citenamefont{Hung},\ and\ \citenamefont{Sher}}]{Frampton:1999xi}%
  \BibitemOpen
  \bibfield{author}{%
  \bibinfo {author} {\bibnamefont{Frampton}, \bibfnamefont{P.~H.}}, \bibinfo
  {author} {\bibfnamefont{P.~Q.}\ \bibnamefont{Hung}},\ and\ \bibinfo {author}
  {\bibfnamefont{M.}~\bibnamefont{Sher}}}%
  , \bibinfo {year} {2000},\ \bibfield{journal}{%
  \Doi{10.1016/S0370-1573(99)00095-2}{\bibinfo {journal} {Phys. Rept.}}\ }%
  \textbf{\bibinfo {volume} {330}},\ \bibinfo {pages} {263},\
  \Eprint{http://arxiv.org/abs/hep-ph/9903387}{arXiv:hep-ph/9903387}%
  \bibAnnoteFile{NoStop}{Frampton:1999xi}%
%%CITATION = HEP-PH/9903387;%%
\bibitem[{\citenamefont{Frederix}\ and\
  \citenamefont{Maltoni}(2009)}]{Frederix:2007gi}%
  \BibitemOpen
  \bibfield{author}{%
  \bibinfo {author} {\bibnamefont{Frederix}, \bibfnamefont{R.}},\ and\ \bibinfo
  {author} {\bibfnamefont{F.}~\bibnamefont{Maltoni}}}%
  , \bibinfo {year} {2009},\ \bibfield{journal}{%
  \Doi{10.1088/1126-6708/2009/01/047}{\bibinfo {journal} {JHEP}}\ }%
  \textbf{\bibinfo {volume} {01}},\ \bibinfo {pages} {047},\
  \Eprint{http://arxiv.org/abs/0712.2355}{arXiv:0712.2355 [hep-ph]}%
  \bibAnnoteFile{NoStop}{Frederix:2007gi}%
%%CITATION = 0712.2355;%%
\bibitem[{\citenamefont{Frixione}\ and\ \citenamefont{Webber}(2002)}]{mcnlo}%
  \BibitemOpen
  \bibfield{author}{%
  \bibinfo {author} {\bibnamefont{Frixione}, \bibfnamefont{S.}},\ and\ \bibinfo
  {author} {\bibfnamefont{B.}~\bibnamefont{Webber}}}%
  , \bibinfo {year} {2002},\ \bibfield{journal}{%
  \Doi{10.1088/1126-6708/2002/06/029}{\bibinfo {journal} {JHEP}}\ }%
  \textbf{\bibinfo {volume} {0206}},\ \bibinfo {pages} {029},\ \bibinfo {note}
  {see also http://www.hep.phy.cam.ac.uk/theory/webber/MCatNLO/},\
  \Eprint{http://arxiv.org/abs/0204244}{arXiv:0204244 [hep-ph]}%
  \bibAnnoteFile{NoStop}{mcnlo}%
\bibitem[{\citenamefont{Galea}(2007)}]{Galea:2006mi}%
  \BibitemOpen
  \bibfield{author}{%
  \bibinfo {author} {\bibnamefont{Galea}, \bibfnamefont{C.}} (\bibinfo
  {collaboration} {D0})}%
  , \bibinfo {year} {2007},\ \bibfield{journal}{%
  \bibinfo {journal} {Acta Phys. Polon.}\ }%
  \textbf{\bibinfo {volume} {B38}},\ \bibinfo {pages} {769}%
  \bibAnnoteFile{NoStop}{Galea:2006mi}%
%%CITATION = APPOA,B38,769;%%
\bibitem[{\citenamefont{Gribov}\ and\ \citenamefont{Lipatov}(1972)}]{dglap}%
  \BibitemOpen
  \bibfield{author}{%
  \bibinfo {author} {\bibnamefont{Gribov}, \bibfnamefont{V.}},\ and\ \bibinfo
  {author} {\bibfnamefont{N.}~\bibnamefont{Lipatov}}}%
  , \bibinfo {year} {1972},\ \bibfield{journal}{%
  \bibinfo {journal} {Sov. J. Nucl. Phys.}\ }%
  \textbf{\bibinfo {volume} {15}},\ \bibinfo {pages} {438},\ \bibinfo {note}
  {also Altarelli G. and Parisi G. (1977), Nucl.Phys. {\bf{B126}}, 298, and
  Dokshitzer Yu.L. (1977), Sov.Phys. JETP, {\bf{46}}, 641}%
  \bibAnnoteFile{NoStop}{dglap}%
\bibitem[{\citenamefont{Grossman}(1994)}]{Grossman:1994jb}%
  \BibitemOpen
  \bibfield{author}{%
  \bibinfo {author} {\bibnamefont{Grossman}, \bibfnamefont{Y.}}}%
  , \bibinfo {year} {1994},\ \bibfield{journal}{%
  \Doi{10.1016/0550-3213(94)90316-6}{\bibinfo {journal} {Nucl. Phys.}}\ }%
  \textbf{\bibinfo {volume} {B426}},\ \bibinfo {pages} {355},\
  \Eprint{http://arxiv.org/abs/hep-ph/9401311}{arXiv:hep-ph/9401311}%
  \bibAnnoteFile{NoStop}{Grossman:1994jb}%
%%CITATION = HEP-PH/9401311;%%
\bibitem[{\citenamefont{Grossman}\ and\
  \citenamefont{Neubert}(2000)}]{Grossman:1999ra}%
  \BibitemOpen
  \bibfield{author}{%
  \bibinfo {author} {\bibnamefont{Grossman}, \bibfnamefont{Y.}},\ and\ \bibinfo
  {author} {\bibfnamefont{M.}~\bibnamefont{Neubert}}}%
  , \bibinfo {year} {2000},\ \bibfield{journal}{%
  \Doi{10.1016/S0370-2693(00)00054-X}{\bibinfo {journal} {Phys. Lett.}}\ }%
  \textbf{\bibinfo {volume} {B474}},\ \bibinfo {pages} {361},\
  \Eprint{http://arxiv.org/abs/hep-ph/9912408}{arXiv:hep-ph/9912408}%
  \bibAnnoteFile{NoStop}{Grossman:1999ra}%
%%CITATION = HEP-PH/9912408;%%
\bibitem[{\citenamefont{Guasch}\ \emph{et~al.}(1995)\citenamefont{Guasch},
  \citenamefont{Jimenez},\ and\ \citenamefont{Sola}}]{Guasch:1995rn}%
  \BibitemOpen
  \bibfield{author}{%
  \bibinfo {author} {\bibnamefont{Guasch}, \bibfnamefont{J.}}, \bibinfo
  {author} {\bibfnamefont{R.~A.}\ \bibnamefont{Jimenez}},\ and\ \bibinfo
  {author} {\bibfnamefont{J.}~\bibnamefont{Sola}}}%
  , \bibinfo {year} {1995},\ \bibfield{journal}{%
  \Doi{10.1016/0370-2693(95)01127-C}{\bibinfo {journal} {Phys.Lett.}}\ }%
  \textbf{\bibinfo {volume} {B360}},\ \bibinfo {pages} {47},\
  \Eprint{http://arxiv.org/abs/hep-ph/9507461}{arXiv:hep-ph/9507461 [hep-ph]}%
  \bibAnnoteFile{NoStop}{Guasch:1995rn}%
\bibitem[{\citenamefont{Gunion}\ \emph{et~al.}(1989)\citenamefont{Gunion},
  \citenamefont{Haber}, \citenamefont{Kane},\ and\
  \citenamefont{Dawson}}]{Gunion:1989we}%
  \BibitemOpen
  \bibfield{author}{%
  \bibinfo {author} {\bibnamefont{Gunion}, \bibfnamefont{J.~F.}}, \bibinfo
  {author} {\bibfnamefont{H.~E.}\ \bibnamefont{Haber}}, \bibinfo {author}
  {\bibfnamefont{G.~L.}\ \bibnamefont{Kane}},\ and\ \bibinfo {author}
  {\bibfnamefont{S.}~\bibnamefont{Dawson}}}%
  , \bibinfo {year} {1989},\ \emph{\bibinfo {title} {{THE HIGGS HUNTER'S
  GUIDE}}},\ \bibinfo {note} {sCIPP-89/13}%
  \bibAnnoteFile{NoStop}{Gunion:1989we}%
\bibitem[{\citenamefont{Haber}\ and\ \citenamefont{Kane}(1985)}]{Haber:1984rc}%
  \BibitemOpen
  \bibfield{author}{%
  \bibinfo {author} {\bibnamefont{Haber}, \bibfnamefont{H.~E.}},\ and\ \bibinfo
  {author} {\bibfnamefont{G.~L.}\ \bibnamefont{Kane}}}%
  , \bibinfo {year} {1985},\ \bibfield{journal}{%
  \Doi{10.1016/0370-1573(85)90051-1}{\bibinfo {journal} {Phys. Rept.}}\ }%
  \textbf{\bibinfo {volume} {117}},\ \bibinfo {pages} {75}%
  \bibAnnoteFile{NoStop}{Haber:1984rc}%
%%CITATION = PRPLC,117,75;%%
\bibitem[{\citenamefont{Han}\ \emph{et~al.}(2003)\citenamefont{Han},
  \citenamefont{Logan}, \citenamefont{McElrath},\ and\
  \citenamefont{Wang}}]{Han:2003gf}%
  \BibitemOpen
  \bibfield{author}{%
  \bibinfo {author} {\bibnamefont{Han}, \bibfnamefont{T.}}, \bibinfo {author}
  {\bibfnamefont{H.~E.}\ \bibnamefont{Logan}}, \bibinfo {author}
  {\bibfnamefont{B.}~\bibnamefont{McElrath}},\ and\ \bibinfo {author}
  {\bibfnamefont{L.-T.}\ \bibnamefont{Wang}}}%
  , \bibinfo {year} {2003},\ \bibfield{journal}{%
  \Doi{10.1016/j.physletb.2004.10.021}{\bibinfo {journal} {Phys. Lett.}}\ }%
  \textbf{\bibinfo {volume} {B563}},\ \bibinfo {pages} {191},\
  \Eprint{http://arxiv.org/abs/hep-ph/0302188}{arXiv:hep-ph/0302188}%
  \bibAnnoteFile{NoStop}{Han:2003gf}%
%%CITATION = HEP-PH/0302188;%%
\bibitem[{\citenamefont{He}\ \emph{et~al.}(2001)\citenamefont{He},
  \citenamefont{Polonsky},\ and\ \citenamefont{Su}}]{He:2001tp}%
  \BibitemOpen
  \bibfield{author}{%
  \bibinfo {author} {\bibnamefont{He}, \bibfnamefont{H.-J.}}, \bibinfo {author}
  {\bibfnamefont{N.}~\bibnamefont{Polonsky}},\ and\ \bibinfo {author}
  {\bibfnamefont{S.-f.}\ \bibnamefont{Su}}}%
  , \bibinfo {year} {2001},\ \bibfield{journal}{%
  \Doi{10.1103/PhysRevD.64.053004}{\bibinfo {journal} {Phys. Rev.}}\ }%
  \textbf{\bibinfo {volume} {D64}},\ \bibinfo {pages} {053004},\
  \Eprint{http://arxiv.org/abs/hep-ph/0102144}{arXiv:hep-ph/0102144}%
  \bibAnnoteFile{NoStop}{He:2001tp}%
%%CITATION = HEP-PH/0102144;%%
\bibitem[{\citenamefont{Heinson}(2010)}]{Heinson:2010}%
  \BibitemOpen
  \bibfield{author}{%
  \bibinfo {author} {\bibnamefont{Heinson}, \bibfnamefont{A.}}}%
  , \bibinfo {year} {2010},\ \bibfield{journal}{%
  \Doi{10.1142/S0217732310032871}{\bibinfo {journal} {Mod. Phys. Lett.}}\ }%
  \textbf{\bibinfo {volume} {A25}},\ \bibinfo {pages} {309},\
  \Eprint{http://arxiv.org/abs/1002.4167}{arXiv:1002.4167 [hep-ex]}%
  \bibAnnoteFile{NoStop}{Heinson:2010}%
%%CITATION = 1002.4167;%%
\bibitem[{\citenamefont{Heinson}\ and\
  \citenamefont{Junk}(2011)}]{Heinson:2011}%
  \BibitemOpen
  \bibfield{author}{%
  \bibinfo {author} {\bibnamefont{Heinson}, \bibfnamefont{A.}},\ and\ \bibinfo
  {author} {\bibfnamefont{T.}~\bibnamefont{Junk}}}%
  , \bibinfo {year} {2011}\ \bibinfo {note} {to appear in Ann. Rev. Nucl. Part.
  Sci. 61},\ \Eprint{http://arxiv.org/abs/1101.1275}{arXiv:1101.1275 [hep-ex]}%
  \bibAnnoteFile{NoStop}{Heinson:2011}%
%%CITATION = 1101.1275;%%
\bibitem[{\citenamefont{Herb}\ \emph{et~al.}(1977)\citenamefont{Herb}
  \emph{et~al.}}]{Bottom1977}%
  \BibitemOpen
  \bibfield{author}{%
  \bibinfo {author} {\bibnamefont{Herb}, \bibfnamefont{S.}}, \emph{et~al.}}%
  , \bibinfo {year} {1977},\ \bibfield{journal}{%
  \Doi{10.1103/PhysRevLett.39.252}{\bibinfo {journal} {Phys. Rev. Lett.}}\ }%
  \textbf{\bibinfo {volume} {39}},\ \bibinfo {pages} {252}%
  \bibAnnoteFile{NoStop}{Bottom1977}%
\bibitem[{\citenamefont{Hewett}\ \emph{et~al.}(2002)\citenamefont{Hewett},
  \citenamefont{Petriello},\ and\ \citenamefont{Rizzo}}]{Hewett:2002fe}%
  \BibitemOpen
  \bibfield{author}{%
  \bibinfo {author} {\bibnamefont{Hewett}, \bibfnamefont{J.~L.}}, \bibinfo
  {author} {\bibfnamefont{F.~J.}\ \bibnamefont{Petriello}},\ and\ \bibinfo
  {author} {\bibfnamefont{T.~G.}\ \bibnamefont{Rizzo}}}%
  , \bibinfo {year} {2002},\ \bibfield{journal}{%
  \bibinfo {journal} {JHEP}\ }%
  \textbf{\bibinfo {volume} {09}},\ \bibinfo {pages} {030},\
  \Eprint{http://arxiv.org/abs/hep-ph/0203091}{arXiv:hep-ph/0203091}%
  \bibAnnoteFile{NoStop}{Hewett:2002fe}%
%%CITATION = HEP-PH/0203091;%%
\bibitem[{\citenamefont{Hill}(1991)}]{Hill:1991at}%
  \BibitemOpen
  \bibfield{author}{%
  \bibinfo {author} {\bibnamefont{Hill}, \bibfnamefont{C.~T.}}}%
  , \bibinfo {year} {1991},\ \bibfield{journal}{%
  \Doi{10.1016/0370-2693(91)91061-Y}{\bibinfo {journal} {Phys. Lett.}}\ }%
  \textbf{\bibinfo {volume} {B266}},\ \bibinfo {pages} {419}%
  \bibAnnoteFile{NoStop}{Hill:1991at}%
%%CITATION = PHLTA,B266,419;%%
\bibitem[{\citenamefont{Hill}(1995)}]{Hill:1994hp}%
  \BibitemOpen
  \bibfield{author}{%
  \bibinfo {author} {\bibnamefont{Hill}, \bibfnamefont{C.~T.}}}%
  , \bibinfo {year} {1995},\ \bibfield{journal}{%
  \Doi{10.1016/0370-2693(94)01660-5}{\bibinfo {journal} {Phys. Lett.}}\ }%
  \textbf{\bibinfo {volume} {B345}},\ \bibinfo {pages} {483},\
  \Eprint{http://arxiv.org/abs/hep-ph/9411426}{arXiv:hep-ph/9411426}%
  \bibAnnoteFile{NoStop}{Hill:1994hp}%
%%CITATION = HEP-PH/9411426;%%
\bibitem[{\citenamefont{Hill}\ and\ \citenamefont{Parke}(1994)}]{Hill:1993hs}%
  \BibitemOpen
  \bibfield{author}{%
  \bibinfo {author} {\bibnamefont{Hill}, \bibfnamefont{C.~T.}},\ and\ \bibinfo
  {author} {\bibfnamefont{S.~J.}\ \bibnamefont{Parke}}}%
  , \bibinfo {year} {1994},\ \bibfield{journal}{%
  \Doi{10.1103/PhysRevD.49.4454}{\bibinfo {journal} {Phys. Rev.}}\ }%
  \textbf{\bibinfo {volume} {D49}},\ \bibinfo {pages} {4454},\
  \Eprint{http://arxiv.org/abs/hep-ph/9312324}{arXiv:hep-ph/9312324}%
  \bibAnnoteFile{NoStop}{Hill:1993hs}%
%%CITATION = HEP-PH/9312324;%%
\bibitem[{\citenamefont{Hoang}\ and\
  \citenamefont{Stewart}(2008)}]{Hoang:2008xm}%
  \BibitemOpen
  \bibfield{author}{%
  \bibinfo {author} {\bibnamefont{Hoang}, \bibfnamefont{A.~H.}},\ and\ \bibinfo
  {author} {\bibfnamefont{I.~W.}\ \bibnamefont{Stewart}}}%
  , \bibinfo {year} {2008},\ \bibfield{journal}{%
  \Doi{10.1016/j.nuclphysbps.2008.10.028}{\bibinfo {journal} {Nucl. Phys. Proc.
  Suppl.}}\ }%
  \textbf{\bibinfo {volume} {185}},\ \bibinfo {pages} {220},\
  \Eprint{http://arxiv.org/abs/0808.0222}{arXiv:0808.0222 [hep-ph]}%
  \bibAnnoteFile{NoStop}{Hoang:2008xm}%
%%CITATION = 0808.0222;%%
\bibitem[{\citenamefont{Hocker}\ and\
  \citenamefont{Kartvelishvili}(1996)}]{Hocker:1995kb}%
  \BibitemOpen
  \bibfield{author}{%
  \bibinfo {author} {\bibnamefont{Hocker}, \bibfnamefont{A.}},\ and\ \bibinfo
  {author} {\bibfnamefont{V.}~\bibnamefont{Kartvelishvili}}}%
  , \bibinfo {year} {1996},\ \bibfield{journal}{%
  \Doi{10.1016/0168-9002(95)01478-0}{\bibinfo {journal} {Nucl. Instrum.
  Meth.}}\ }%
  \textbf{\bibinfo {volume} {A372}},\ \bibinfo {pages} {469},\
  \Eprint{http://arxiv.org/abs/hep-ph/9509307}{arXiv:hep-ph/9509307}%
  \bibAnnoteFile{NoStop}{Hocker:1995kb}%
%%CITATION = HEP-PH/9509307;%%
\bibitem[{\citenamefont{Holdom}\ \emph{et~al.}(2009)\citenamefont{Holdom},
  \citenamefont{Hou}, \citenamefont{Hurth}, \citenamefont{Mangano},
  \citenamefont{Sultansoy} \emph{et~al.}}]{Holdom:2009rf}%
  \BibitemOpen
  \bibfield{author}{%
  \bibinfo {author} {\bibnamefont{Holdom}, \bibfnamefont{B.}}, \bibinfo
  {author} {\bibfnamefont{W.}~\bibnamefont{Hou}}, \bibinfo {author}
  {\bibfnamefont{T.}~\bibnamefont{Hurth}}, \bibinfo {author}
  {\bibfnamefont{M.}~\bibnamefont{Mangano}}, \bibinfo {author}
  {\bibfnamefont{S.}~\bibnamefont{Sultansoy}}, \emph{et~al.}}%
  , \bibinfo {year} {2009},\ \bibfield{journal}{%
  \Doi{10.1186/1754-0410-3-4}{\bibinfo {journal} {PMC Phys.}}\ }%
  \textbf{\bibinfo {volume} {A3}},\ \bibinfo {pages} {4},\
  \Eprint{http://arxiv.org/abs/0904.4698}{arXiv:0904.4698 [hep-ph]}%
  \bibAnnoteFile{NoStop}{Holdom:2009rf}%
\bibitem[{\citenamefont{Huber}\ and\
  \citenamefont{Shafi}(2001)}]{Huber:2000fh}%
  \BibitemOpen
  \bibfield{author}{%
  \bibinfo {author} {\bibnamefont{Huber}, \bibfnamefont{S.~J.}},\ and\ \bibinfo
  {author} {\bibfnamefont{Q.}~\bibnamefont{Shafi}}}%
  , \bibinfo {year} {2001},\ \bibfield{journal}{%
  \Doi{10.1103/PhysRevD.63.045010}{\bibinfo {journal} {Phys. Rev.}}\ }%
  \textbf{\bibinfo {volume} {D63}},\ \bibinfo {pages} {045010},\
  \Eprint{http://arxiv.org/abs/hep-ph/0005286}{arXiv:hep-ph/0005286}%
  \bibAnnoteFile{NoStop}{Huber:2000fh}%
%%CITATION = HEP-PH/0005286;%%
\bibitem[{\citenamefont{Incandela}\ \emph{et~al.}(2009)\citenamefont{Incandela}
  \emph{et~al.}}]{Incandela:2009}%
  \BibitemOpen
  \bibfield{author}{%
  \bibinfo {author} {\bibnamefont{Incandela}, \bibfnamefont{J.}},
  \emph{et~al.}}%
  , \bibinfo {year} {2009},\ \bibfield{journal}{%
  \Doi{10.1016/j.ppnp.2009.08.001}{\bibinfo {journal} {Prog.Part.Nucl.Phys.}}\
  }%
  \textbf{\bibinfo {volume} {63}},\ \bibinfo {pages} {239},\
  \Eprint{http://arxiv.org/abs/0904.2499}{arXiv:0904.2499 [hep-ex]}%
  \bibAnnoteFile{NoStop}{Incandela:2009}%
%%CITATION = 0904.2499;%%
\bibitem[{\citenamefont{Jezabek}\ and\
  \citenamefont{K{\"{u}}hn}(1993)}]{Kuhn1993}%
  \BibitemOpen
  \bibfield{author}{%
  \bibinfo {author} {\bibnamefont{Jezabek}, \bibfnamefont{M.}},\ and\ \bibinfo
  {author} {\bibfnamefont{J.}~\bibnamefont{K{\"{u}}hn}}}%
  , \bibinfo {year} {1993},\ \bibfield{journal}{%
  \Doi{10.1103/PhysRevD.48.R1910}{\bibinfo {journal} {Phys. Rev.}}\ }%
  \textbf{\bibinfo {volume} {D48}},\ \bibinfo {pages} {1910},\
  \Eprint{http://arxiv.org/abs/9302295}{arXiv:9302295 [hep-ph]}%
  \bibAnnoteFile{NoStop}{Kuhn1993}%
\bibitem[{\citenamefont{Jezabek}\ and\
  \citenamefont{K{\"{u}}hn}(1994)}]{Kuhn1994}%
  \BibitemOpen
  \bibfield{author}{%
  \bibinfo {author} {\bibnamefont{Jezabek}, \bibfnamefont{M.}},\ and\ \bibinfo
  {author} {\bibfnamefont{J.}~\bibnamefont{K{\"{u}}hn}}}%
  , \bibinfo {year} {1994},\ \bibfield{journal}{%
  \Doi{10.1103/PhysRevD.49.4970}{\bibinfo {journal} {Phys. Rev.}}\ }%
  \textbf{\bibinfo {volume} {D48}},\ \bibinfo {pages} {4970}%
  \bibAnnoteFile{NoStop}{Kuhn1994}%
\bibitem[{\citenamefont{Jezabek}\ and\
  \citenamefont{Kuhn}(1989)}]{Jezabek:1988iv}%
  \BibitemOpen
  \bibfield{author}{%
  \bibinfo {author} {\bibnamefont{Jezabek}, \bibfnamefont{M.}},\ and\ \bibinfo
  {author} {\bibfnamefont{J.~H.}\ \bibnamefont{Kuhn}}}%
  , \bibinfo {year} {1989},\ \bibfield{journal}{%
  \Doi{10.1016/0550-3213(89)90108-9}{\bibinfo {journal} {Nucl. Phys.}}\ }%
  \textbf{\bibinfo {volume} {B314}},\ \bibinfo {pages} {1}%
  \bibAnnoteFile{NoStop}{Jezabek:1988iv}%
%%CITATION = NUPHA,B314,1;%%
\bibitem[{\citenamefont{Juste}(1999)}]{Juste:1998rw}%
  \BibitemOpen
  \bibfield{author}{%
  \bibinfo {author} {\bibnamefont{Juste}, \bibfnamefont{A.}}}%
  , \bibinfo {year} {1999},\ \enquote{\bibinfo {title} {{Measurement of the W
  mass in e+ e- annihilation}},}\ \bibinfo {note} {CERN-THESIS-99-012}%
  \bibAnnoteFile{NoStop}{Juste:1998rw}%
\bibitem[{\citenamefont{Kane}\ \emph{et~al.}(1992)\citenamefont{Kane},
  \citenamefont{Ladinsky},\ and\ \citenamefont{Yuan}}]{Kane:1991bg}%
  \BibitemOpen
  \bibfield{author}{%
  \bibinfo {author} {\bibnamefont{Kane}, \bibfnamefont{G.~L.}}, \bibinfo
  {author} {\bibfnamefont{G.~A.}\ \bibnamefont{Ladinsky}},\ and\ \bibinfo
  {author} {\bibfnamefont{C.~P.}\ \bibnamefont{Yuan}}}%
  , \bibinfo {year} {1992},\ \bibfield{journal}{%
  \Doi{10.1103/PhysRevD.45.124}{\bibinfo {journal} {Phys. Rev.}}\ }%
  \textbf{\bibinfo {volume} {D45}},\ \bibinfo {pages} {124}%
  \bibAnnoteFile{NoStop}{Kane:1991bg}%
%%CITATION = PHRVA,D45,124;%%
\bibitem[{\citenamefont{Kane}\ and\ \citenamefont{Mrenna}(1996)}]{Kane:1996ny}%
  \BibitemOpen
  \bibfield{author}{%
  \bibinfo {author} {\bibnamefont{Kane}, \bibfnamefont{G.~L.}},\ and\ \bibinfo
  {author} {\bibfnamefont{S.}~\bibnamefont{Mrenna}}}%
  , \bibinfo {year} {1996},\ \bibfield{journal}{%
  \Doi{10.1103/PhysRevLett.77.3502}{\bibinfo {journal} {Phys. Rev. Lett.}}\ }%
  \textbf{\bibinfo {volume} {77}},\ \bibinfo {pages} {3502},\
  \Eprint{http://arxiv.org/abs/hep-ph/9605351}{arXiv:hep-ph/9605351}%
  \bibAnnoteFile{NoStop}{Kane:1996ny}%
%%CITATION = HEP-PH/9605351;%%
\bibitem[{\citenamefont{Kehoe}\ \emph{et~al.}(2008)\citenamefont{Kehoe},
  \citenamefont{Narain},\ and\ \citenamefont{Kumar}}]{Kehoe:2008}%
  \BibitemOpen
  \bibfield{author}{%
  \bibinfo {author} {\bibnamefont{Kehoe}, \bibfnamefont{R.}}, \bibinfo {author}
  {\bibfnamefont{M.}~\bibnamefont{Narain}},\ and\ \bibinfo {author}
  {\bibfnamefont{A.}~\bibnamefont{Kumar}}}%
  , \bibinfo {year} {2008},\ \bibfield{journal}{%
  \Doi{10.1142/S0217751X08039293}{\bibinfo {journal} {Int.J.Mod.Phys.}}\ }%
  \textbf{\bibinfo {volume} {A23}},\ \bibinfo {pages} {353},\
  \Eprint{http://arxiv.org/abs/0712.2733}{arXiv:0712.2733 [hep-ex]}%
  \bibAnnoteFile{NoStop}{Kehoe:2008}%
%%CITATION = 0712.2733;%%
\bibitem[{\citenamefont{Kidonakis}\ and\
  \citenamefont{Vogt}(2008)}]{Kidonakis:2008mu}%
  \BibitemOpen
  \bibfield{author}{%
  \bibinfo {author} {\bibnamefont{Kidonakis}, \bibfnamefont{N.}},\ and\
  \bibinfo {author} {\bibfnamefont{R.}~\bibnamefont{Vogt}}}%
  , \bibinfo {year} {2008},\ \bibfield{journal}{%
  \Doi{10.1103/PhysRevD.78.074005}{\bibinfo {journal} {Phys. Rev.}}\ }%
  \textbf{\bibinfo {volume} {D78}},\ \bibinfo {pages} {074005},\
  \Eprint{http://arxiv.org/abs/0805.3844}{arXiv:0805.3844 [hep-ph]}%
  \bibAnnoteFile{NoStop}{Kidonakis:2008mu}%
%%CITATION = 0805.3844;%%
\bibitem[{\citenamefont{Kobayashi}\ and\
  \citenamefont{Maskawa}(1973)}]{KM1973}%
  \BibitemOpen
  \bibfield{author}{%
  \bibinfo {author} {\bibnamefont{Kobayashi}, \bibfnamefont{M.}},\ and\
  \bibinfo {author} {\bibfnamefont{T.}~\bibnamefont{Maskawa}}}%
  , \bibinfo {year} {1973},\ \bibfield{journal}{%
  \bibinfo {journal} {Prog. Theor. Phys.}\ }%
  \textbf{\bibinfo {volume} {49}},\ \bibinfo {pages} {652}%
  \bibAnnoteFile{NoStop}{KM1973}%
\bibitem[{\citenamefont{Kondo}\ \emph{et~al.}(1993)\citenamefont{Kondo},
  \citenamefont{Chikamatsu},\ and\ \citenamefont{Kim}}]{Kondo:1993in}%
  \BibitemOpen
  \bibfield{author}{%
  \bibinfo {author} {\bibnamefont{Kondo}, \bibfnamefont{K.}}, \bibinfo {author}
  {\bibfnamefont{T.}~\bibnamefont{Chikamatsu}},\ and\ \bibinfo {author}
  {\bibfnamefont{S.~H.}\ \bibnamefont{Kim}}}%
  , \bibinfo {year} {1993},\ \bibfield{journal}{%
  \Doi{10.1143/JPSJ.62.1177}{\bibinfo {journal} {J. Phys. Soc. Jap.}}\ }%
  \textbf{\bibinfo {volume} {62}},\ \bibinfo {pages} {1177}%
  \bibAnnoteFile{NoStop}{Kondo:1993in}%
%%CITATION = JUPSA,62,1177;%%
\bibitem[{\citenamefont{Kribs}\ \emph{et~al.}(2007)\citenamefont{Kribs},
  \citenamefont{Plehn}, \citenamefont{Spannowsky},\ and\
  \citenamefont{Tait}}]{Kribs:2007nz}%
  \BibitemOpen
  \bibfield{author}{%
  \bibinfo {author} {\bibnamefont{Kribs}, \bibfnamefont{G.~D.}}, \bibinfo
  {author} {\bibfnamefont{T.}~\bibnamefont{Plehn}}, \bibinfo {author}
  {\bibfnamefont{M.}~\bibnamefont{Spannowsky}},\ and\ \bibinfo {author}
  {\bibfnamefont{T.~M.}\ \bibnamefont{Tait}}}%
  , \bibinfo {year} {2007},\ \bibfield{journal}{%
  \Doi{10.1103/PhysRevD.76.075016}{\bibinfo {journal} {Phys.Rev.}}\ }%
  \textbf{\bibinfo {volume} {D76}},\ \bibinfo {pages} {075016},\
  \Eprint{http://arxiv.org/abs/0706.3718}{arXiv:0706.3718 [hep-ph]}%
  \bibAnnoteFile{NoStop}{Kribs:2007nz}%
\bibitem[{\citenamefont{Kronfeld}\ and\
  \citenamefont{Quigg}(2010)}]{Kronfeld:2010bx}%
  \BibitemOpen
  \bibfield{author}{%
  \bibinfo {author} {\bibnamefont{Kronfeld}, \bibfnamefont{A.~S.}},\ and\
  \bibinfo {author} {\bibfnamefont{C.}~\bibnamefont{Quigg}}}%
  , \bibinfo {year} {2010}\
  \Eprint{http://arxiv.org/abs/1002.5032}{arXiv:1002.5032 [hep-ph]}%
  \bibAnnoteFile{NoStop}{Kronfeld:2010bx}%
%%CITATION = 1002.5032;%%
\bibitem[{\citenamefont{Kuhn}\ and\
  \citenamefont{Rodrigo}(1998)}]{Kuhn:1998jr}%
  \BibitemOpen
  \bibfield{author}{%
  \bibinfo {author} {\bibnamefont{Kuhn}, \bibfnamefont{J.~H.}},\ and\ \bibinfo
  {author} {\bibfnamefont{G.}~\bibnamefont{Rodrigo}}}%
  , \bibinfo {year} {1998},\ \bibfield{journal}{%
  \Doi{10.1103/PhysRevLett.81.49}{\bibinfo {journal} {Phys. Rev. Lett.}}\ }%
  \textbf{\bibinfo {volume} {81}},\ \bibinfo {pages} {49},\
  \Eprint{http://arxiv.org/abs/hep-ph/9802268}{arXiv:hep-ph/9802268}%
  \bibAnnoteFile{NoStop}{Kuhn:1998jr}%
%%CITATION = HEP-PH/9802268;%%
\bibitem[{\citenamefont{Kuhn}\ and\
  \citenamefont{Rodrigo}(1999)}]{Kuhn:1998kw}%
  \BibitemOpen
  \bibfield{author}{%
  \bibinfo {author} {\bibnamefont{Kuhn}, \bibfnamefont{J.~H.}},\ and\ \bibinfo
  {author} {\bibfnamefont{G.}~\bibnamefont{Rodrigo}}}%
  , \bibinfo {year} {1999},\ \bibfield{journal}{%
  \Doi{10.1103/PhysRevD.59.054017}{\bibinfo {journal} {Phys. Rev.}}\ }%
  \textbf{\bibinfo {volume} {D59}},\ \bibinfo {pages} {054017},\
  \Eprint{http://arxiv.org/abs/hep-ph/9807420}{arXiv:hep-ph/9807420}%
  \bibAnnoteFile{NoStop}{Kuhn:1998kw}%
%%CITATION = HEP-PH/9807420;%%
\bibitem[{\citenamefont{Lane}\ and\
  \citenamefont{Eichten}(1995)}]{Lane:1995gw}%
  \BibitemOpen
  \bibfield{author}{%
  \bibinfo {author} {\bibnamefont{Lane}, \bibfnamefont{K.~D.}},\ and\ \bibinfo
  {author} {\bibfnamefont{E.}~\bibnamefont{Eichten}}}%
  , \bibinfo {year} {1995},\ \bibfield{journal}{%
  \Doi{10.1016/0370-2693(95)00482-Z}{\bibinfo {journal} {Phys. Lett.}}\ }%
  \textbf{\bibinfo {volume} {B352}},\ \bibinfo {pages} {382},\
  \Eprint{http://arxiv.org/abs/hep-ph/9503433}{arXiv:hep-ph/9503433}%
  \bibAnnoteFile{NoStop}{Lane:1995gw}%
%%CITATION = HEP-PH/9503433;%%
\bibitem[{\citenamefont{Langacker}(1981)}]{Langacker:1980js}%
  \BibitemOpen
  \bibfield{author}{%
  \bibinfo {author} {\bibnamefont{Langacker}, \bibfnamefont{P.}}}%
  , \bibinfo {year} {1981},\ \bibfield{journal}{%
  \Doi{10.1016/0370-1573(81)90059-4}{\bibinfo {journal} {Phys. Rept.}}\ }%
  \textbf{\bibinfo {volume} {72}},\ \bibinfo {pages} {185}%
  \bibAnnoteFile{NoStop}{Langacker:1980js}%
%%CITATION = PRPLC,72,185;%%
\bibitem[{\citenamefont{Lange}\ and\ \citenamefont{Ryd}(1998)}]{evtgen}%
  \BibitemOpen
  \bibfield{author}{%
  \bibinfo {author} {\bibnamefont{Lange}, \bibfnamefont{D.}},\ and\ \bibinfo
  {author} {\bibfnamefont{A.}~\bibnamefont{Ryd}}}%
  , \bibinfo {year} {1998},\ \enquote{\bibinfo {title} {{The EvtGen Event
  Generator Package}},}\ \bibinfo {note} {See
  http://hep.ucsb.edu/people/lange/EvtGen/}%
  \bibAnnoteFile{NoStop}{evtgen}%
\bibitem[{\citenamefont{Langenfeld}\
  \emph{et~al.}(2009)\citenamefont{Langenfeld}, \citenamefont{Moch},\ and\
  \citenamefont{Uwer}}]{Langenfeld:2009wd}%
  \BibitemOpen
  \bibfield{author}{%
  \bibinfo {author} {\bibnamefont{Langenfeld}, \bibfnamefont{U.}}, \bibinfo
  {author} {\bibfnamefont{S.}~\bibnamefont{Moch}},\ and\ \bibinfo {author}
  {\bibfnamefont{P.}~\bibnamefont{Uwer}}}%
  , \bibinfo {year} {2009},\ \bibfield{journal}{%
  \Doi{10.1103/PhysRevD.80.054009}{\bibinfo {journal} {Phys. Rev.}}\ }%
  \textbf{\bibinfo {volume} {D80}},\ \bibinfo {pages} {054009},\
  \Eprint{http://arxiv.org/abs/0906.5273}{arXiv:0906.5273 [hep-ph]}%
  \bibAnnoteFile{NoStop}{Langenfeld:2009wd}%
%%CITATION = 0906.5273;%%
\bibitem[{\citenamefont{Larios}\ \emph{et~al.}(2006)\citenamefont{Larios},
  \citenamefont{Martinez},\ and\ \citenamefont{Perez}}]{Larios:2006pb}%
  \BibitemOpen
  \bibfield{author}{%
  \bibinfo {author} {\bibnamefont{Larios}, \bibfnamefont{F.}}, \bibinfo
  {author} {\bibfnamefont{R.}~\bibnamefont{Martinez}},\ and\ \bibinfo {author}
  {\bibfnamefont{M.~A.}\ \bibnamefont{Perez}}}%
  , \bibinfo {year} {2006},\ \bibfield{journal}{%
  \Doi{10.1142/S0217751X06033039}{\bibinfo {journal} {Int. J. Mod. Phys.}}\ }%
  \textbf{\bibinfo {volume} {A21}},\ \bibinfo {pages} {3473},\
  \Eprint{http://arxiv.org/abs/hep-ph/0605003}{arXiv:hep-ph/0605003}%
  \bibAnnoteFile{NoStop}{Larios:2006pb}%
%%CITATION = HEP-PH/0605003;%%
\bibitem[{\citenamefont{Leike}(1999)}]{Leike:1998wr}%
  \BibitemOpen
  \bibfield{author}{%
  \bibinfo {author} {\bibnamefont{Leike}, \bibfnamefont{A.}}}%
  , \bibinfo {year} {1999},\ \bibfield{journal}{%
  \Doi{10.1016/S0370-1573(98)00133-1}{\bibinfo {journal} {Phys. Rept.}}\ }%
  \textbf{\bibinfo {volume} {317}},\ \bibinfo {pages} {143},\
  \Eprint{http://arxiv.org/abs/hep-ph/9805494}{arXiv:hep-ph/9805494}%
  \bibAnnoteFile{NoStop}{Leike:1998wr}%
%%CITATION = HEP-PH/9805494;%%
\bibitem[{\citenamefont{Lester}\ and\
  \citenamefont{Summers}(1999)}]{Lester:1999}%
  \BibitemOpen
  \bibfield{author}{%
  \bibinfo {author} {\bibnamefont{Lester}, \bibfnamefont{C.}},\ and\ \bibinfo
  {author} {\bibfnamefont{D.}~\bibnamefont{Summers}}}%
  , \bibinfo {year} {1999},\ \bibfield{journal}{%
  \Doi{10.1016/S0370-2693(99)00945-4}{\bibinfo {journal} {Phys. Lett.}}\ }%
  \textbf{\bibinfo {volume} {B463}},\ \bibinfo {pages} {99},\
  \Eprint{http://arxiv.org/abs/9906349}{arXiv:9906349 [hep-ph]}%
  \bibAnnoteFile{NoStop}{Lester:1999}%
%%CITATION = hep-ph/9906349;%%
\bibitem[{\citenamefont{Lillie}\ \emph{et~al.}(2007)\citenamefont{Lillie},
  \citenamefont{Randall},\ and\ \citenamefont{Wang}}]{Lillie:2007yh}%
  \BibitemOpen
  \bibfield{author}{%
  \bibinfo {author} {\bibnamefont{Lillie}, \bibfnamefont{B.}}, \bibinfo
  {author} {\bibfnamefont{L.}~\bibnamefont{Randall}},\ and\ \bibinfo {author}
  {\bibfnamefont{L.-T.}\ \bibnamefont{Wang}}}%
  , \bibinfo {year} {2007},\ \bibfield{journal}{%
  \Doi{10.1088/1126-6708/2007/09/074}{\bibinfo {journal} {JHEP}}\ }%
  \textbf{\bibinfo {volume} {09}},\ \bibinfo {pages} {074},\
  \Eprint{http://arxiv.org/abs/hep-ph/0701166}{arXiv:hep-ph/0701166}%
  \bibAnnoteFile{NoStop}{Lillie:2007yh}%
%%CITATION = HEP-PH/0701166;%%
\bibitem[{\citenamefont{Loader}(1999)}]{Loader:1999}%
  \BibitemOpen
  \bibfield{author}{%
  \bibinfo {author} {\bibnamefont{Loader}, \bibfnamefont{C.}}}%
  , \bibinfo {year} {1999},\ \emph{\bibinfo {title} {{Local regression and
  likelihood}}}\ (\bibinfo {publisher} {Springer})%
  \bibAnnoteFile{NoStop}{Loader:1999}%
\bibitem[{\citenamefont{Lyons}\ \emph{et~al.}(1988)\citenamefont{Lyons},
  \citenamefont{Gibaut},\ and\ \citenamefont{Clifford}}]{Lyons:1988rp}%
  \BibitemOpen
  \bibfield{author}{%
  \bibinfo {author} {\bibnamefont{Lyons}, \bibfnamefont{L.}}, \bibinfo {author}
  {\bibfnamefont{D.}~\bibnamefont{Gibaut}},\ and\ \bibinfo {author}
  {\bibfnamefont{P.}~\bibnamefont{Clifford}}}%
  , \bibinfo {year} {1988},\ \bibfield{journal}{%
  \Doi{10.1016/0168-9002(88)90018-6}{\bibinfo {journal} {Nucl. Instrum.
  Meth.}}\ }%
  \textbf{\bibinfo {volume} {A270}},\ \bibinfo {pages} {110}%
  \bibAnnoteFile{NoStop}{Lyons:1988rp}%
%%CITATION = NUIMA,A270,110;%%
\bibitem[{\citenamefont{Mahlon}\ and\
  \citenamefont{Parke}(1996)}]{Mahlon:1995zn}%
  \BibitemOpen
  \bibfield{author}{%
  \bibinfo {author} {\bibnamefont{Mahlon}, \bibfnamefont{G.}},\ and\ \bibinfo
  {author} {\bibfnamefont{S.~J.}\ \bibnamefont{Parke}}}%
  , \bibinfo {year} {1996},\ \bibfield{journal}{%
  \Doi{10.1103/PhysRevD.53.4886}{\bibinfo {journal} {Phys. Rev.}}\ }%
  \textbf{\bibinfo {volume} {D53}},\ \bibinfo {pages} {4886},\
  \Eprint{http://arxiv.org/abs/hep-ph/9512264}{arXiv:hep-ph/9512264}%
  \bibAnnoteFile{NoStop}{Mahlon:1995zn}%
%%CITATION = HEP-PH/9512264;%%
\bibitem[{\citenamefont{Mahlon}\ and\
  \citenamefont{Parke}(1997)}]{Mahlon:1997uc}%
  \BibitemOpen
  \bibfield{author}{%
  \bibinfo {author} {\bibnamefont{Mahlon}, \bibfnamefont{G.}},\ and\ \bibinfo
  {author} {\bibfnamefont{S.~J.}\ \bibnamefont{Parke}}}%
  , \bibinfo {year} {1997},\ \bibfield{journal}{%
  \Doi{10.1016/S0370-2693(97)00987-8}{\bibinfo {journal} {Phys. Lett.}}\ }%
  \textbf{\bibinfo {volume} {B411}},\ \bibinfo {pages} {173},\
  \Eprint{http://arxiv.org/abs/hep-ph/9706304}{arXiv:hep-ph/9706304}%
  \bibAnnoteFile{NoStop}{Mahlon:1997uc}%
%%CITATION = HEP-PH/9706304;%%
\bibitem[{\citenamefont{Mahlon}\ and\
  \citenamefont{Parke}(2010)}]{Mahlon:2010gw}%
  \BibitemOpen
  \bibfield{author}{%
  \bibinfo {author} {\bibnamefont{Mahlon}, \bibfnamefont{G.}},\ and\ \bibinfo
  {author} {\bibfnamefont{S.~J.}\ \bibnamefont{Parke}}}%
  , \bibinfo {year} {2010},\ \bibfield{journal}{%
  \Doi{10.1103/PhysRevD.81.074024}{\bibinfo {journal} {Phys. Rev.}}\ }%
  \textbf{\bibinfo {volume} {D81}},\ \bibinfo {pages} {074024},\
  \Eprint{http://arxiv.org/abs/1001.3422}{arXiv:1001.3422 [hep-ph]}%
  \bibAnnoteFile{NoStop}{Mahlon:2010gw}%
%%CITATION = 1001.3422;%%
\bibitem[{\citenamefont{Malkawi}\ \emph{et~al.}(1996)\citenamefont{Malkawi},
  \citenamefont{Tait},\ and\ \citenamefont{Yuan}}]{Malkawi:1996fs}%
  \BibitemOpen
  \bibfield{author}{%
  \bibinfo {author} {\bibnamefont{Malkawi}, \bibfnamefont{E.}}, \bibinfo
  {author} {\bibfnamefont{T.~M.~P.}\ \bibnamefont{Tait}},\ and\ \bibinfo
  {author} {\bibfnamefont{C.~P.}\ \bibnamefont{Yuan}}}%
  , \bibinfo {year} {1996},\ \bibfield{journal}{%
  \Doi{10.1016/0370-2693(96)00859-3}{\bibinfo {journal} {Phys. Lett.}}\ }%
  \textbf{\bibinfo {volume} {B385}},\ \bibinfo {pages} {304},\
  \Eprint{http://arxiv.org/abs/hep-ph/9603349}{arXiv:hep-ph/9603349}%
  \bibAnnoteFile{NoStop}{Malkawi:1996fs}%
%%CITATION = HEP-PH/9603349;%%
\bibitem[{\citenamefont{Mangano}\ \emph{et~al.}(2003)\citenamefont{Mangano},
  \citenamefont{Moretti}, \citenamefont{Piccinini}, \citenamefont{Pittau},\
  and\ \citenamefont{Polosa}}]{alpgen}%
  \BibitemOpen
  \bibfield{author}{%
  \bibinfo {author} {\bibnamefont{Mangano}, \bibfnamefont{M.}}, \bibinfo
  {author} {\bibfnamefont{M.}~\bibnamefont{Moretti}}, \bibinfo {author}
  {\bibfnamefont{F.}~\bibnamefont{Piccinini}}, \bibinfo {author}
  {\bibfnamefont{R.}~\bibnamefont{Pittau}},\ and\ \bibinfo {author}
  {\bibfnamefont{A.}~\bibnamefont{Polosa}}}%
  , \bibinfo {year} {2003},\ \bibfield{journal}{%
  \Doi{10.1088/1126-6708/2003/07/001}{\bibinfo {journal} {JHEP}}\ }%
  \textbf{\bibinfo {volume} {0307}},\ \bibinfo {pages} {001},\ \bibinfo {note}
  {see also http://mlm.web.cern.ch/mlm/alpgen/},\
  \Eprint{http://arxiv.org/abs/0206293}{arXiv:0206293 [hep-ph]}%
  \bibAnnoteFile{NoStop}{alpgen}%
\bibitem[{\citenamefont{Martin}\ \emph{et~al.}(2007)\citenamefont{Martin},
  \citenamefont{Roberts}, \citenamefont{Stirling},\ and\
  \citenamefont{Watt}}]{mrst}%
  \BibitemOpen
  \bibfield{author}{%
  \bibinfo {author} {\bibnamefont{Martin}, \bibfnamefont{A.}}, \bibinfo
  {author} {\bibfnamefont{R.}~\bibnamefont{Roberts}}, \bibinfo {author}
  {\bibfnamefont{W.}~\bibnamefont{Stirling}},\ and\ \bibinfo {author}
  {\bibfnamefont{G.}~\bibnamefont{Watt}} (\bibinfo {collaboration}
  {MRST/MRSW})}%
  , \bibinfo {year} {2007},\ \bibfield{journal}{%
  \Doi{10.1016/j.physletb.2007.07.040}{\bibinfo {journal} {Phys.Lett.}}\ }%
  \textbf{\bibinfo {volume} {B652}},\ \bibinfo {pages} {292},\ \bibinfo {note}
  {see also http://durpdg.dur.ac.uk/hepdata/mrs.html},\
  \Eprint{http://arxiv.org/abs/0706.0459}{arXiv:0706.0459 [hep-ph]}%
  \bibAnnoteFile{NoStop}{mrst}%
\bibitem[{\citenamefont{Martin}(1997)}]{Martin:1997ns}%
  \BibitemOpen
  \bibfield{author}{%
  \bibinfo {author} {\bibnamefont{Martin}, \bibfnamefont{S.~P.}}}%
  , \bibinfo {year} {1997}\
  \Eprint{http://arxiv.org/abs/hep-ph/9709356}{arXiv:hep-ph/9709356}%
  \bibAnnoteFile{NoStop}{Martin:1997ns}%
%%CITATION = HEP-PH/9709356;%%
\bibitem[{\citenamefont{Moch}\ and\ \citenamefont{Uwer}(2008)}]{Moch:2008qy}%
  \BibitemOpen
  \bibfield{author}{%
  \bibinfo {author} {\bibnamefont{Moch}, \bibfnamefont{S.}},\ and\ \bibinfo
  {author} {\bibfnamefont{P.}~\bibnamefont{Uwer}}}%
  , \bibinfo {year} {2008},\ \bibfield{journal}{%
  \Doi{10.1103/PhysRevD.78.034003}{\bibinfo {journal} {Phys. Rev.}}\ }%
  \textbf{\bibinfo {volume} {D78}},\ \bibinfo {pages} {034003},\
  \Eprint{http://arxiv.org/abs/0804.1476}{arXiv:0804.1476 [hep-ph]}%
  \bibAnnoteFile{NoStop}{Moch:2008qy}%
%%CITATION = 0804.1476;%%
\bibitem[{\citenamefont{Nadolsky}\ \emph{et~al.}(2008)\citenamefont{Nadolsky}
  \emph{et~al.}}]{Nadolsky:2008zw}%
  \BibitemOpen
  \bibfield{author}{%
  \bibinfo {author} {\bibnamefont{Nadolsky}, \bibfnamefont{P.~M.}},
  \emph{et~al.}}%
  , \bibinfo {year} {2008},\ \bibfield{journal}{%
  \Doi{10.1103/PhysRevD.78.013004}{\bibinfo {journal} {Phys. Rev.}}\ }%
  \textbf{\bibinfo {volume} {D78}},\ \bibinfo {pages} {013004},\
  \Eprint{http://arxiv.org/abs/0802.0007}{arXiv:0802.0007 [hep-ph]}%
  \bibAnnoteFile{NoStop}{Nadolsky:2008zw}%
%%CITATION = 0802.0007;%%
\bibitem[{\citenamefont{Nakamura}\ \emph{et~al.}(2010)\citenamefont{Nakamura}
  \emph{et~al.}}]{PDG}%
  \BibitemOpen
  \bibfield{author}{%
  \bibinfo {author} {\bibnamefont{Nakamura}, \bibfnamefont{K.}},
  \emph{et~al.}}%
  , \bibinfo {year} {2010},\ \bibfield{journal}{%
  \bibinfo {journal} {J. Phys.}\ }%
  \textbf{\bibinfo {volume} {G 37}},\ \bibinfo {pages} {075021}%
  \bibAnnoteFile{NoStop}{PDG}%
\bibitem[{\citenamefont{Nam}(2002)}]{Nam:2002rq}%
  \BibitemOpen
  \bibfield{author}{%
  \bibinfo {author} {\bibnamefont{Nam}, \bibfnamefont{S.-h.}}}%
  , \bibinfo {year} {2002},\ \bibfield{journal}{%
  \Doi{10.1103/PhysRevD.66.055008}{\bibinfo {journal} {Phys. Rev.}}\ }%
  \textbf{\bibinfo {volume} {D66}},\ \bibinfo {pages} {055008},\
  \Eprint{http://arxiv.org/abs/hep-ph/0206037}{arXiv:hep-ph/0206037}%
  \bibAnnoteFile{NoStop}{Nam:2002rq}%
%%CITATION = HEP-PH/0206037;%%
\bibitem[{\citenamefont{Nilles}(1984)}]{Nilles:1983ge}%
  \BibitemOpen
  \bibfield{author}{%
  \bibinfo {author} {\bibnamefont{Nilles}, \bibfnamefont{H.~P.}}}%
  , \bibinfo {year} {1984},\ \bibfield{journal}{%
  \Doi{10.1016/0370-1573(84)90008-5}{\bibinfo {journal} {Phys. Rept.}}\ }%
  \textbf{\bibinfo {volume} {110}},\ \bibinfo {pages} {1}%
  \bibAnnoteFile{NoStop}{Nilles:1983ge}%
%%CITATION = PRPLC,110,1;%%
\bibitem[{\citenamefont{Parke}\ and\
  \citenamefont{Shadmi}(1996)}]{Parke:1996pr}%
  \BibitemOpen
  \bibfield{author}{%
  \bibinfo {author} {\bibnamefont{Parke}, \bibfnamefont{S.~J.}},\ and\ \bibinfo
  {author} {\bibfnamefont{Y.}~\bibnamefont{Shadmi}}}%
  , \bibinfo {year} {1996},\ \bibfield{journal}{%
  \Doi{10.1016/0370-2693(96)00998-7}{\bibinfo {journal} {Phys. Lett.}}\ }%
  \textbf{\bibinfo {volume} {B387}},\ \bibinfo {pages} {199},\
  \Eprint{http://arxiv.org/abs/hep-ph/9606419}{arXiv:hep-ph/9606419}%
  \bibAnnoteFile{NoStop}{Parke:1996pr}%
%%CITATION = HEP-PH/9606419;%%
\bibitem[{\citenamefont{Perl}\ \emph{et~al.}(1975)\citenamefont{Perl}
  \emph{et~al.}}]{Tau1975}%
  \BibitemOpen
  \bibfield{author}{%
  \bibinfo {author} {\bibnamefont{Perl}, \bibfnamefont{M.}}, \emph{et~al.}}%
  , \bibinfo {year} {1975},\ \bibfield{journal}{%
  \Doi{10.1103/PhysRevLett.35.1489}{\bibinfo {journal} {Phys. Rev. Lett.}}\ }%
  \textbf{\bibinfo {volume} {35}},\ \bibinfo {pages} {1489}%
  \bibAnnoteFile{NoStop}{Tau1975}%
\bibitem[{\citenamefont{Pleier}(2009)}]{Pleier:2009}%
  \BibitemOpen
  \bibfield{author}{%
  \bibinfo {author} {\bibnamefont{Pleier}, \bibfnamefont{M.-A.}}}%
  , \bibinfo {year} {2009},\ \bibfield{journal}{%
  \Doi{10.1142/S0217751X09044541}{\bibinfo {journal} {Int.J.Mod.Phys.}}\ }%
  \textbf{\bibinfo {volume} {A24}},\ \bibinfo {pages} {2899},\
  \Eprint{http://arxiv.org/abs/0810.5226}{arXiv:0810.5226 [hep-ex]}%
  \bibAnnoteFile{NoStop}{Pleier:2009}%
%%CITATION = 0810.5226;%%
\bibitem[{\citenamefont{Pomarol}(2000)}]{Pomarol:1999ad}%
  \BibitemOpen
  \bibfield{author}{%
  \bibinfo {author} {\bibnamefont{Pomarol}, \bibfnamefont{A.}}}%
  , \bibinfo {year} {2000},\ \bibfield{journal}{%
  \Doi{10.1016/S0370-2693(00)00737-1}{\bibinfo {journal} {Phys. Lett.}}\ }%
  \textbf{\bibinfo {volume} {B486}},\ \bibinfo {pages} {153},\
  \Eprint{http://arxiv.org/abs/hep-ph/9911294}{arXiv:hep-ph/9911294}%
  \bibAnnoteFile{NoStop}{Pomarol:1999ad}%
%%CITATION = HEP-PH/9911294;%%
\bibitem[{\citenamefont{Pumplin}\ \emph{et~al.}(2002)\citenamefont{Pumplin},
  \citenamefont{Stump}, \citenamefont{Huston}, \citenamefont{Lai},
  \citenamefont{Nadolsky},\ and\ \citenamefont{Tung}}]{cteq}%
  \BibitemOpen
  \bibfield{author}{%
  \bibinfo {author} {\bibnamefont{Pumplin}, \bibfnamefont{J.}}, \bibinfo
  {author} {\bibfnamefont{D.}~\bibnamefont{Stump}}, \bibinfo {author}
  {\bibfnamefont{J.}~\bibnamefont{Huston}}, \bibinfo {author}
  {\bibfnamefont{H.}~\bibnamefont{Lai}}, \bibinfo {author}
  {\bibfnamefont{P.}~\bibnamefont{Nadolsky}},\ and\ \bibinfo {author}
  {\bibfnamefont{W.}~\bibnamefont{Tung}} (\bibinfo {collaboration} {CTEQ})}%
  , \bibinfo {year} {2002},\ \bibfield{journal}{%
  \Doi{10.1088/1126-6708/2002/07/012}{\bibinfo {journal} {JHEP}}\ }%
  \textbf{\bibinfo {volume} {07}},\ \bibinfo {pages} {012},\ \bibinfo {note}
  {see also http://www.phys.psu.edu/~cteq},\
  \Eprint{http://arxiv.org/abs/0201195}{arXiv:0201195 [hep-ph]}%
  \bibAnnoteFile{NoStop}{cteq}%
\bibitem[{\citenamefont{Quadt}(2006)}]{Quadt:2007jk}%
  \BibitemOpen
  \bibfield{author}{%
  \bibinfo {author} {\bibnamefont{Quadt}, \bibfnamefont{A.}}}%
  , \bibinfo {year} {2006},\ \bibfield{journal}{%
  \Doi{10.1140/epjc/s2006-02631-6}{\bibinfo {journal} {Eur. Phys. J.}}\ }%
  \textbf{\bibinfo {volume} {C48}},\ \bibinfo {pages} {835}%
  \bibAnnoteFile{NoStop}{Quadt:2007jk}%
%%CITATION = EPHJA,C48,835;%%
\bibitem[{\citenamefont{Quigg}(2009)}]{Quigg2009}%
  \BibitemOpen
  \bibfield{author}{%
  \bibinfo {author} {\bibnamefont{Quigg}, \bibfnamefont{C.}}}%
  , \bibinfo {year} {2009},\ \bibfield{journal}{%
  \Doi{10.1146/annurev.nucl.010909.083126}{\bibinfo {journal}
  {Ann.Rev.Nucl.Part.Sci.}}\ }%
  \textbf{\bibinfo {volume} {59}},\ \bibinfo {pages} {505},\
  \Eprint{http://arxiv.org/abs/0905.3187}{arXiv:0905.3187 [hep-ph]}%
  \bibAnnoteFile{NoStop}{Quigg2009}%
\bibitem[{\citenamefont{Randall}\ and\
  \citenamefont{Schwartz}(2001)}]{Randall:2001gb}%
  \BibitemOpen
  \bibfield{author}{%
  \bibinfo {author} {\bibnamefont{Randall}, \bibfnamefont{L.}},\ and\ \bibinfo
  {author} {\bibfnamefont{M.~D.}\ \bibnamefont{Schwartz}}}%
  , \bibinfo {year} {2001},\ \bibfield{journal}{%
  \bibinfo {journal} {JHEP}\ }%
  \textbf{\bibinfo {volume} {11}},\ \bibinfo {pages} {003},\
  \Eprint{http://arxiv.org/abs/hep-th/0108114}{arXiv:hep-th/0108114}%
  \bibAnnoteFile{NoStop}{Randall:2001gb}%
%%CITATION = HEP-TH/0108114;%%
\bibitem[{\citenamefont{Randall}\ and\
  \citenamefont{Schwartz}(2002)}]{Randall:2001gc}%
  \BibitemOpen
  \bibfield{author}{%
  \bibinfo {author} {\bibnamefont{Randall}, \bibfnamefont{L.}},\ and\ \bibinfo
  {author} {\bibfnamefont{M.~D.}\ \bibnamefont{Schwartz}}}%
  , \bibinfo {year} {2002},\ \bibfield{journal}{%
  \Doi{10.1103/PhysRevLett.88.081801}{\bibinfo {journal} {Phys. Rev. Lett.}}\
  }%
  \textbf{\bibinfo {volume} {88}},\ \bibinfo {pages} {081801},\
  \Eprint{http://arxiv.org/abs/hep-th/0108115}{arXiv:hep-th/0108115}%
  \bibAnnoteFile{NoStop}{Randall:2001gc}%
%%CITATION = HEP-TH/0108115;%%
\bibitem[{\citenamefont{Randall}\ and\
  \citenamefont{Sundrum}(1999)}]{Randall:1999ee}%
  \BibitemOpen
  \bibfield{author}{%
  \bibinfo {author} {\bibnamefont{Randall}, \bibfnamefont{L.}},\ and\ \bibinfo
  {author} {\bibfnamefont{R.}~\bibnamefont{Sundrum}}}%
  , \bibinfo {year} {1999},\ \bibfield{journal}{%
  \Doi{10.1103/PhysRevLett.83.3370}{\bibinfo {journal} {Phys. Rev. Lett.}}\ }%
  \textbf{\bibinfo {volume} {83}},\ \bibinfo {pages} {3370},\
  \Eprint{http://arxiv.org/abs/hep-ph/9905221}{arXiv:hep-ph/9905221}%
  \bibAnnoteFile{NoStop}{Randall:1999ee}%
%%CITATION = HEP-PH/9905221;%%
\bibitem[{\citenamefont{Read}(1999)}]{Read:1999kh}%
  \BibitemOpen
  \bibfield{author}{%
  \bibinfo {author} {\bibnamefont{Read}, \bibfnamefont{A.~L.}}}%
  , \bibinfo {year} {1999},\ \bibfield{journal}{%
  \Doi{10.1016/S0168-9002(98)01347-3}{\bibinfo {journal} {Nucl. Instrum.
  Meth.}}\ }%
  \textbf{\bibinfo {volume} {A425}},\ \bibinfo {pages} {357}%
  \bibAnnoteFile{NoStop}{Read:1999kh}%
%%CITATION = NUIMA,A425,357;%%
\bibitem[{\citenamefont{Rizzo}(2001)}]{Rizzo:2001sd}%
  \BibitemOpen
  \bibfield{author}{%
  \bibinfo {author} {\bibnamefont{Rizzo}, \bibfnamefont{T.~G.}}}%
  , \bibinfo {year} {2001},\ \bibfield{journal}{%
  \Doi{10.1103/PhysRevD.64.095010}{\bibinfo {journal} {Phys. Rev.}}\ }%
  \textbf{\bibinfo {volume} {D64}},\ \bibinfo {pages} {095010},\
  \Eprint{http://arxiv.org/abs/hep-ph/0106336}{arXiv:hep-ph/0106336}%
  \bibAnnoteFile{NoStop}{Rizzo:2001sd}%
%%CITATION = HEP-PH/0106336;%%
\bibitem[{\citenamefont{Rizzo}\ and\
  \citenamefont{Wells}(2000)}]{Rizzo:1999br}%
  \BibitemOpen
  \bibfield{author}{%
  \bibinfo {author} {\bibnamefont{Rizzo}, \bibfnamefont{T.~G.}},\ and\ \bibinfo
  {author} {\bibfnamefont{J.~D.}\ \bibnamefont{Wells}}}%
  , \bibinfo {year} {2000},\ \bibfield{journal}{%
  \Doi{10.1103/PhysRevD.61.016007}{\bibinfo {journal} {Phys. Rev.}}\ }%
  \textbf{\bibinfo {volume} {D61}},\ \bibinfo {pages} {016007},\
  \Eprint{http://arxiv.org/abs/hep-ph/9906234}{arXiv:hep-ph/9906234}%
  \bibAnnoteFile{NoStop}{Rizzo:1999br}%
%%CITATION = HEP-PH/9906234;%%
\bibitem[{\citenamefont{Ross}(1984)}]{Ross:1985ai}%
  \BibitemOpen
  \bibfield{author}{%
  \bibinfo {author} {\bibnamefont{Ross}, \bibfnamefont{G.~G.}}}%
  , \bibinfo {year} {1984},\ \emph{\bibinfo {title} {{Grand Unified
  Theories}}},\ \bibinfo {note} {reading, USA: Benjamin/cummings ( Frontiers In
  Physics, 60)}%
  \bibAnnoteFile{NoStop}{Ross:1985ai}%
\bibitem[{\citenamefont{Schaile}\ and\
  \citenamefont{Zerwas}(1992)}]{Schaile1992}%
  \BibitemOpen
  \bibfield{author}{%
  \bibinfo {author} {\bibnamefont{Schaile}, \bibfnamefont{D.}},\ and\ \bibinfo
  {author} {\bibfnamefont{P.}~\bibnamefont{Zerwas}}}%
  , \bibinfo {year} {1992},\ \bibfield{journal}{%
  \Doi{10.1103/PhysRevD.45.3262}{\bibinfo {journal} {Phys. Rev.}}\ }%
  \textbf{\bibinfo {volume} {D45}},\ \bibinfo {pages} {3262}%
  \bibAnnoteFile{NoStop}{Schaile1992}%
\bibitem[{\citenamefont{Scott}(1992)}]{Scott:1992}%
  \BibitemOpen
  \bibfield{author}{%
  \bibinfo {author} {\bibnamefont{Scott}, \bibfnamefont{D.~W.}}}%
  , \bibinfo {year} {1992},\ \emph{\bibinfo {title} {{Multivariate density
  estimation: theory, practice, and visualization}}}\ (\bibinfo {publisher}
  {Wiley-Interscience})%
  \bibAnnoteFile{NoStop}{Scott:1992}%
\bibitem[{\citenamefont{Sehgal}\ and\
  \citenamefont{Wanninger}(1988)}]{Sehgal:1987wi}%
  \BibitemOpen
  \bibfield{author}{%
  \bibinfo {author} {\bibnamefont{Sehgal}, \bibfnamefont{L.~M.}},\ and\
  \bibinfo {author} {\bibfnamefont{M.}~\bibnamefont{Wanninger}}}%
  , \bibinfo {year} {1988},\ \bibfield{journal}{%
  \Doi{10.1016/0370-2693(88)91138-0}{\bibinfo {journal} {Phys. Lett.}}\ }%
  \textbf{\bibinfo {volume} {B200}},\ \bibinfo {pages} {211}%
  \bibAnnoteFile{NoStop}{Sehgal:1987wi}%
%%CITATION = PHLTA,B200,211;%%
\bibitem[{\citenamefont{Shu}\ \emph{et~al.}(2010)\citenamefont{Shu},
  \citenamefont{Tait},\ and\ \citenamefont{Wang}}]{Shu:2009xf}%
  \BibitemOpen
  \bibfield{author}{%
  \bibinfo {author} {\bibnamefont{Shu}, \bibfnamefont{J.}}, \bibinfo {author}
  {\bibfnamefont{T.~M.~P.}\ \bibnamefont{Tait}},\ and\ \bibinfo {author}
  {\bibfnamefont{K.}~\bibnamefont{Wang}}}%
  , \bibinfo {year} {2010},\ \bibfield{journal}{%
  \Doi{10.1103/PhysRevD.81.034012}{\bibinfo {journal} {Phys. Rev.}}\ }%
  \textbf{\bibinfo {volume} {D81}},\ \bibinfo {pages} {034012},\
  \Eprint{http://arxiv.org/abs/0911.3237}{arXiv:0911.3237 [hep-ph]}%
  \bibAnnoteFile{NoStop}{Shu:2009xf}%
%%CITATION = 0911.3237;%%
\bibitem[{\citenamefont{Silverman}(1998)}]{Silverman:1998}%
  \BibitemOpen
  \bibfield{author}{%
  \bibinfo {author} {\bibnamefont{Silverman}, \bibfnamefont{B.~W.}}}%
  , \bibinfo {year} {1998},\ \emph{\bibinfo {title} {{Density estimation for
  statistics and data analysis}}}\ (\bibinfo {publisher} {Chapman and Hall})%
  \bibAnnoteFile{NoStop}{Silverman:1998}%
\bibitem[{\citenamefont{Simmons}(1997)}]{Simmons:1996fz}%
  \BibitemOpen
  \bibfield{author}{%
  \bibinfo {author} {\bibnamefont{Simmons}, \bibfnamefont{E.~H.}}}%
  , \bibinfo {year} {1997},\ \bibfield{journal}{%
  \Doi{10.1103/PhysRevD.55.1678}{\bibinfo {journal} {Phys. Rev.}}\ }%
  \textbf{\bibinfo {volume} {D55}},\ \bibinfo {pages} {1678},\
  \Eprint{http://arxiv.org/abs/hep-ph/9608269}{arXiv:hep-ph/9608269}%
  \bibAnnoteFile{NoStop}{Simmons:1996fz}%
%%CITATION = HEP-PH/9608269;%%
\bibitem[{\citenamefont{Sinervo}(2003)}]{Sinervo:2003wm}%
  \BibitemOpen
  \bibfield{author}{%
  \bibinfo {author} {\bibnamefont{Sinervo}, \bibfnamefont{P.}}}%
  , \bibinfo {year} {2003}\ \bibinfo {note} {prepared for PHYSTAT2003:
  Statistical Problems in Particle Physics, Astrophysics, and Cosmology, Menlo
  Park, California, 8-11 Sep 2003}%
  \bibAnnoteFile{NoStop}{Sinervo:2003wm}%
%%CITATION = ECONF,C030908,TUAT004;%%
\bibitem[{\citenamefont{Sj{\"{o}}strand}\
  \emph{et~al.}(2001)\citenamefont{Sj{\"{o}}strand}, \citenamefont{Eden},
  \citenamefont{Friberg}, \citenamefont{L{\"{o}}nnblad}, \citenamefont{Miu},
  \citenamefont{Mrenna},\ and\ \citenamefont{Norrbin}}]{pythia}%
  \BibitemOpen
  \bibfield{author}{%
  \bibinfo {author} {\bibnamefont{Sj{\"{o}}strand}, \bibfnamefont{T.}},
  \bibinfo {author} {\bibfnamefont{P.}~\bibnamefont{Eden}}, \bibinfo {author}
  {\bibfnamefont{C.}~\bibnamefont{Friberg}}, \bibinfo {author}
  {\bibfnamefont{L.}~\bibnamefont{L{\"{o}}nnblad}}, \bibinfo {author}
  {\bibfnamefont{G.}~\bibnamefont{Miu}}, \bibinfo {author}
  {\bibfnamefont{S.}~\bibnamefont{Mrenna}},\ and\ \bibinfo {author}
  {\bibfnamefont{E.}~\bibnamefont{Norrbin}}}%
  , \bibinfo {year} {2001},\ \bibfield{journal}{%
  \Doi{10.1016/S0010-4655(00)00236-8}{\bibinfo {journal} {Comput. Phys.
  Commun.}}\ }%
  \textbf{\bibinfo {volume} {135}},\ \bibinfo {pages} {238},\ \bibinfo {note}
  {see also http://home.thep.lu.se/~torbjorn/Pythia.html},\
  \Eprint{http://arxiv.org/abs/0010017}{arXiv:0010017 [hep-ph]}%
  \bibAnnoteFile{NoStop}{pythia}%
\bibitem[{\citenamefont{Smith}\ and\
  \citenamefont{Willenbrock}(1997)}]{Smith:1996xz}%
  \BibitemOpen
  \bibfield{author}{%
  \bibinfo {author} {\bibnamefont{Smith}, \bibfnamefont{M.~C.}},\ and\ \bibinfo
  {author} {\bibfnamefont{S.~S.}\ \bibnamefont{Willenbrock}}}%
  , \bibinfo {year} {1997},\ \bibfield{journal}{%
  \Doi{10.1103/PhysRevLett.79.3825}{\bibinfo {journal} {Phys. Rev. Lett.}}\ }%
  \textbf{\bibinfo {volume} {79}},\ \bibinfo {pages} {3825},\
  \Eprint{http://arxiv.org/abs/hep-ph/9612329}{arXiv:hep-ph/9612329}%
  \bibAnnoteFile{NoStop}{Smith:1996xz}%
%%CITATION = HEP-PH/9612329;%%
\bibitem[{\citenamefont{Stanley}\ and\
  \citenamefont{Miikkulainen}(2002)}]{Stanley:2002zz}%
  \BibitemOpen
  \bibfield{author}{%
  \bibinfo {author} {\bibnamefont{Stanley}, \bibfnamefont{K.~O.}},\ and\
  \bibinfo {author} {\bibfnamefont{R.}~\bibnamefont{Miikkulainen}}}%
  , \bibinfo {year} {2002},\ \bibfield{journal}{%
  \bibinfo {journal} {Evol. Comput.}\ }%
  \textbf{\bibinfo {volume} {10}},\ \bibinfo {pages} {99}%
  \bibAnnoteFile{NoStop}{Stanley:2002zz}%
%%CITATION = EOCME,10,99;%%
\bibitem[{\citenamefont{Valassi}(2003)}]{Valassi:2003mu}%
  \BibitemOpen
  \bibfield{author}{%
  \bibinfo {author} {\bibnamefont{Valassi}, \bibfnamefont{A.}}}%
  , \bibinfo {year} {2003},\ \bibfield{journal}{%
  \Doi{10.1016/S0168-9002(03)00329-2}{\bibinfo {journal} {Nucl. Instrum.
  Meth.}}\ }%
  \textbf{\bibinfo {volume} {A500}},\ \bibinfo {pages} {391}%
  \bibAnnoteFile{NoStop}{Valassi:2003mu}%
%%CITATION = NUIMA,A500,391;%%
\bibitem[{\citenamefont{Wagner}(2010)}]{Wagner:2010}%
  \BibitemOpen
  \bibfield{author}{%
  \bibinfo {author} {\bibnamefont{Wagner}, \bibfnamefont{W.}}}%
  , \bibinfo {year} {2010}\ \bibinfo {note} {submitted to Mod. Phys. Lett. A},\
  \Eprint{http://arxiv.org/abs/1003.4359}{arXiv:1003.4359 [hep-ex]}%
  \bibAnnoteFile{NoStop}{Wagner:2010}%
%%CITATION = 1003.4359;%%
\bibitem[{\citenamefont{Wicke}(2010)}]{Wicke:2010}%
  \BibitemOpen
  \bibfield{author}{%
  \bibinfo {author} {\bibnamefont{Wicke}, \bibfnamefont{D.}}}%
  , \bibinfo {year} {2010}\ \bibinfo {note} {habilitation thesis, Univ.
  Wuppertal},\ \Eprint{http://arxiv.org/abs/1005.2460}{arXiv:1005.2460
  [hep-ex]}%
  \bibAnnoteFile{NoStop}{Wicke:2010}%
%%CITATION = 1005.2460;%%
\bibitem[{\citenamefont{Yuan}(1993)}]{Yuan:1993ck}%
  \BibitemOpen
  \bibfield{author}{%
  \bibinfo {author} {\bibnamefont{Yuan}, \bibfnamefont{C.~P.}}}%
  , \bibinfo {year} {1993}\
  \Eprint{http://arxiv.org/abs/hep-ph/9308240}{arXiv:hep-ph/9308240}%
  \bibAnnoteFile{NoStop}{Yuan:1993ck}%
%%CITATION = HEP-PH/9308240;%%
\bibitem[{\citenamefont{Yuan}(1995)}]{Yuan:1995cm}%
  \BibitemOpen
  \bibfield{author}{%
  \bibinfo {author} {\bibnamefont{Yuan}, \bibfnamefont{C.~P.}}}%
  , \bibinfo {year} {1995}\
  \Eprint{http://arxiv.org/abs/hep-ph/9509209}{arXiv:hep-ph/9509209}%
  \bibAnnoteFile{NoStop}{Yuan:1995cm}%
%%CITATION = HEP-PH/9509209;%%
\bibitem[{\citenamefont{Zhang}\ \emph{et~al.}(2000)\citenamefont{Zhang},
  \citenamefont{Wang}, \citenamefont{Kuang},\ and\
  \citenamefont{Zhou}}]{Zhang:1999qy}%
  \BibitemOpen
  \bibfield{author}{%
  \bibinfo {author} {\bibnamefont{Zhang}, \bibfnamefont{L.}}, \bibinfo {author}
  {\bibfnamefont{X.-L.}\ \bibnamefont{Wang}}, \bibinfo {author}
  {\bibfnamefont{Y.-P.}\ \bibnamefont{Kuang}},\ and\ \bibinfo {author}
  {\bibfnamefont{H.-Y.}\ \bibnamefont{Zhou}}}%
  , \bibinfo {year} {2000},\ \bibfield{journal}{%
  \Doi{10.1103/PhysRevD.61.115007}{\bibinfo {journal} {Phys. Rev.}}\ }%
  \textbf{\bibinfo {volume} {D61}},\ \bibinfo {pages} {115007},\
  \Eprint{http://arxiv.org/abs/hep-ph/9910265}{arXiv:hep-ph/9910265}%
  \bibAnnoteFile{NoStop}{Zhang:1999qy}%
%%CITATION = HEP-PH/9910265;%%
\end{thebibliography}%

% ============================================================================
% --- GO HOME
% ============================================================================
\end{document}